%% file: sample-journal.tex
\documentclass[siggraph]{acmart}

\usepackage{booktabs} % For formal tables

\setcitestyle{square}

\usepackage{subfigure}
\usepackage{amsmath}
\usepackage{amssymb}
\usepackage{hyperref}
\usepackage{url}

\usepackage[ruled]{algorithm2e} % For algorithms

\SetAlFnt{\small}
\SetAlCapFnt{\small}
\SetAlCapNameFnt{\small}
\SetAlCapHSkip{0pt}
\IncMargin{-\parindent}

\setcopyright{none}
\renewcommand\footnotetextcopyrightpermission[1]{} % removes footnote with conference information in first column
\usepackage{soul}
\usepackage{color}

\usepackage{color}

\newcommand{\normN}[1]{\lVert#1\rVert}
\newcommand{\normScalar}[1]{|#1|}
\newcommand{\tabincell}[2]{\begin{tabular}{@{}#1@{}}#2\end{tabular}}
\newcommand{\tablefont}{\fontsize{8pt}{\baselineskip}\selectfont}
\newcommand{\eqfont}{\fontsize{7pt}{\baselineskip}\selectfont}

\hypersetup{draft}
\begin{document}
% Title portion
\title{Low Rank Matrix Approximation for Geometry Filtering}
%\xuequan{F$^2$G: A Feature-aware Approach for Geometry Filtering[previous title: A Two-step Approach for Geometry Processing]} }

\author{Xuequan Lu}
\orcid{0000-0003-0959-408X}
\affiliation{%
  \institution{Nanyang Technological University}
  %\streetaddress{104 Jamestown Rd}
  %\city{Singapore}
  %\state{Singapore}
  %\postcode{23185}
  \country{Singapore}}
\email{xuequanlu@ntu.edu.sg}
\author{Scott Schaefer}
\affiliation{%
  \institution{Texas A\&M University}
  \city{College Station}
  \state{Texas}
  \country{USA}
}
\email{schaefer@cs.tamu.edu}
\author{Jun Luo}
\affiliation{%
 \institution{Nanyang Technological University}
 %\streetaddress{Rono-Hills}
 %\city{Doimukh}
 %\state{Arunachal Pradesh}
 \country{Singapore}}
\email{junluo@ntu.edu.sg}
\author{Lizhuang Ma}
\affiliation{%
  \institution{Shanghai Jiao Tong University}
  %\streetaddress{30 Shuangqing Rd}
  \city{Shanghai}
  %\state{}
  \country{China}
}
\email{ma-lz@cs.sjtu.edu.cn}
\author{Ying He}
\affiliation{%
  \institution{Nanyang Technological University}
  \country{Singapore}}
\email{YHe@ntu.edu.sg}
% \affiliation{%
%   \institution{University of Minnesota}
% \email{tinghe@uva.edu}
% \author{Chengdu Huang}
% \author{John A. Stankovic}
% \author{Tarek F. Abdelzaher}
% \affiliation{%
%   \institution{University of Virginia}
%   \department{School of Engineering}
%   \city{Charlottesville}
%   \state{VA}
%   \postcode{22903}

\renewcommand\shortauthors{Lu, X. et al}

\begin{abstract}
We propose a robust normal estimation method for both point clouds and meshes using a low rank matrix approximation algorithm. First, we compute a local isotropic structure for each point and find its similar, non-local structures that we organize into a matrix.  We then show that a low rank matrix approximation algorithm can robustly estimate normals for both point clouds and meshes.  Furthermore, we provide a new filtering method for point cloud data to smooth the position data to fit the estimated normals.  We show applications of our method to point cloud filtering, point set upsampling, surface reconstruction, mesh denoising, and geometric texture removal.  Our experiments show that our method outperforms current methods in both visual quality and accuracy.

\end{abstract}





\maketitle

%-----------------------------------------------
\input{paper/introduction}

\input{paper/relatedwork}

\input{paper/methodology}

\input{paper/results}

\input{paper/discussion}

\input{paper/conclusion}

%-----------------------------------------------

\bibliographystyle{ACM-Reference-Format}
\bibliography{sample-bibliography} 

\input{paper/appendix}

\end{document}

%% file: paper/introduction.tex
\section{Introduction}\label{sec:introduction}
Normal estimation for point cloud models or mesh shapes is important since it is often the first step in a geometry processing pipeline.  This estimation is often followed by a filtering process to update the position data and remove noise~\cite{Sun2007}.
A variety of computer graphics applications, such as point cloud filtering \cite{Avron2010,Sun2015,Lu2017tvcg}, point set upsampling \cite{Huang2013}, surface reconstruction \cite{Oztireli2009}, mesh denoising \cite{Sun2007,Zheng2011,Zhang2015} and geometric texture removal \cite{Wang2015} rely heavily on the quality of estimated normals and subsequent filtering of position data.  

Current state of the art techniques in mesh denoising \cite{Sun2007,Zheng2011,Zhang2015} and geometric texture removal \cite{Wang2015} can achieve impressive results. However, these methods are still limited in their ability to recover sharp edges in challenging regions.
Normal estimation for point clouds has been an active area of research in recent years \cite{Boulch2012,Huang2013,Boulch2016}. However, these methods perform suboptimally when estimating normals in noisy point clouds. Specifically, \cite{Boulch2012,Boulch2016} are less robust in the presence of considerable noise. The bilateral filter can preserve geometric features but sometimes may fail due to the locality of its computations and lack of self-adaption of parameters. 

Updating point positions using the estimated normals in point clouds has received sparse treatment so far \cite{Avron2010,Sun2015}. However, those position update approaches using the $L_0$ or $L_1$ norms are complex to solve and hard to implement. Moreover, they restrict each point to only move along its normal orientation potentially leading to suboptimal results or slow convergence. 

\begin{figure}[htbp]
%\vspace{-0.0cm}
\centering
\begin{minipage}[b]{0.95\linewidth}
%\subfigure[]
{\label{}
\includegraphics[width=1\linewidth]{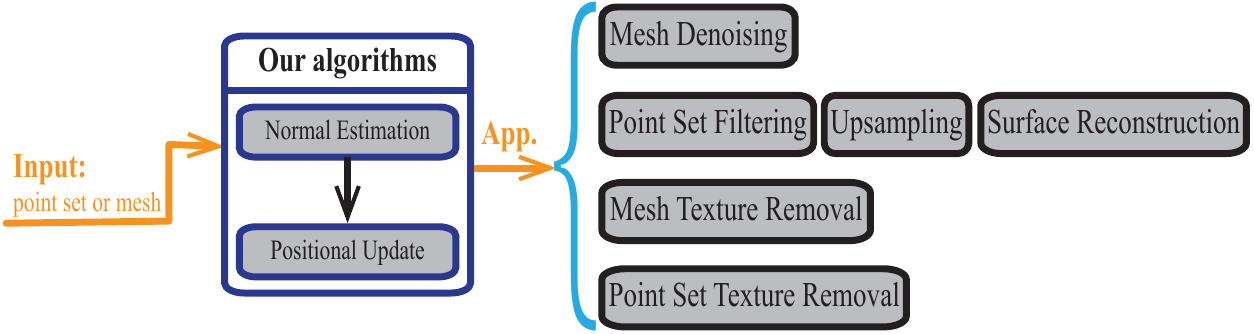}}
\end{minipage}
\caption{Overview of our approach and its applications. Our method can be applied to various geometry processing tasks directly or indirectly.}
\label{fig:overview}
%\vspace{-0.65cm}
\end{figure}

To address the issues shown above, we propose a new normal estimation method for both meshes and point clouds and a new position update algorithm for point clouds.  Our method benefits various geometry processing applications, directly or indirectly, such as point cloud filtering, point set upsampling, surface reconstruction, mesh denoising, and geometric texture removal (Figure \ref{fig:overview}). Given a point cloud or mesh as input, our method first estimates point or face normals, then updates the positions of points or vertices using the estimated normals. We observe that: (1) non-local methods could be more accurate than local techniques; (2) there usually exist similar structures of each local isotropic structure (Section \ref{sec:nonlocalstructures}) in geometry shapes; (3) the matrix constructed by similar structures should be low-rank. Motivated by these observations, we propose a novel normal estimation technique which consists of two sub-steps: (i) non-local similar structures location and (ii) weighted nuclear norm minimization. We adopt the former to find similar structures of each local isotropic structure. We employ the latter \cite{Gu2014} to handle the problem of recovering low-rank matrices. We also present a fast and effective point update algorithm for point clouds to filter the point positions to better match the estimated normals.

The \textbf{main contributions} of this paper are:
\begin{itemize}
\item a novel normal estimation technique for both point cloud shapes and mesh models;
\item a new position update algorithm for point cloud data;
\item analysis of the convergence of the proposed normal estimation technique and point update algorithm, experimentally or theoretically.
\end{itemize}

Extensive experiments and comparisons show that our method outperforms  
current methods in terms of visual quality and accuracy.

% method overview

%% file: paper/relatedwork.tex
\section{Related Work}
\label{sec:relatedwork}
In this section, we only review the research works that are most related to this work. We first review the previous research on normal estimation. Then we review some previous works which employed the nuclear norm minimization or its weighted version. 

\subsection{Normal Estimation}
\label{sec:normalestimaitonresearch}
Normal estimation for geometric shapes can be classified into two types: (1) normal estimation for point clouds, and (2) normal estimation for mesh shapes.

\textbf{Normal estimation for point clouds.} Hoppe et al. \cite{Hoppe1992} estimated normals by computing the tangent plane at each data point using principal component analysis (PCA) of the local neighborhood. Later, a variety of variants of PCA have been proposed \cite{Alexa2001,Pauly2002,Mitra2003,Lange2005,Huang2009} to estimate normals. Nevertheless, the normals estimated by these techniques tend to smear sharp features. Researchers also estimate normals using Voronoi cells or Voronoi-PCA \cite{Dey2004,Alliez2007}. Minimizing the $L_1$ or $L_0$ norm can preserve sharp features as these norms can be used to measure sparsity in the derivative of the normal field \cite{Avron2010,Sun2015}. Yet, the solutions are complex and computationally expensive. Li et al. \cite{Li2010} estimated normals by using robust statistics to detect the best local tangent plane for each point. Another set of techniques attempted to better estimate normals near edges and corners by point clustering in a neighborhood \cite{Zhang2013,Liu2015}. Later they presented a  pair consistency voting scheme which outputs multiple normals per feature point \cite{Zhang2018}. Boulch and Marlet \cite{Boulch2012} use a robust randomized Hough transform to estimate point normals. Convolutional neural networks have recently been applied to estimate normals in point clouds \cite{Boulch2016}. Such estimation methods are usually less robust for point clouds with considerable amount of noise. Bilateral smoothing of PCA normals \cite{Oztireli2009,Huang2013} is simple and effective, but it suffers from inaccuracy due to the locality of its computations and may blur edges with small dihedral angles. 

\textbf{Normal estimation for mesh shapes.} Most methods focus on the estimation of face normals in mesh shapes. One simple, direct way is to compute the face normals by the cross product of two edges in a triangle face. However, such normals can deviate from the true normals significantly even in the presence of small position noise. There exist a considerable amount of research work to smooth these face normals. One approach uses the bilateral filter \cite{Lee2005,Wang2006,Zheng2011}, inspired by the founding works \cite{Jones2003,Fleishman2003}. Mean, median and alpha-trimming methods \cite{Yagou2002,Yagou2003,Shen2004} are also used to estimate face normals. Sun et al. \cite{Sun2007,Sun2008} present two different methods to filter face normals. Recently, researchers have presented filtering methods \cite{Solomon2014,Zhang2015,Zhang2015cgf,Lu2017,Yadav2017} based on mean shift, total variation, guided normals, $L_1$ median, and normal voting tensor. Wang et al. \cite{Wang2016} estimated face normals via cascaded normal regression.

\subsection{Nonlocal Methods for Point Clouds and Nuclear Norm Minimization }
\label{sec:nnmresearch}
Previous researchers proposed non-local methods for point clouds. For example, Zheng et al. \cite{Zheng2010} applied non-local filtering to  3D buildings that exhibit large scale repetitions and self-similarities. Digne presented a non-local denoising framework to unorganized point clouds by building an intrinsic descriptor \cite{Digne2012}, and recently proposed a shape analysis approach with colleagues based on the non-local analysis of local shape variations \cite{Digne2018}. 

The nuclear norm of a matrix is defined as the sum of the absolute values of its singular values (see Eq. \eqref{eq:nuclearnorm}). It has been proved that most low-rank matrices can be recovered by minimizing their nuclear norm \cite{Candes2009}. Cai et al. \cite{Cai2010} provided a simple solution to the low-rank matrix approximation problem by minimizing the nuclear norm. The nuclear norm minimization has been broadly employed to matrix completion \cite{Candes2009,Cai2010}, robust principle component analysis \cite{Wright2009}, low-rank representation for subspace clustering \cite{Liu2010} and low-rank textures \cite{Zhang2012}. Gu et al. \cite{Gu2014,Gu2017} presented a weighted version of the nuclear norm minimization, which has been adopted to image processing applications such as image denoising, background subtraction and image inpainting.

%% file: paper/methodology.tex
\section{Normal Estimation}
\label{sec:normalestimation}
In this section, we take point clouds, consisting of positions as well as normals, as input and further extend to meshes later. First of all, we present an algorithm to locate and construct non-local similar structures for each local isotropic structure of a point (Section \ref{sec:nonlocalstructures}). We then describe how to estimate normals via weighted nuclear norm minimization on non-local similar structures (Section \ref{sec:wnnm}).

\subsection{Non-local Similar Structures}
\label{sec:nonlocalstructures}
\textbf{Local structure.} We define each point $\mathbf{p}_i$ has a local structure $S_i$ which consists of $k_{local}$ nearest neighbors. Locating structures similar to a specific local structure is difficult due to the irregularity of points. 

\textbf{Tensor voting.} We assume each local structure embeds a representative orientation. To do so, we first define the tensor at a point $\mathbf{p}_i$ as
\begin{equation}\label{eq:singletensor}
\begin{aligned}
\mathbf{T}_{ij} = \eta(\normN{\mathbf{p}_i-\mathbf{p}_j}) \phi(\theta_{ij}) \mathbf{n}_j^T\mathbf{n}_j,
\end{aligned}
\end{equation}
where $\mathbf{p}_j$ ($1\times3$ vector) is one of the $k_{local}$ nearest neighbors of $\mathbf{p}_i$, which we denote as $j\in S_i$, and $\mathbf{n}_j$ ($1\times3$ vector) is the normal of $\mathbf{p}_j$. $\eta$ and $\phi$ are the weights induced by spatial distances and intersection angles ($\theta_{ij}$) of two neighboring normals, which are given by \cite{Huang2013,Zheng2011}: $\eta(x)=e^{-(\frac{x}{\sigma_p})^2}$, $\phi(\theta)=e^{-(\frac{1-\cos(\theta)}{1-\cos(\sigma_{\theta})})^2}$. $\sigma_{p}$ and $\sigma_{\theta}$ are the scaling parameters, which are empirically set to two times the maximal distance between any two points in the $k_{local}$ nearest neighbors within the local structure and $30^\circ$, respectively. 

For each local structure $S_i$, we can derive the accumulated tensor by aggregating all the induced tensor votes $\{i,j\in S_i | \mathbf{T}_{ij}\}$. This final tensor encodes the local structure, which provides a reliable, representative orientation that will be later used to compute the local isotropic structure and locate similar structures.

\begin{equation}\label{eq:overalltensor}
\begin{aligned}
\mathbf{T}_{i} = \sum_{j\in S_i}\mathbf{T}_{ij}
\end{aligned}
\end{equation}

Let $\lambda_{i1} \geq \lambda_{i2} \geq \lambda_{i3}$ be the eigenvalues of  $\mathbf{T}_i$ with the corresponding eigenvectors $\mathbf{e}_{i1}$, $\mathbf{e}_{i2}$ and $\mathbf{e}_{i3}$. In tensor voting \cite{Wu2012}, $\lambda_{i1}-\lambda_{i2}$ indicates surface saliency with a normal direction $\mathbf{e}_{i1}$; $\lambda_{i2}-\lambda_{i3}$ indicates curve saliency with a tangent orientation $\mathbf{e}_{i3}$; $\lambda_{i3}$ denotes junction saliency. Therefore, we take $\mathbf{e}_{i1}$ as the representative orientation for the local structure $S_i$ of point $\mathbf{p}_i$.

%illustration of local structure, local isotropic structure and similar structures
\begin{figure}[t]
%\vspace{-0.0cm}
\centering
\begin{minipage}[b]{0.32\linewidth}
\subfigure[Local structure]{\label{}\includegraphics[width=1\linewidth]{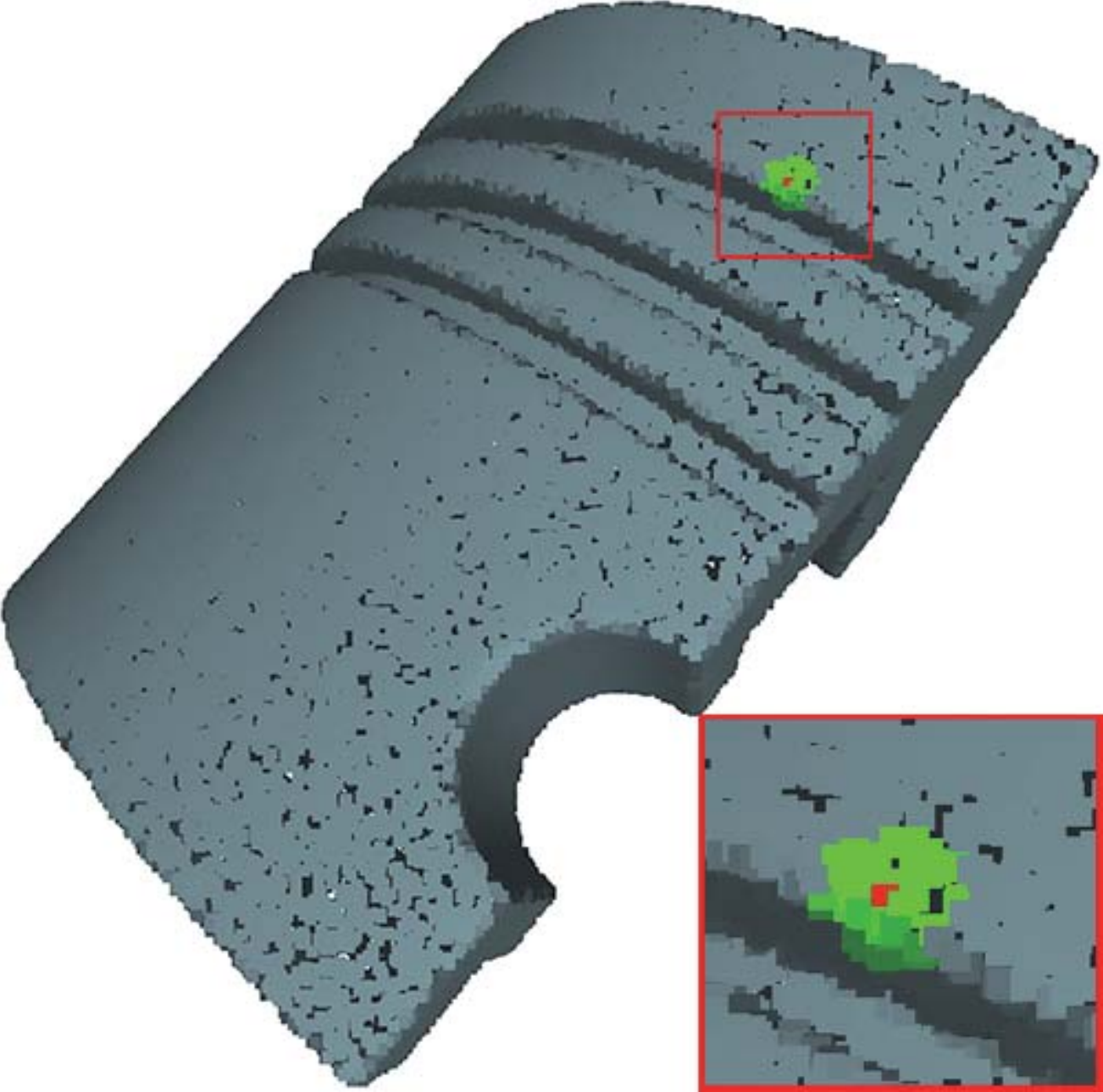}}
\end{minipage}
\begin{minipage}[b]{0.32\linewidth}
\subfigure[Local isotropic structure]{\label{}\includegraphics[width=1\linewidth]{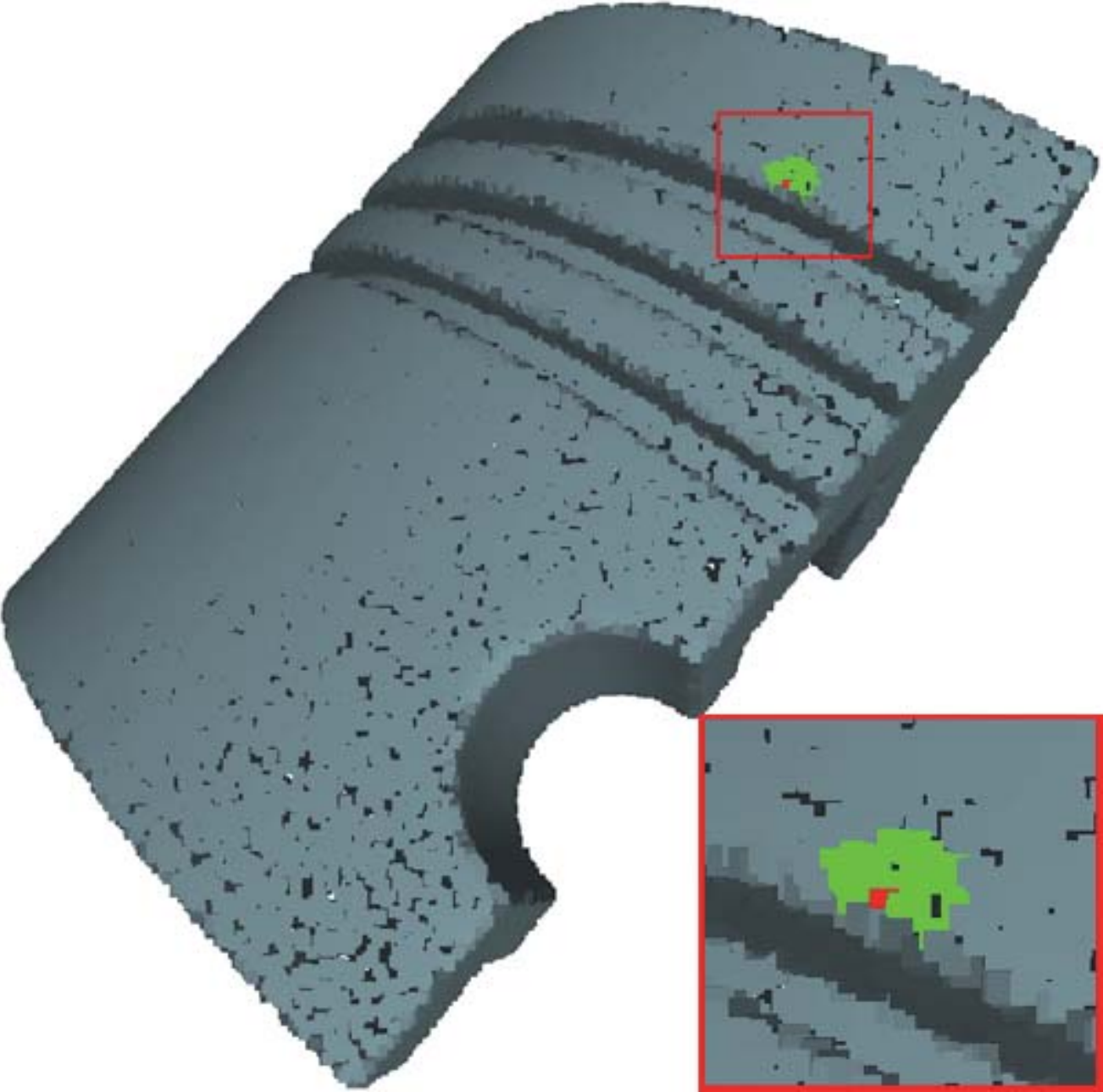}}
\end{minipage}
\begin{minipage}[b]{0.32\linewidth}
\subfigure[Similar structures]{\label{}\includegraphics[width=1\linewidth]{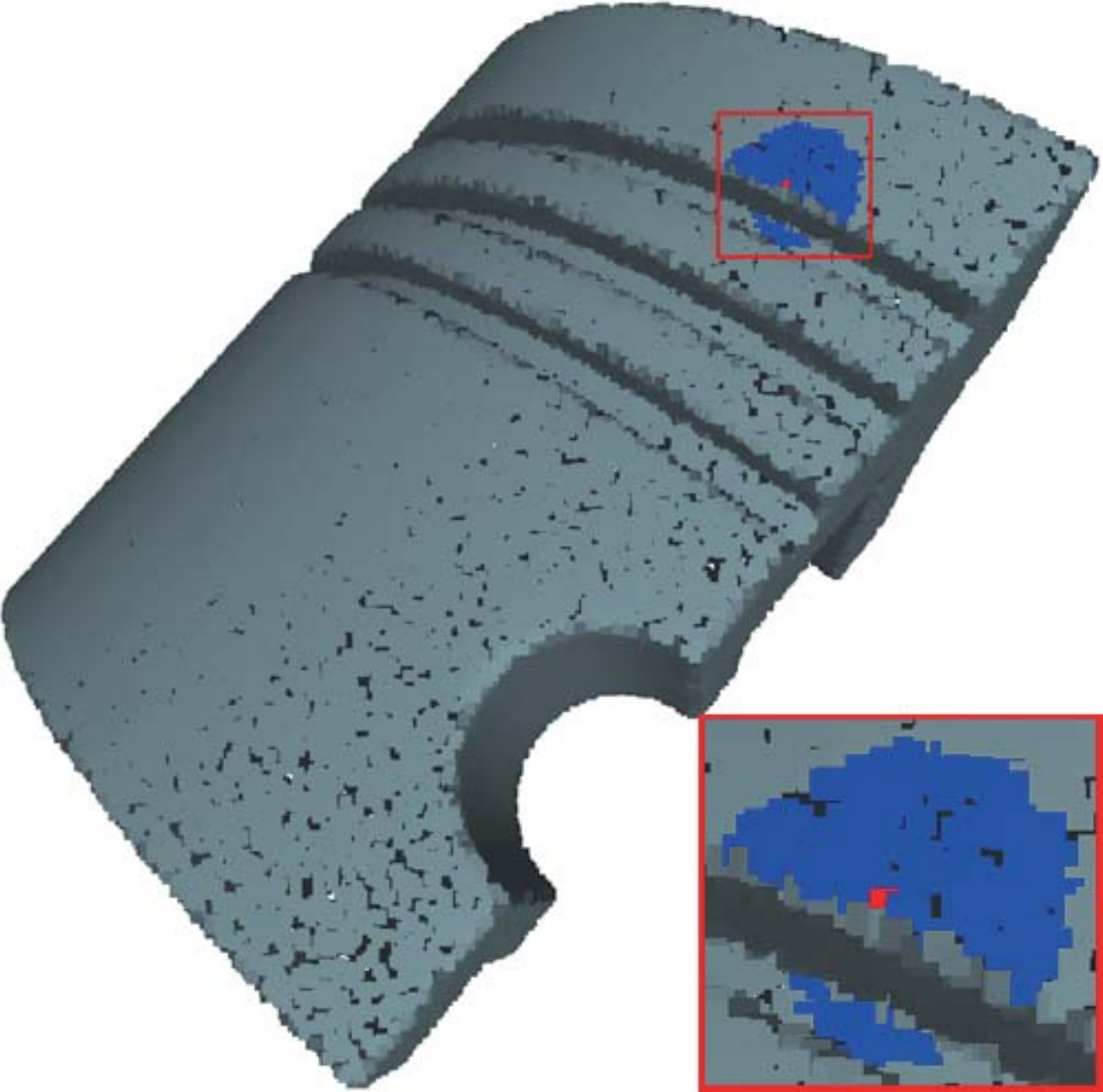}}
\end{minipage}
\caption{(a) The local structure (green points) of the centered red point. (b) The local isotropic structure (green) of the red point. (c) The similar local isotropic structures of the local isotropic structure denoted by the red point. Each blue point denotes its isotropic structure.  }
\label{fig:localstructure}
%\vspace{-0.65cm}
\end{figure}

\textbf{Local isotropic structure.} We assume that each local structure has a subset of points that are on the same isotropic surface with the representative orientation. We call this subset of points the local isotropic structure. Now we describe how to obtain a local isotropic structure $S_i^{iso}$ from a local structure $S_i$ and locate similar local isotropic structures for $S_i^{iso}$. We present a simple and effective way to achieve this: comparing the intersection angles of two normals. Specifically, to obtain $S_i^{iso}$, we
\begin{itemize}
\item compute the intersection angles between each point normal and the representative orientation within a local structure $S_i$;
\item add the current point to $S_i^{iso}$ if the corresponding intersection angle is less than or equal to a threshold, $\theta_{th}$. 
\end{itemize}

For simplicity, we will refer to similar local isotropic structures as similar structures. Given an isotropic structure $S_i^{iso}$, we identify its non-local similar structures by comparing the angles between the representative orientation of each structure.  If the angle is less than or equal to $\theta_{th}$ (we use the same threshold for simplicity), we define the two isotropic structures to be similar. The underlying rationale of our similarity search is: the point normals in a local isotropic structure are bounded by the representative orientation, indicating these points are on the same isotropic surface; the similar structures search is also bounded by the representative orientations, implying the similar structures are on the similar isotropic surfaces. These similar structures will often overlap on the same isotropic surface as shown in Figure~\ref{fig:localstructure}.  In the figure, we show the local structure (a), the local isotropic structure (b), and the similar structures (c).

\textbf{Remark 1.} We do not consider other types of similarity such as rotation-invariant similarity which is often sensitive to noise. Moreover, the rotation quantities have to be estimated in the presence of noise. It could also introduce significant computational overhead to both similarity search and the following SVD operation. Even though our method does not have such a property, we observe it works very well for a wide range of models (see Section \ref{sec:results}).

\subsection{Weighted Nuclear Norm Minimization}
\label{sec:wnnm}

For each non-local similar structure $S_l^{iso}$ for the isotropic structure $S_i^{iso}$ associated with the point $p_i$, we append the point normals of $S_l^{iso}$ as rows to a matrix $\mathbf{M}$.  Note that the dimensions of this matrix are $\hat{r} \times 3$.  This matrix already has a maximal rank of 3 and is a low rank matrix.  To make the low rank matrix approximation meaningful, we reshape the matrix $\mathbf{M}$ to be close to a square matrix.

We do so by finding dimensions $r$ and $c$ of a new matrix $\mathbf{Z}'$ where $\hat{r} \times 3 = r \times c$ and we minimize $|r-c|$.  Given that the structure in $\mathbf{M}$ is isotropic, removing one or more points does not affect this structure significantly.  Therefore, we find $r$ and $c$ iteratively by measuring if $|r-c|\ge 6$.  If so, we remove a point normal from $\mathbf{M}$ and solve for $r$,$c$ again.  We repeat such a process until $|r-c|<6$ ($r$ is not constrained to be a multiple of three).  Then we simply copy the column entries in $\mathbf{M}$ to $\mathbf{Z}'$ filling each column of $\mathbf{Z}'$ before continuing to the next column.  The resulting matrix $\mathbf{Z}'$ should be low rank since all point normals come from similar isotropic structures.  

We then cast the normal estimation problem as a low-rank matrix approximation problem. We attempt to recover a low-rank matrix $\mathbf{Z}$ from $\mathbf{Z}'$ using nuclear norm minimization.  We first present some fundamental knowledge about nuclear norm minimization and then show how we estimate normals with weighted nuclear norm minimization.

%matrix ordering
\begin{figure}[htbp]
%\vspace{-0.0cm}
\centering
\begin{minipage}[b]{0.3\linewidth}
\subfigure[]{\label{}\includegraphics[width=1\linewidth]{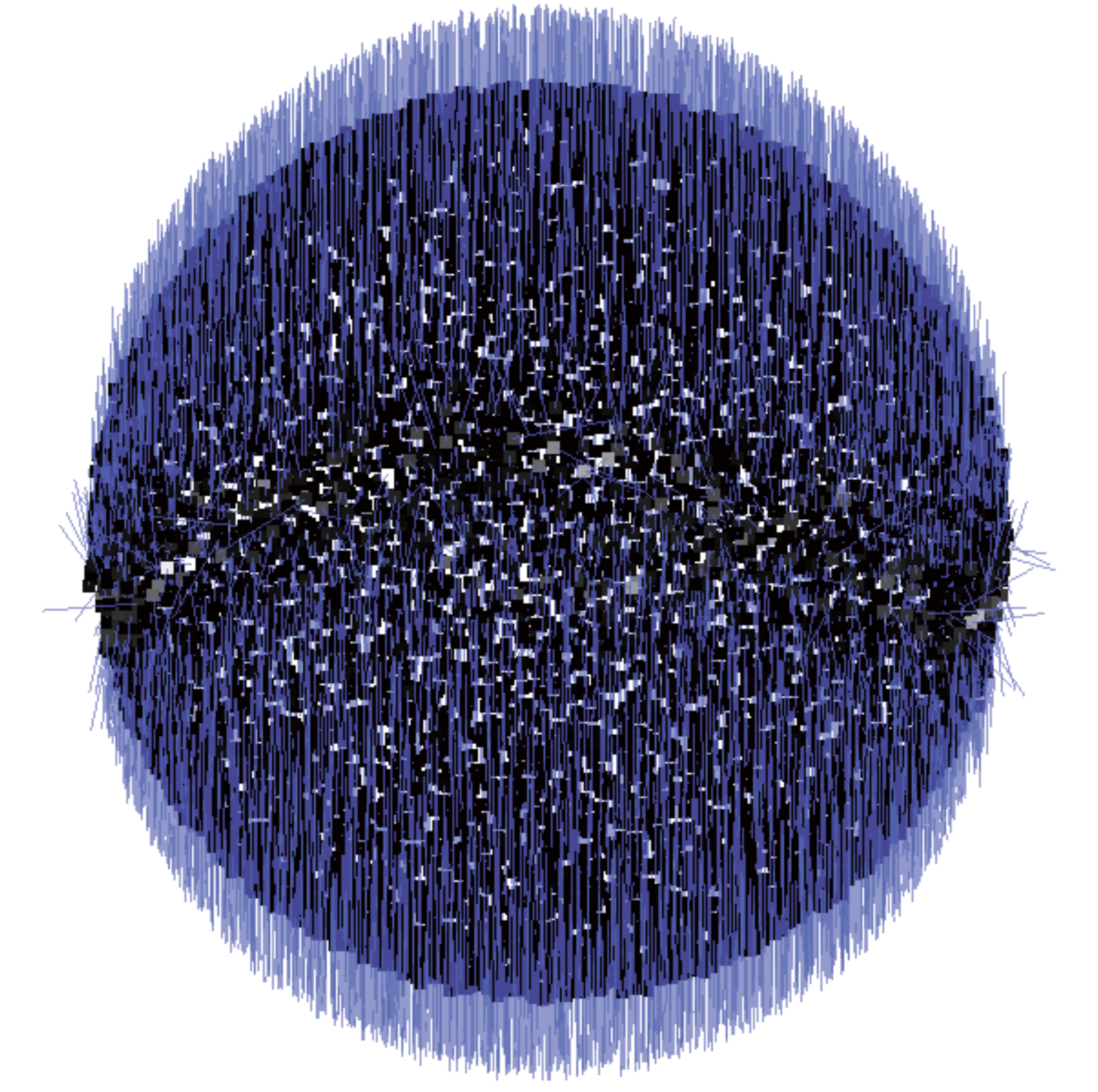}}
\end{minipage}
\begin{minipage}[b]{0.3\linewidth}
\subfigure[]{\label{}\includegraphics[width=1\linewidth]{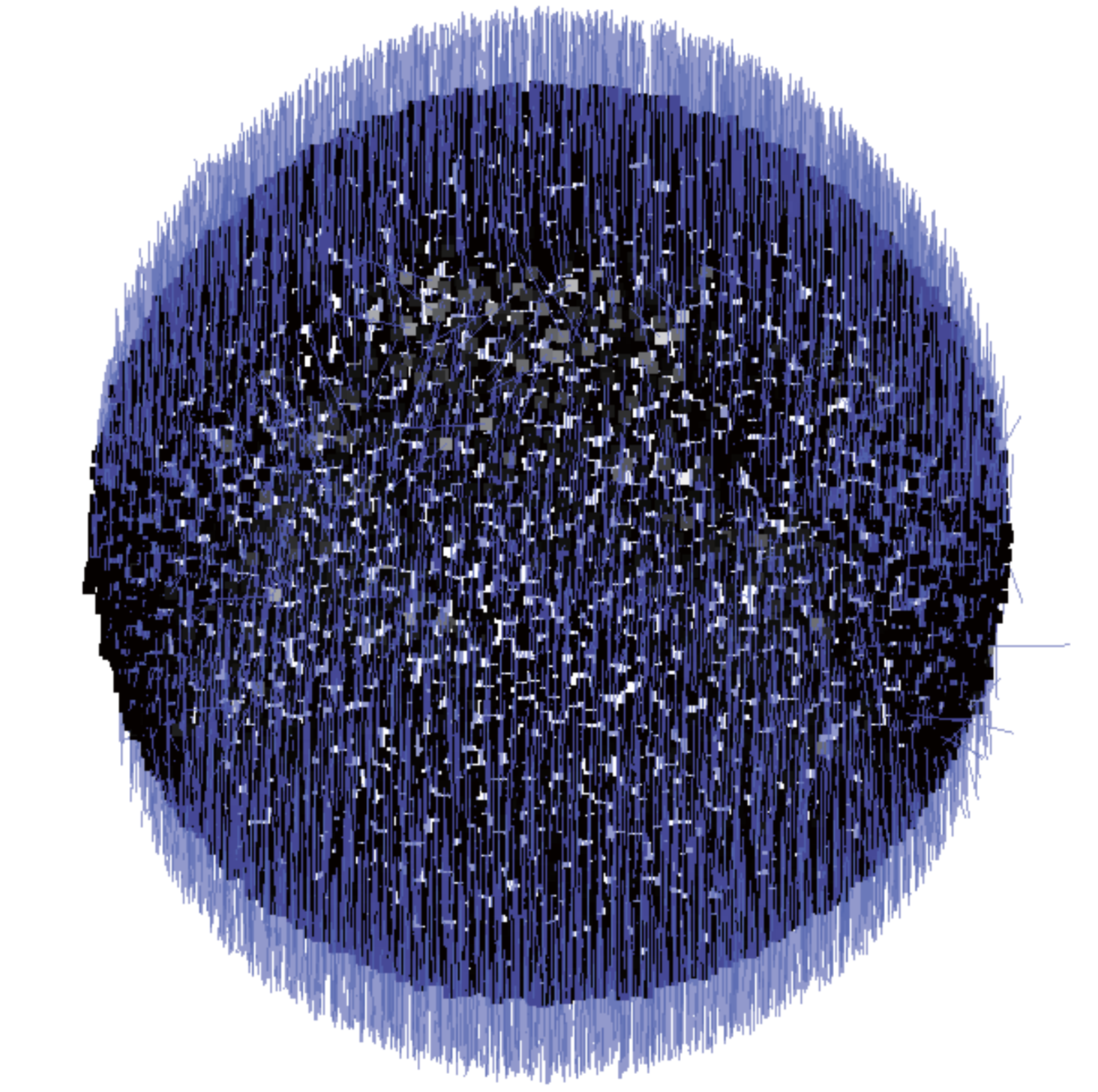}}
\end{minipage}
\begin{minipage}[b]{0.3\linewidth}
\subfigure[]{\label{}\includegraphics[width=1\linewidth]{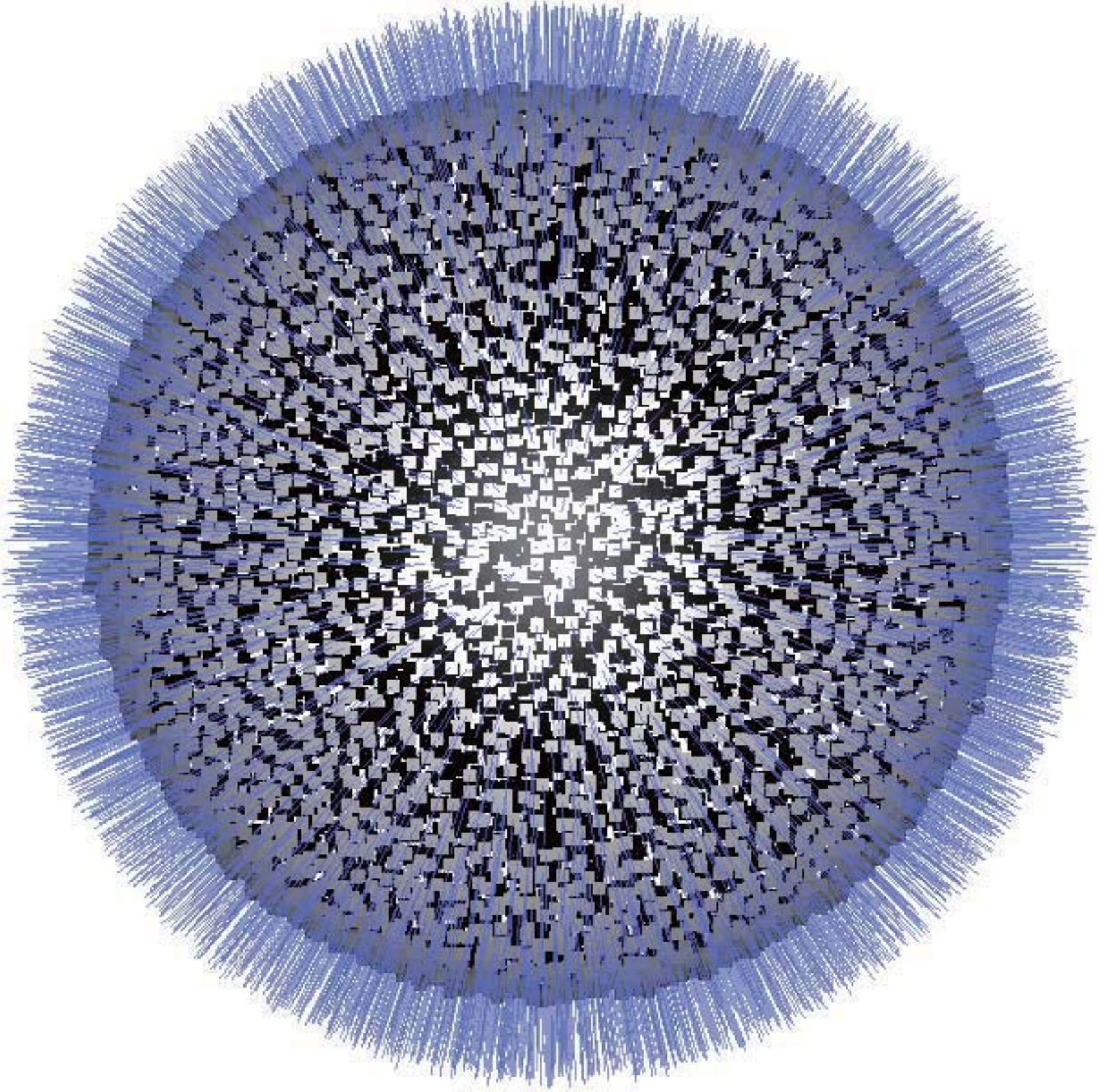}}
\end{minipage}
\caption{Three ways of matrix construction: (a) random permutation, (b) permute matrix $M$ with the order of $x$, $y$ and $z$ of one by one normal, (c) ours: permute matrix $M$ with the order of $x$ of all normals, then $y$ of all normals and finally $z$. Blue lines indicate point normals.   }
\label{fig:matrixordering}
%\vspace{-0.65cm}
\end{figure}

%rank-related: patch & final
\begin{figure}[htt]
%\vspace{-0.0cm}
\centering
\begin{minipage}[b]{0.3\linewidth}
\subfigure[Average]{\label{}\includegraphics[width=1\linewidth]{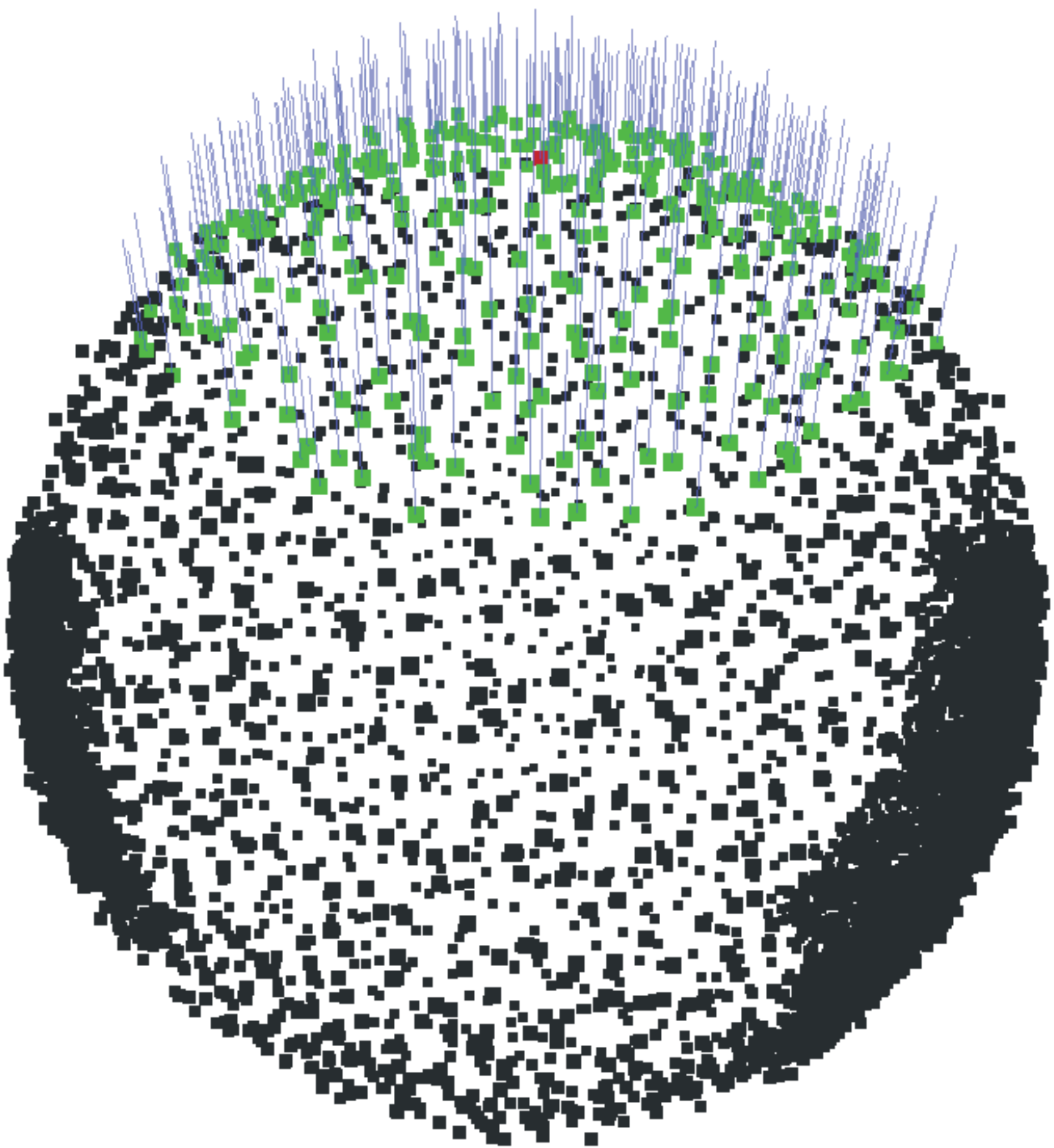}}
\end{minipage}
\begin{minipage}[b]{0.3\linewidth}
\subfigure[$\beta=20$]{\label{}\includegraphics[width=1\linewidth]{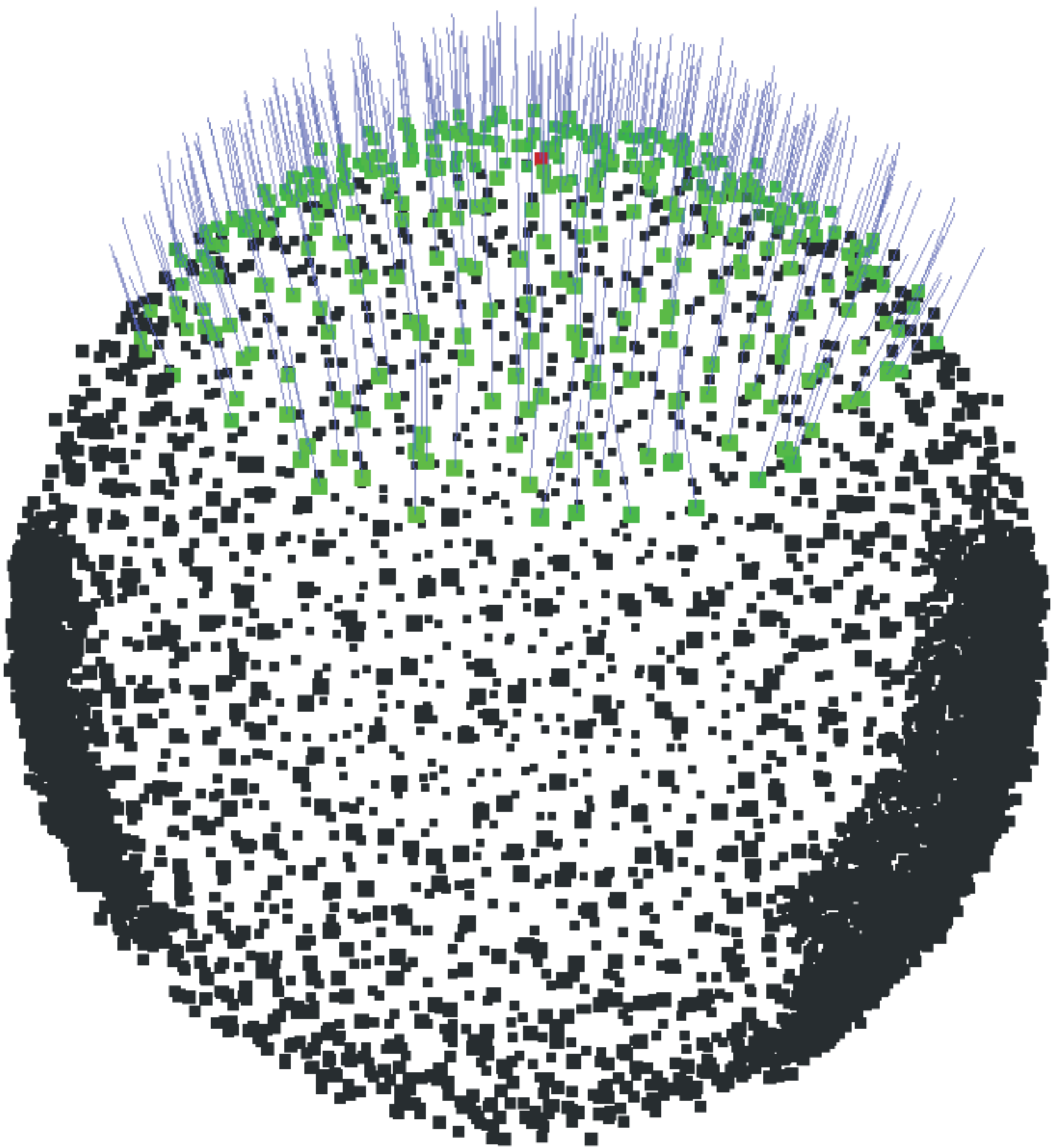}}
\end{minipage}
\begin{minipage}[b]{0.3\linewidth}
\subfigure[$\beta=1$]{\label{}\includegraphics[width=1\linewidth]{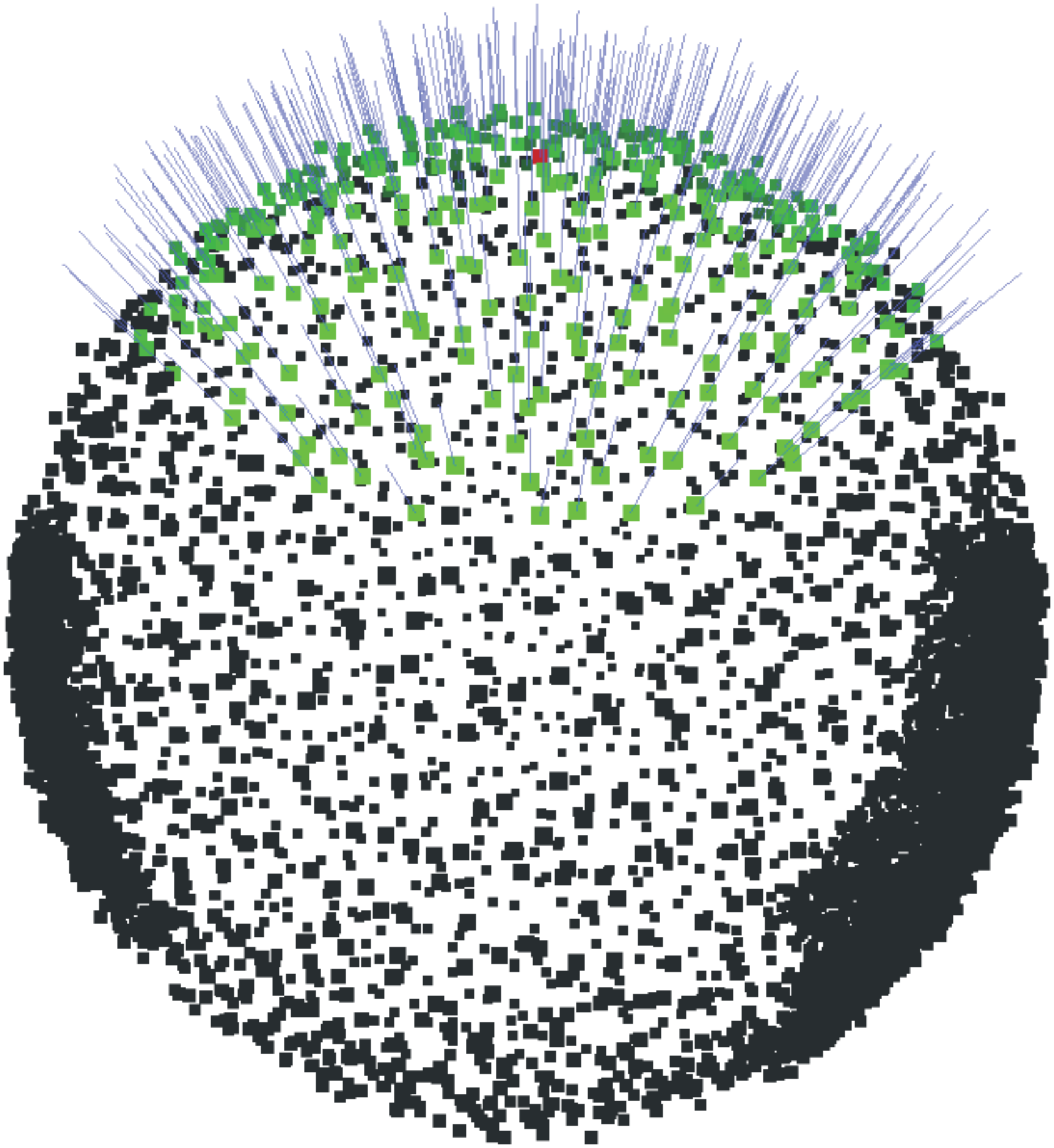}}
\end{minipage}\\
\begin{minipage}[b]{0.3\linewidth}
\subfigure[Average]{\label{}\includegraphics[width=1\linewidth]{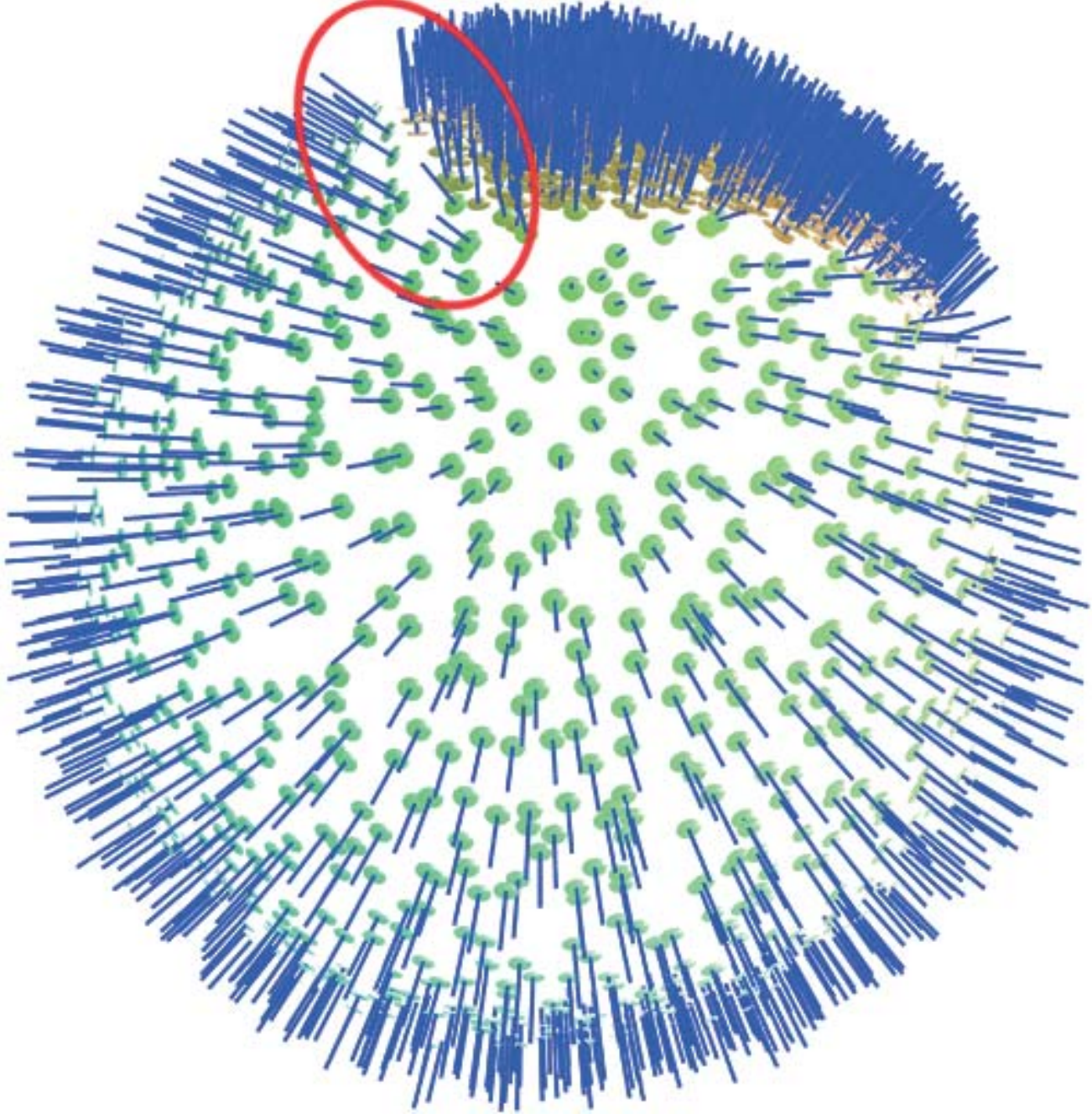}}
\end{minipage}
\begin{minipage}[b]{0.3\linewidth}
\subfigure[$\beta=20$]{\label{}\includegraphics[width=1\linewidth]{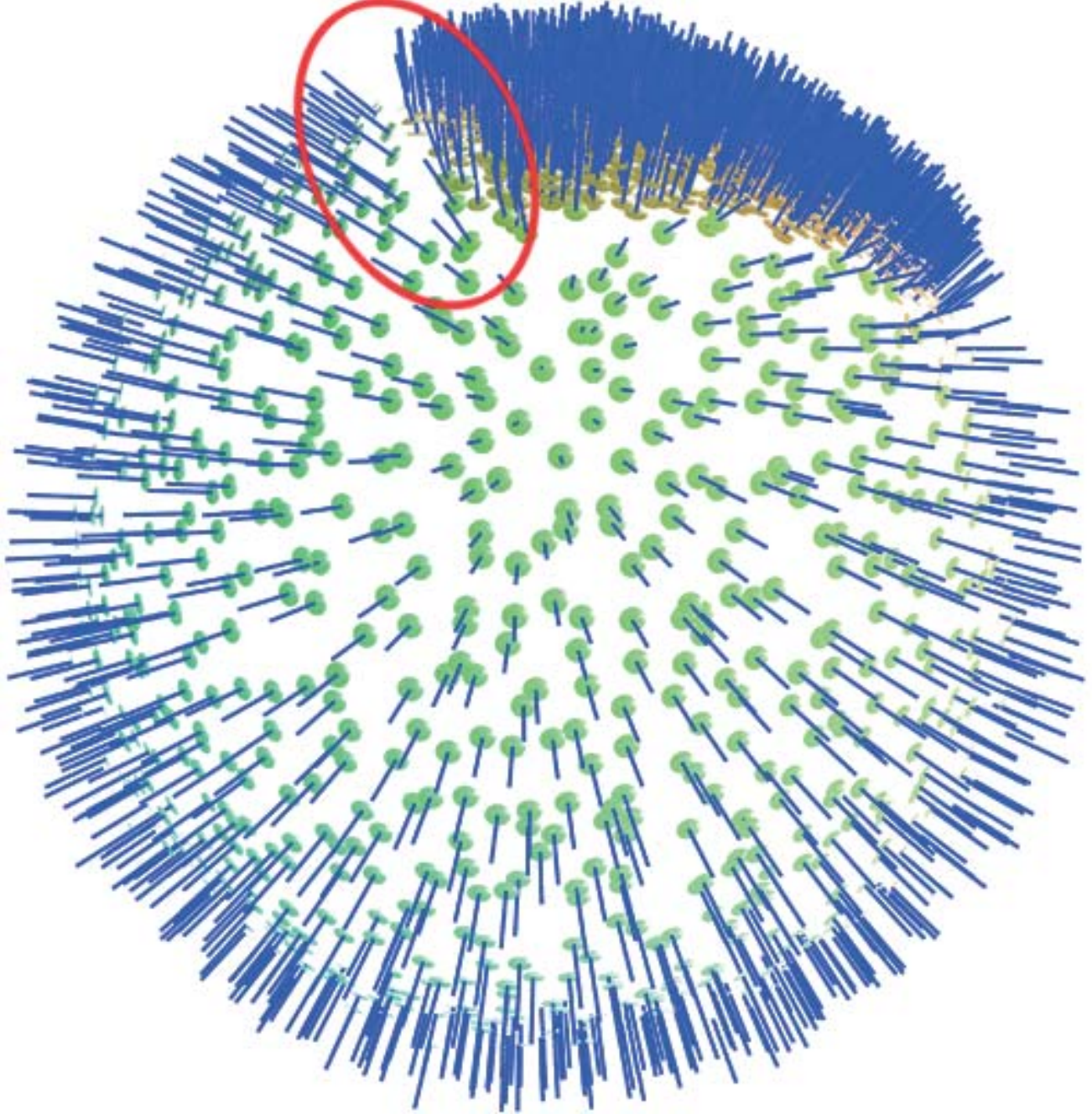}}
\end{minipage}
\begin{minipage}[b]{0.3\linewidth}
\subfigure[$\beta=1$]{\label{}\includegraphics[width=1\linewidth]{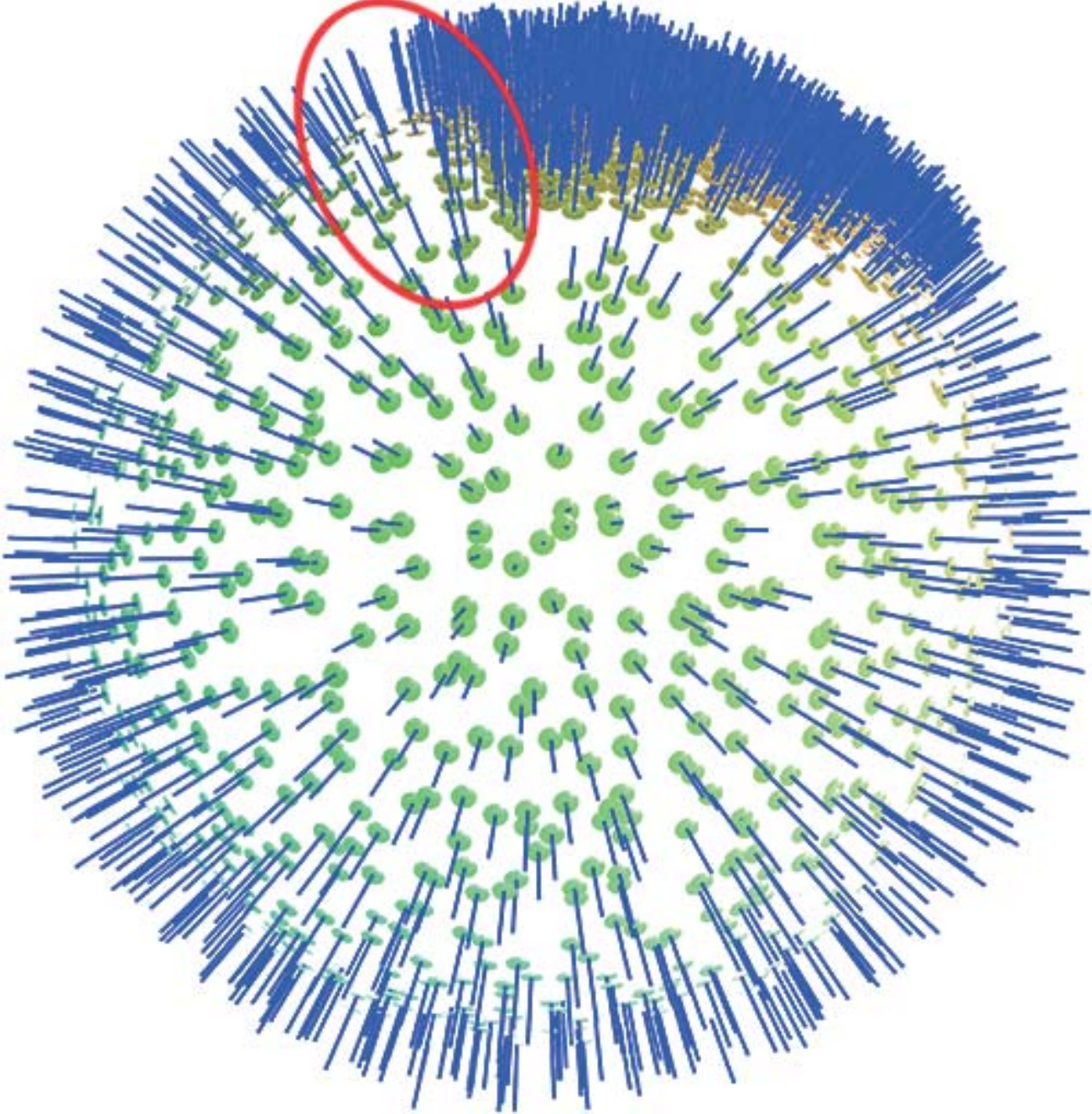}}
\end{minipage}
\caption{First row: recovered normals of a matrix constructed by similar local isotropic structures, with averaging and different $\beta$. Second row: results after performing several normal estimation iterations on the same input. The mean square angular errors for (d-f) are ($\times10^{-2}$): $3.50$, $2.58$ and $0.41$, respectively. }
\label{fig:rankrelated}
%\vspace{-0.65cm}
\end{figure}

\textbf{Nuclear norm.} The nuclear norm of a matrix is defined as the sum of the absolute values of its singular values, shown in Eq. \eqref{eq:nuclearnorm}.
\begin{equation}\label{eq:nuclearnorm}
\begin{aligned}
\normN{\mathbf{Z}}_{*} = \sum_{m} {|\delta_m|},
\end{aligned}
\end{equation}
where $\delta_m$ is the $m$-th singular value of matrix $\mathbf{Z}$. $\normN{\mathbf{Z}}_{*}$ indicates the nuclear norm of $\mathbf{Z}$.

\textbf{Nuclear norm minimization.} Nuclear norm minimization is frequently used to approximate the known matrix, $\mathbf{Z}'$, by a low-rank matrix, $\mathbf{Z}$, while minimizing the nuclear norm of $\mathbf{Z}$. Cai et al. \cite{Cai2010} demonstrated that the low-rank matrix $\mathbf{Z}$ can be easily solved by adding a Frobenius-norm data term.
\begin{equation}\label{eq:nnm}
\begin{aligned}
\min_{\mathbf{Z}} \alpha\normN{\mathbf{Z}}_{*} + \normN{\mathbf{Z}'-\mathbf{Z}}_F^2,
\end{aligned}
\end{equation}
where $\alpha$ is the weighting parameter. The minimizing matrix $\mathbf{Z}$ is then
\begin{equation}\label{eq:solutionnnm}
\begin{aligned}
\mathbf{Z} = \mathbf{U}\psi(\mathbf{S},\alpha)\mathbf{V}^T, %why put a hat on Z?
\end{aligned}
\end{equation}
where $\mathbf{Z}'=\mathbf{U}\mathbf{S}\mathbf{V}^T$ denotes the SVD of $\mathbf{Z}'$ and $\mathbf{S}_{m,m}$ is the $m$-th diagonal element in $\mathbf{S}$. $\psi$ is the soft-thresholding function on $\mathbf{S}$ and the parameter $\alpha$, i.e., $\psi(\mathbf{S}_{m,m},\alpha) = \max(0,\mathbf{S}_{m,m}-\alpha)$.

Nuclear norm minimization treats and shrinks each singular value equally. However, in general, larger singular values should be shrunk less to better approximate the known matrix and preserve the major components. The weighted nuclear norm minimization solves this issue \cite{Gu2014}.

\textbf{Weighted nuclear norm minimization.} The weighted nuclear norm of a matrix $\mathbf{Z}$ is
\begin{equation}\label{eq:weightednuclearnorm}
\begin{aligned}
\normN{\mathbf{Z}}_{*,\mathbf{w}} = \sum_{m} {|w_m\delta_m|},
\end{aligned}
\end{equation}
where $w_m$ is the non-negative weight imposed on the $m$-th singular value and $\mathbf{w} = \{w_m\}$. We can then write the low-rank matrix approximation problem as
\begin{equation}\label{eq:wnnm}
\begin{aligned}
\min_{\mathbf{Z}} \normN{\mathbf{Z}}_{*,\mathbf{w}} + \normN{\mathbf{Z}'-\mathbf{Z}}_F^2
\end{aligned}
\end{equation}

Suppose the singular values $\{\delta_m\}$ are sorted in a non-ascending order, the corresponding weights $\{w_m\}$ should be in a non-descending order. Hence, we define the weight function as a Gaussian function.
\begin{equation}\label{eq:weightwnnm}
\begin{aligned}
w_m = \beta e^{-(\frac{2\delta_m}{\delta_1})^2}
\end{aligned}
\end{equation}
$\beta$ denotes the regularized coefficient which defaults to $1.0$. $\delta_1$ is the first singular value after sorting $\{\delta_m\}$ in a non-increasing order. We did not use the original weight definition in \cite{Gu2014} since it needs noise variance which should be unknown in normal estimation. Also, we found their weight determination is not suitable for normal-constructed matrices. Then we solve Eq. \eqref{eq:wnnm} by the generalized soft thresholding operation on the singular values with weights \cite{Gu2014}. 
\begin{equation}\label{eq:solutionwnnm}
\begin{aligned}
\mathbf{Z} = \mathbf{U}\psi(\mathbf{S},\{w_m\})\mathbf{V}^T, 
\end{aligned}
\end{equation}
where $\psi(\mathbf{S}_{m,m},w_m)=\max(0, \mathbf{S}_{m,m}-w_m)$. Here $\psi$ changes to the generalized soft-thresholding function by assigning weights to singular values, and Eq. \eqref{eq:solutionwnnm} becomes the weighted version of Eq. \eqref{eq:solutionnnm}. 

\textbf{Remark 2: matrix ordering.} We investigated the ordering of the constructed matrix. We found that the ordering of points significantly influences the minimization result (Figure \ref{fig:matrixordering}). We suspect this is due to the neighboring information and three coordinates of each normal in different ordering matrices, which is more complicated than the regular single-channel grayscale images.

\textbf{Remark 3: $\beta$ and averaging.} We tested the effects of averaging and $\beta$. We found that $\beta$ is related to the number of ranks after the generalized soft-thresholding. A smaller $\beta$ leads to more ranks retained, which also indicates more noise is left behind. For instance, the numbers of ranks for the sphere example (Fig. \ref{fig:rankrelated}) with $\beta=0.05$, $\beta=1.0$ and $\beta=20.0$ are 147, 85 and 1, respectively. We also observed that the number of ranks is related to capturing changes in the surface: a greater $\beta$ captures less surface changes (e.g., Figure \ref{fig:rankrelated}(b,e)). We found no relations between averaging and the most low-rank method (i.e., with the largest rank retained). The most low-rank method may also recover normals of a matrix with different directions (Figure \ref{fig:rankrelated}(a,b)). Figure \ref{fig:rankrelated} (d-f) show that our default $\beta$ is more robust than both averaging and the most low-rank method.

\begin{algorithm}[t]
\SetAlgoNoLine
\KwIn{non-local similar structures of each local isotropic structure}
\KwOut{New matrices $\{\mathbf{Z}\}$ }
\For{each local isotropic structure $S_i^{iso}$}
{
      \textbullet~construct a matrix $\mathbf{Z}'$\\
      \textbullet~compute the SVD of $\mathbf{Z}'$\\
      \textbullet~compute the weights via Eq. \eqref{eq:weightwnnm}\\
      \textbullet~recover $\mathbf{Z}$ via Eq. \eqref{eq:solutionwnnm}
}
\caption{Weighted nuclear norm minimization}
\label{alg:wnnm}
\end{algorithm}

%density difference
\begin{figure}[htbp]
%\vspace{-0.0cm}
\centering
\begin{minipage}[b]{0.3\linewidth}
\subfigure[Input]{\label{}\includegraphics[width=1\linewidth]{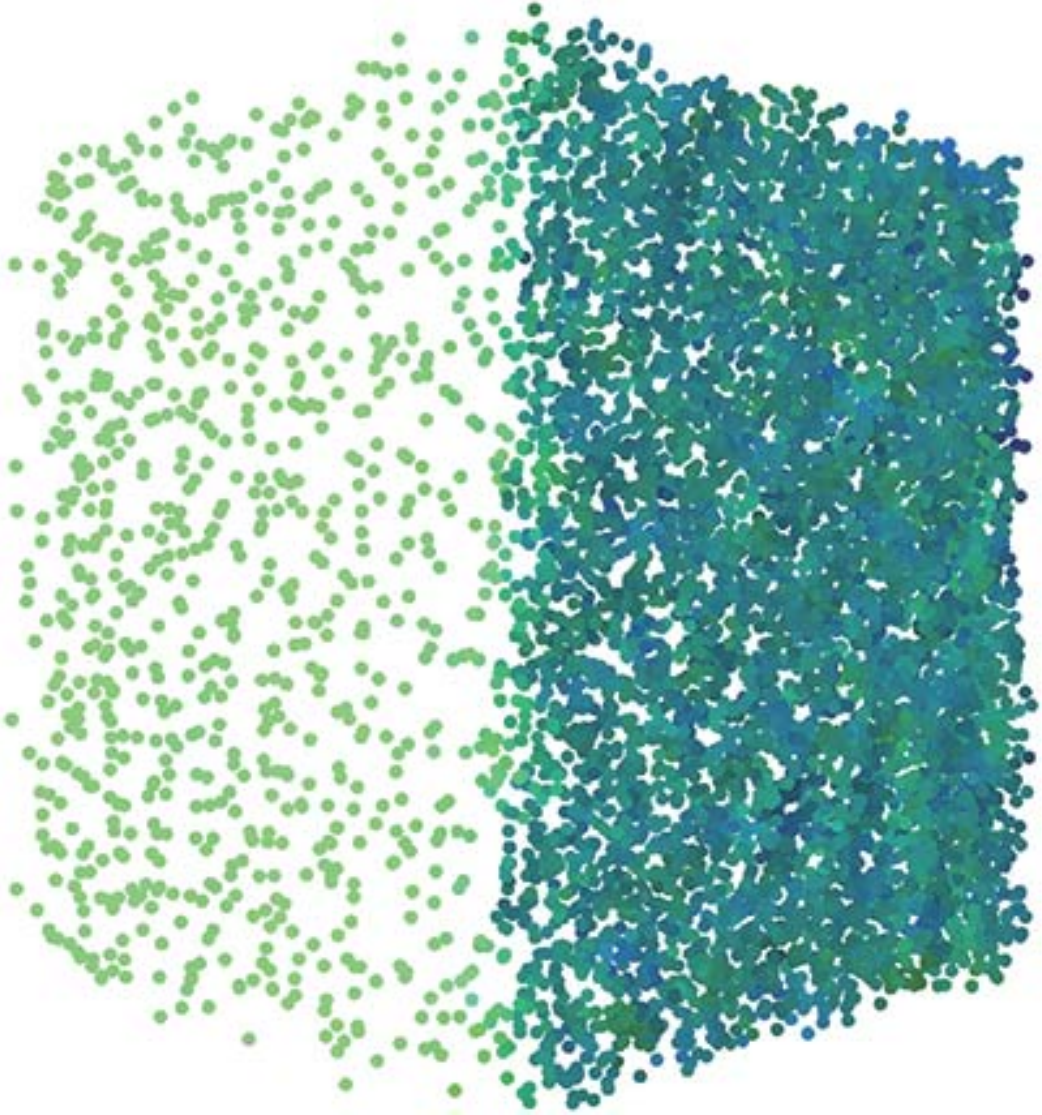}}
\end{minipage}
\begin{minipage}[b]{0.3\linewidth}
\subfigure[\protect\cite{Huang2013}]{\label{}\includegraphics[width=1\linewidth]{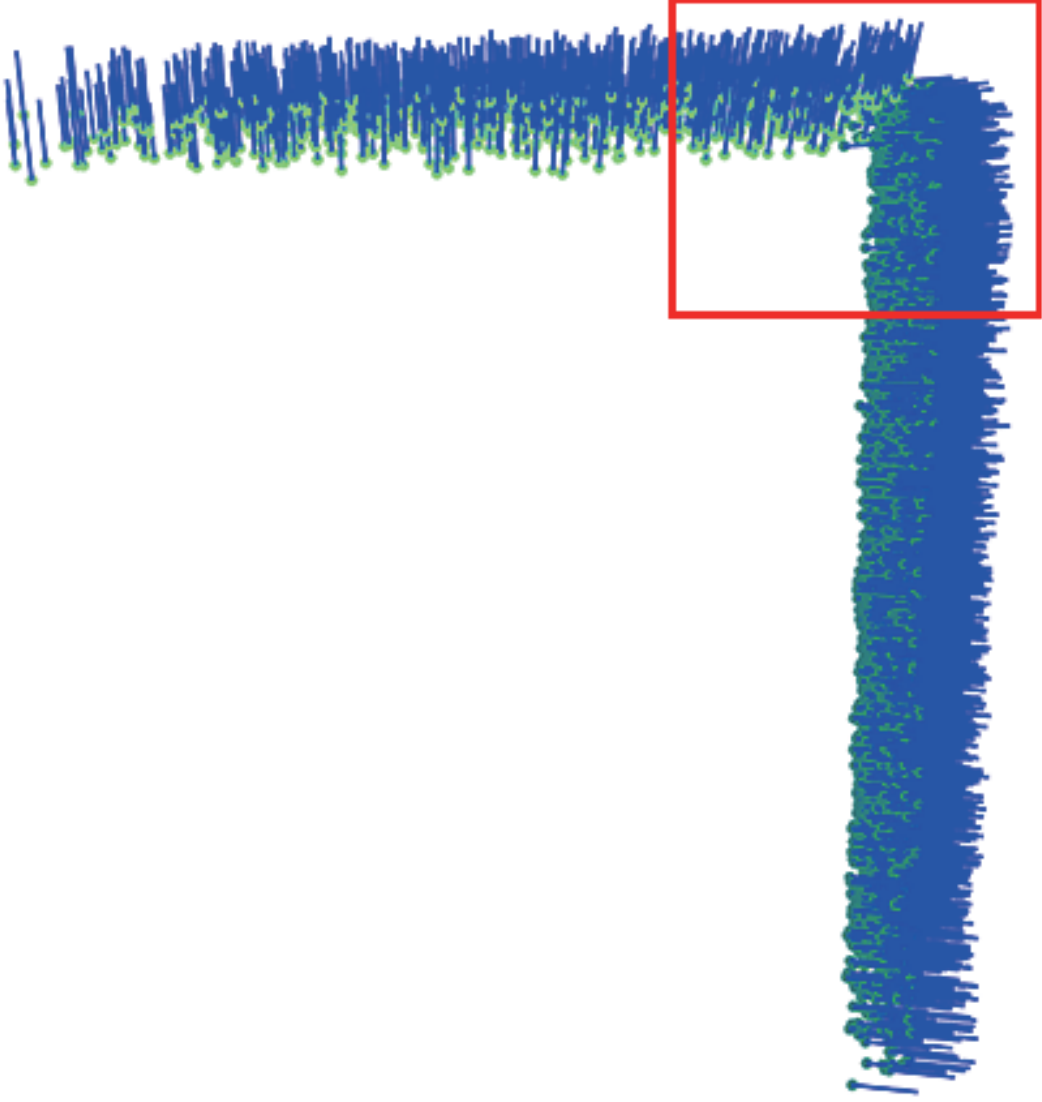}}
\end{minipage}
\begin{minipage}[b]{0.3\linewidth}
\subfigure[Ours]{\label{}\includegraphics[width=1\linewidth]{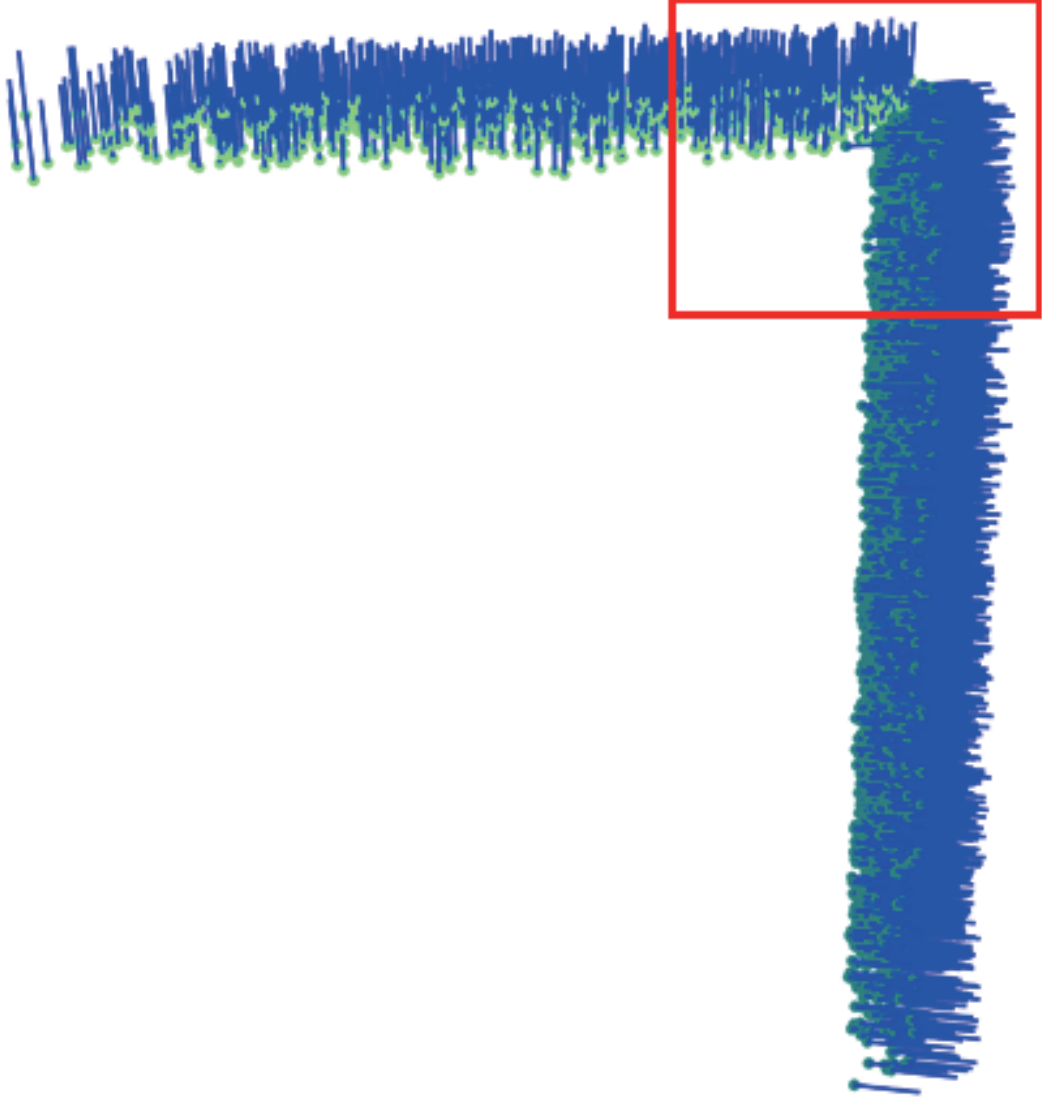}}
\end{minipage}
\caption{Normal estimation for irregularly sampled data. (a) The points on the right side are significantly denser than the points on top. (b) and (c) are shown in a different view from (a). Blue lines indicate point normals. The mean square angular error (MSAE, in radians) of (b) and (c) are $1.87\times 10^{-3}$ and $3.90\times 10^{-4}$, respectively.}
\label{fig:densitycmp}
%\vspace{-0.65cm}
\end{figure}

\subsection{Algorithm}
\label{sec:algorithm}

Each point may have multiple normals in the recovered matrices $\{\mathbf{Z}\}$, as the similar structures often overlap. We compute the final normal of each point by simply averaging the corresponding normals in $\{\mathbf{Z}\}$ after calling Algorithm~\ref{alg:wnnm}. To achieve quality normal estimations, we iterate non-local similar structures clustering (Section~\ref{sec:nonlocalstructures}) and the weighted nuclear norm minimization in Algorithm~\ref{alg:wnnm}. 
Notice that our normal estimation algorithm is feature-aware, because each matrix consists only of similar local isotropic structures. Our normal estimation technique is more accurate and robust than the bilateral filter \cite{Huang2013} when handling irregularly sampled data (Figure \ref{fig:densitycmp}).

\textbf{Extension to mesh models.} Our algorithm can be easily extended to handle mesh models. One natural way is to take the vertices/normals of a mesh as points/normals in a point cloud. However, to achieve desired results, face normals are frequently used to update vertex positions \cite{Sun2007,Zheng2011,Zhang2015}. Hence, we use the centers of faces and the corresponding normals as points.  Moreover, we use the mesh topology to compute neighbors in Section~\ref{sec:nonlocalstructures}.

%%point clouds
%cube
\begin{figure*}[htbp]
%\vspace{-0.0cm}
\centering
\begin{minipage}[b]{0.16\linewidth}
{\label{}\includegraphics[width=1\linewidth]{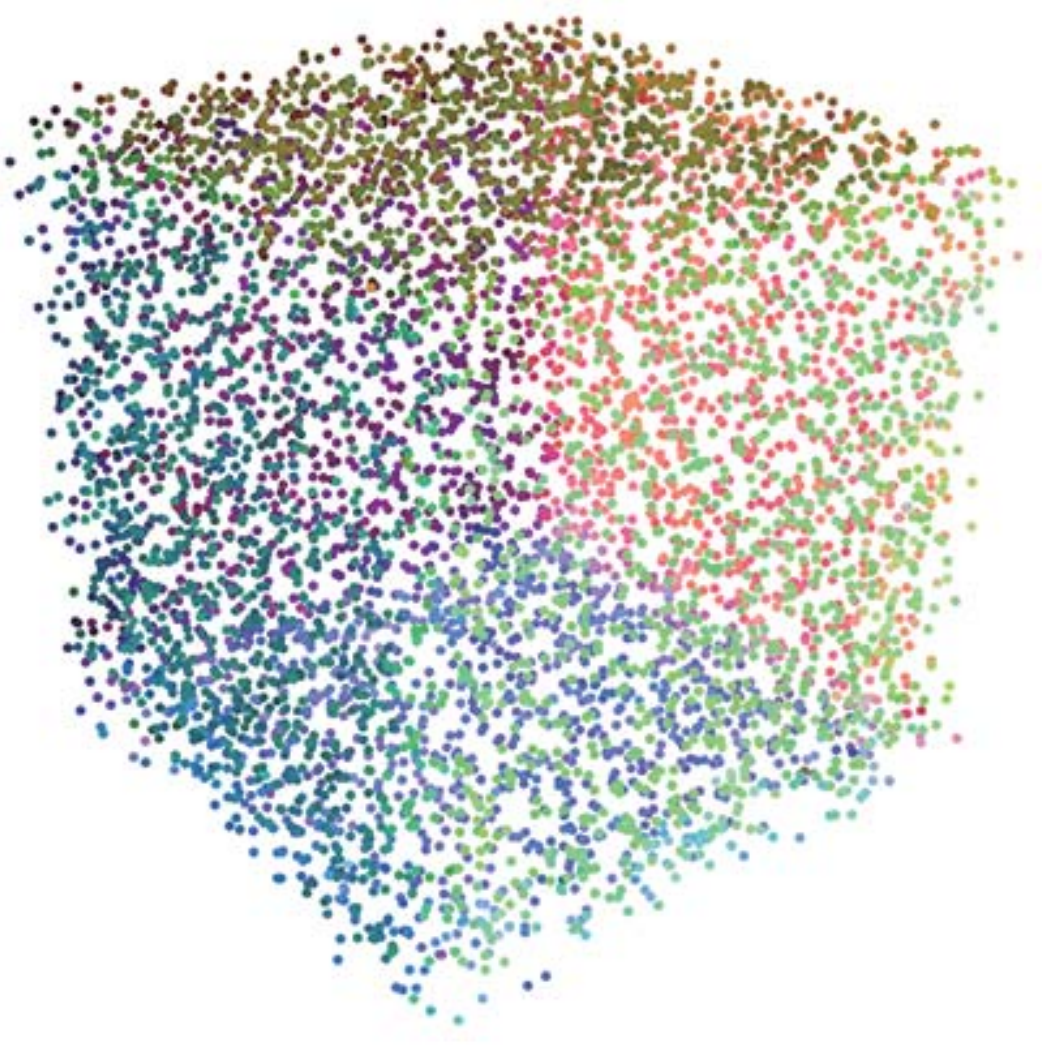}}
\end{minipage}
\begin{minipage}[b]{0.16\linewidth}
{\label{}\includegraphics[width=1\linewidth]{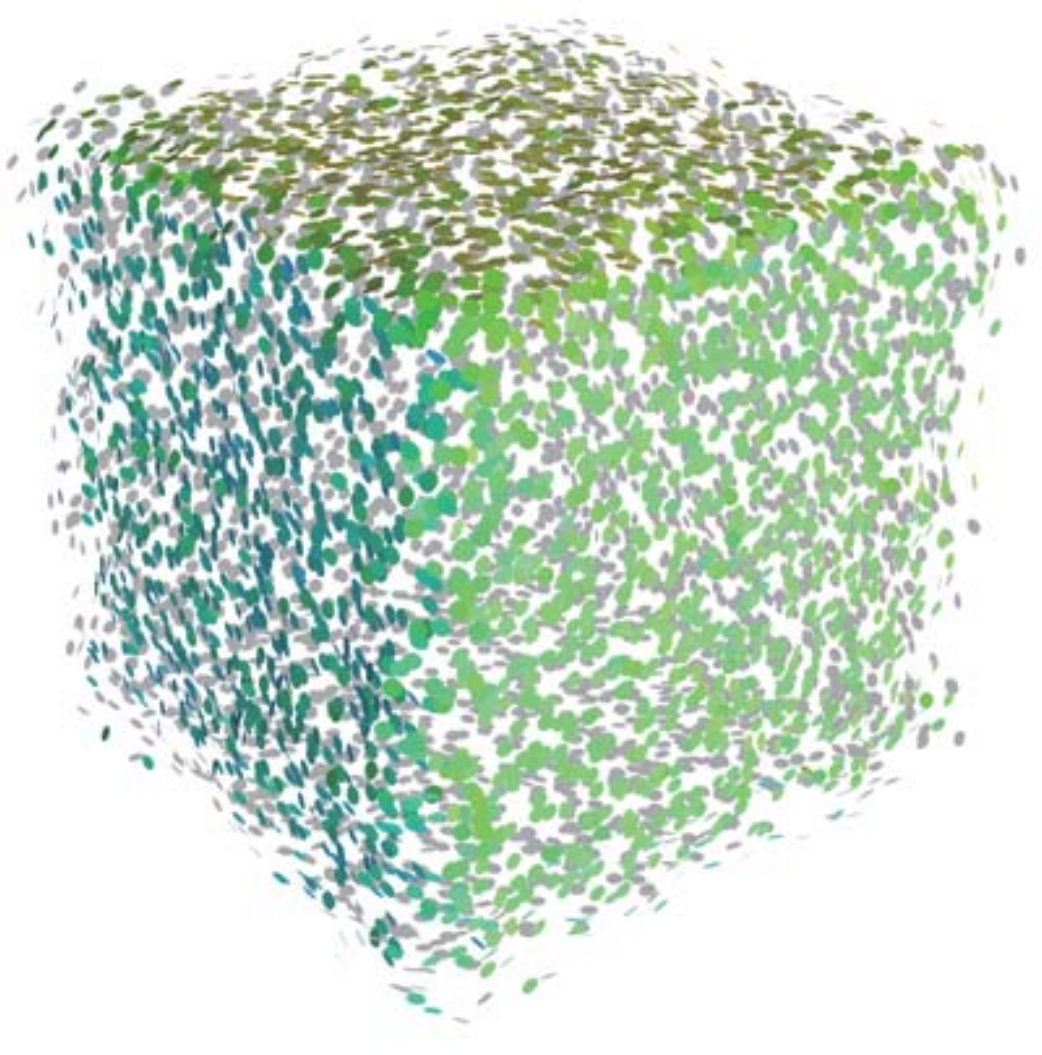}}
\end{minipage}
\begin{minipage}[b]{0.16\linewidth}
{\label{}\includegraphics[width=1\linewidth]{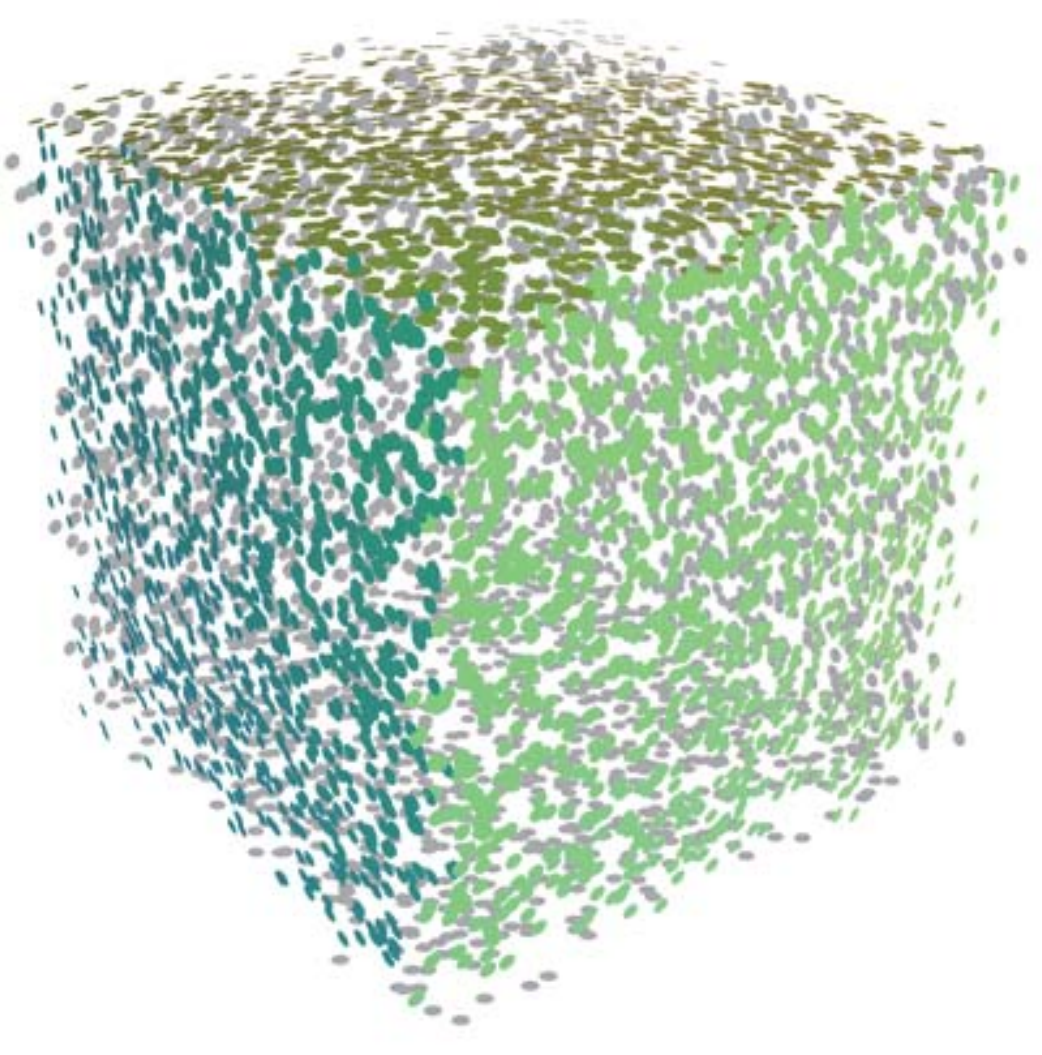}}
\end{minipage}
\begin{minipage}[b]{0.16\linewidth}
{\label{}\includegraphics[width=1\linewidth]{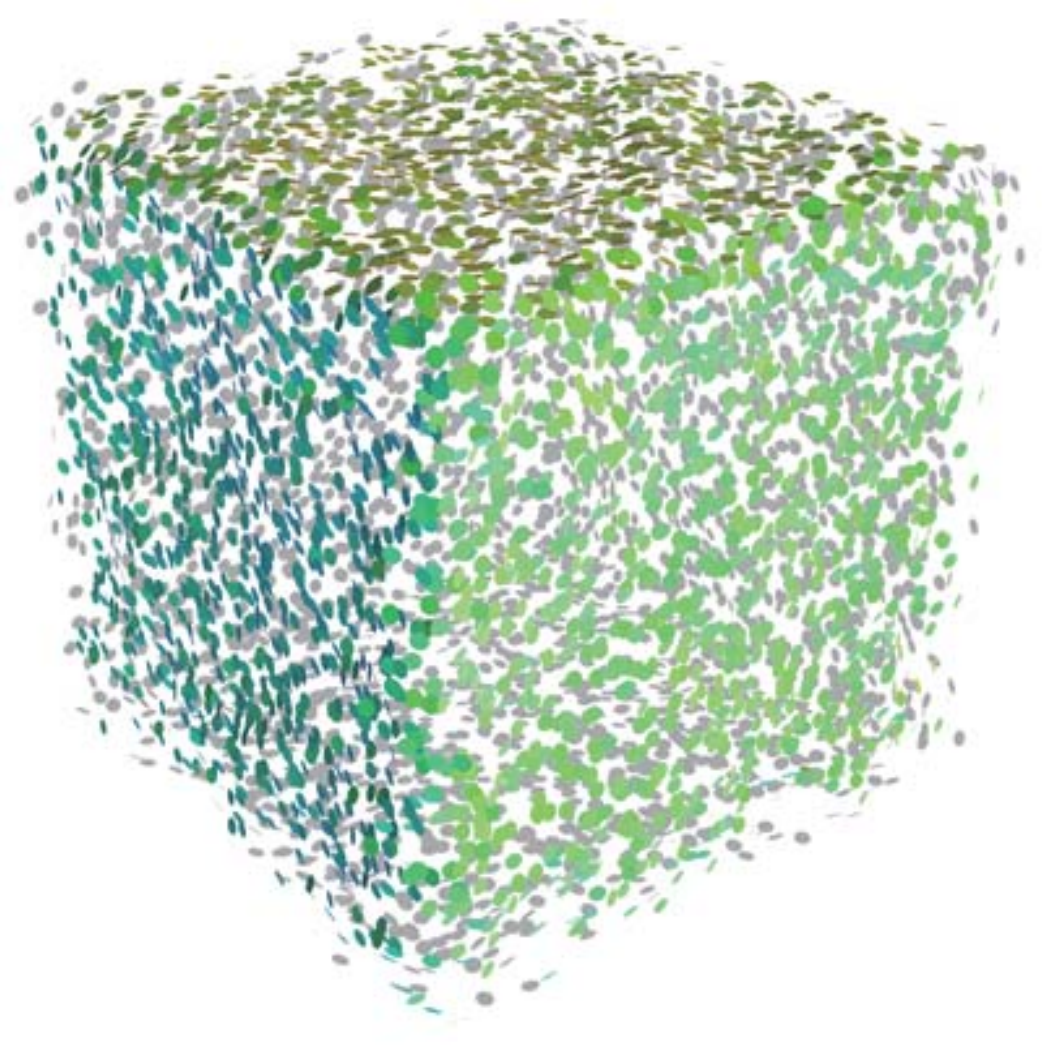}}
\end{minipage}
\begin{minipage}[b]{0.16\linewidth}
{\label{}\includegraphics[width=1\linewidth]{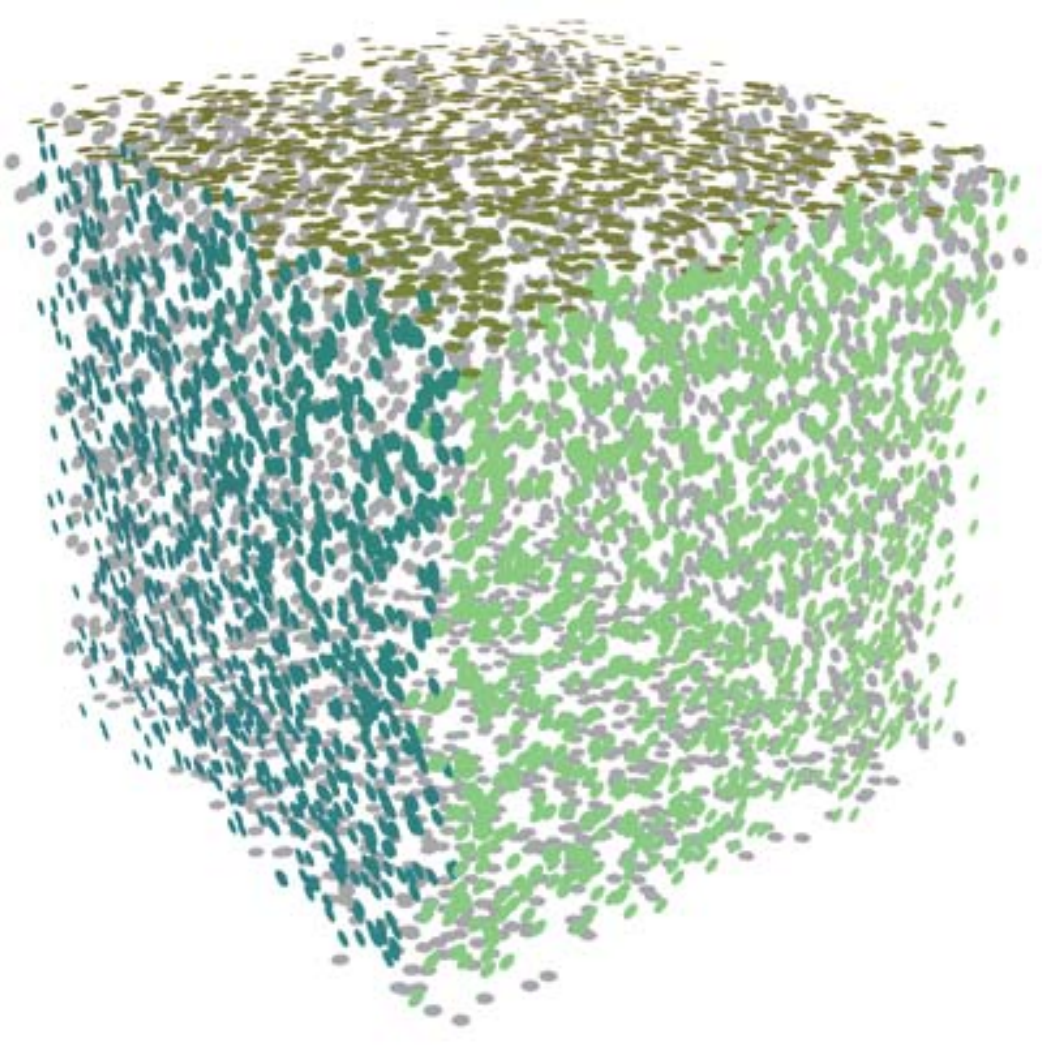}}
\end{minipage}	\\
\begin{minipage}[b]{0.115\linewidth}
{\label{}\includegraphics[width=1\linewidth]{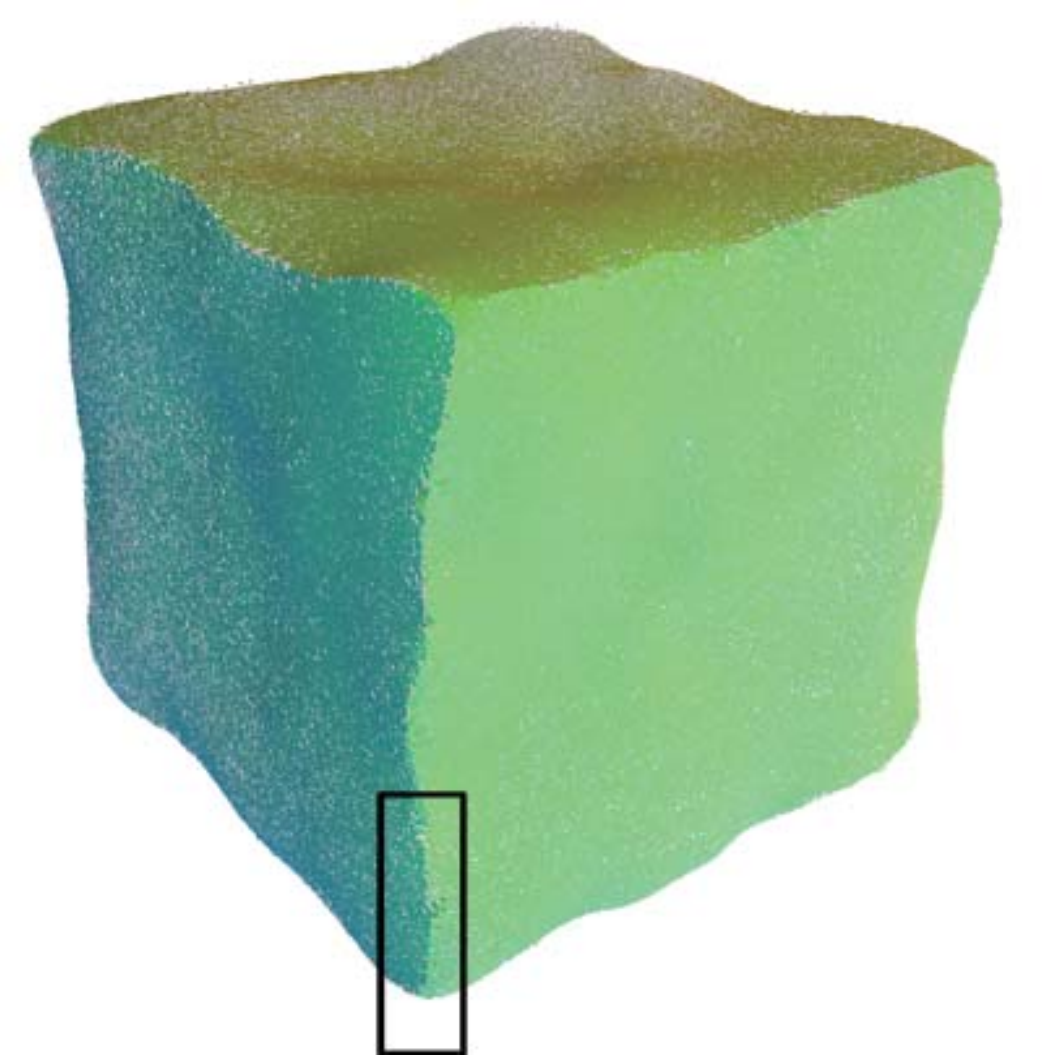}}
\end{minipage}
\begin{minipage}[b]{0.04\linewidth}
{\label{}\includegraphics[width=1\linewidth]{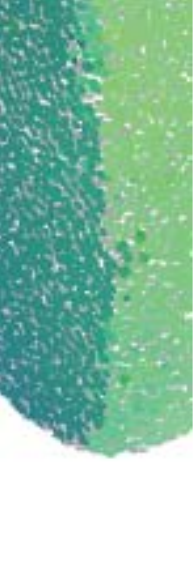}}
\end{minipage}
\begin{minipage}[b]{0.115\linewidth}
{\label{}\includegraphics[width=1\linewidth]{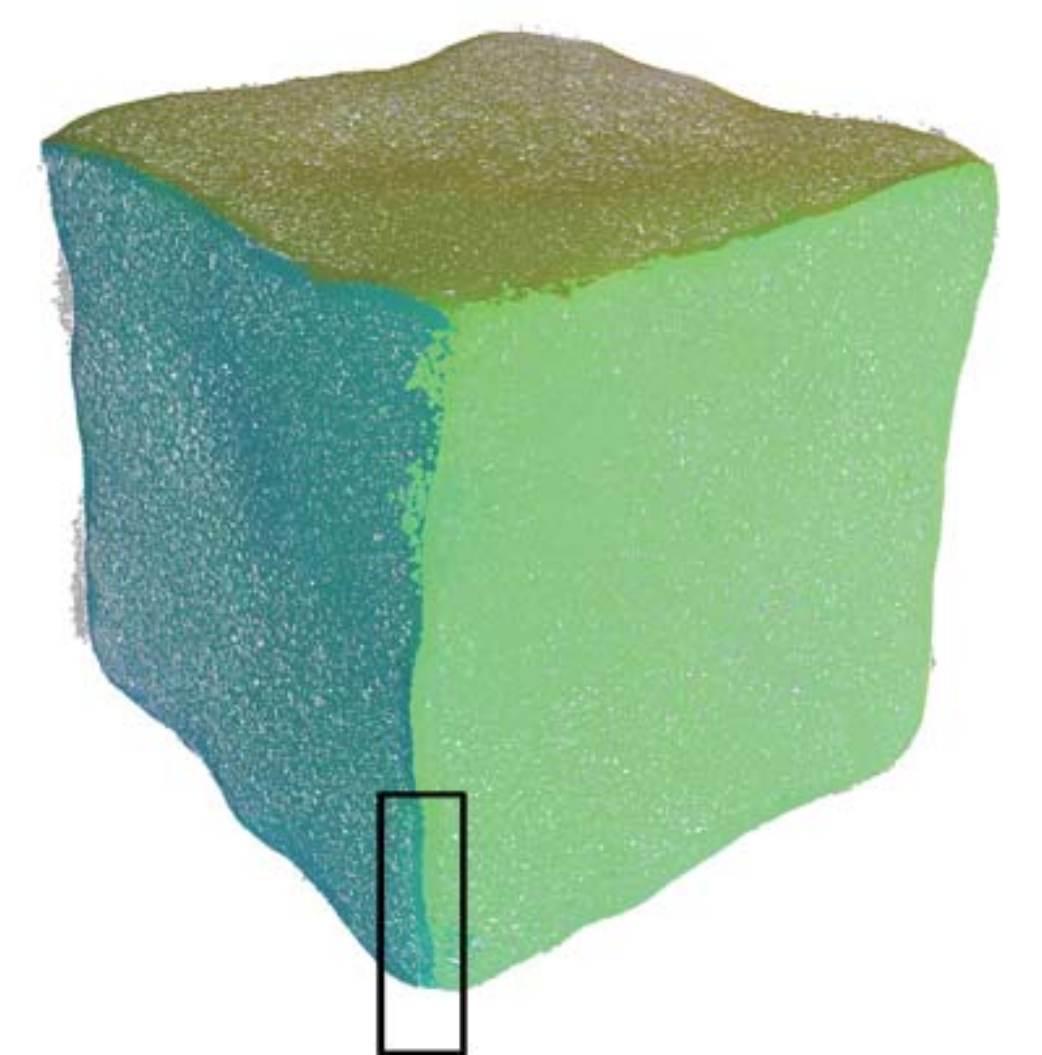}}
\end{minipage}
\begin{minipage}[b]{0.04\linewidth}
{\label{}\includegraphics[width=1\linewidth]{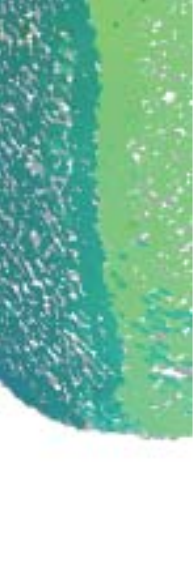}}
\end{minipage}
\begin{minipage}[b]{0.115\linewidth}
{\label{}\includegraphics[width=1\linewidth]{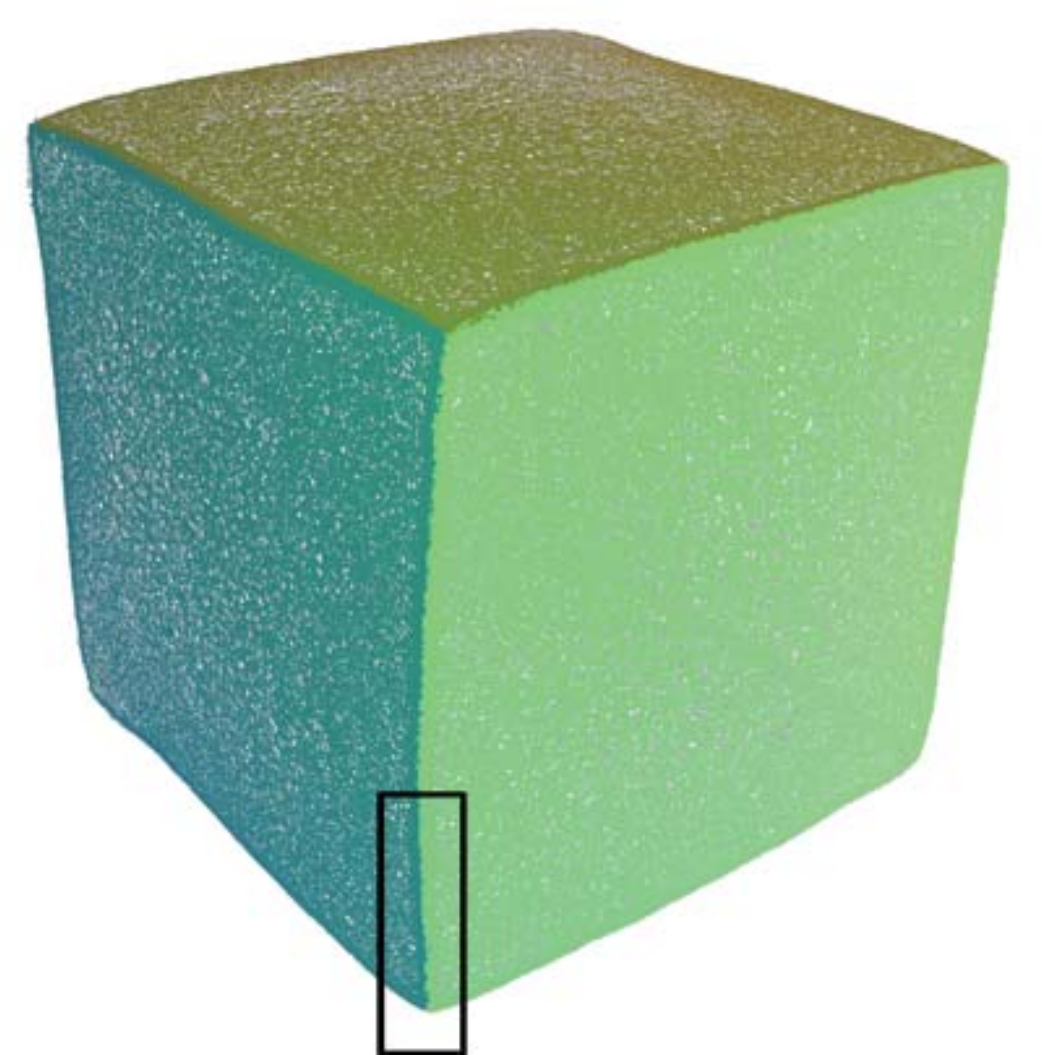}}
\end{minipage}
\begin{minipage}[b]{0.04\linewidth}
{\label{}\includegraphics[width=1\linewidth]{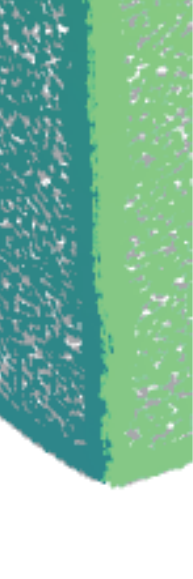}}
\end{minipage}
\begin{minipage}[b]{0.115\linewidth}
{\label{}\includegraphics[width=1\linewidth]{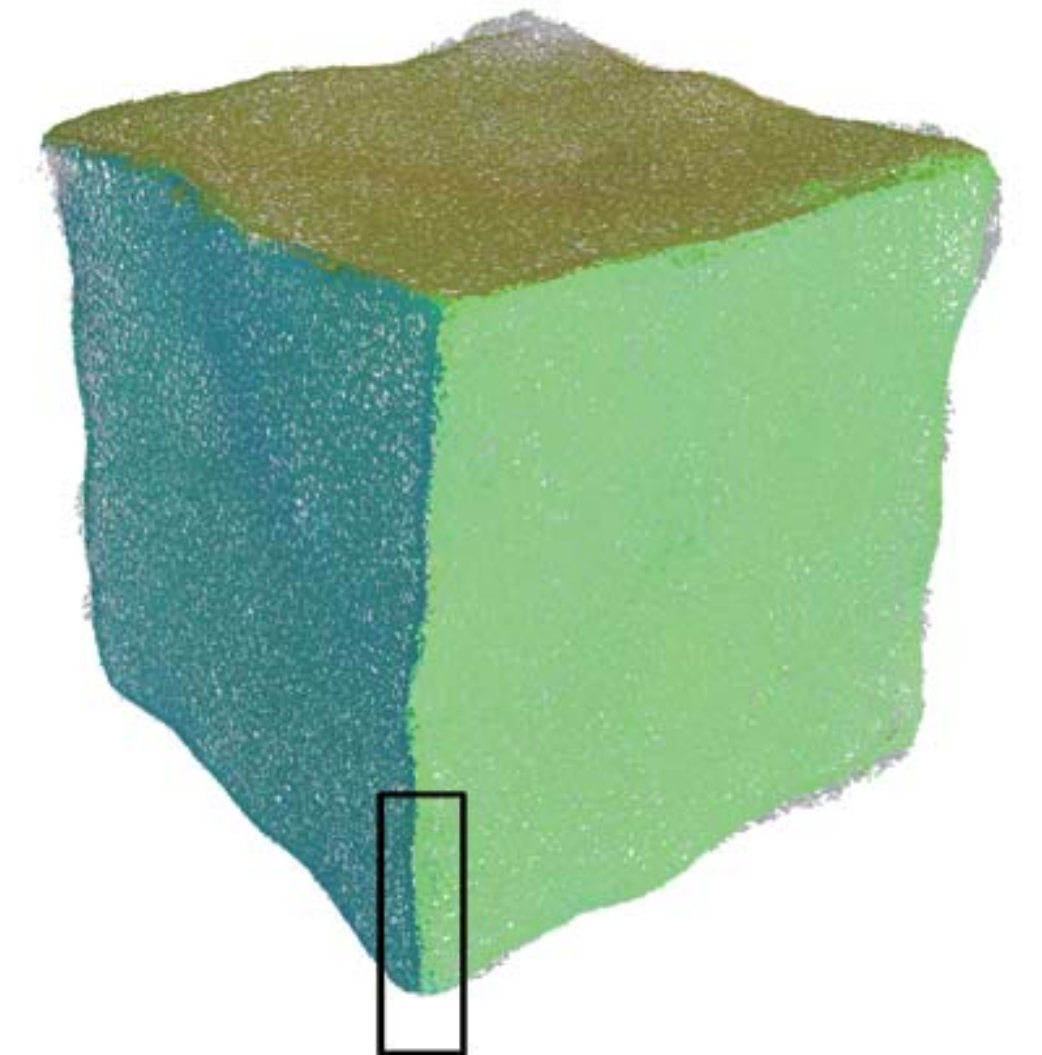}}
\end{minipage}
\begin{minipage}[b]{0.04\linewidth}
{\label{}\includegraphics[width=1\linewidth]{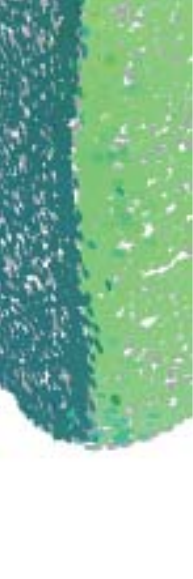}}
\end{minipage}
\begin{minipage}[b]{0.115\linewidth}
{\label{}\includegraphics[width=1\linewidth]{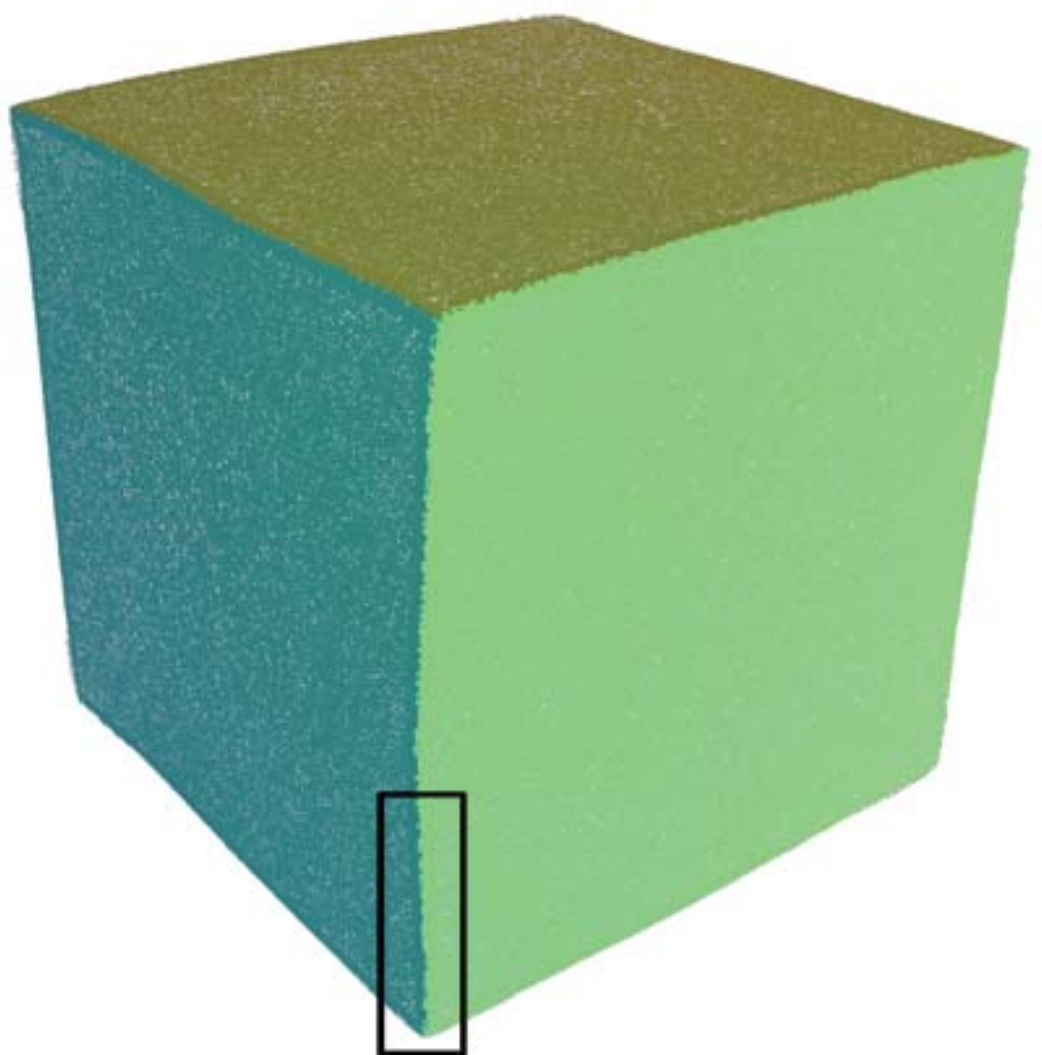}}
\end{minipage}
\begin{minipage}[b]{0.04\linewidth}
{\label{}\includegraphics[width=1\linewidth]{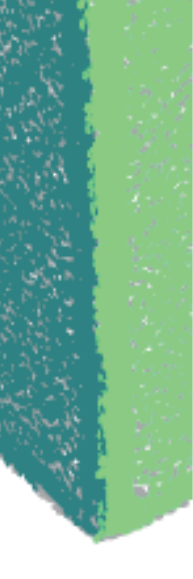}}
\end{minipage}	\\
\begin{minipage}[b]{0.16\linewidth}
\subfigure[\protect\cite{Hoppe1992}]{\label{}\includegraphics[width=1\linewidth]{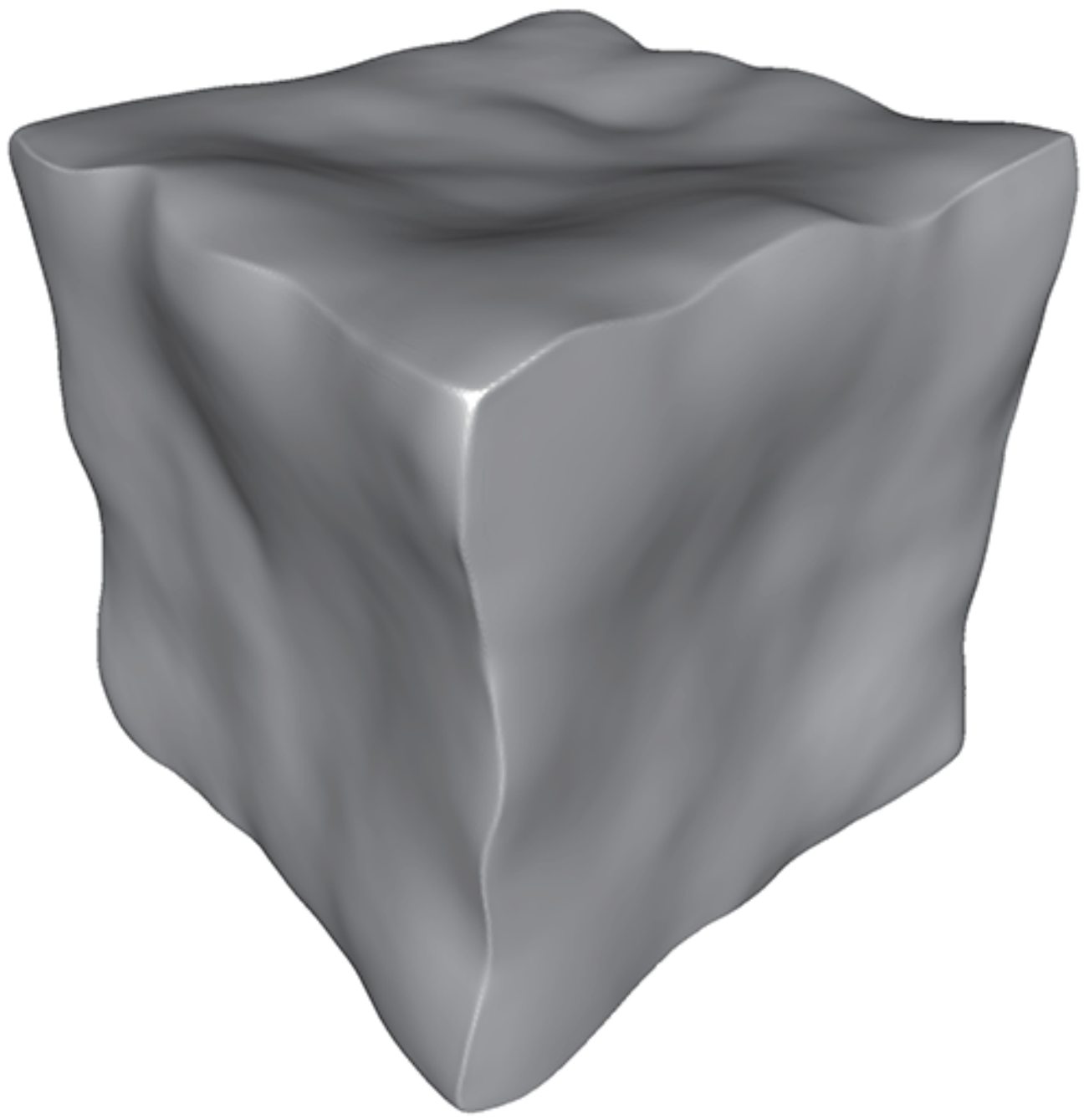}}
\end{minipage}
\begin{minipage}[b]{0.16\linewidth}
\subfigure[\protect\cite{Boulch2012}]{\label{}\includegraphics[width=1\linewidth]{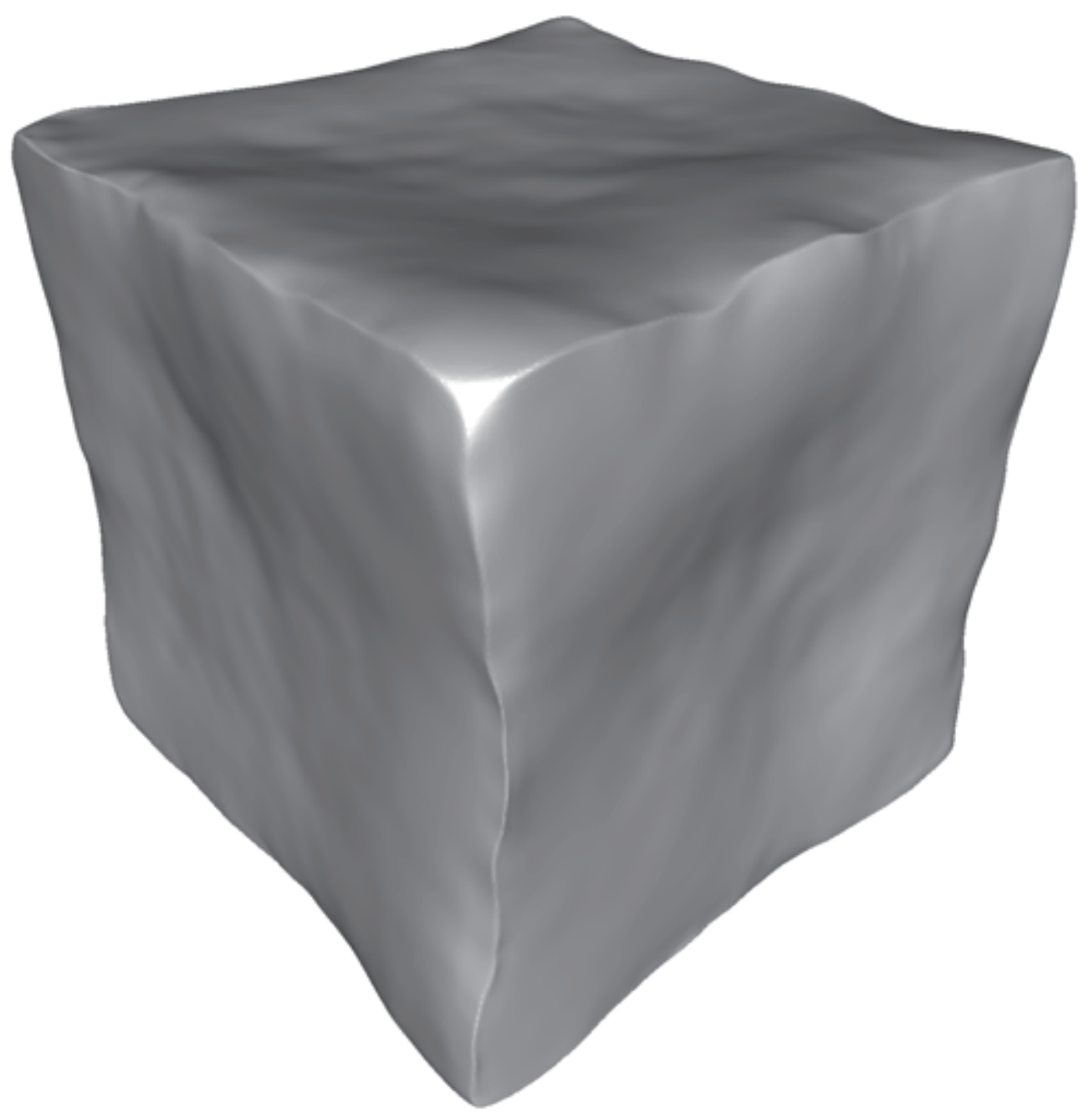}}
\end{minipage}
\begin{minipage}[b]{0.16\linewidth}
\subfigure[\protect\cite{Huang2013}]{\label{}\includegraphics[width=1\linewidth]{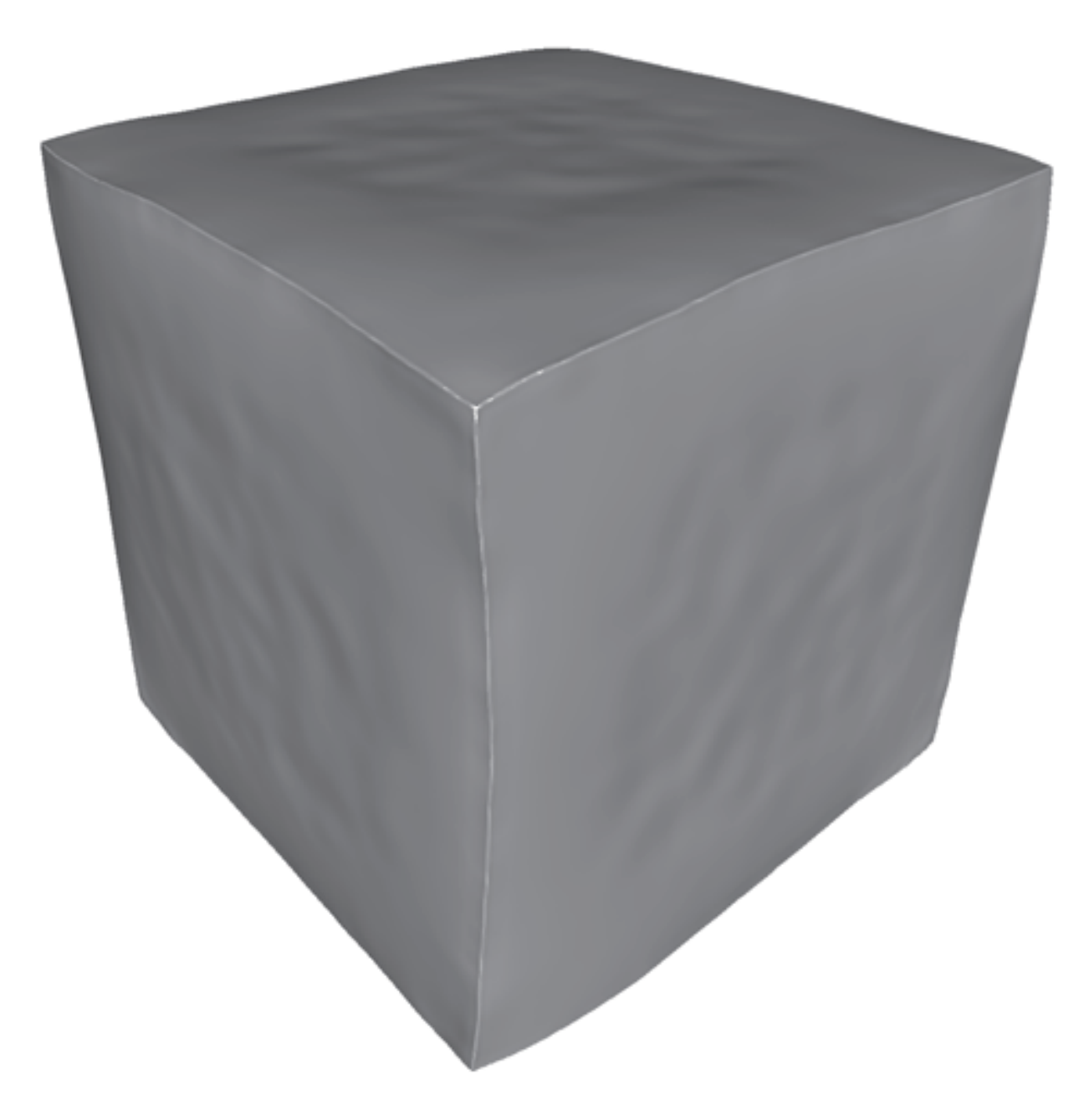}}
\end{minipage}
\begin{minipage}[b]{0.16\linewidth}
\subfigure[\protect\cite{Boulch2016}]{\label{}\includegraphics[width=1\linewidth]{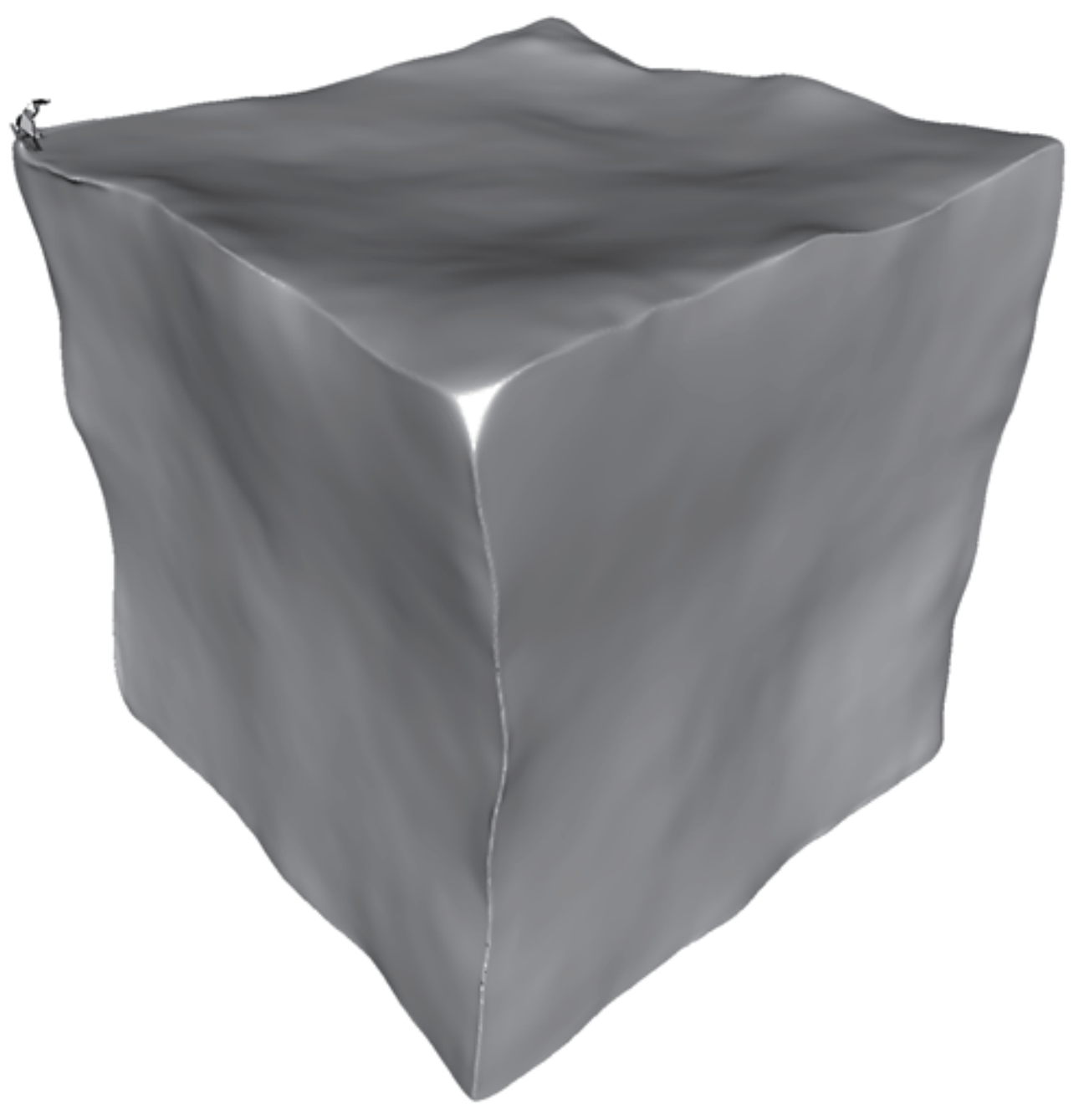}}
\end{minipage}
\begin{minipage}[b]{0.16\linewidth}
\subfigure[Ours]{\label{}\includegraphics[width=1\linewidth]{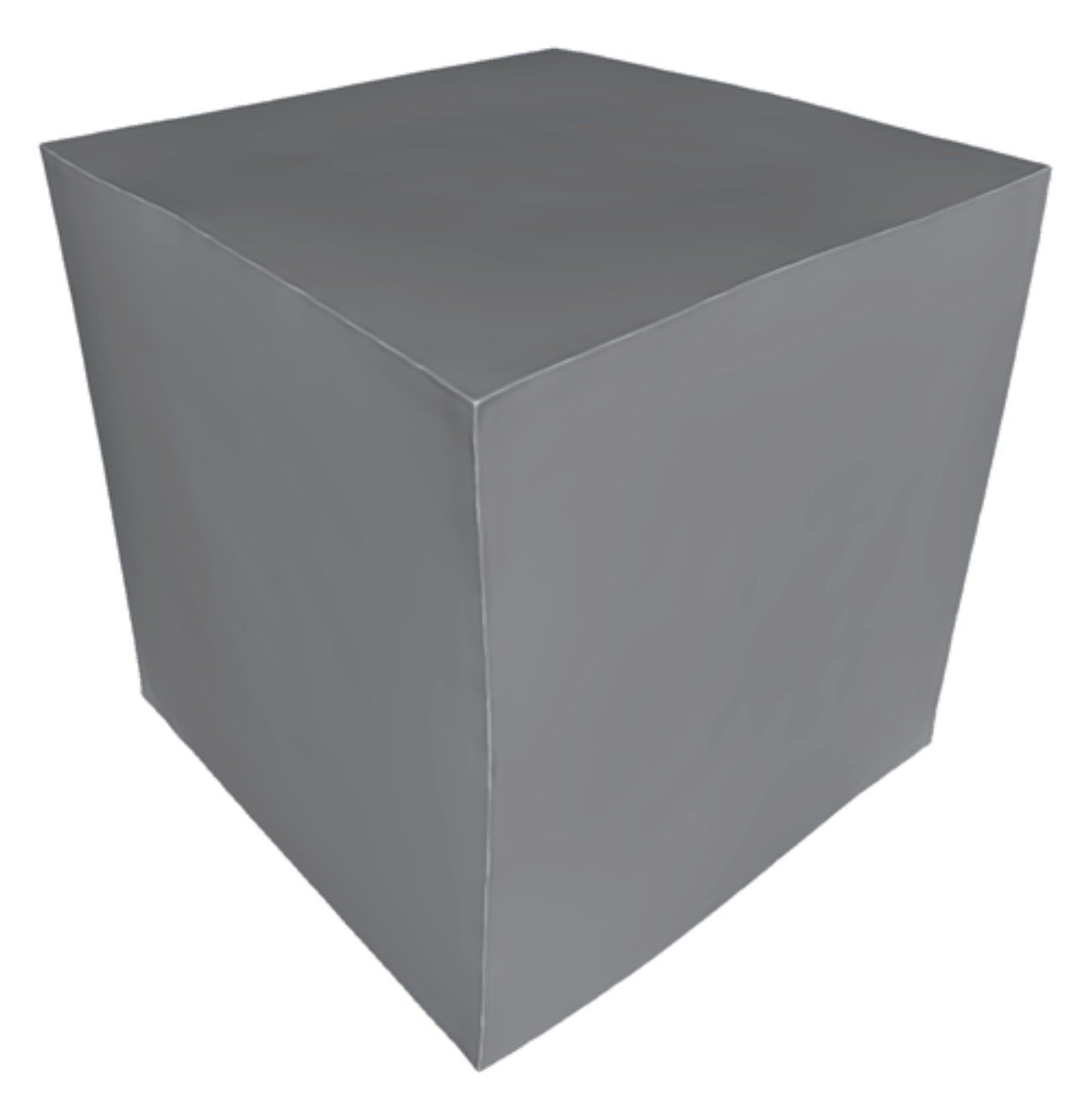}}
\end{minipage}
\caption{The first row: normal results of the Cube point cloud (synthetic noise). The second row: upsampling results of the filtered results by updating position with the normals in the first row. The third row: the corresponding surface reconstruction results.}
\label{fig:cube_point}
%\vspace{-0.65cm}
\end{figure*}

%normal non-convergence
\begin{figure}[htbp]
%\vspace{-0.0cm}
\centering
\begin{minipage}[b]{0.47\linewidth}
\subfigure[]{\label{}\includegraphics[width=1\linewidth]{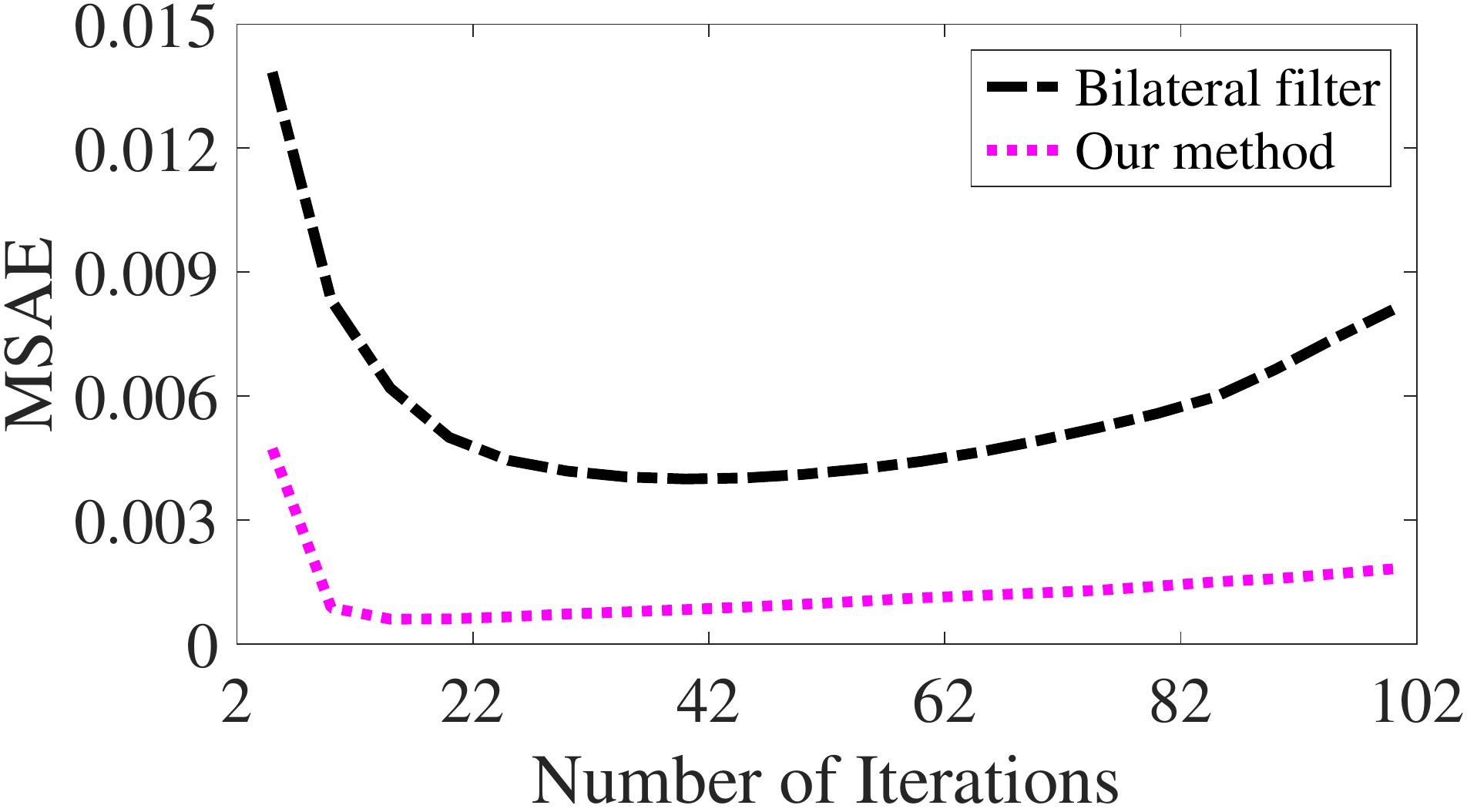}}
\end{minipage}
\begin{minipage}[b]{0.45\linewidth}
\subfigure[]{\label{}\includegraphics[width=1\linewidth]{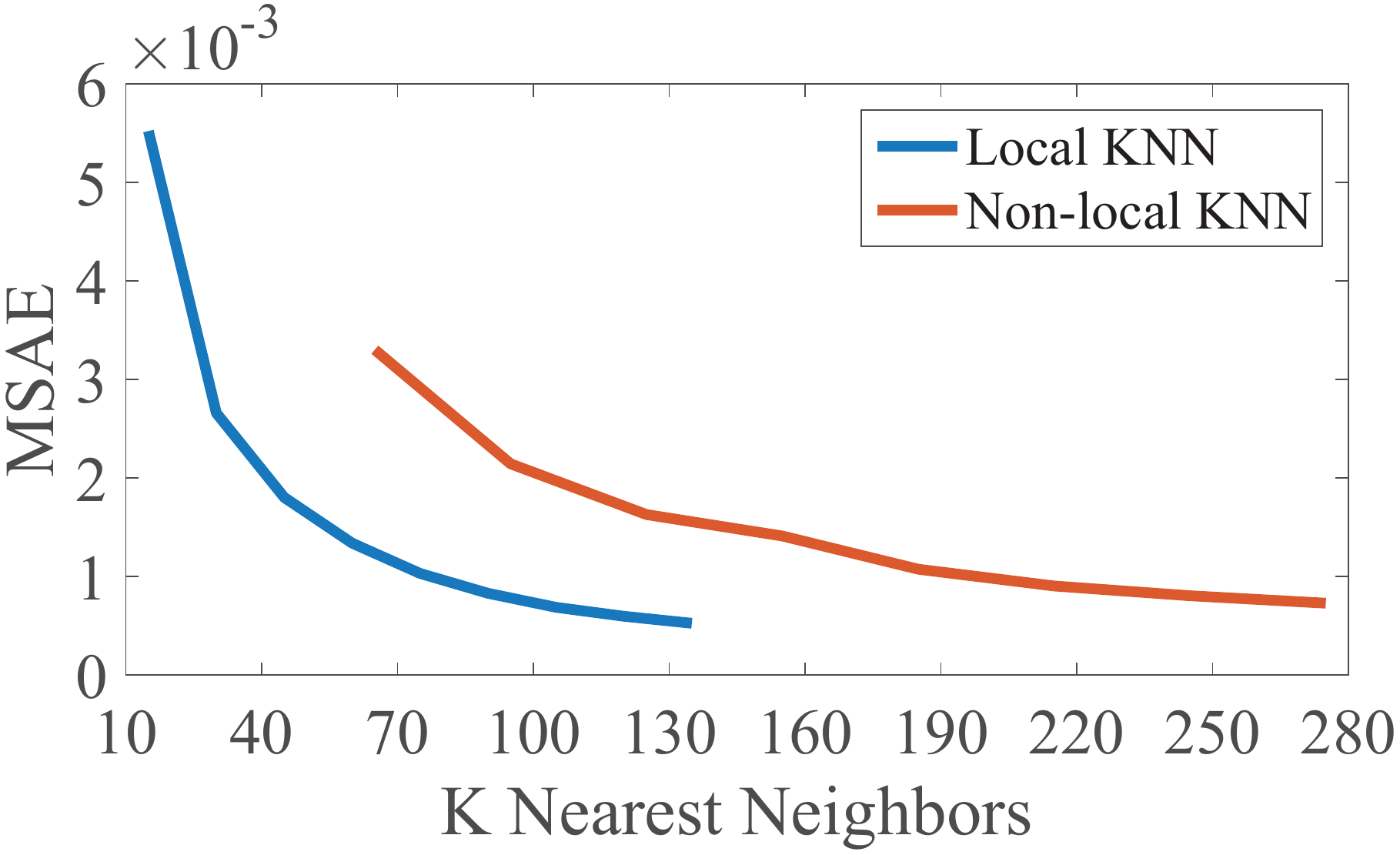}}
\end{minipage}
\caption{(a) Normal errors of our method and the bilateral filter \cite{Huang2013}. (b) Greater $k_{local}$ or $k_{non}$ leads to smaller normal errors. }
\label{fig:normalnonconvergeandKNN}
%\vspace{-0.65cm}
\end{figure}

% with or without updating local knn info
\begin{figure}[thbp]
%\vspace{-0.0cm}
\centering
\begin{minipage}[b]{0.2\linewidth}
\subfigure[]{\label{}\includegraphics[width=1\linewidth]{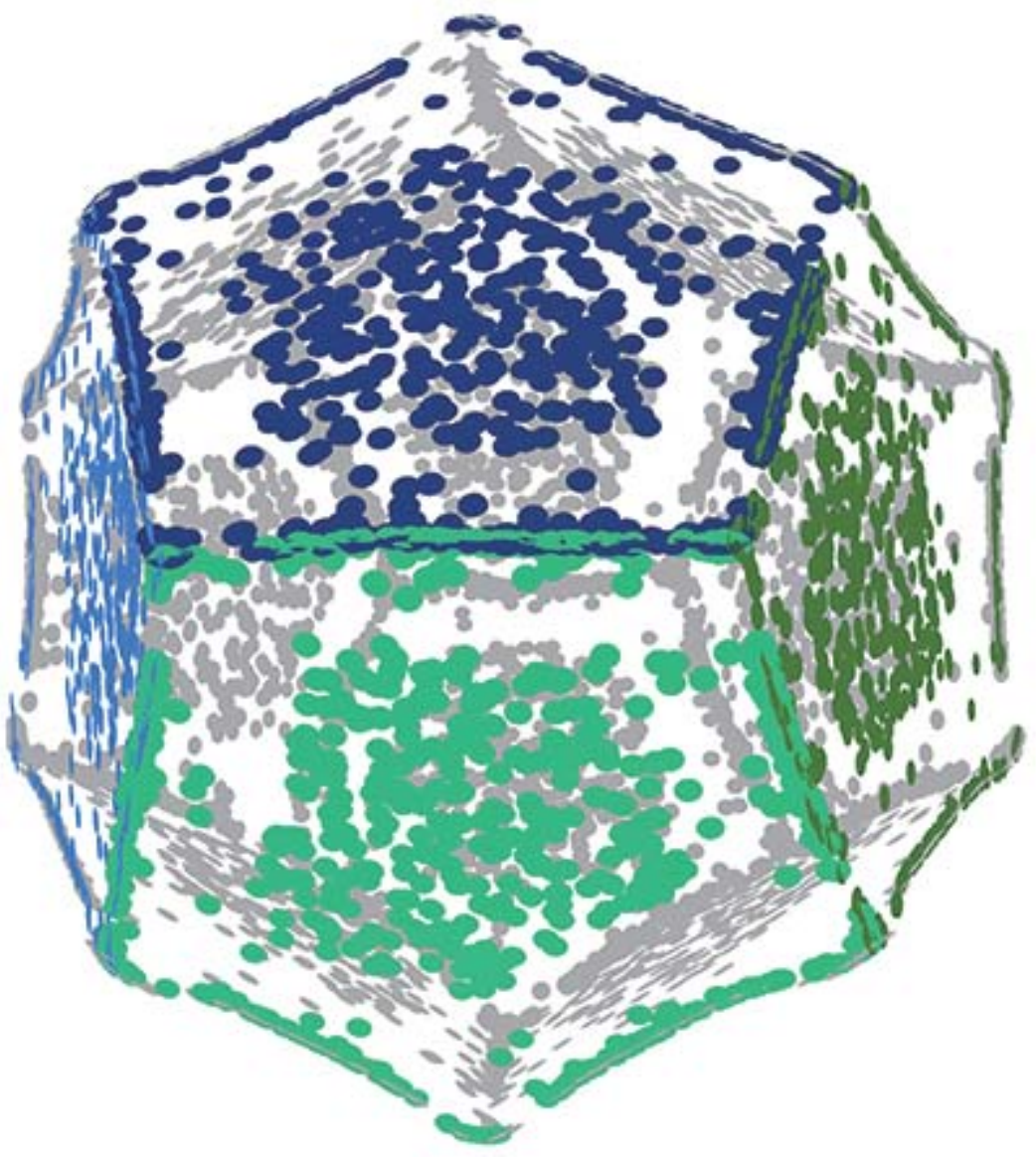}}
\end{minipage}
\begin{minipage}[b]{0.2\linewidth}
\subfigure[]{\label{}\includegraphics[width=1\linewidth]{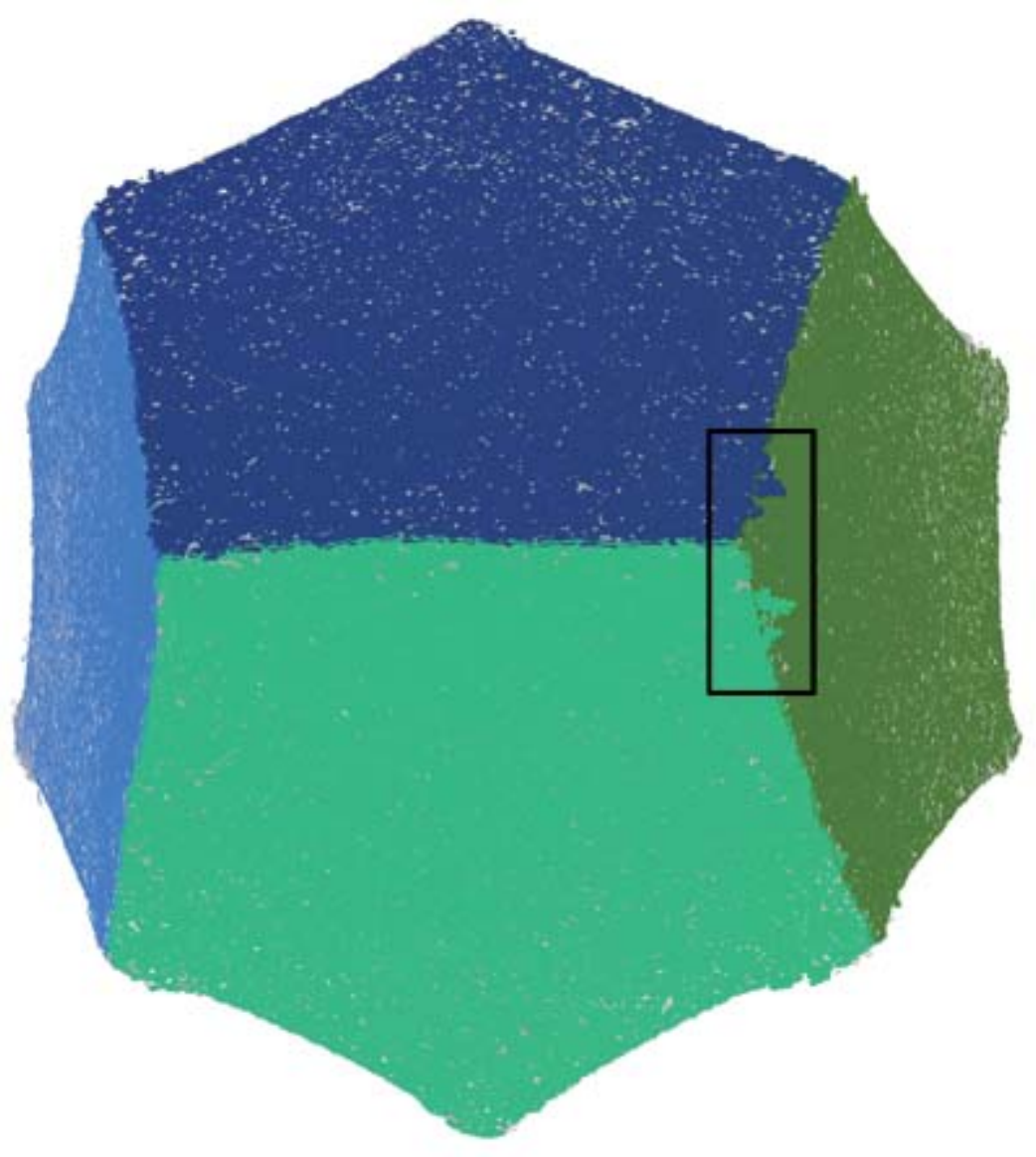}}
\end{minipage}
\begin{minipage}[b]{0.075\linewidth}
{\label{}\includegraphics[width=1\linewidth]{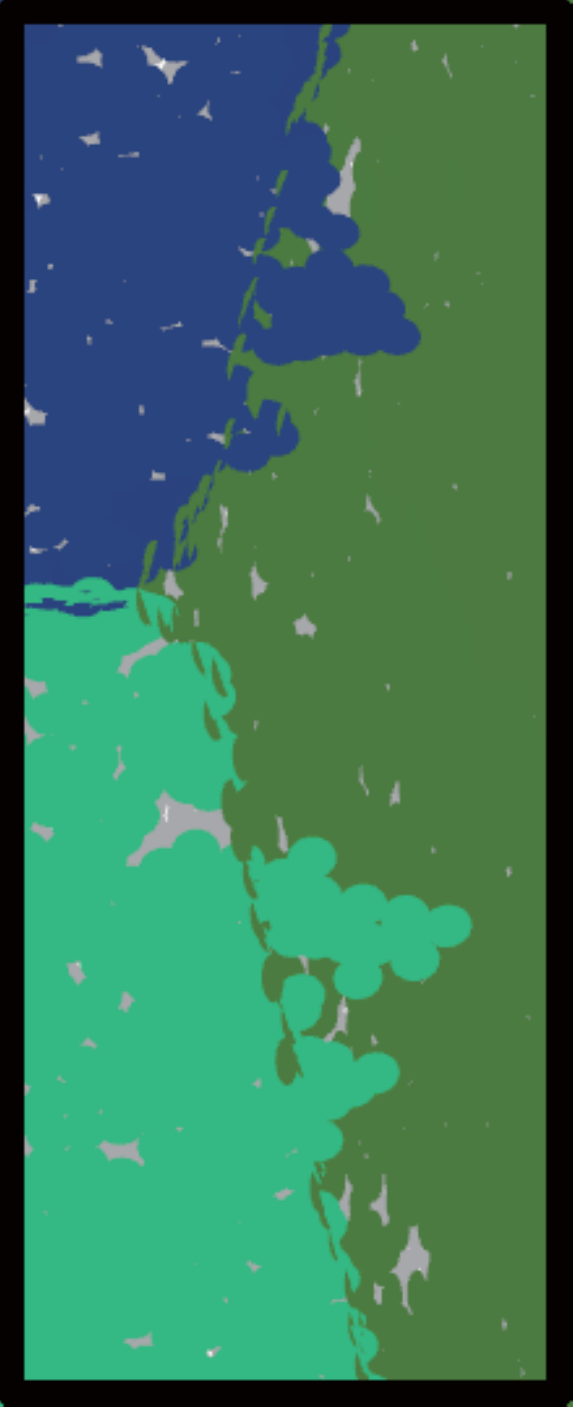}}
\end{minipage}
\begin{minipage}[b]{0.2\linewidth}
\subfigure[]{\label{}\includegraphics[width=1\linewidth]{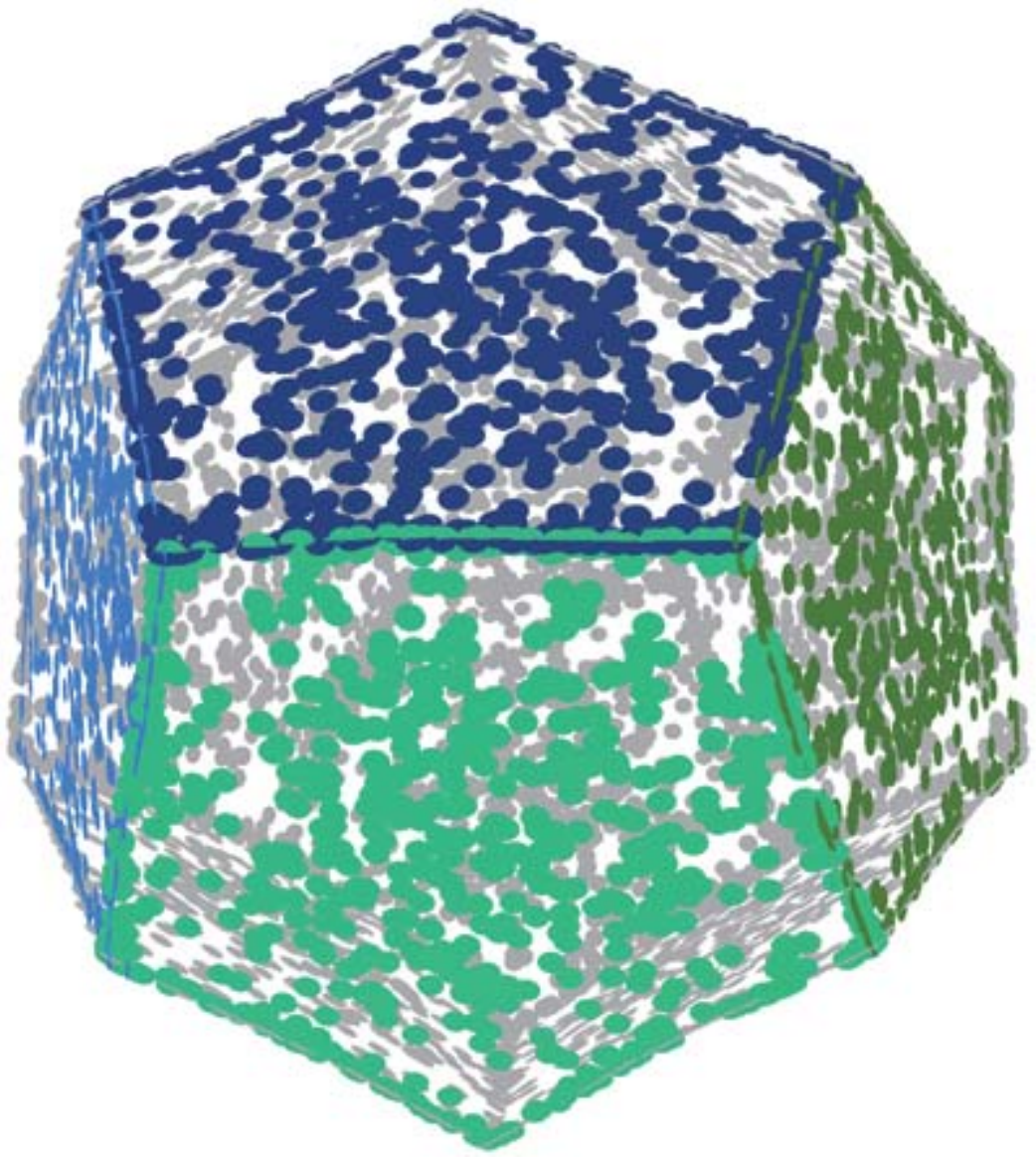}}
\end{minipage}
\begin{minipage}[b]{0.2\linewidth}
\subfigure[]{\label{}\includegraphics[width=1\linewidth]{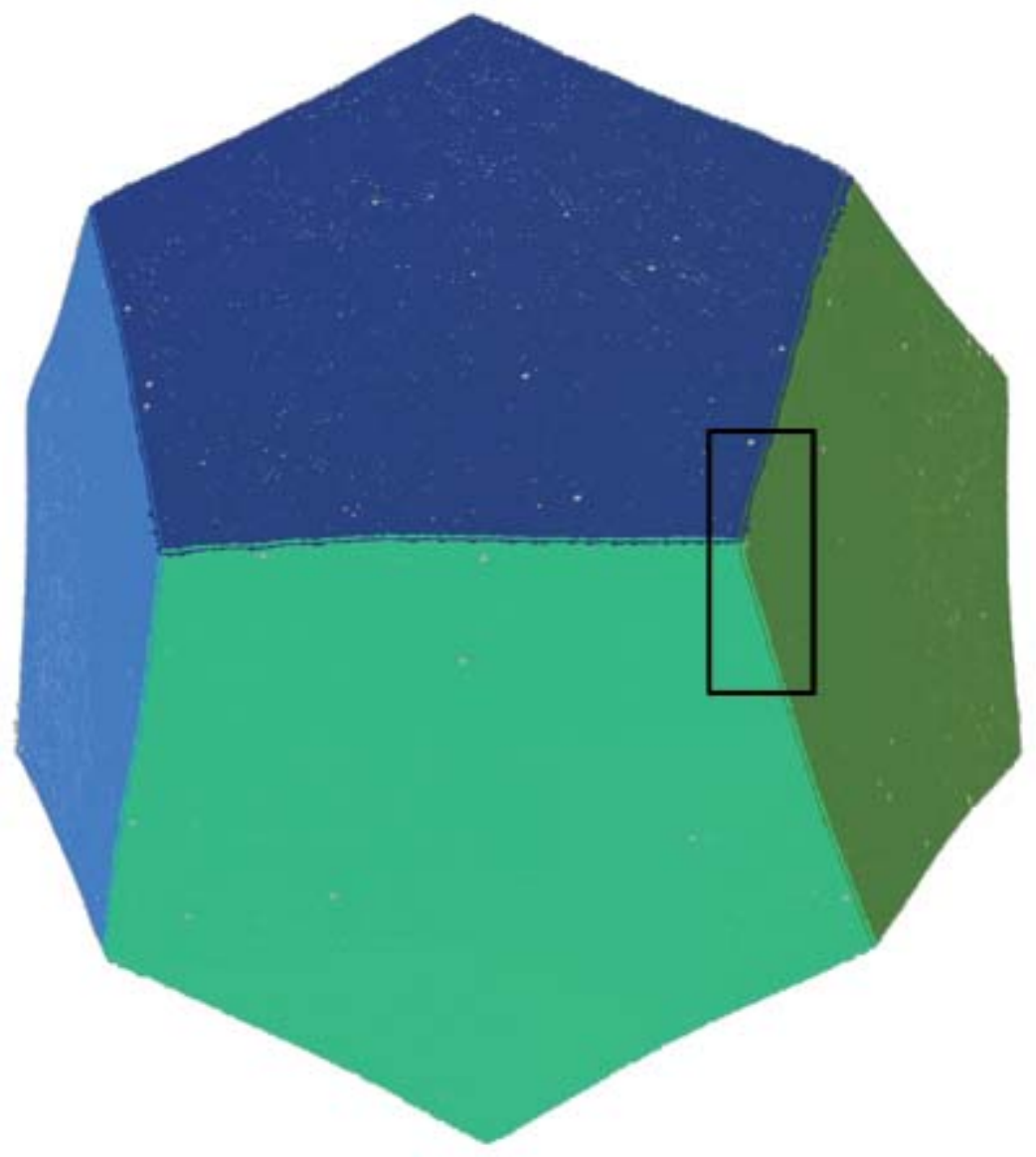}}
\end{minipage}
\begin{minipage}[b]{0.075\linewidth}
{\label{}\includegraphics[width=1\linewidth]{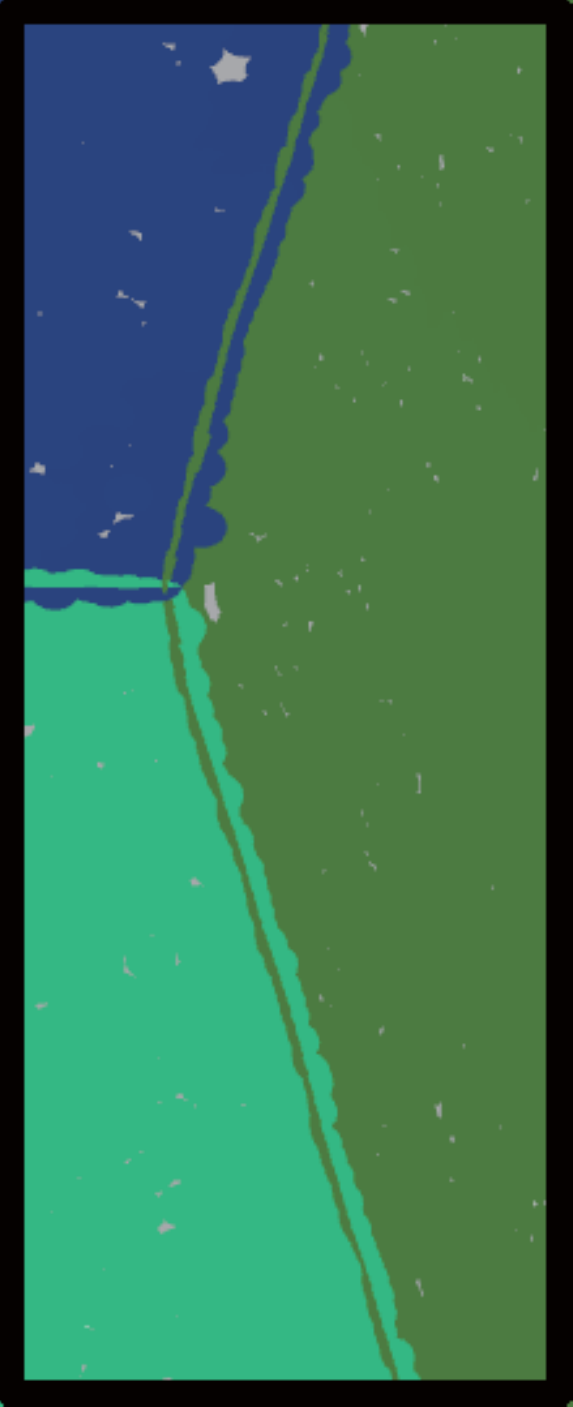}}
\end{minipage}
\caption{Comparison of with and without updating neighboring information in each iteration. (a) Position update with updating neighboring information. (b) The upsampling of (a). (c) Position update without updating neighboring information. (d) The upsampling of (c).}
\label{fig:dod_knnupdate}
%\vspace{-0.65cm}
\end{figure}

% dod
\begin{figure*}[htbp]
%\vspace{-0.0cm}
\centering
\begin{minipage}[b]{0.16\linewidth}
{\label{}\includegraphics[width=1\linewidth]{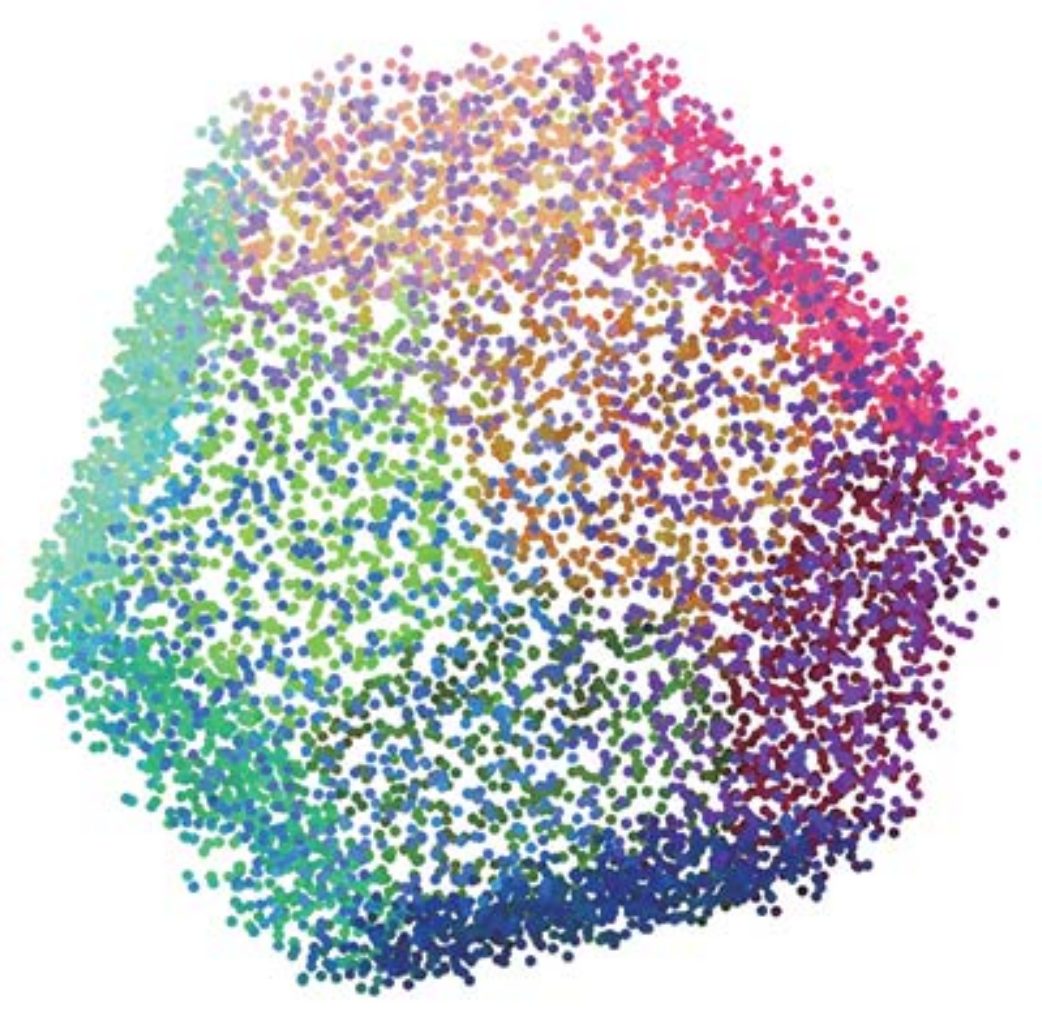}}
\end{minipage}
\begin{minipage}[b]{0.16\linewidth}
{\label{}\includegraphics[width=1\linewidth]{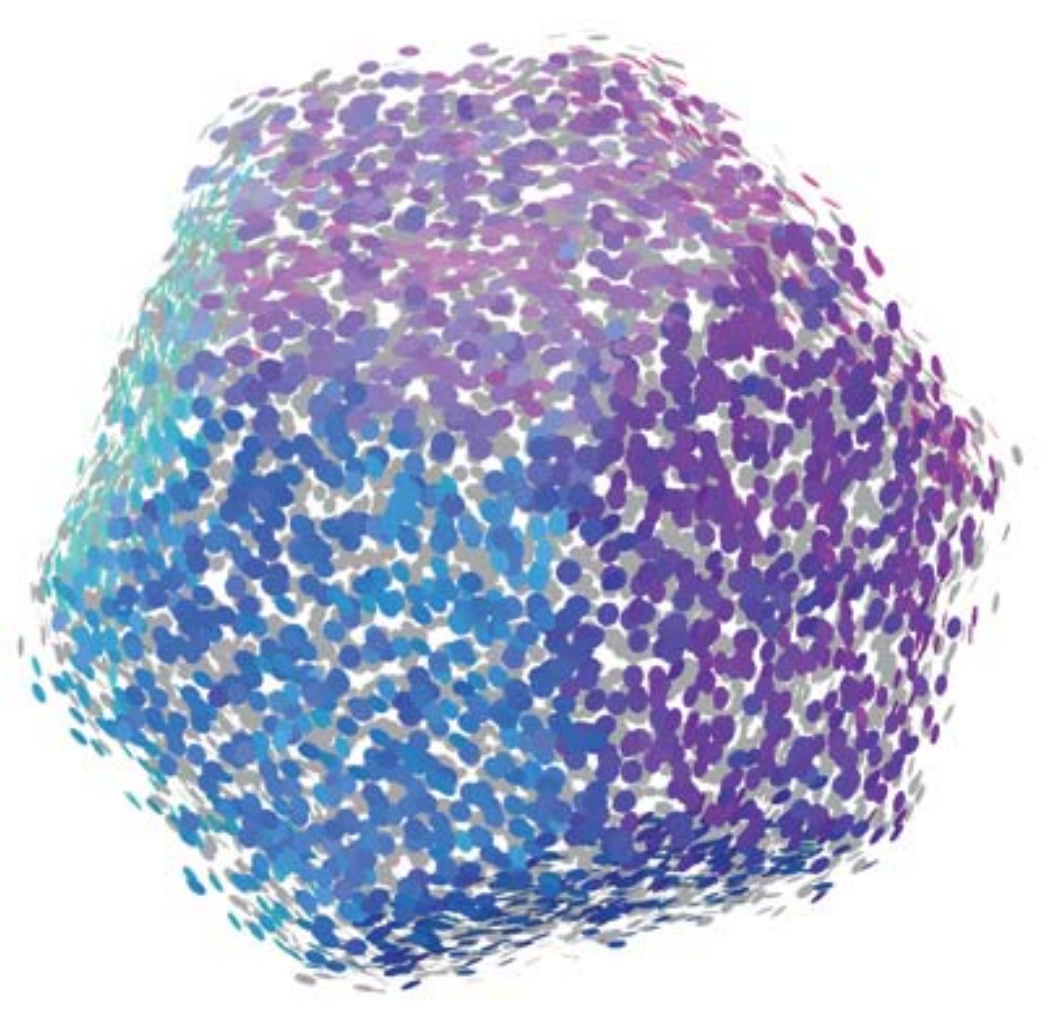}}
\end{minipage}
\begin{minipage}[b]{0.16\linewidth}
{\label{}\includegraphics[width=1\linewidth]{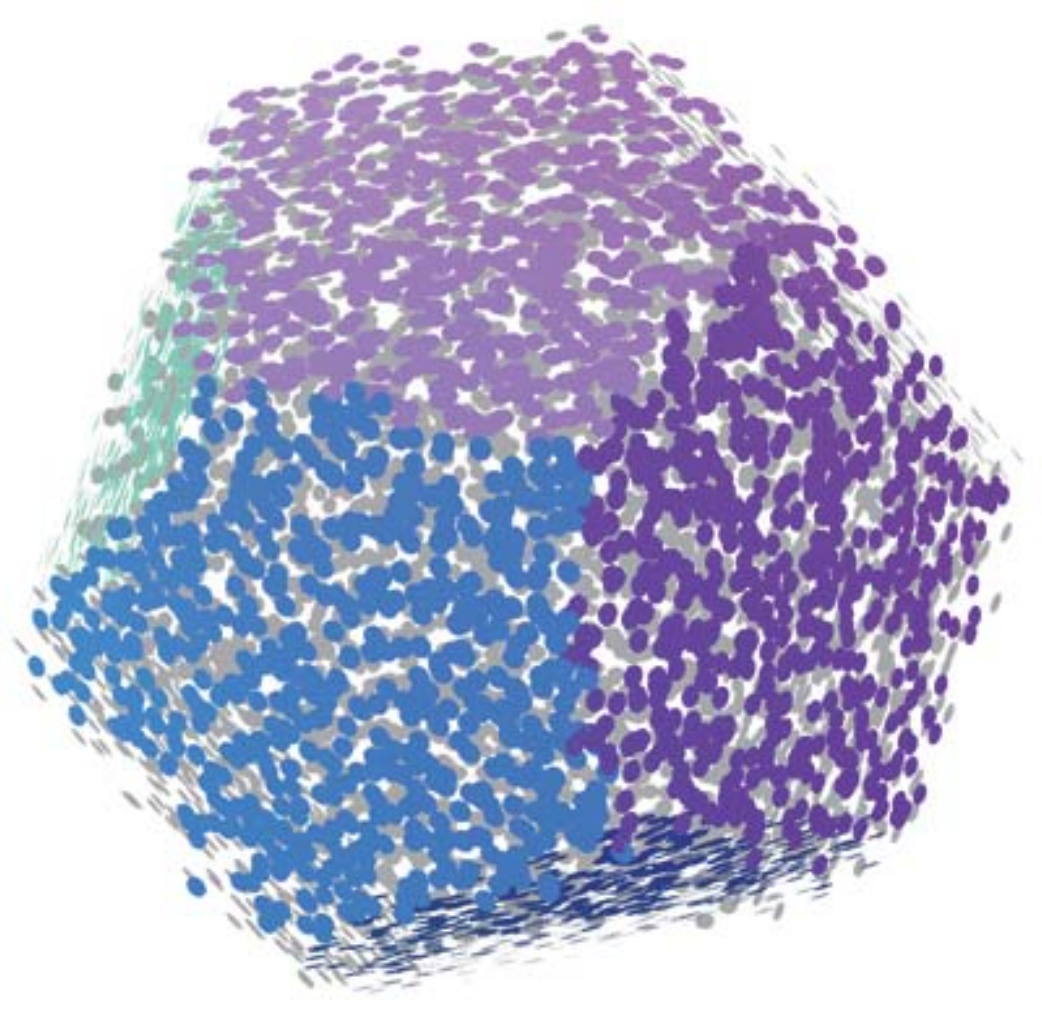}}
\end{minipage}
\begin{minipage}[b]{0.16\linewidth}
{\label{}\includegraphics[width=1\linewidth]{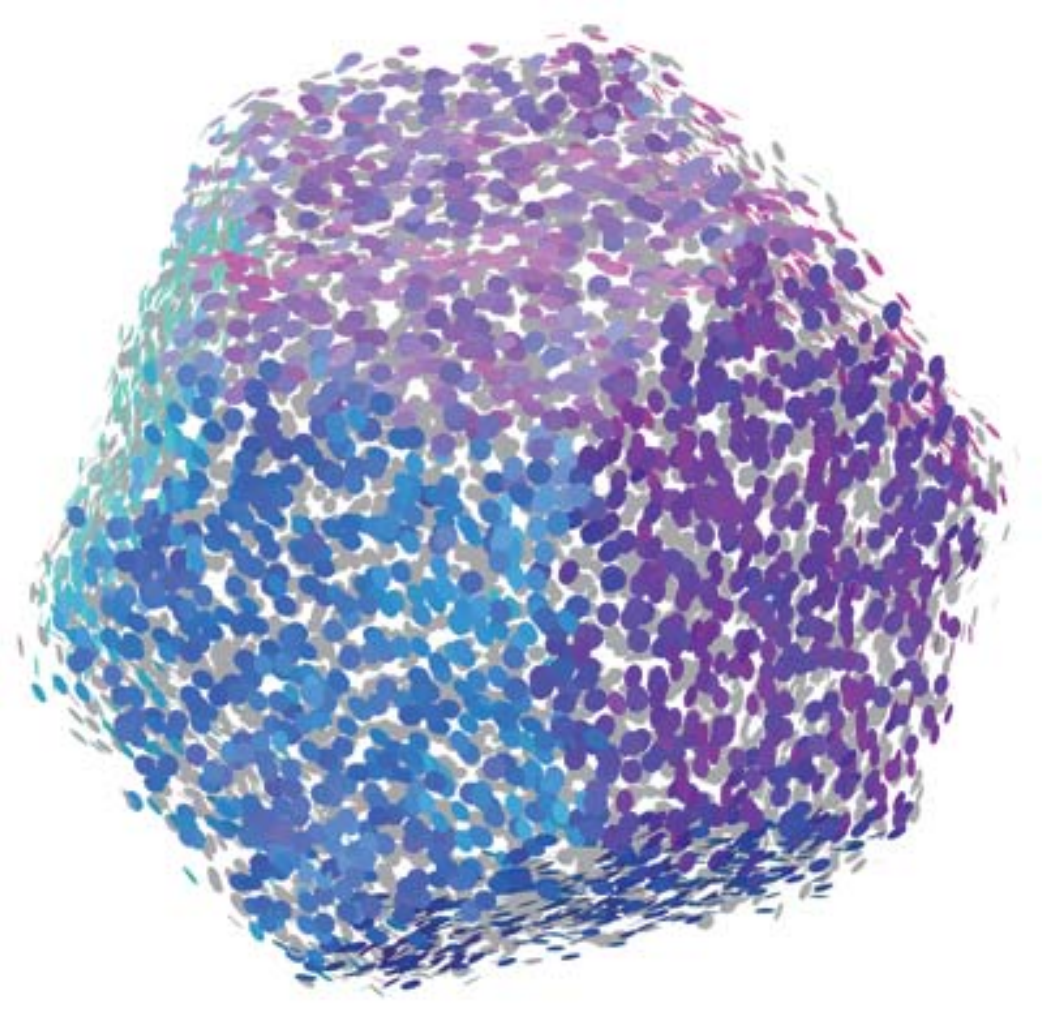}}
\end{minipage}
\begin{minipage}[b]{0.16\linewidth}
{\label{}\includegraphics[width=1\linewidth]{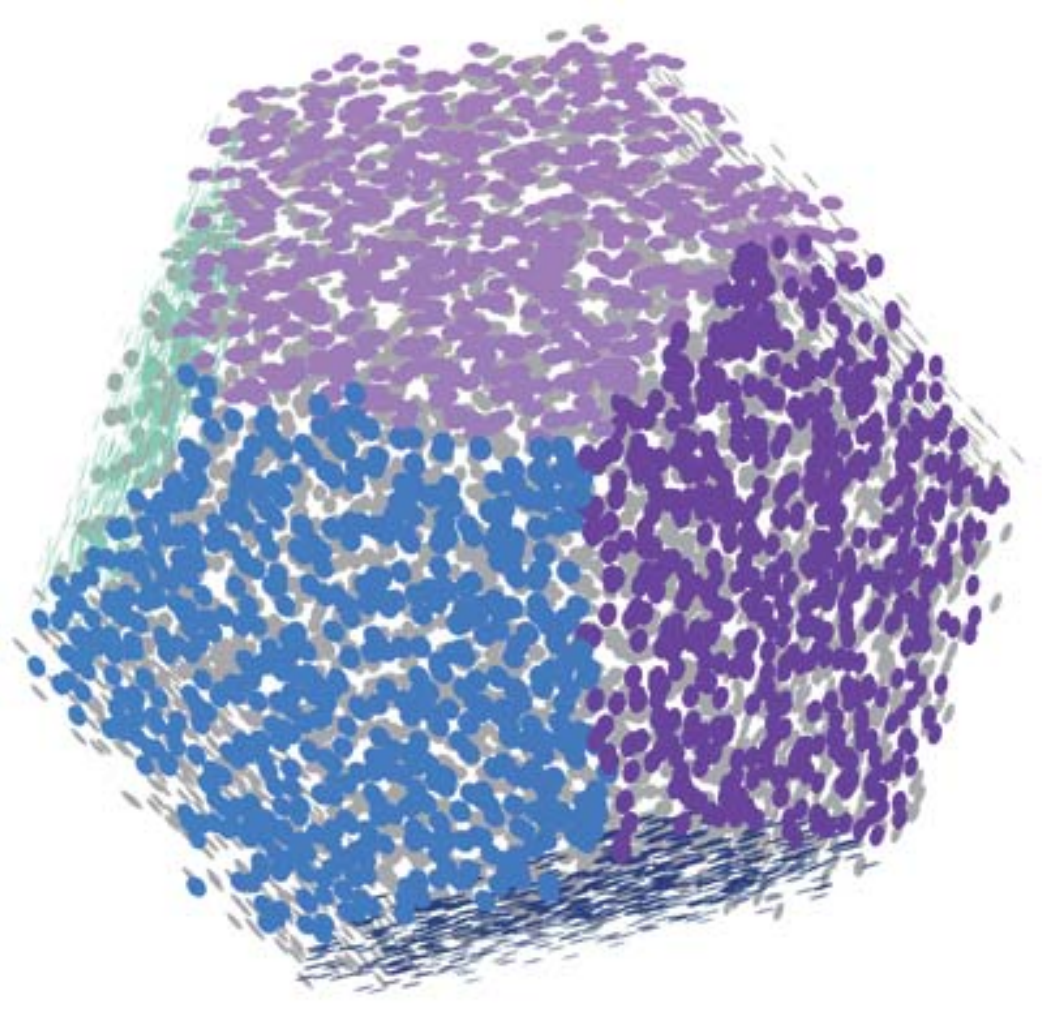}}
\end{minipage}	\\
\begin{minipage}[b]{0.16\linewidth}
{\label{}\includegraphics[width=1\linewidth]{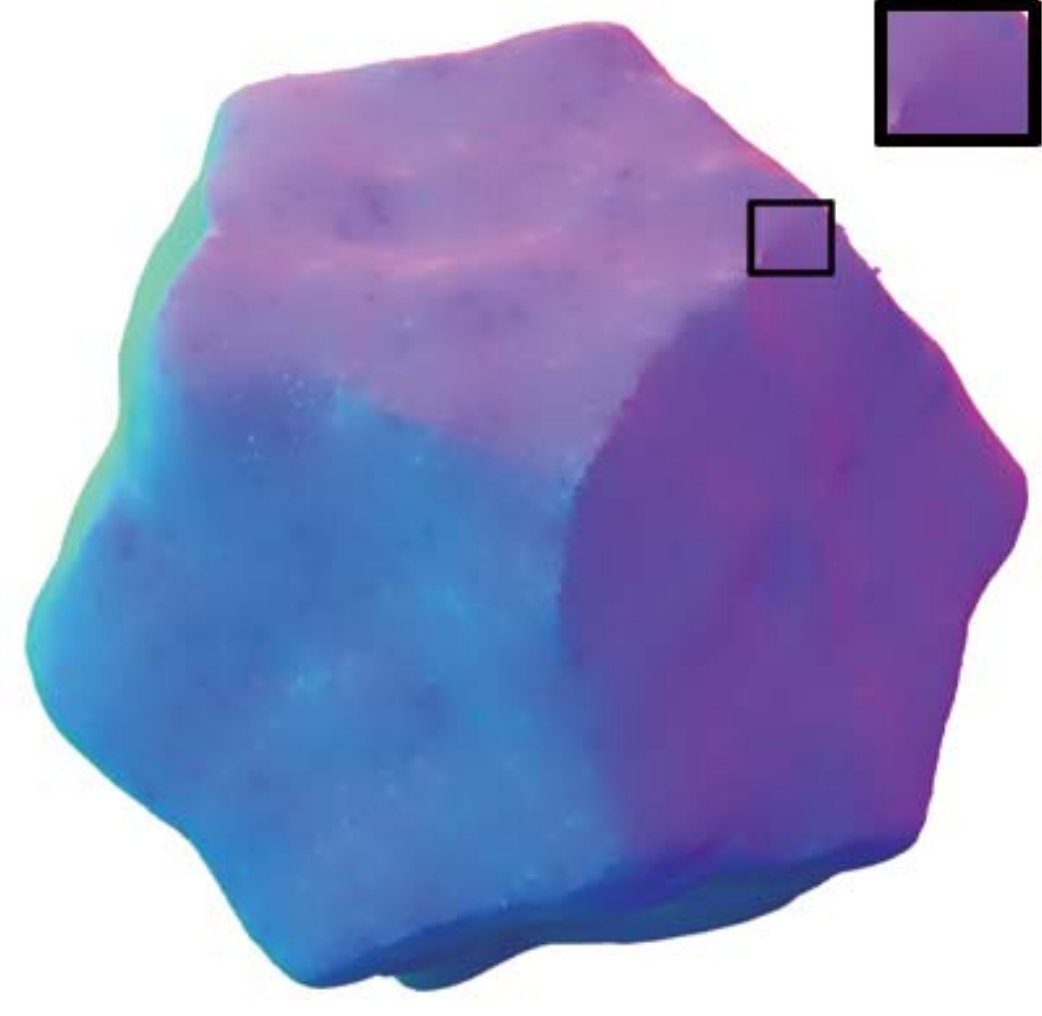}}
\end{minipage}
\begin{minipage}[b]{0.16\linewidth}
{\label{}\includegraphics[width=1\linewidth]{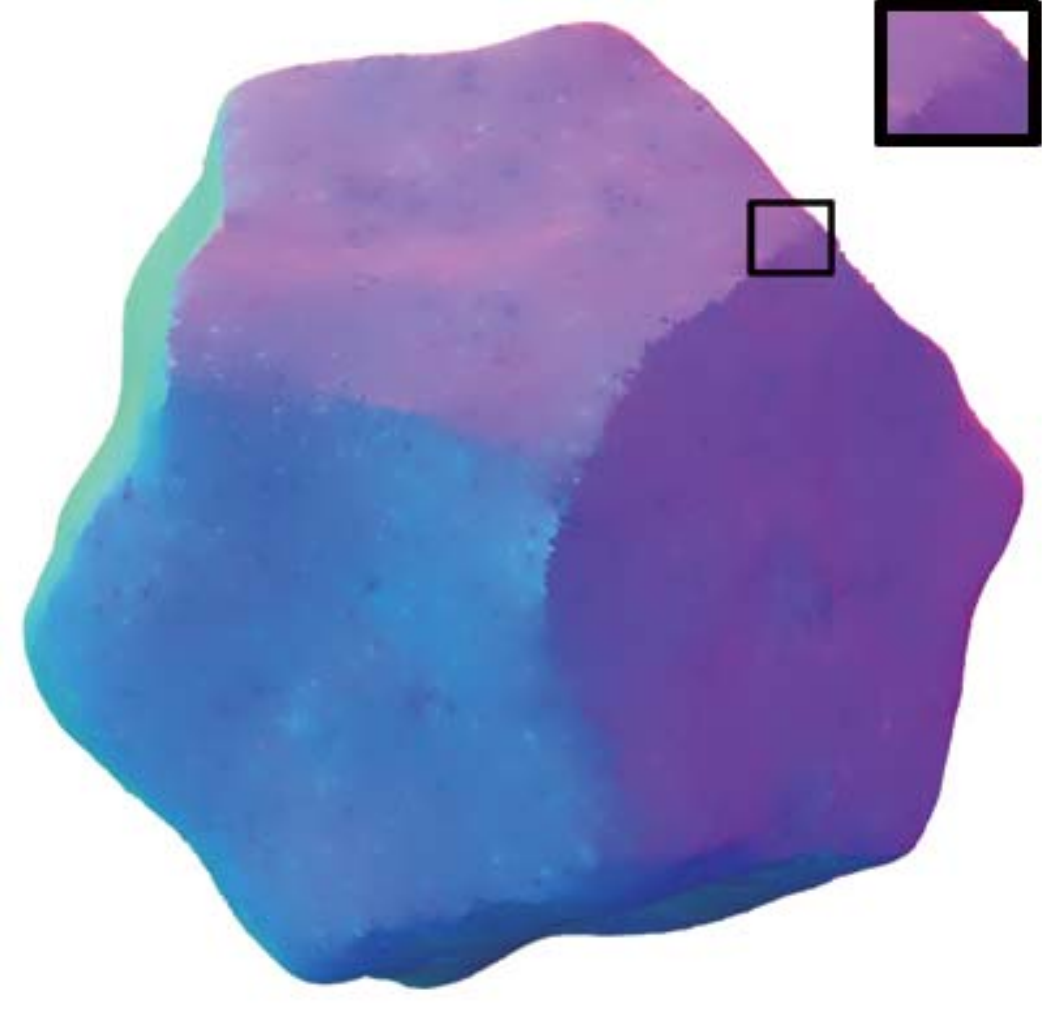}}
\end{minipage}
\begin{minipage}[b]{0.16\linewidth}
{\label{}\includegraphics[width=1\linewidth]{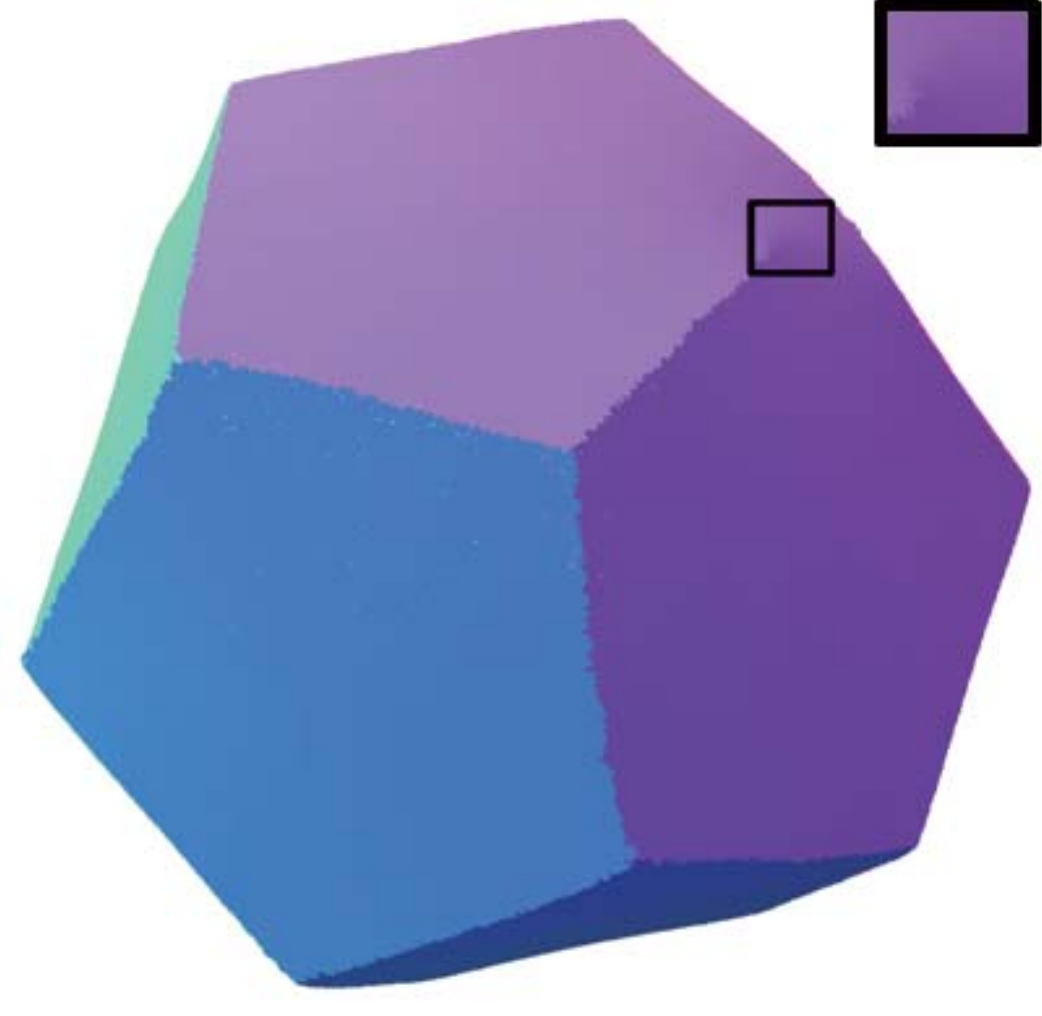}}
\end{minipage}
\begin{minipage}[b]{0.16\linewidth}
{\label{}\includegraphics[width=1\linewidth]{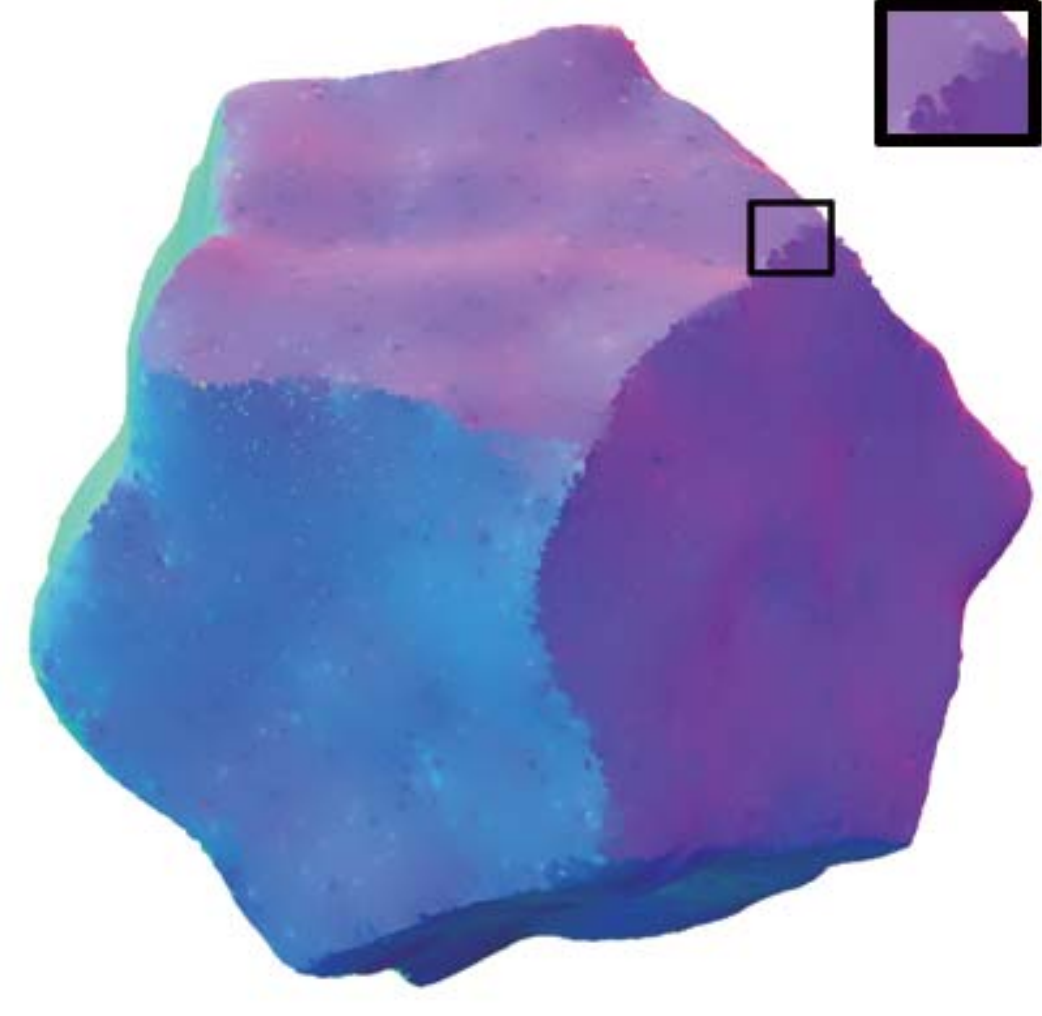}}
\end{minipage}
\begin{minipage}[b]{0.16\linewidth}
{\label{}\includegraphics[width=1\linewidth]{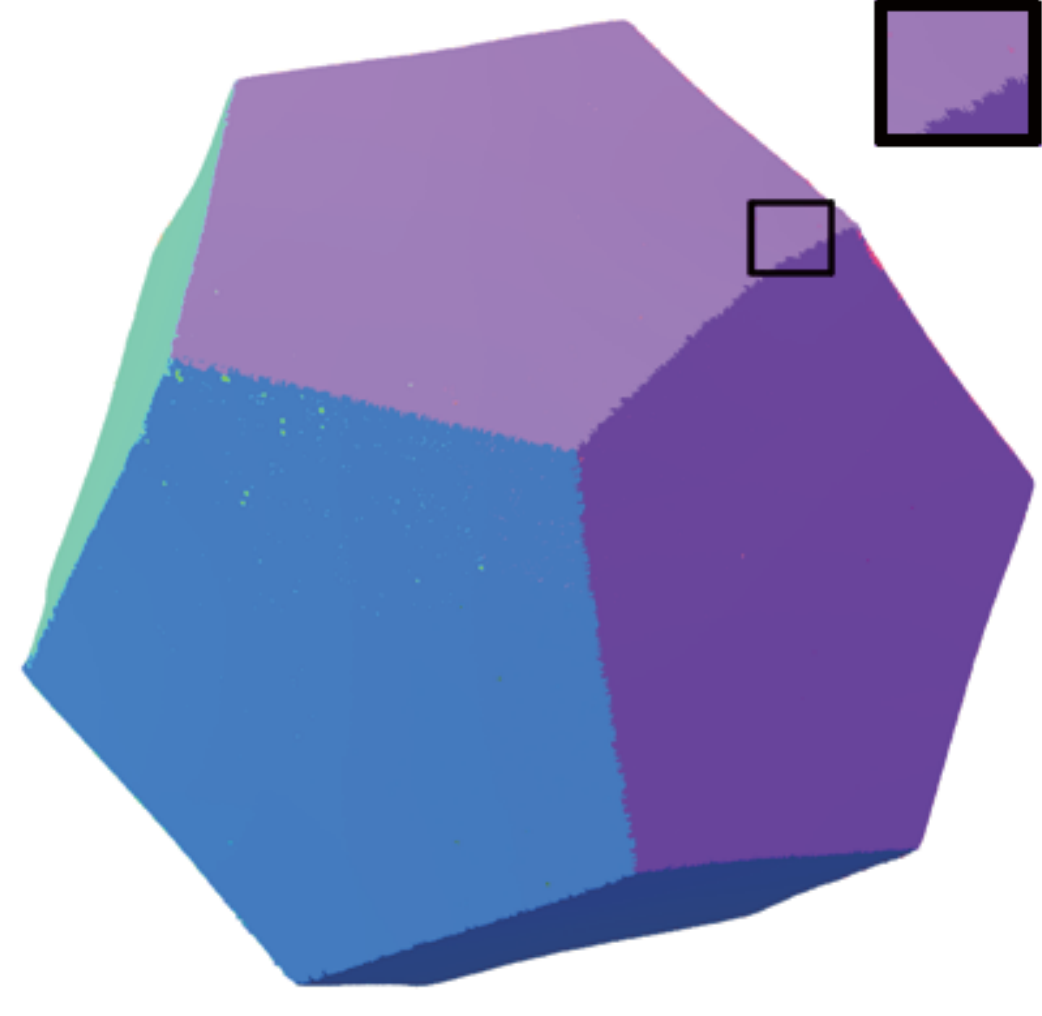}}
\end{minipage}	\\
\begin{minipage}[b]{0.16\linewidth}
\subfigure[\protect\cite{Hoppe1992}]{\label{}\includegraphics[width=1\linewidth]{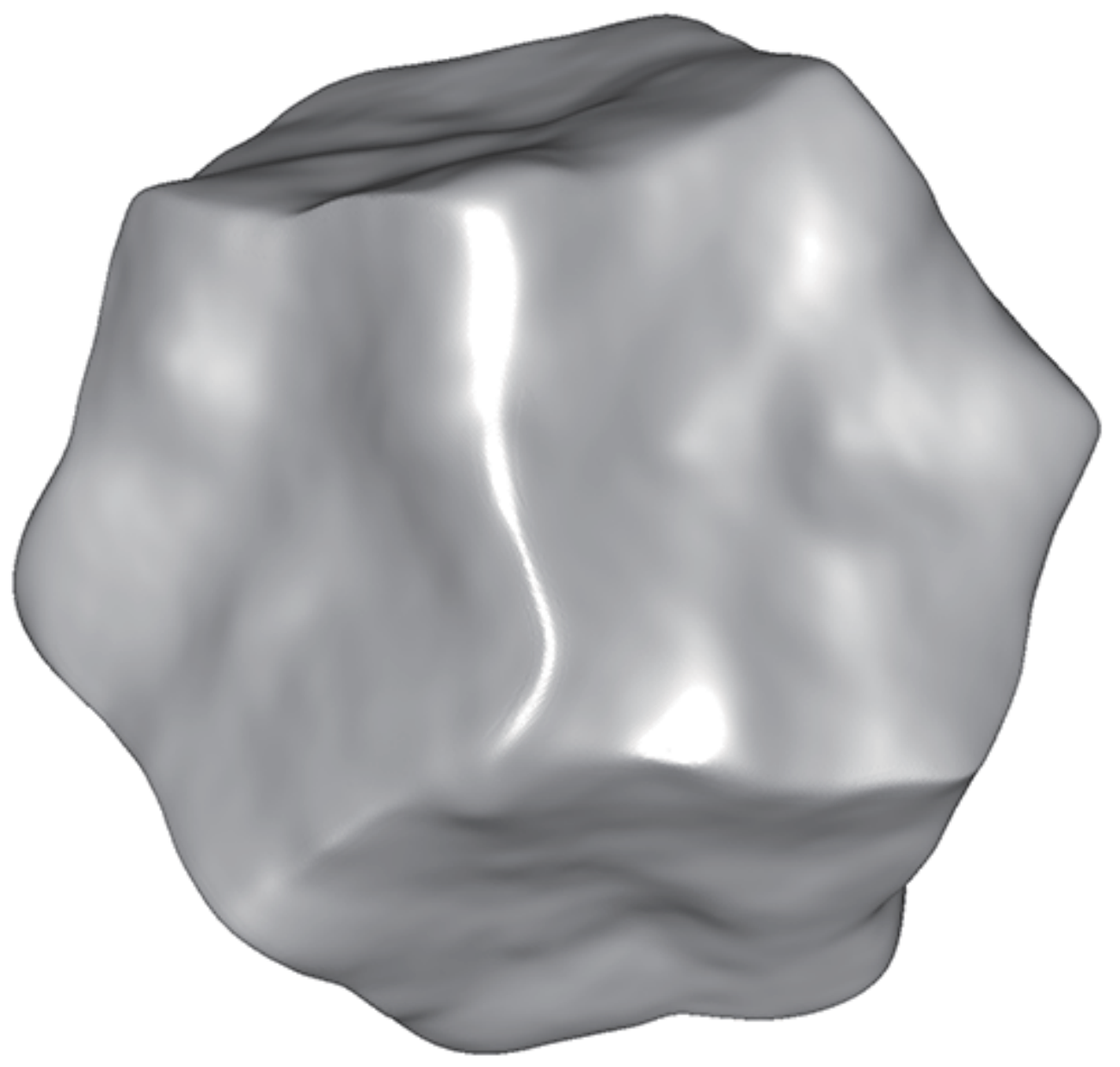}}
\end{minipage}
\begin{minipage}[b]{0.16\linewidth}
\subfigure[\protect\cite{Boulch2012}]{\label{}\includegraphics[width=1\linewidth]{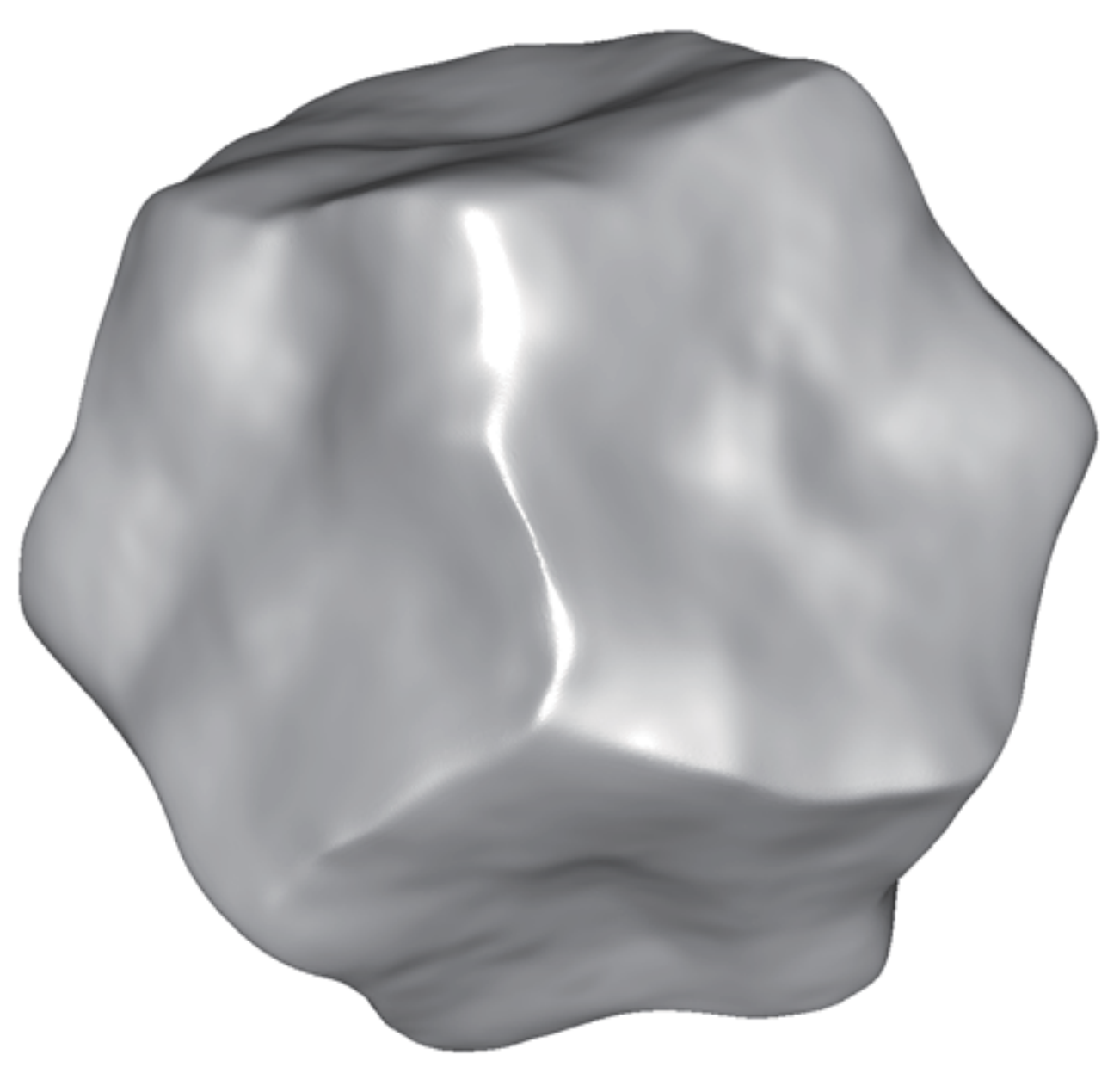}}
\end{minipage}
\begin{minipage}[b]{0.16\linewidth}
\subfigure[\protect\cite{Huang2013}]{\label{}\includegraphics[width=1\linewidth]{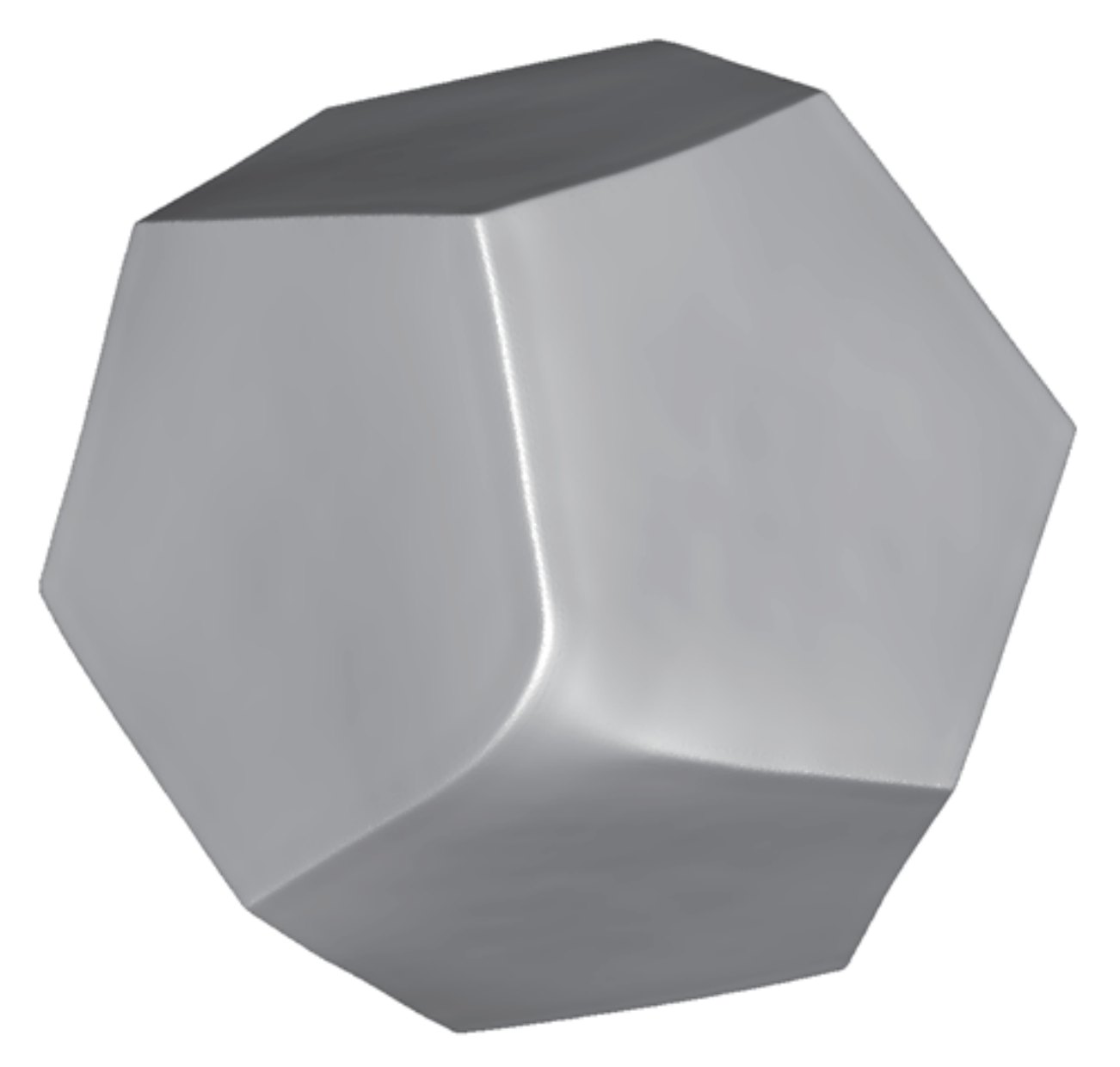}}
\end{minipage}
\begin{minipage}[b]{0.16\linewidth}
\subfigure[\protect\cite{Boulch2016}]{\label{}\includegraphics[width=1\linewidth]{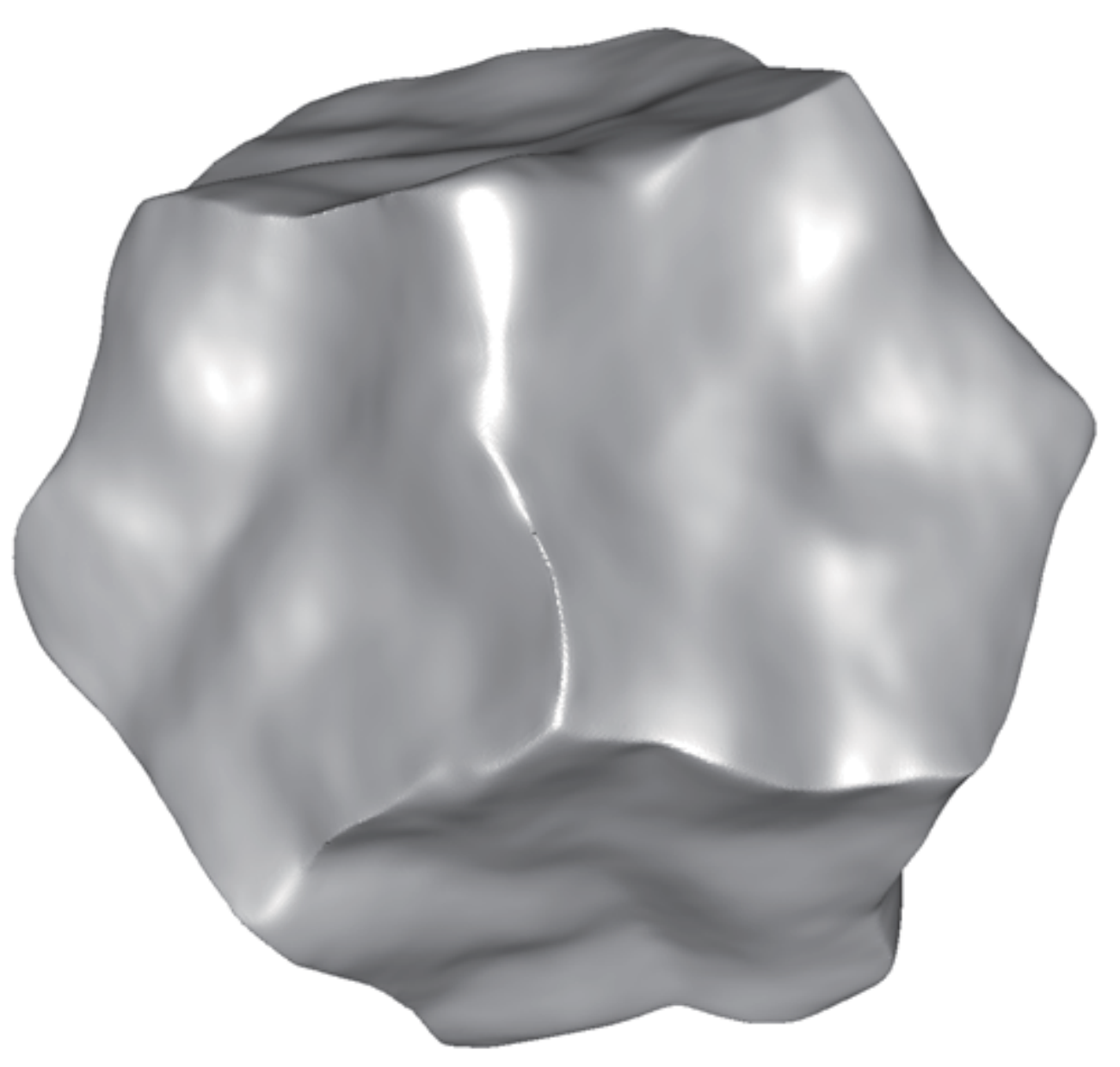}}
\end{minipage}
\begin{minipage}[b]{0.16\linewidth}
\subfigure[Ours]{\label{}\includegraphics[width=1\linewidth]{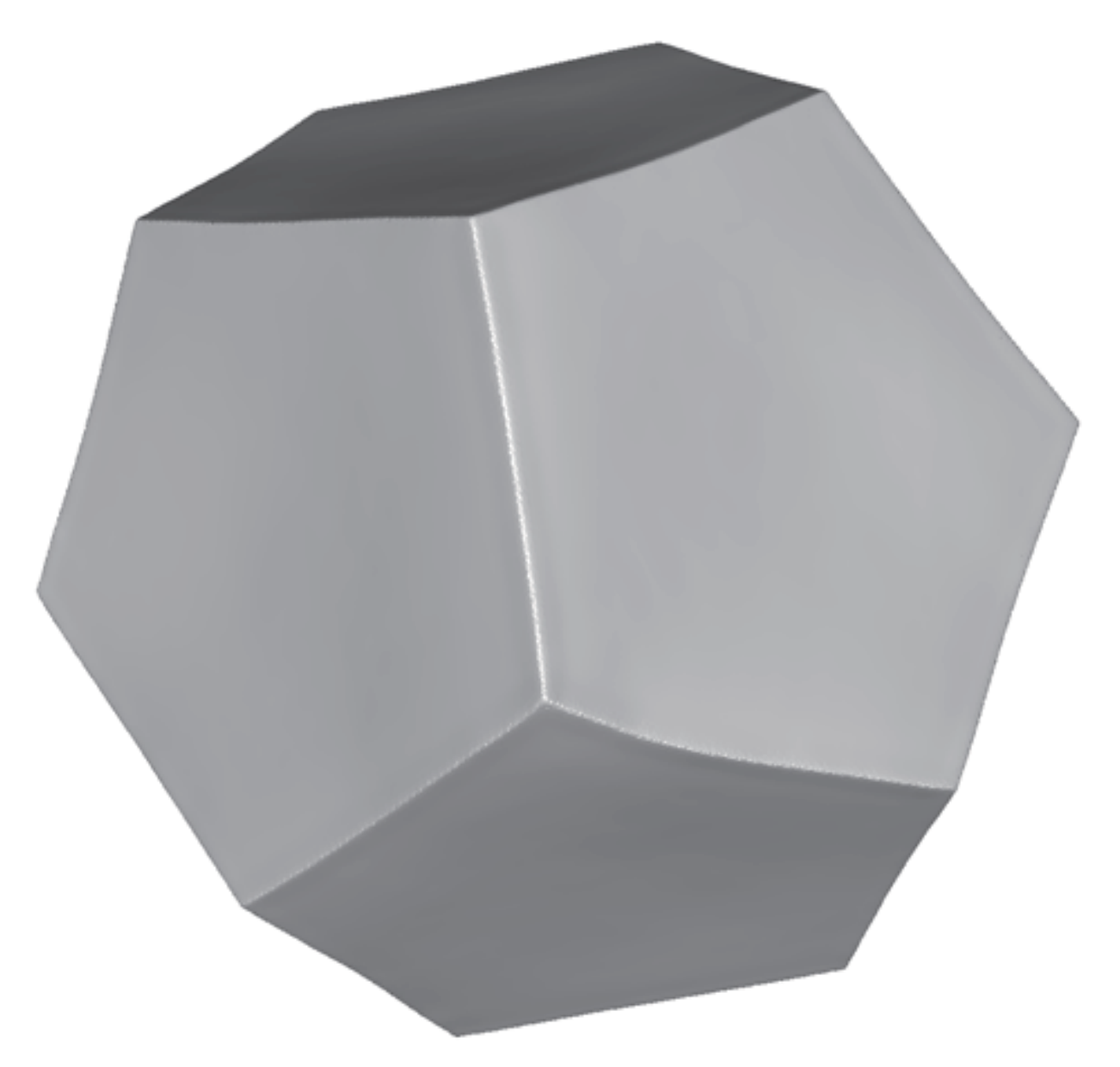}}
\end{minipage}
\caption{The first row: normal results of the Dodecahedron point cloud (synthetic noise). The second row: upsampling results of the filtered results by updating position with the normals in the first row. The third row: the corresponding surface reconstruction results. }
\label{fig:dod_point}
%\vspace{-0.65cm}
\end{figure*}

%scanned: car
\begin{figure*}[htbp]
%\vspace{-0.0cm}
\centering
\begin{minipage}[b]{0.16\linewidth}
{\label{}\includegraphics[width=1\linewidth]{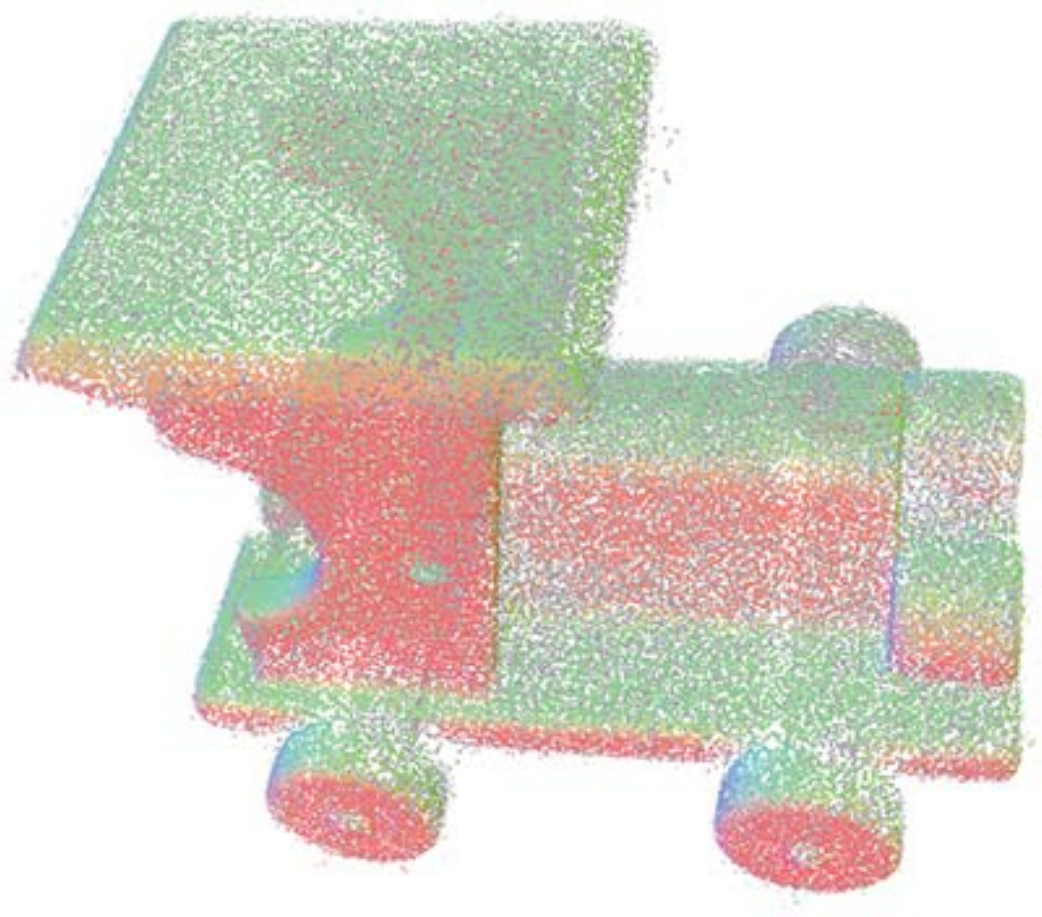}}
\end{minipage}
\begin{minipage}[b]{0.16\linewidth}
{\label{}\includegraphics[width=1\linewidth]{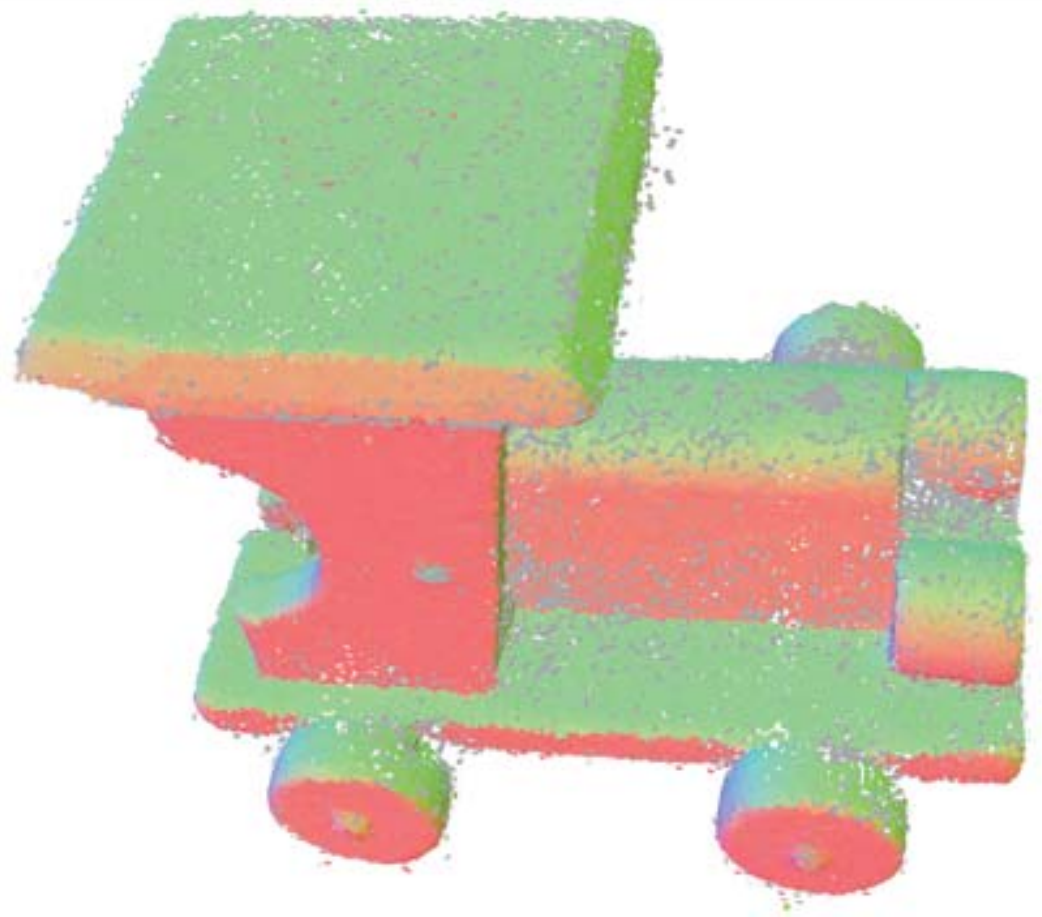}}
\end{minipage}
\begin{minipage}[b]{0.16\linewidth}
{\label{}\includegraphics[width=1\linewidth]{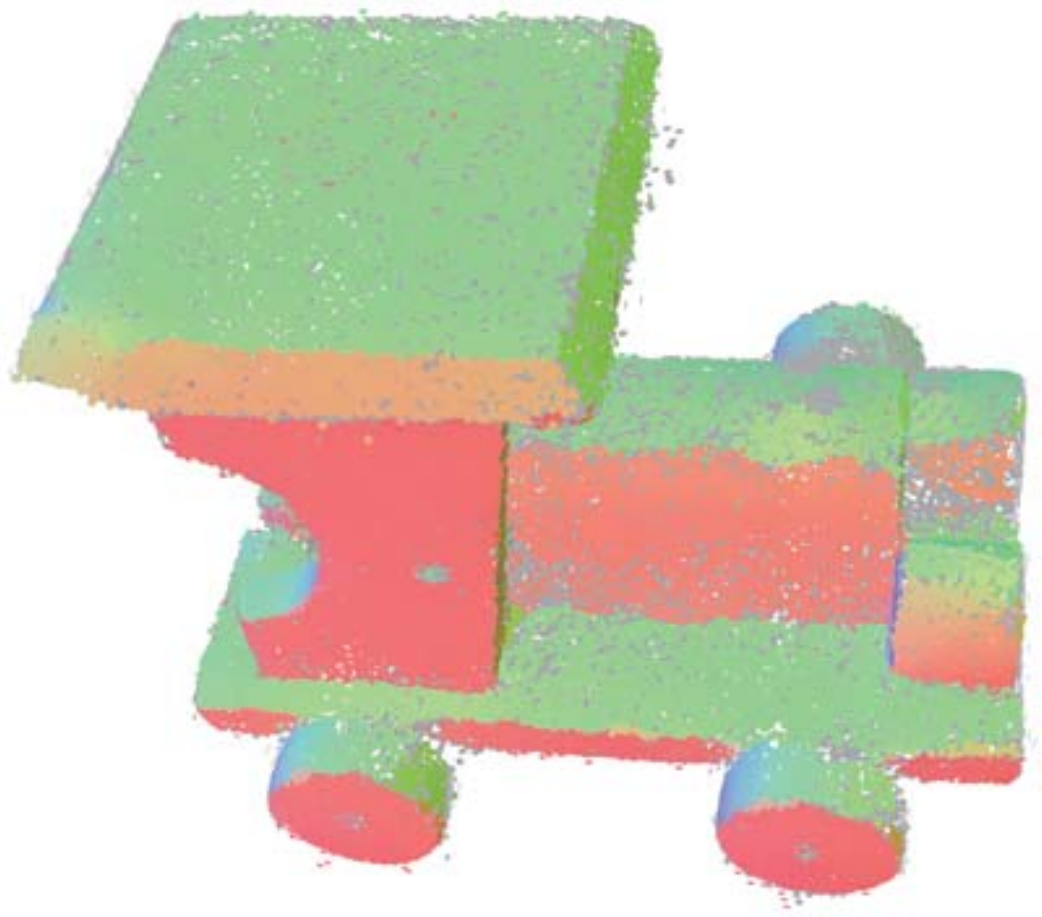}}
\end{minipage}
\begin{minipage}[b]{0.16\linewidth}
{\label{}\includegraphics[width=1\linewidth]{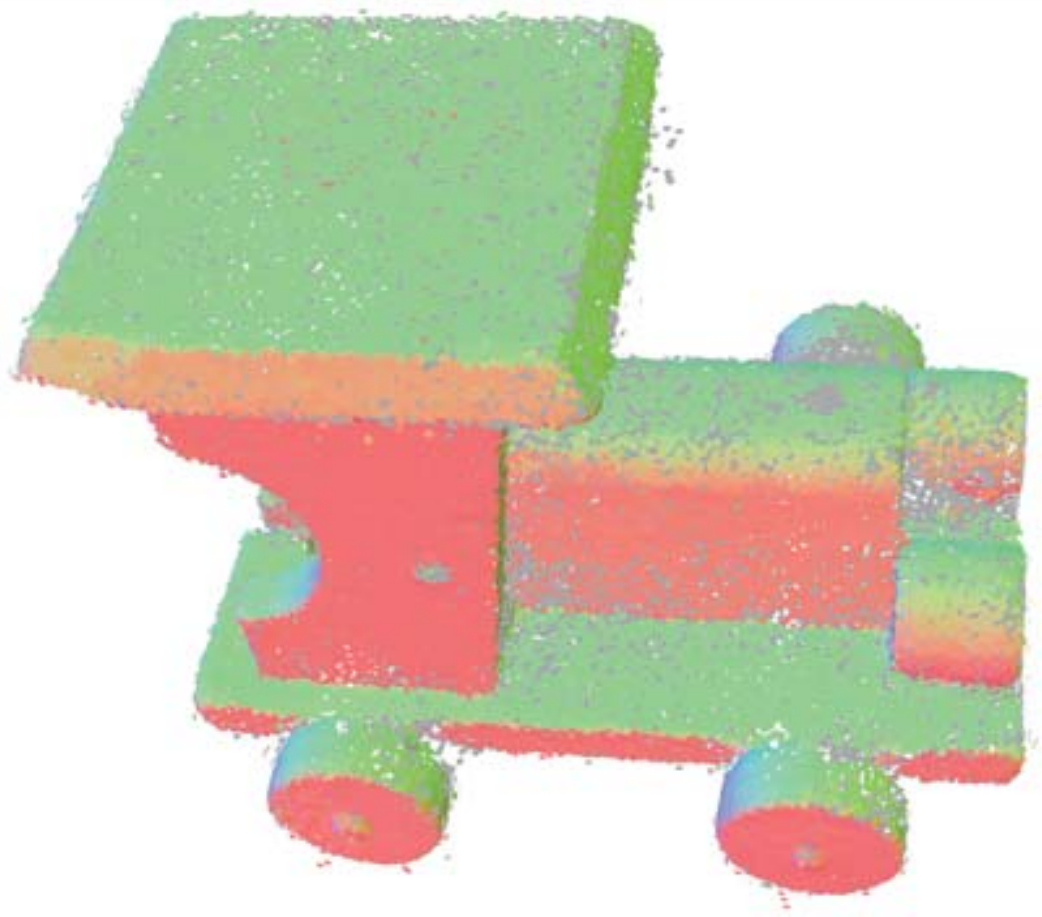}}
\end{minipage}
\begin{minipage}[b]{0.16\linewidth}
{\label{}\includegraphics[width=1\linewidth]{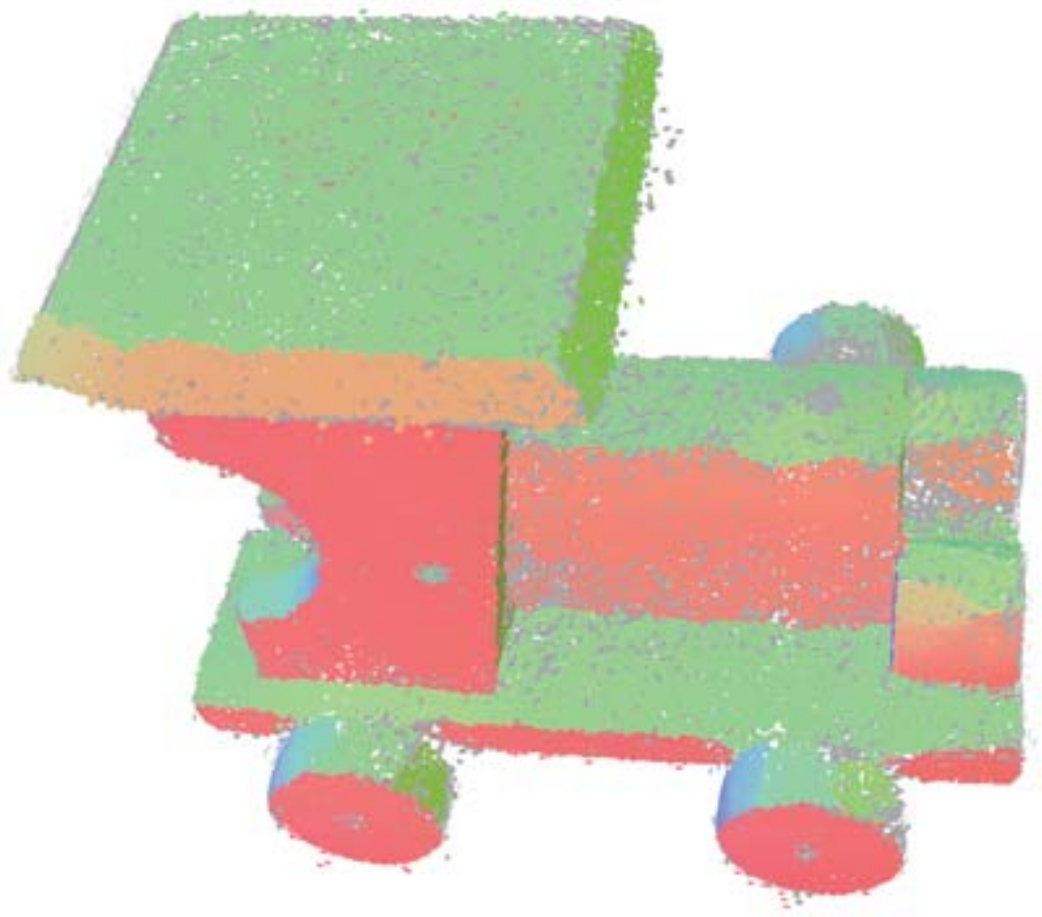}}
\end{minipage}	\\
\begin{minipage}[b]{0.16\linewidth}
{\label{}\includegraphics[width=1\linewidth]{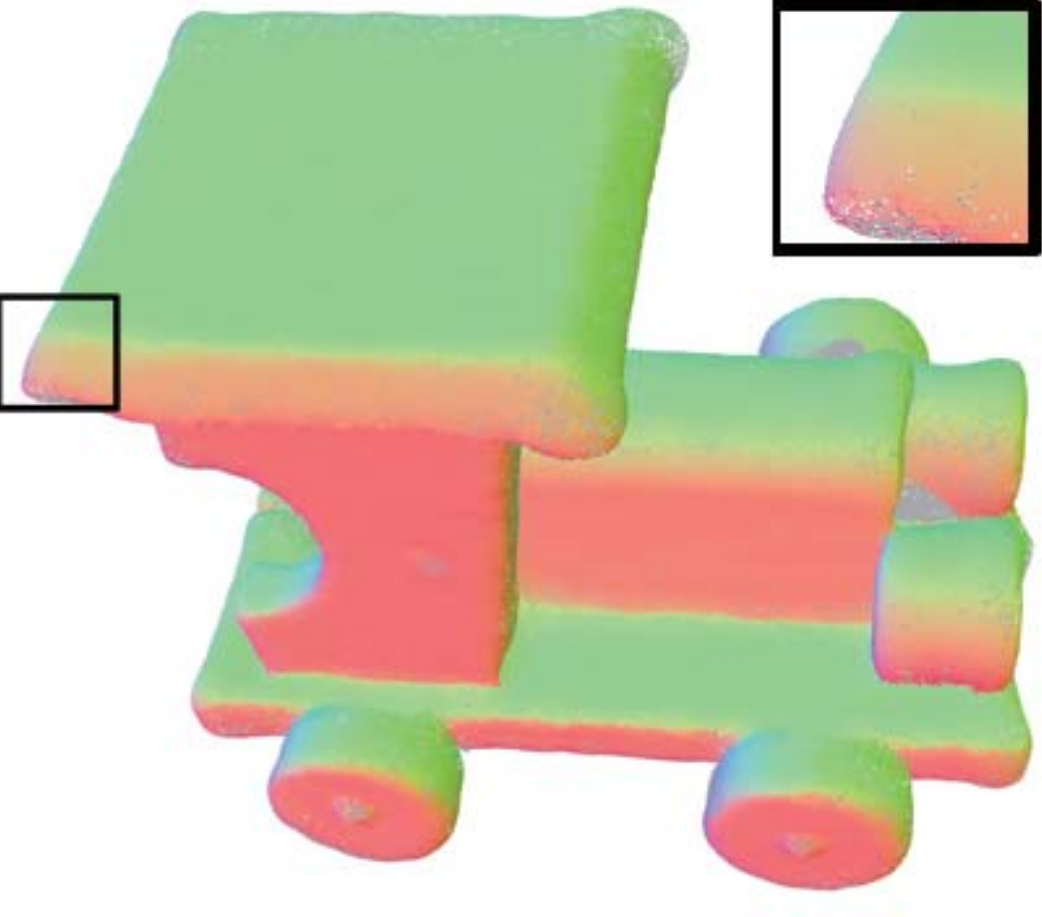}}
\end{minipage}
\begin{minipage}[b]{0.16\linewidth}
{\label{}\includegraphics[width=1\linewidth]{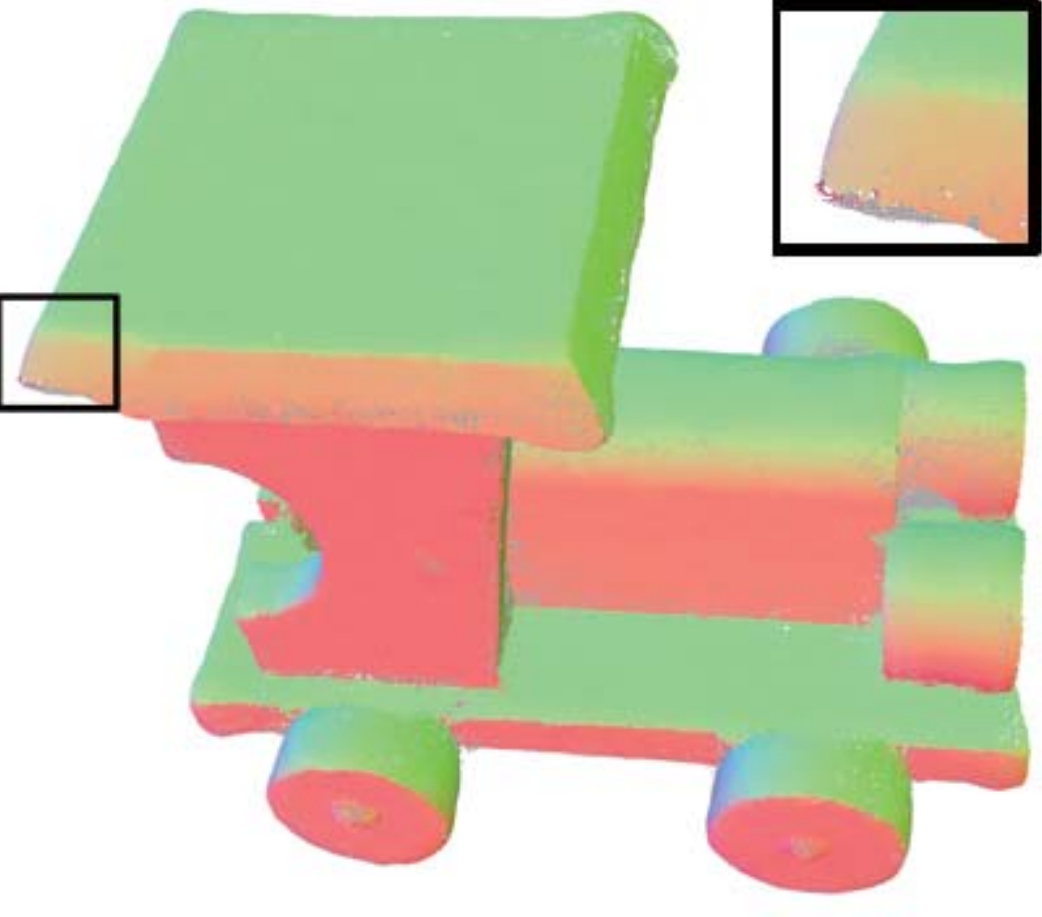}}
\end{minipage}
\begin{minipage}[b]{0.16\linewidth}
{\label{}\includegraphics[width=1\linewidth]{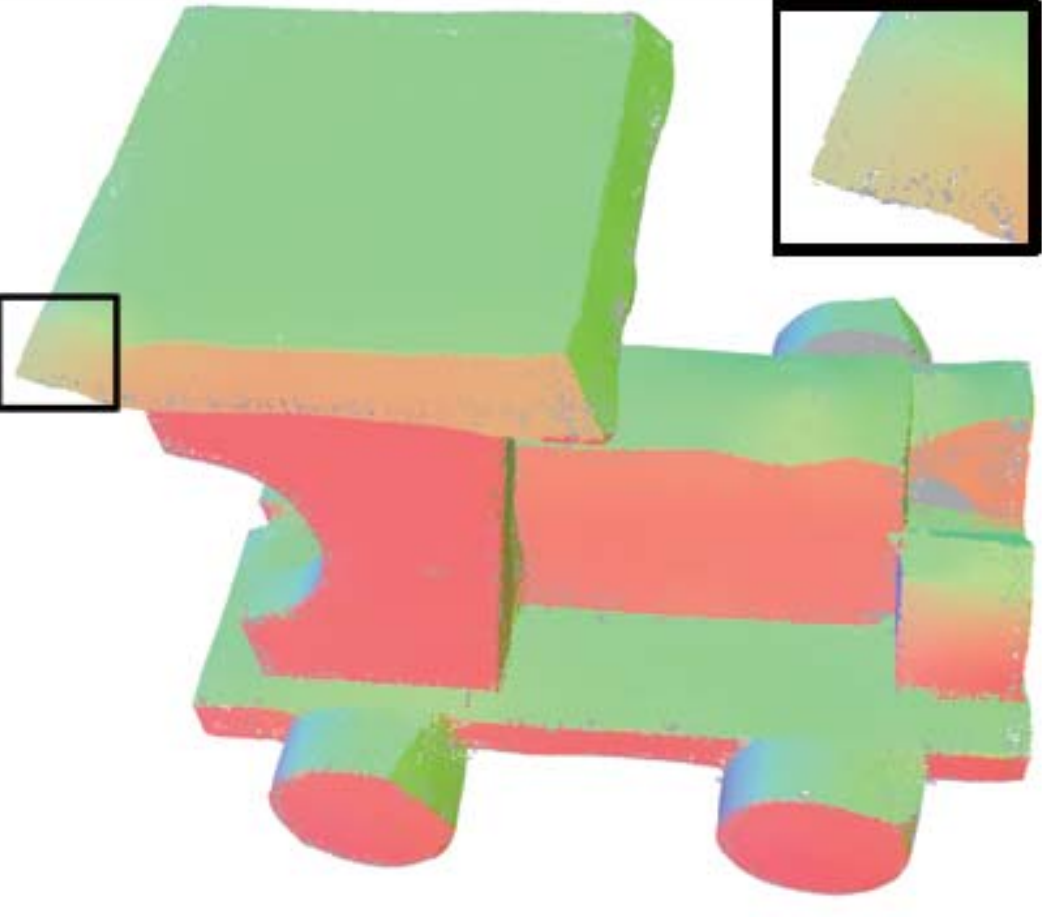}}
\end{minipage}
\begin{minipage}[b]{0.16\linewidth}
{\label{}\includegraphics[width=1\linewidth]{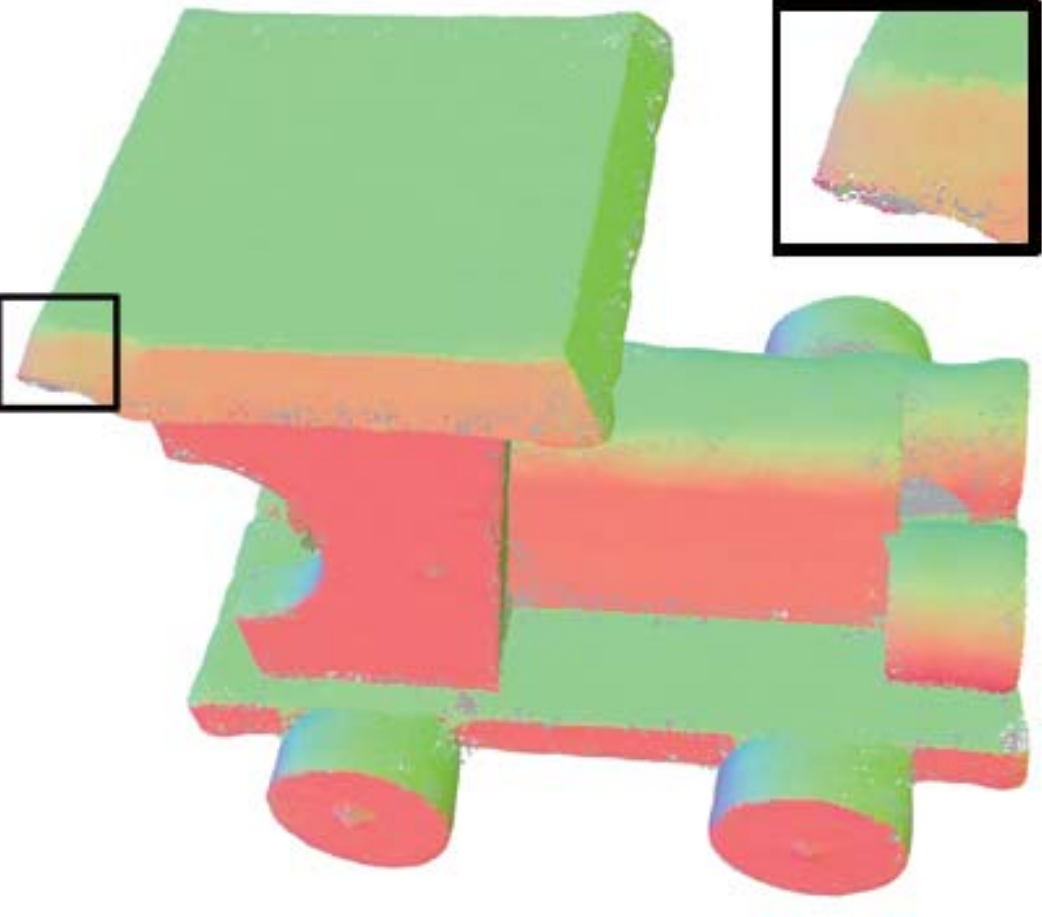}}
\end{minipage}
\begin{minipage}[b]{0.16\linewidth}
{\label{}\includegraphics[width=1\linewidth]{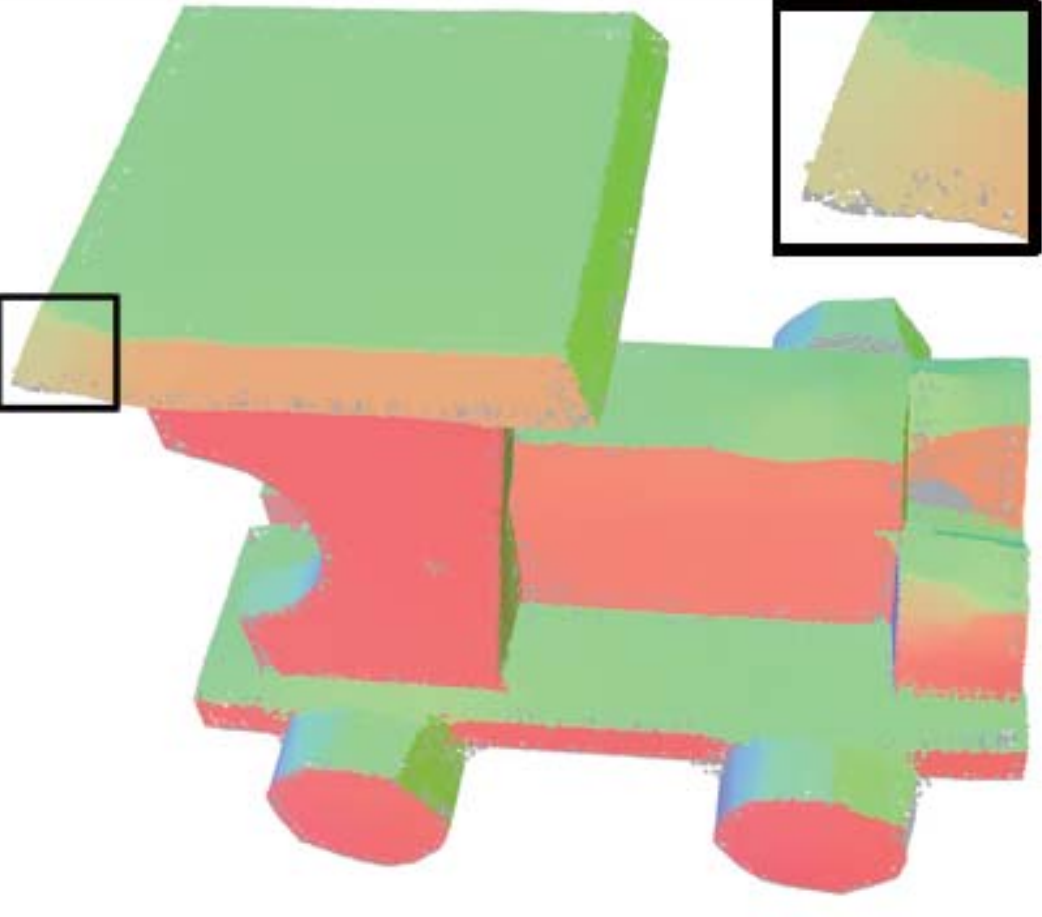}}
\end{minipage}	\\
\begin{minipage}[b]{0.16\linewidth}
\subfigure[\protect\cite{Hoppe1992}]{\label{}\includegraphics[width=1\linewidth]{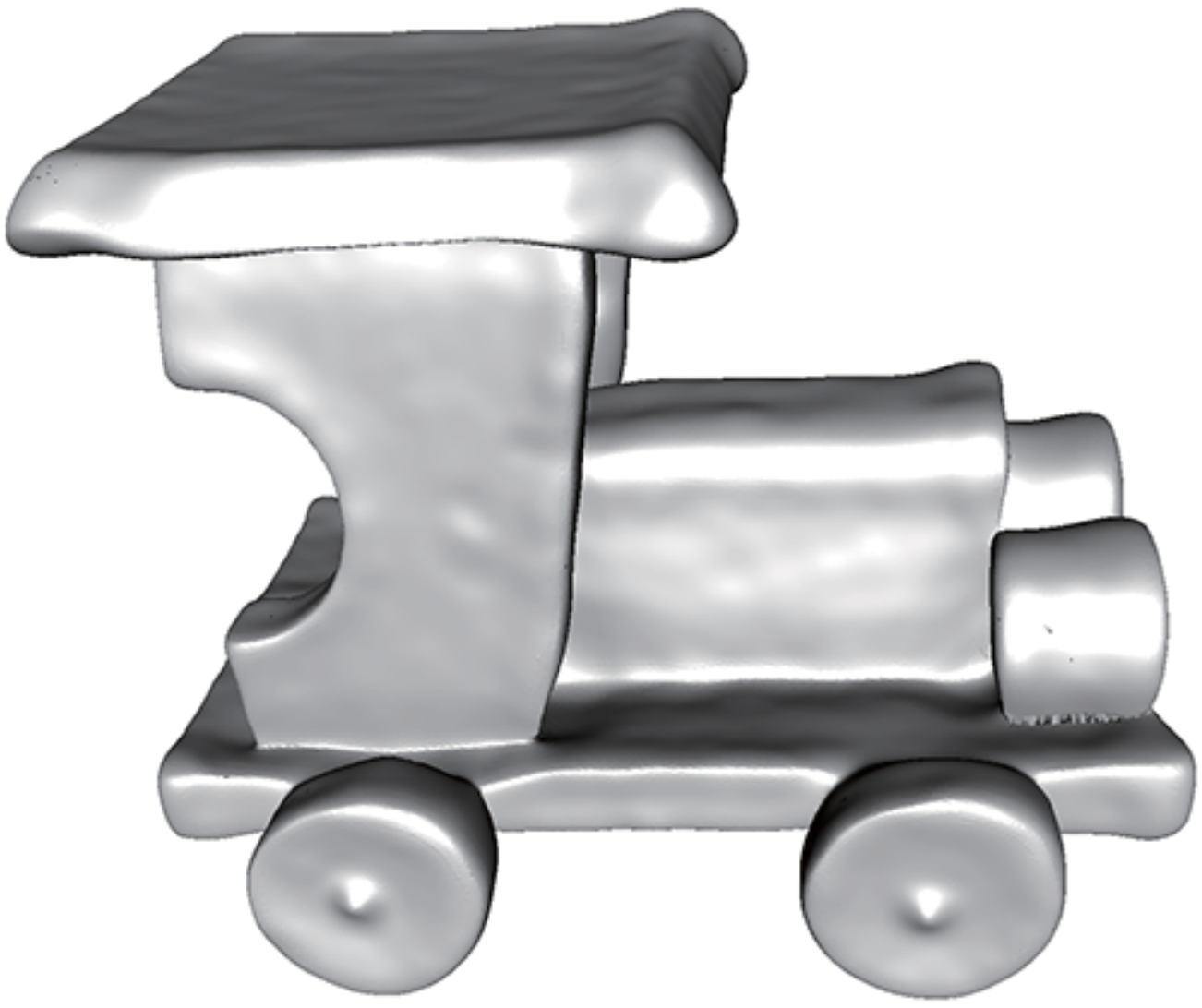}}
\end{minipage}
\begin{minipage}[b]{0.16\linewidth}
\subfigure[\protect\cite{Boulch2012}]{\label{}\includegraphics[width=1\linewidth]{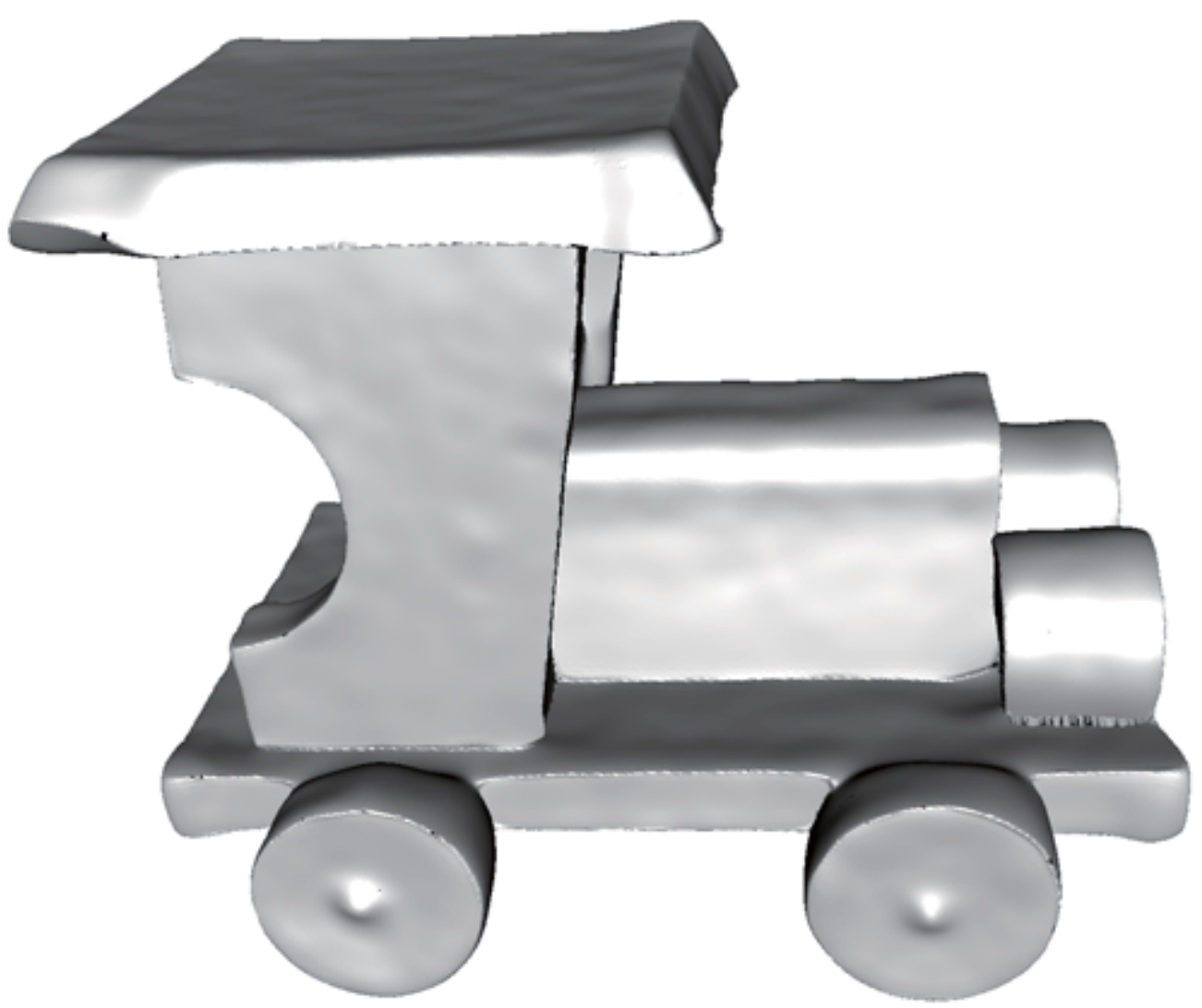}}
\end{minipage}
\begin{minipage}[b]{0.16\linewidth}
\subfigure[\protect\cite{Huang2013}]{\label{}\includegraphics[width=1\linewidth]{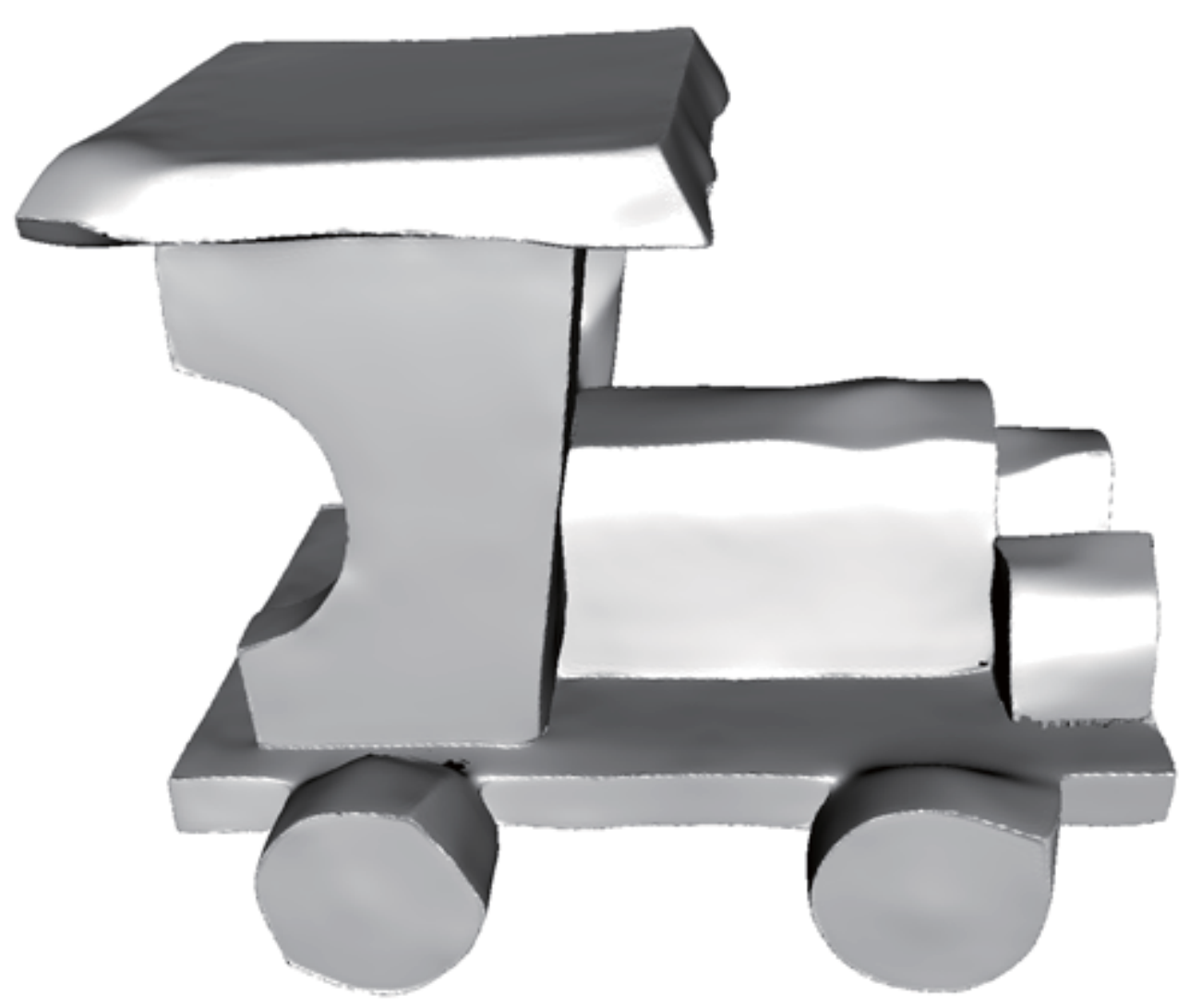}}
\end{minipage}
\begin{minipage}[b]{0.16\linewidth}
\subfigure[\protect\cite{Boulch2016}]{\label{}\includegraphics[width=1\linewidth]{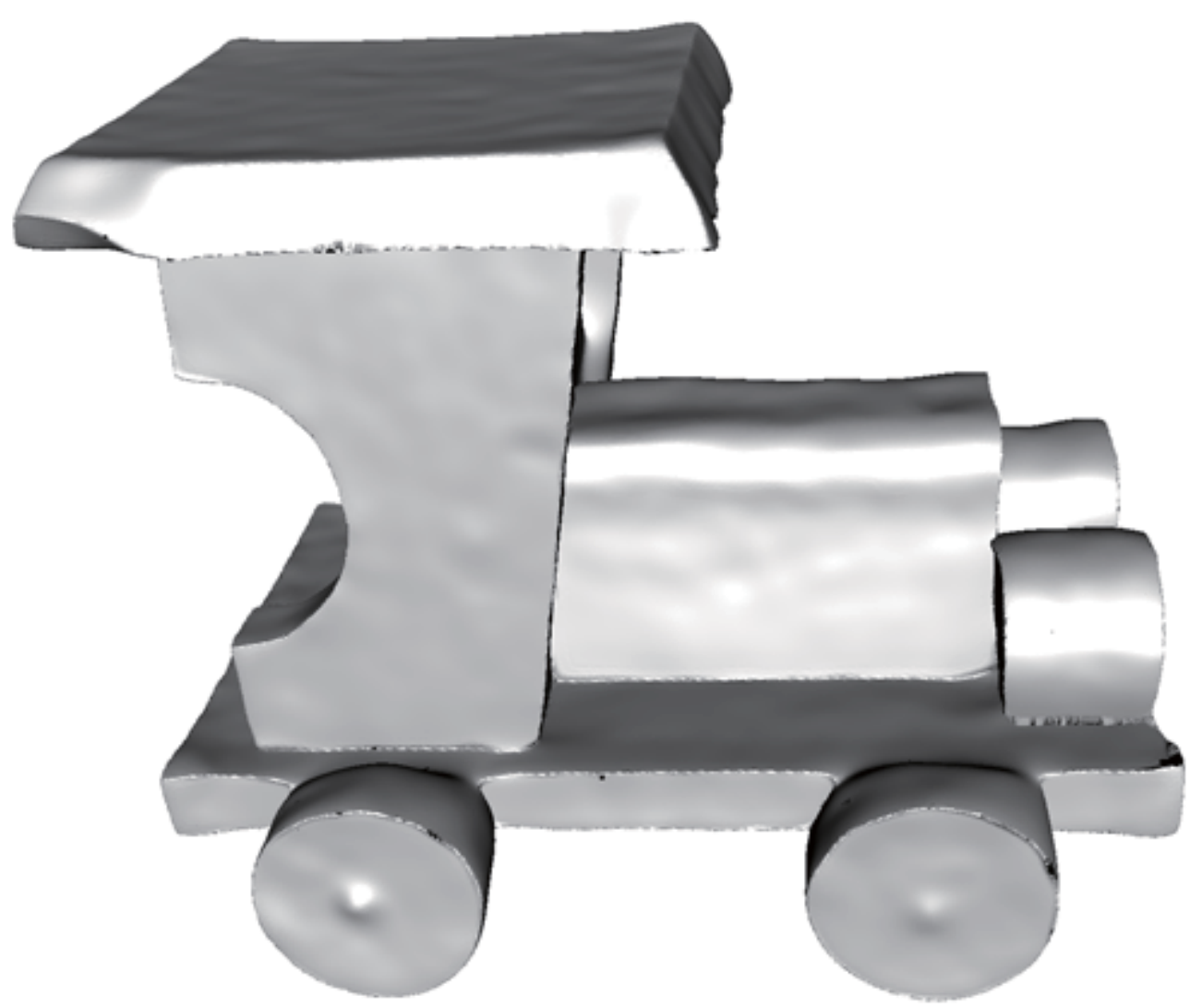}}
\end{minipage}
\begin{minipage}[b]{0.16\linewidth}
\subfigure[Ours]{\label{}\includegraphics[width=1\linewidth]{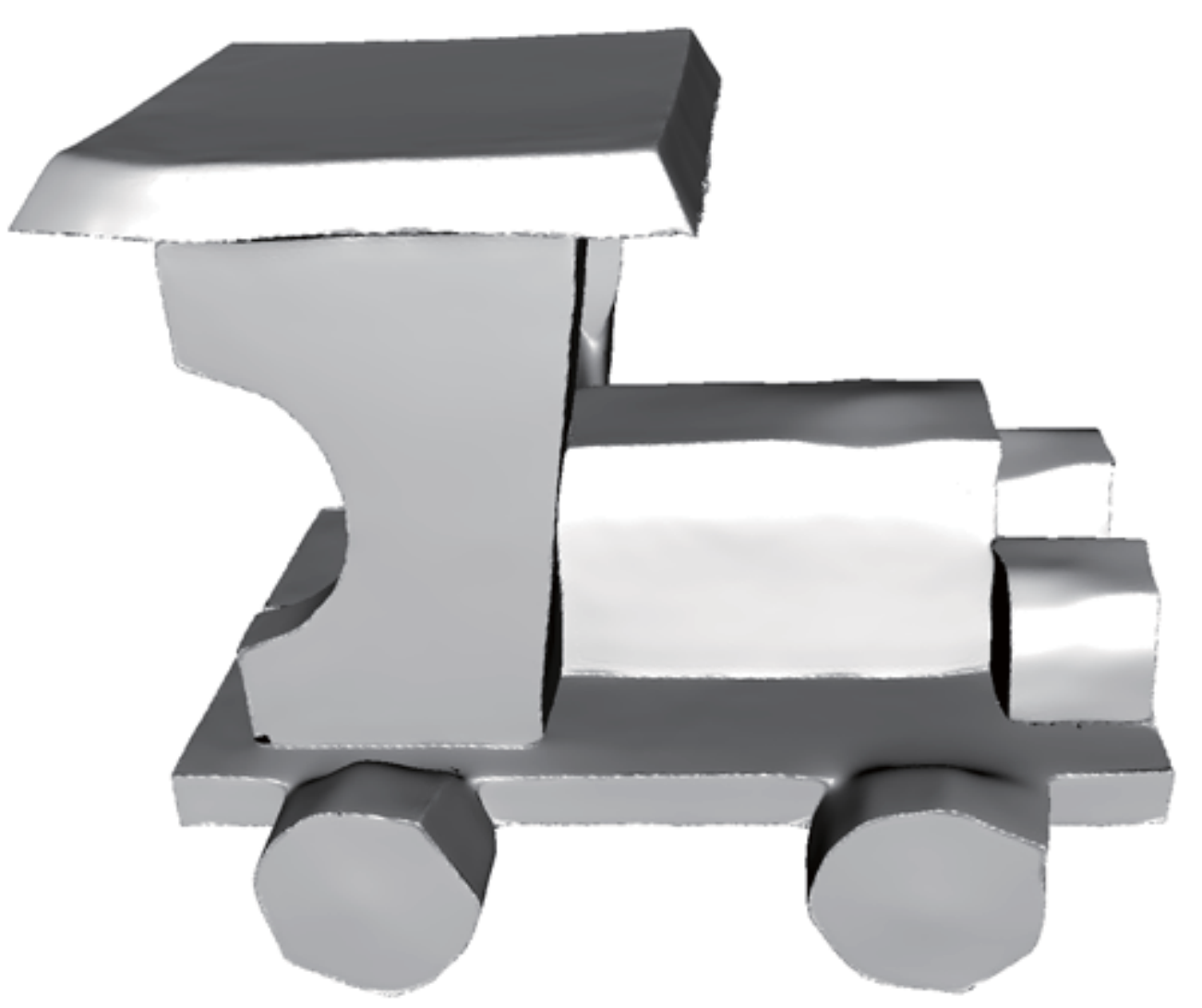}}
\end{minipage}
\caption{The first row: normal results of the scanned Car point cloud. The second row: upsampling results of the filtered results by updating position with the normals in the first row. The third row: the corresponding surface reconstruction. Comparing with other methods, \cite{Huang2013} and our method are better in sharp edges preservation and hereby generate more sharpened results. }
\label{fig:car_point}
%\vspace{-0.65cm}
\end{figure*}

%scanned: house
\begin{figure*}[htbp]
%\vspace{-0.0cm}
\centering
\begin{minipage}[b]{0.16\linewidth}
{\label{}\includegraphics[width=1\linewidth]{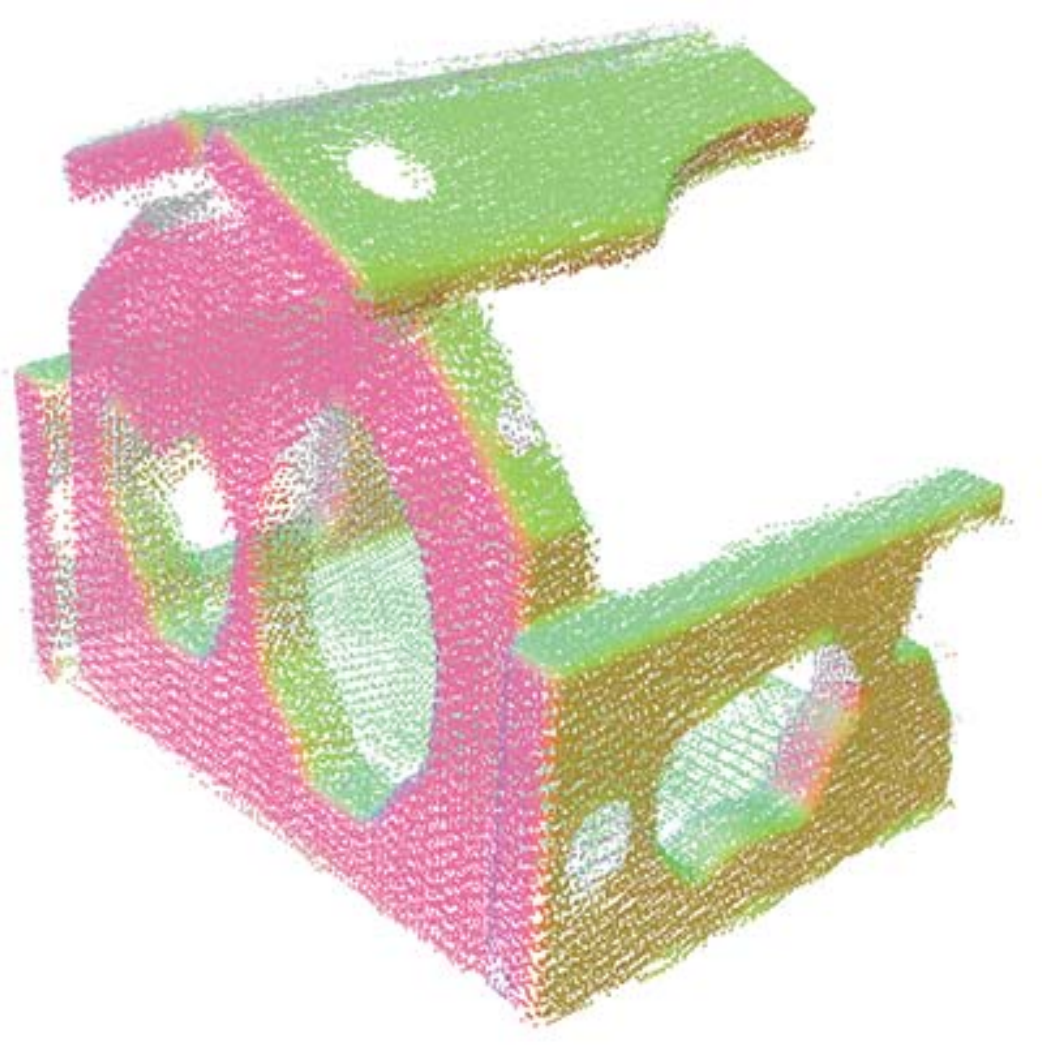}}
\end{minipage}
\begin{minipage}[b]{0.16\linewidth}
{\label{}\includegraphics[width=1\linewidth]{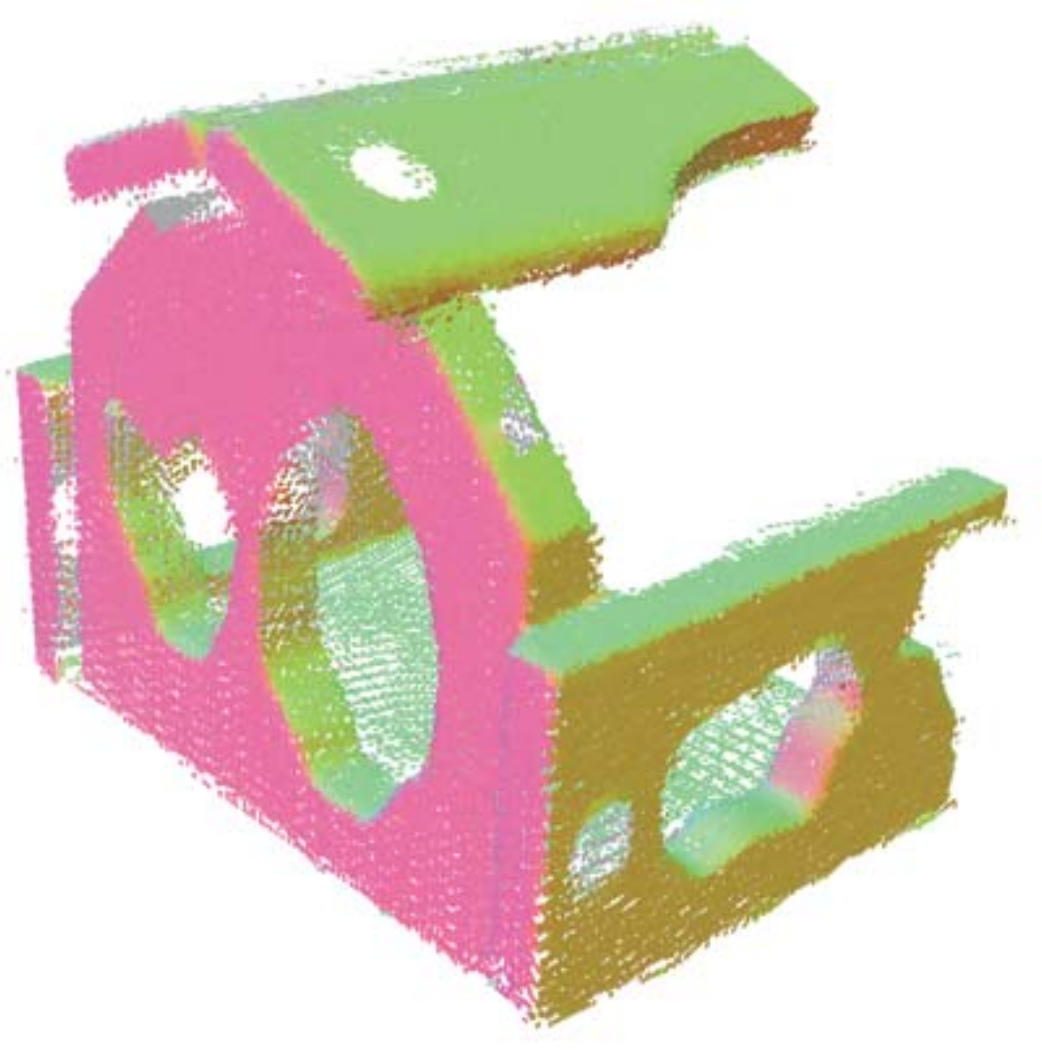}}
\end{minipage}
\begin{minipage}[b]{0.16\linewidth}
{\label{}\includegraphics[width=1\linewidth]{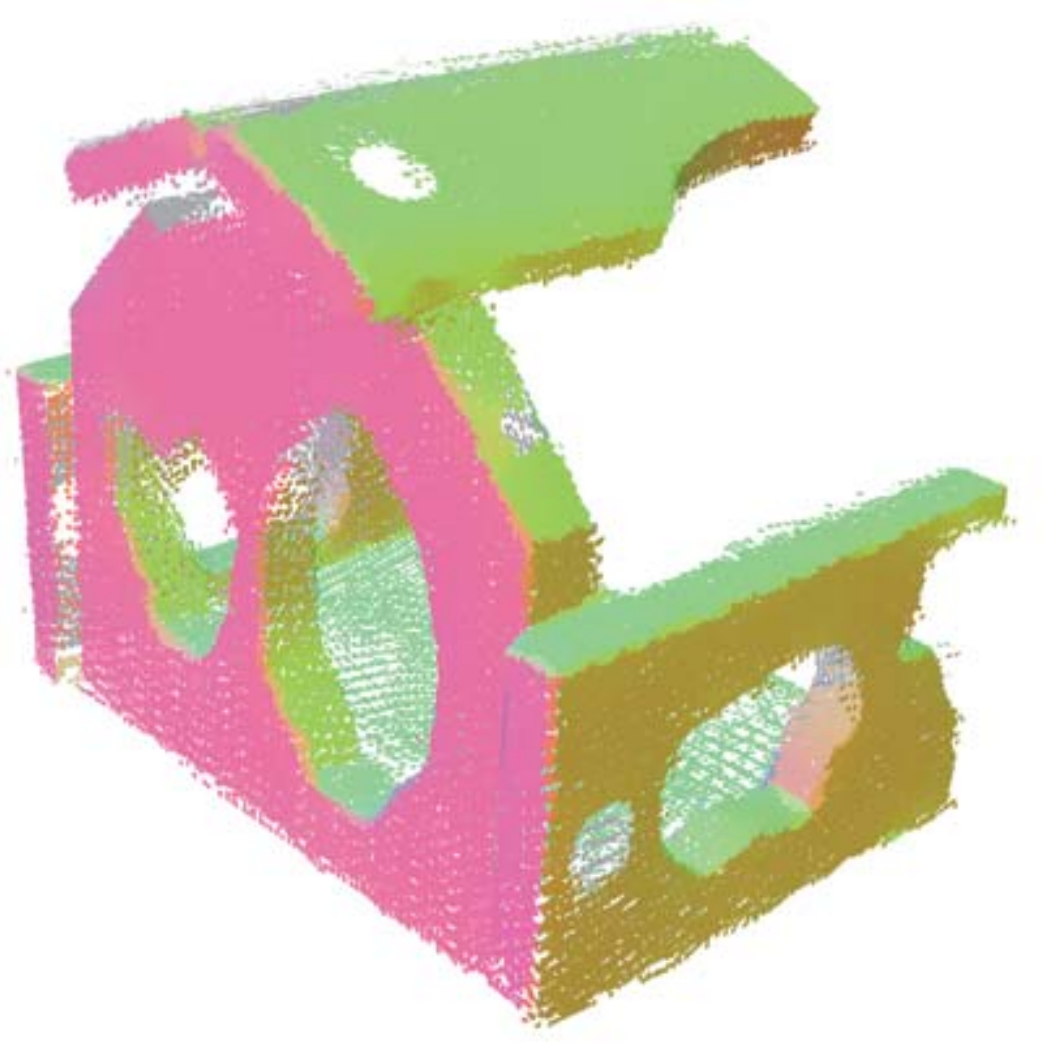}}
\end{minipage}
\begin{minipage}[b]{0.16\linewidth}
{\label{}\includegraphics[width=1\linewidth]{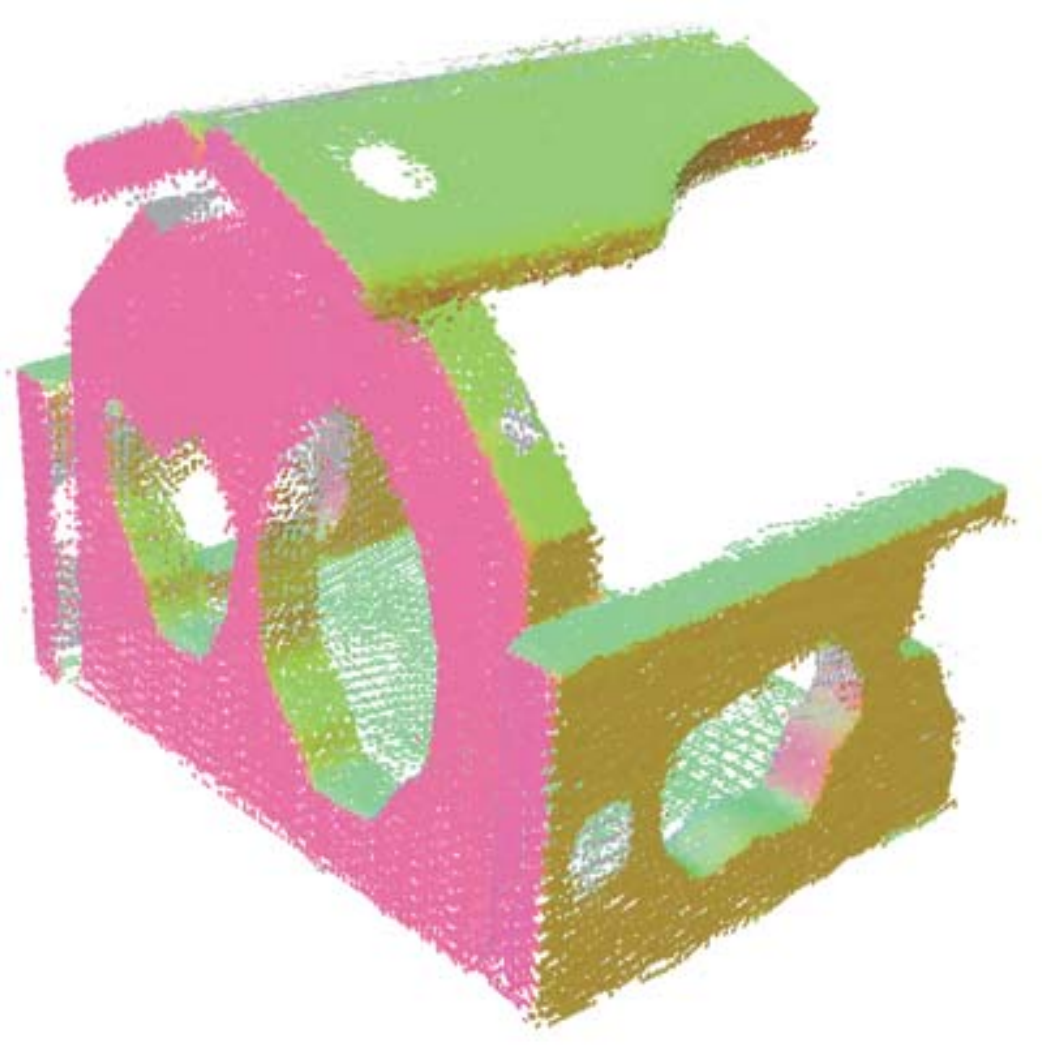}}
\end{minipage}
\begin{minipage}[b]{0.16\linewidth}
{\label{}\includegraphics[width=1\linewidth]{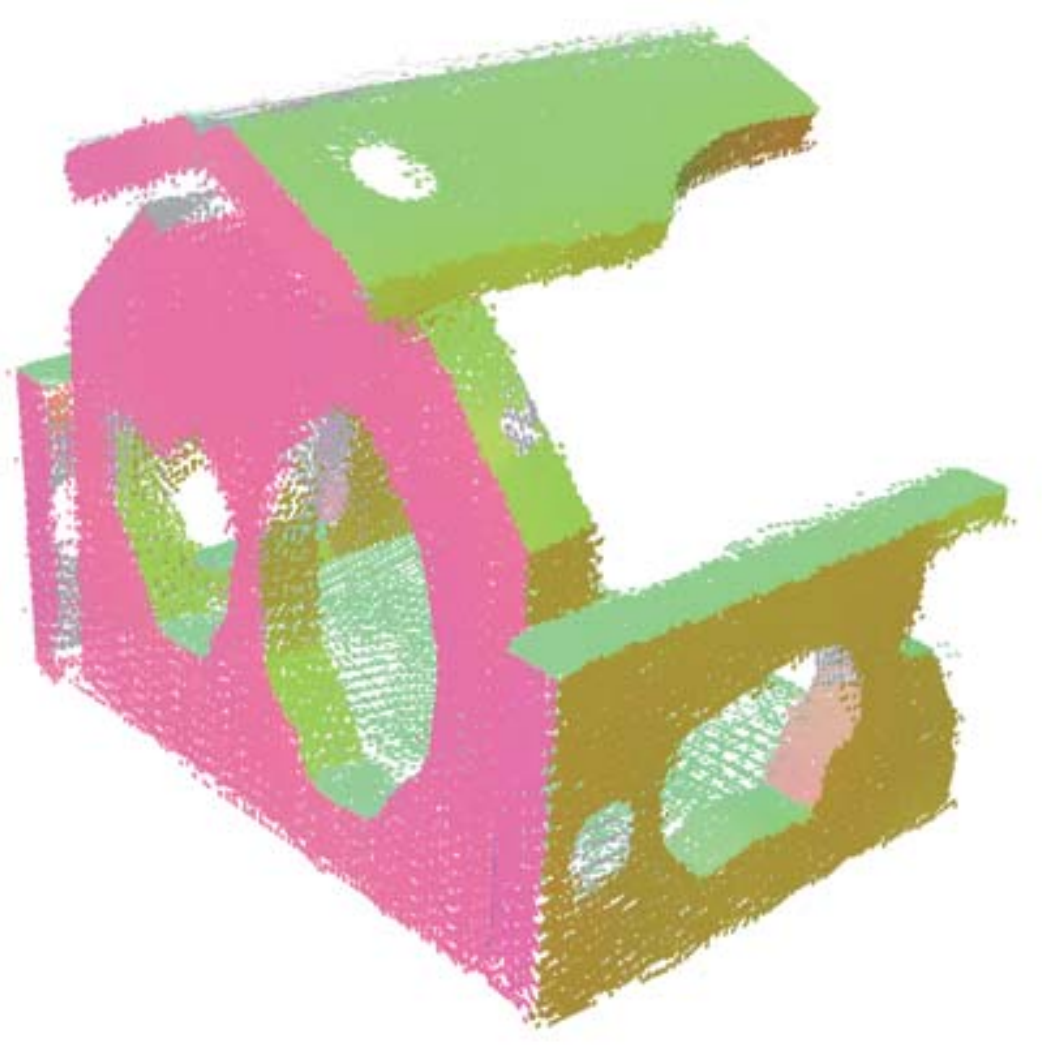}}
\end{minipage}	\\
\begin{minipage}[b]{0.16\linewidth}
{\label{}\includegraphics[width=1\linewidth]{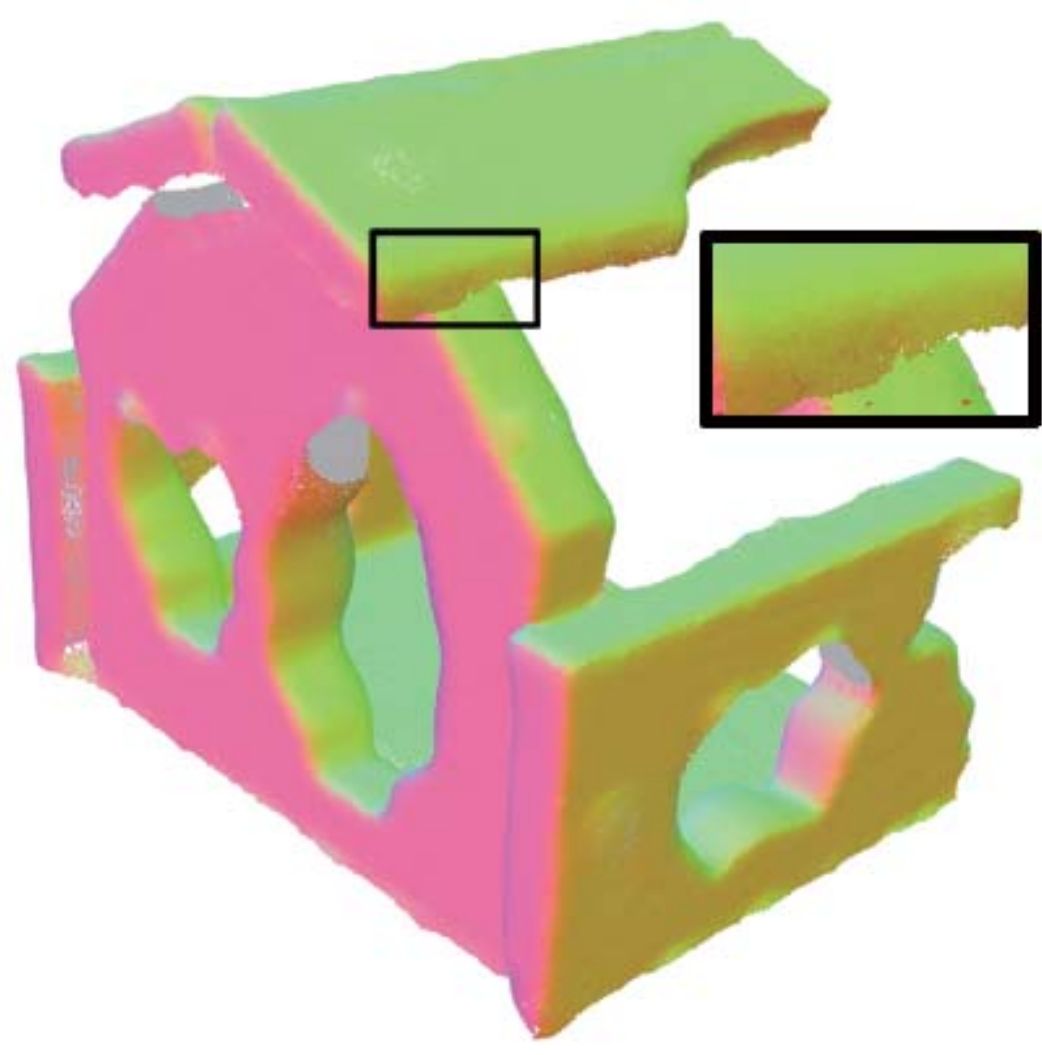}}
\end{minipage}
\begin{minipage}[b]{0.16\linewidth}
{\label{}\includegraphics[width=1\linewidth]{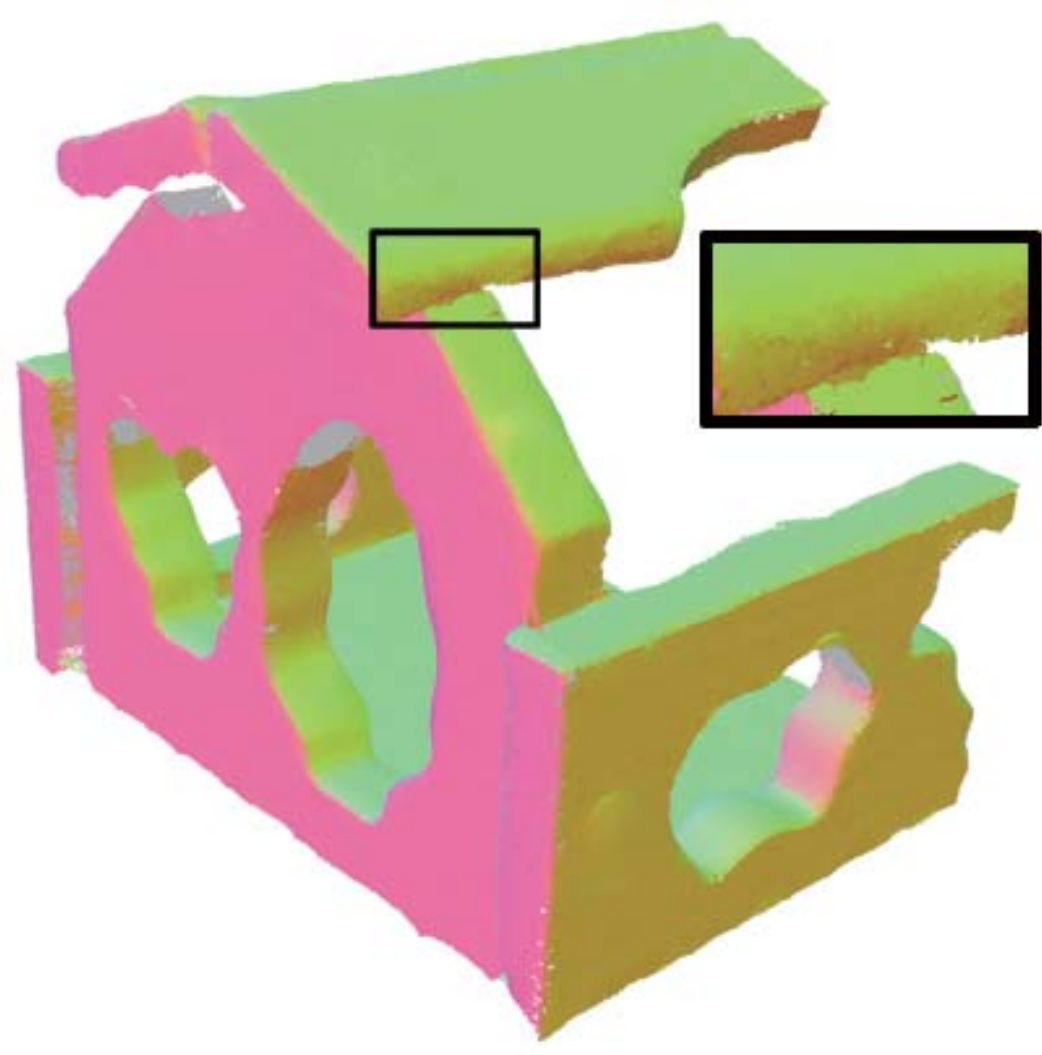}}
\end{minipage}
\begin{minipage}[b]{0.16\linewidth}
{\label{}\includegraphics[width=1\linewidth]{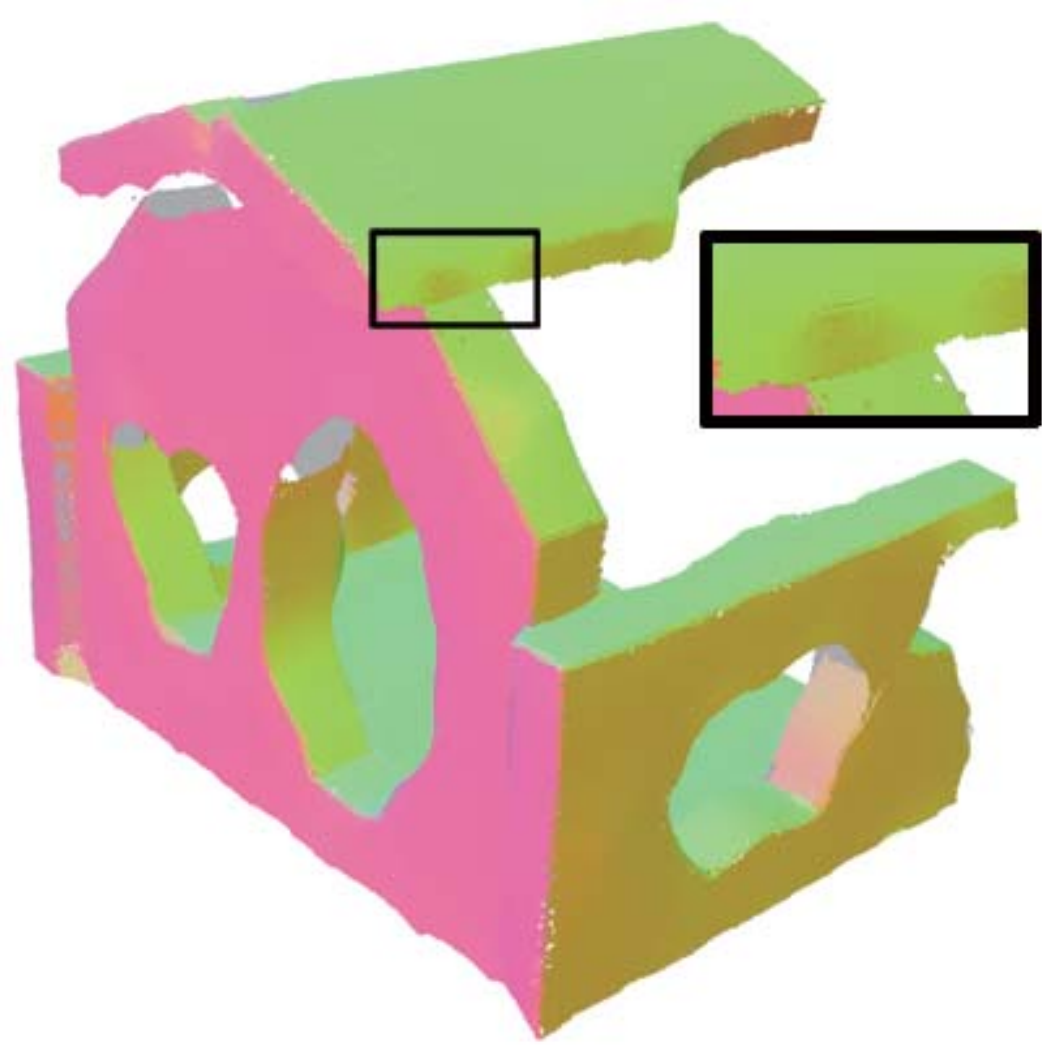}}
\end{minipage}
\begin{minipage}[b]{0.16\linewidth}
{\label{}\includegraphics[width=1\linewidth]{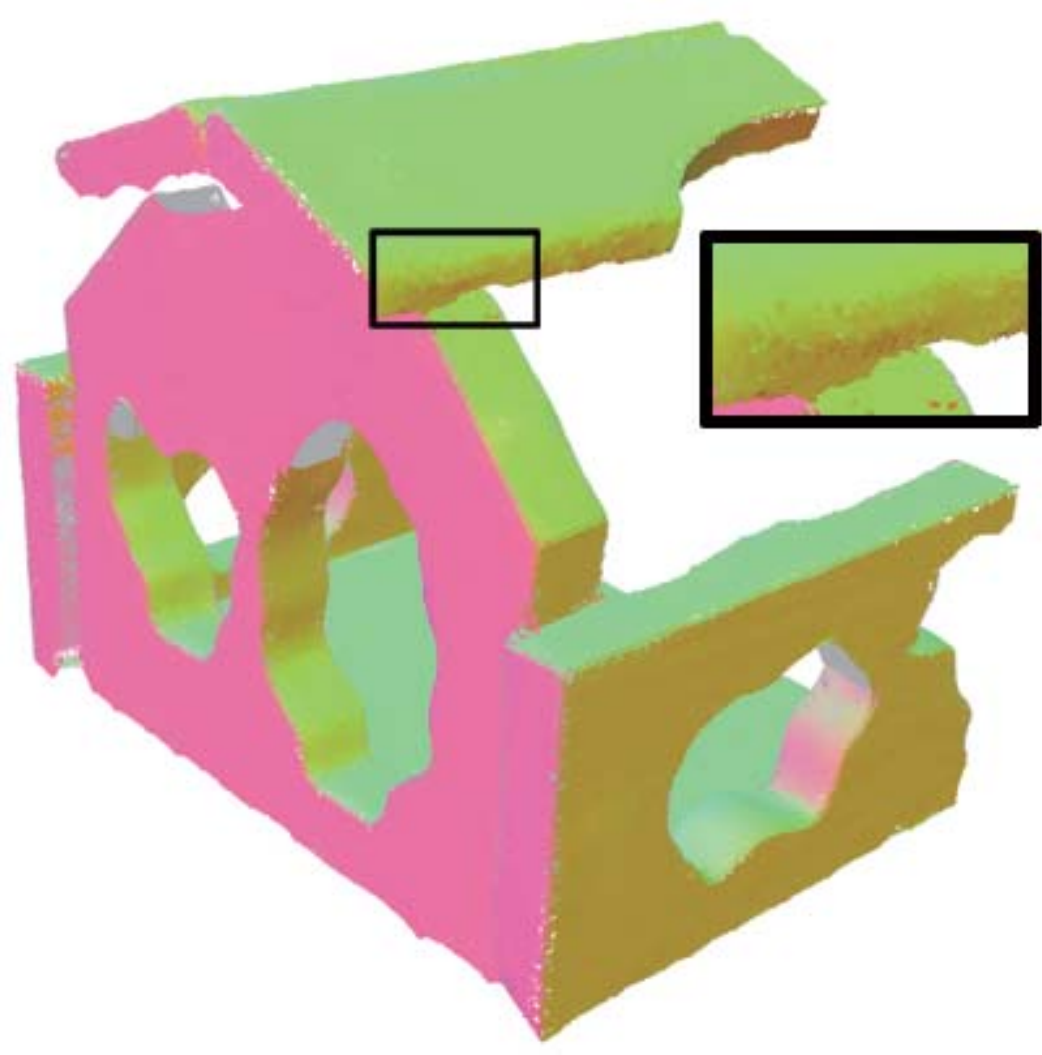}}
\end{minipage}
\begin{minipage}[b]{0.16\linewidth}
{\label{}\includegraphics[width=1\linewidth]{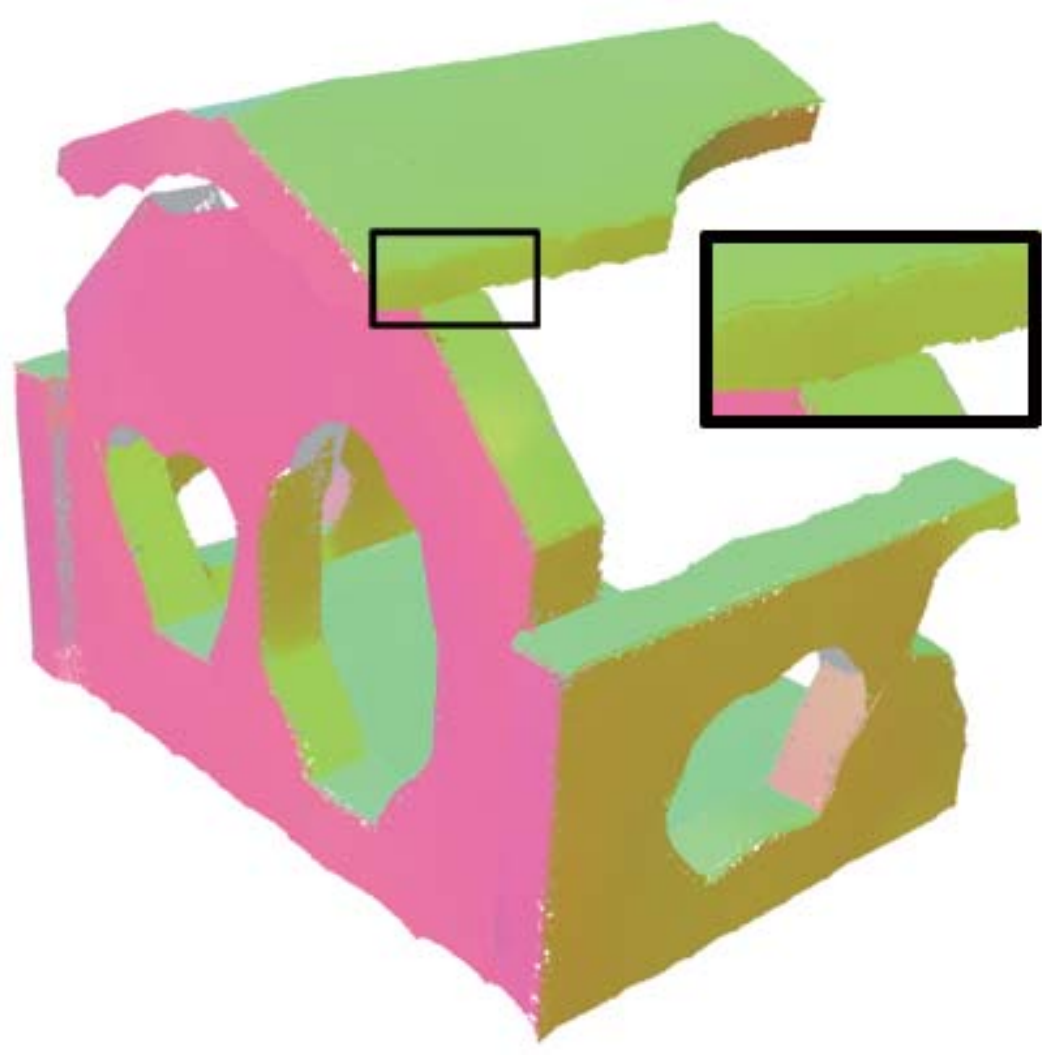}}
\end{minipage}	\\
\begin{minipage}[b]{0.16\linewidth}
\subfigure[\protect\cite{Hoppe1992}]{\label{}\includegraphics[width=1\linewidth]{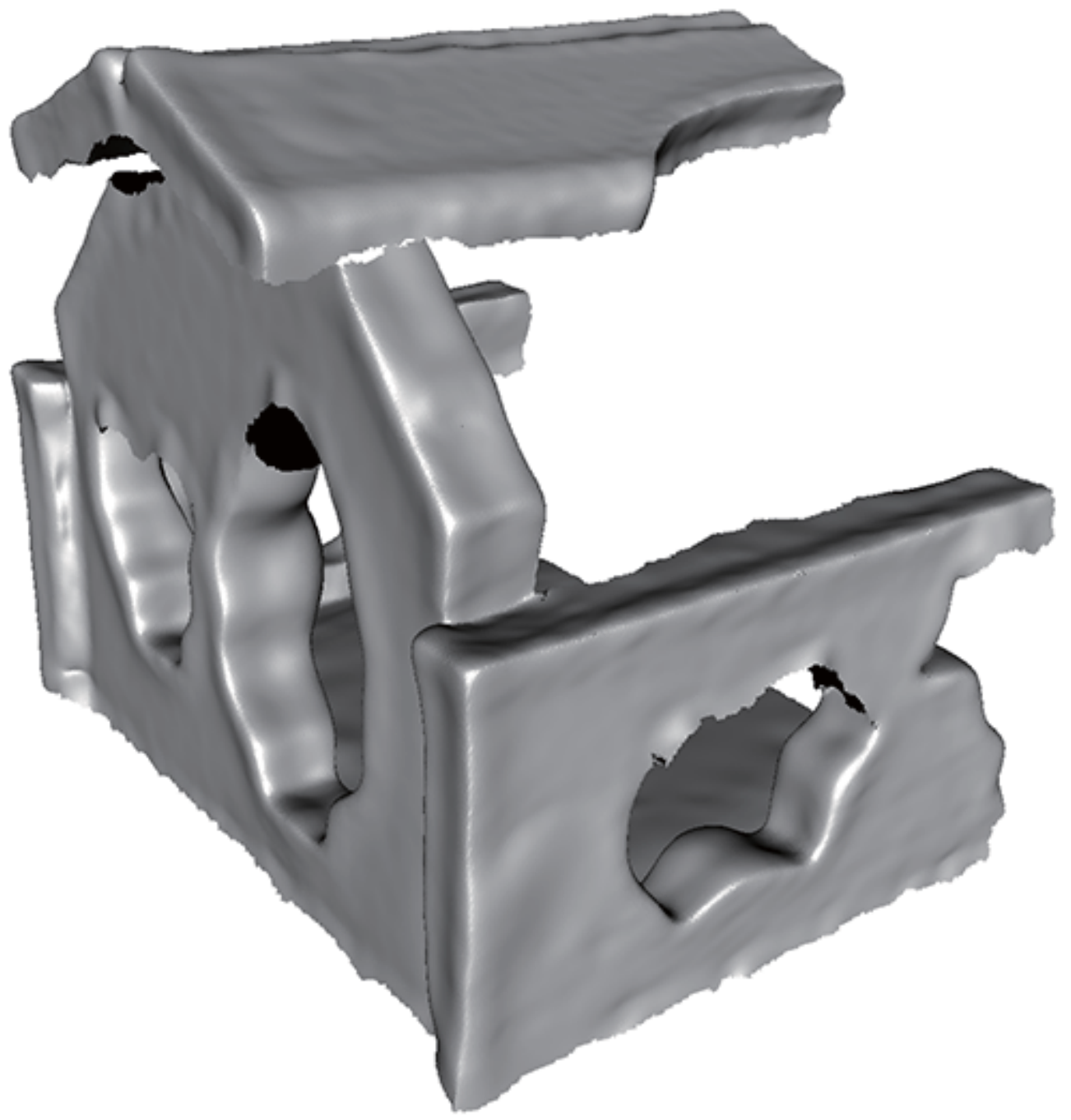}}
\end{minipage}
\begin{minipage}[b]{0.16\linewidth}
\subfigure[\protect\cite{Boulch2012}]{\label{}\includegraphics[width=1\linewidth]{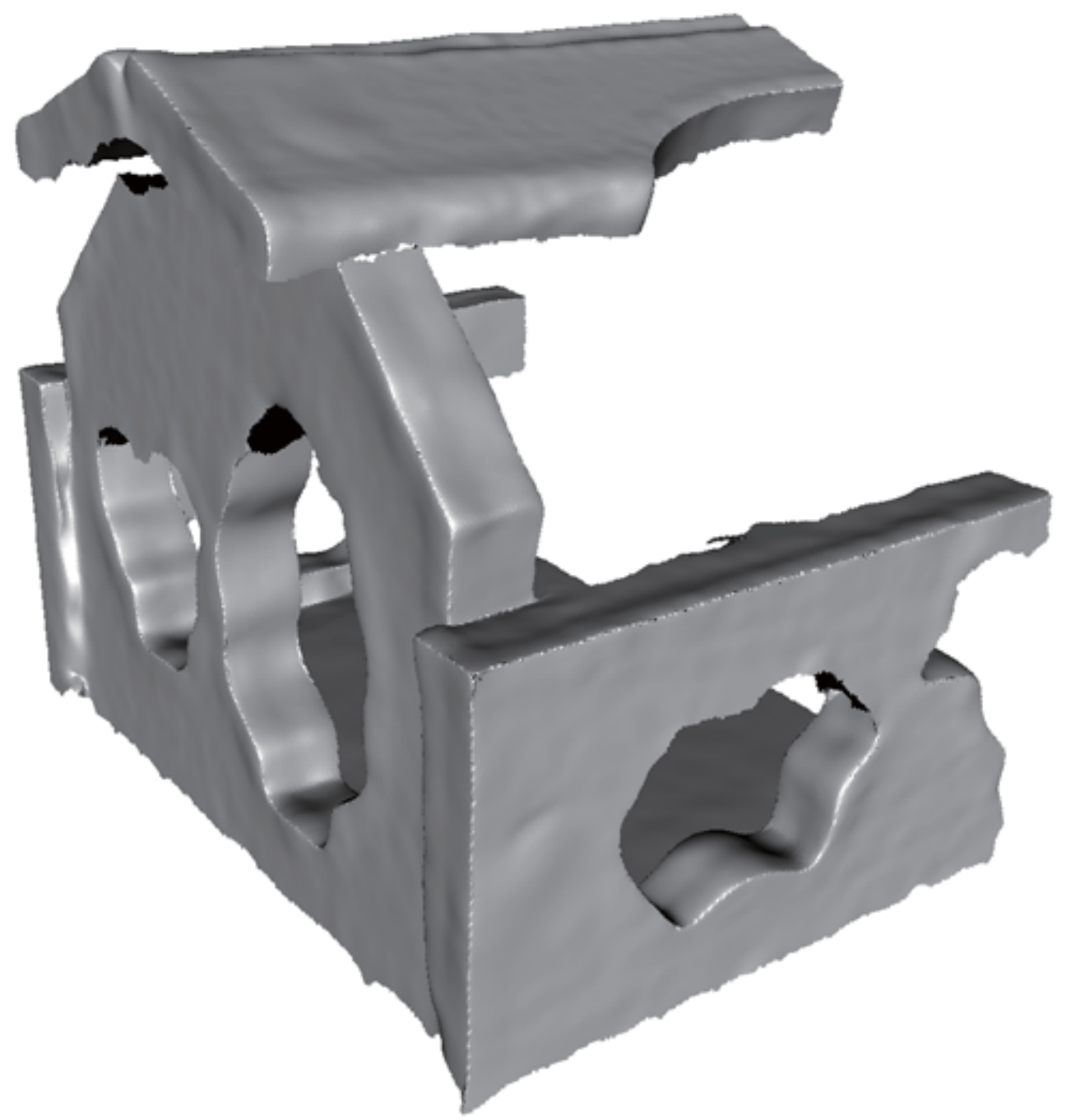}}
\end{minipage}
\begin{minipage}[b]{0.16\linewidth}
\subfigure[\protect\cite{Huang2013}]{\label{}\includegraphics[width=1\linewidth]{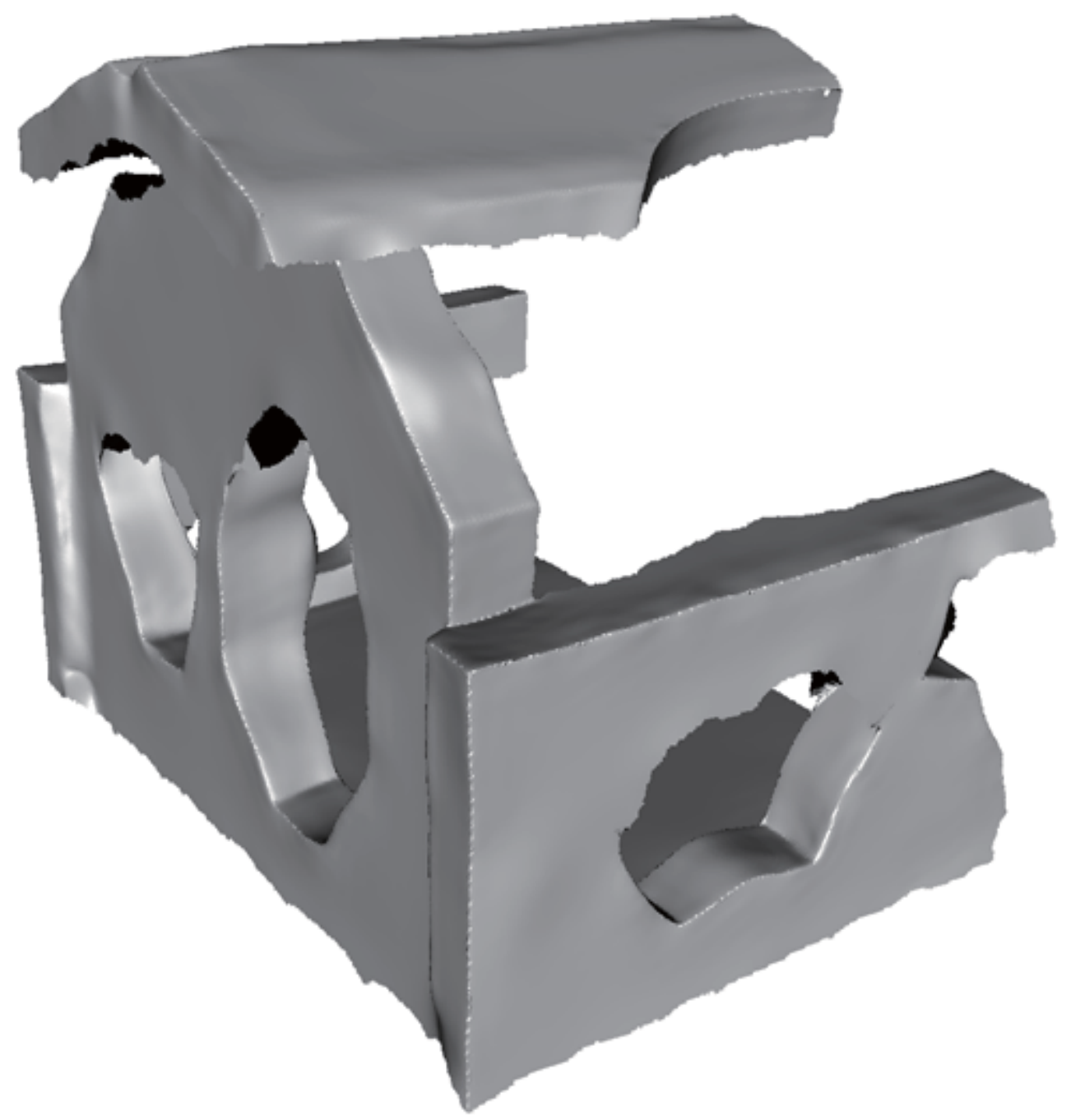}}
\end{minipage}
\begin{minipage}[b]{0.16\linewidth}
\subfigure[\protect\cite{Boulch2016}]{\label{}\includegraphics[width=1\linewidth]{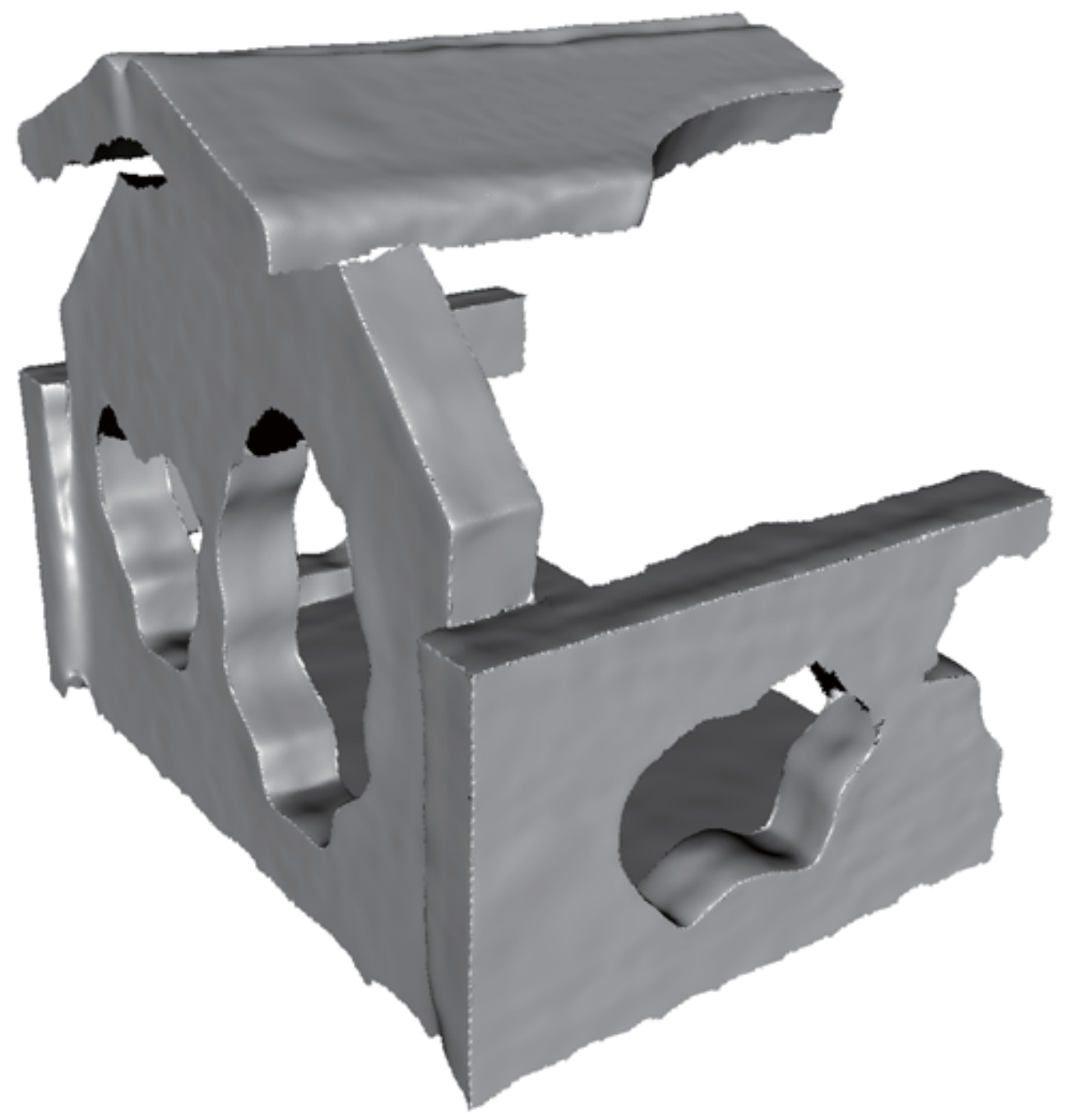}}
\end{minipage}
\begin{minipage}[b]{0.16\linewidth}
\subfigure[Ours]{\label{}\includegraphics[width=1\linewidth]{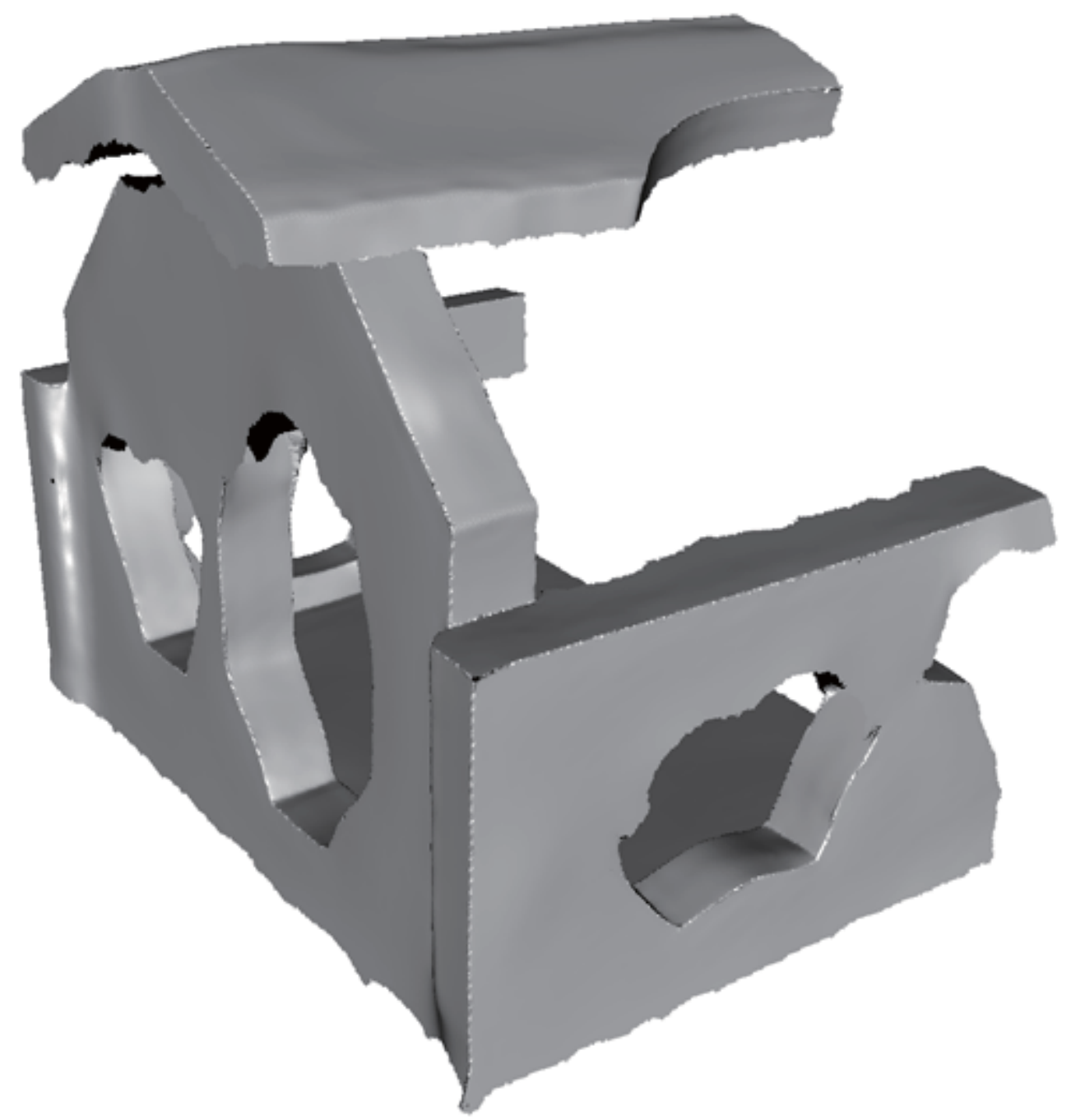}}
\end{minipage}
\caption{The first row: normal results of the scanned House point cloud. The second row: upsampling results of the filtered results by updating position with the normals in the first row. The third row: the corresponding surface reconstruction results.}
\label{fig:house_point}
%\vspace{-0.65cm}
\end{figure*}

%scanned: iron
\begin{figure*}[htbp]
%\vspace{-0.0cm}
\centering
\begin{minipage}[b]{0.16\linewidth}
{\label{}\includegraphics[width=1\linewidth]{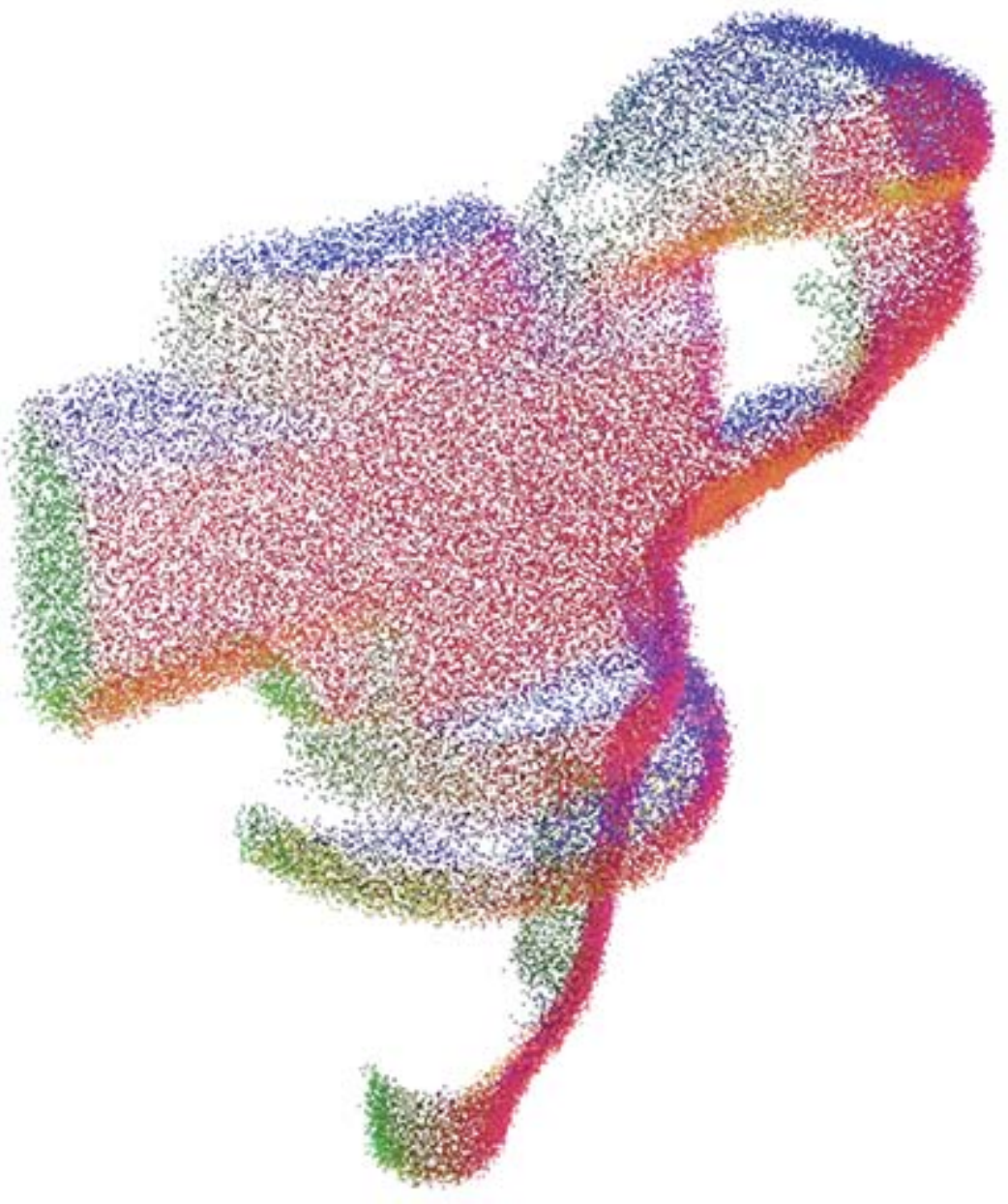}}
\end{minipage}
\begin{minipage}[b]{0.16\linewidth}
{\label{}\includegraphics[width=1\linewidth]{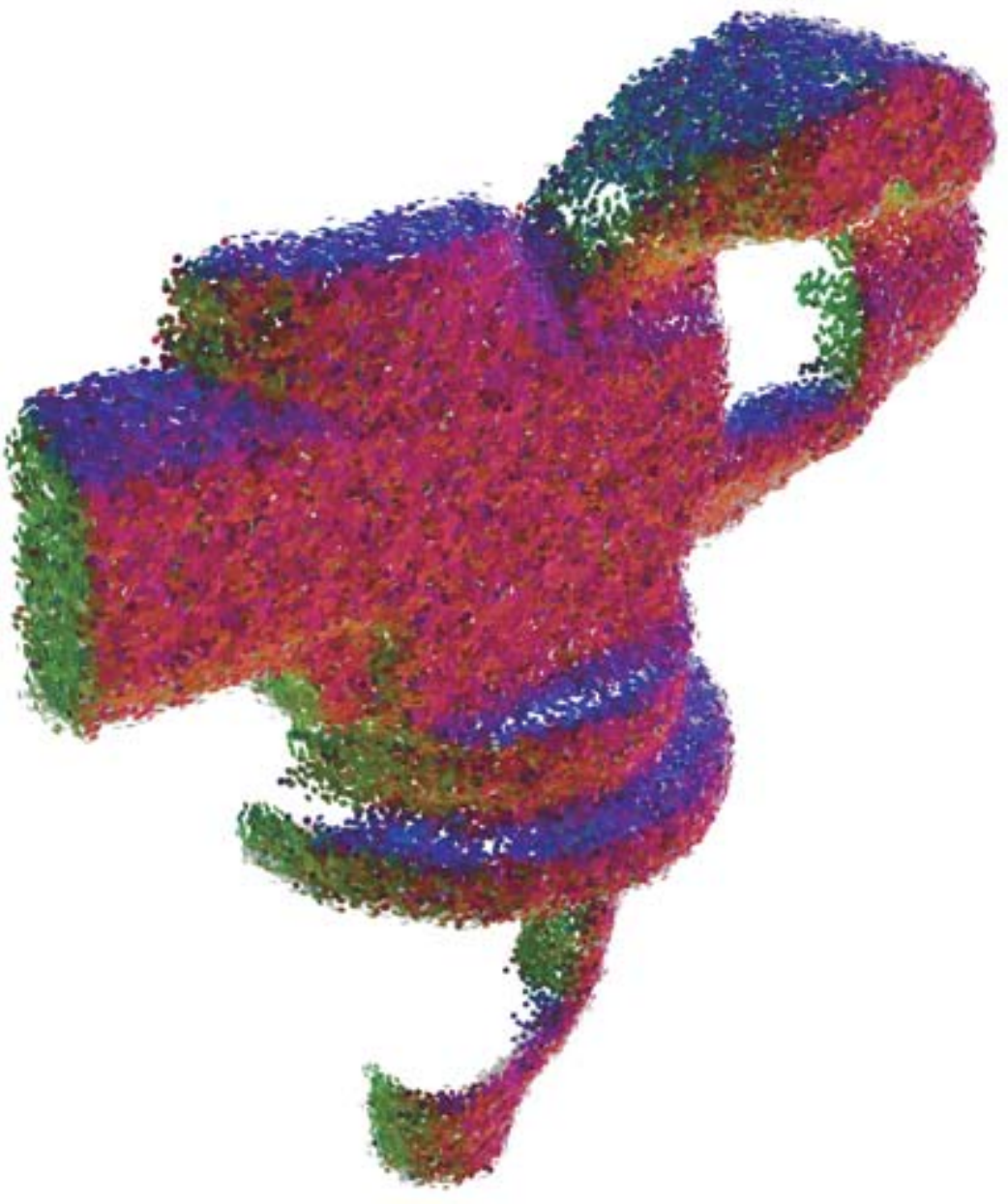}}
\end{minipage}
\begin{minipage}[b]{0.16\linewidth}
{\label{}\includegraphics[width=1\linewidth]{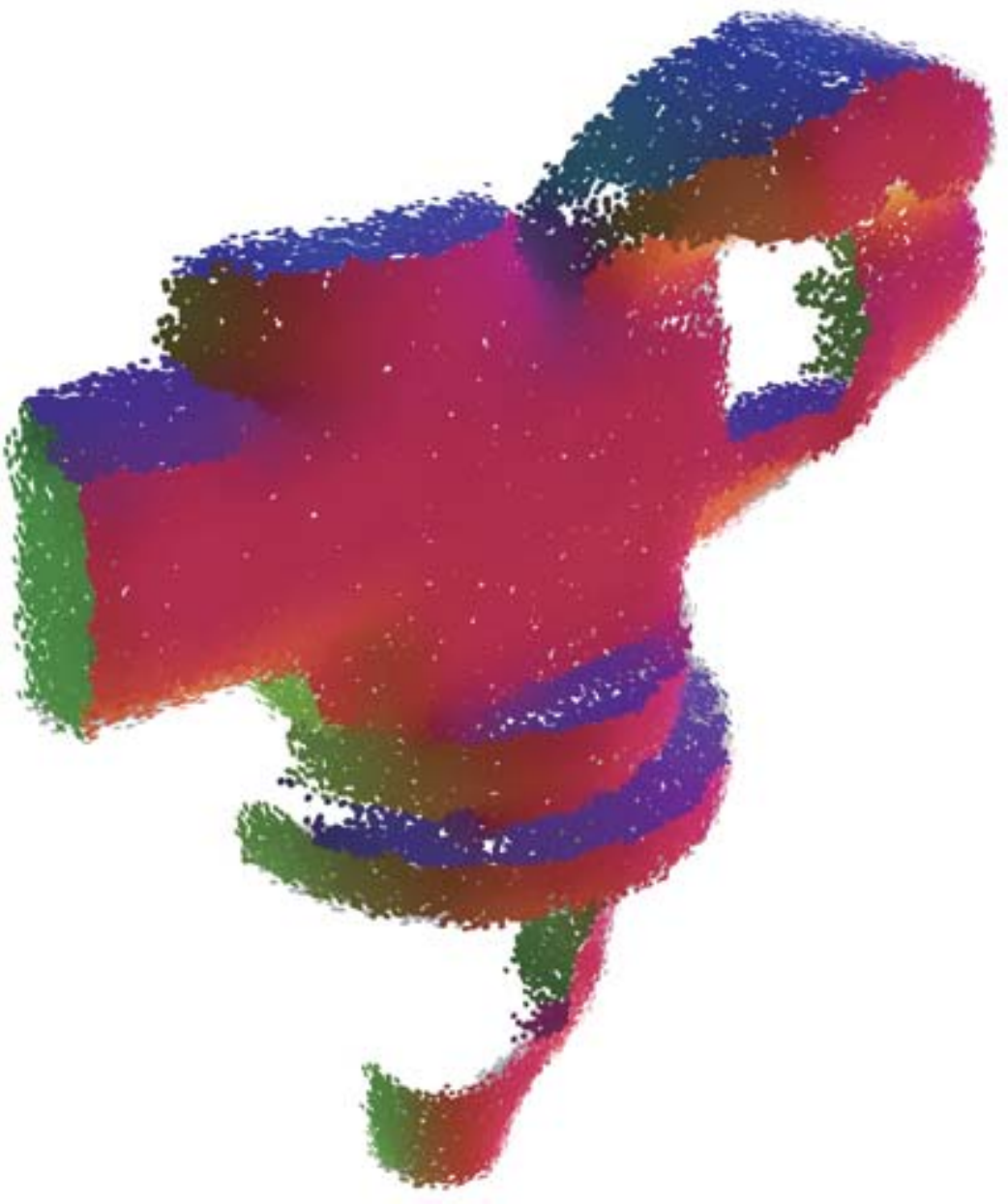}}
\end{minipage}
\begin{minipage}[b]{0.16\linewidth}
{\label{}\includegraphics[width=1\linewidth]{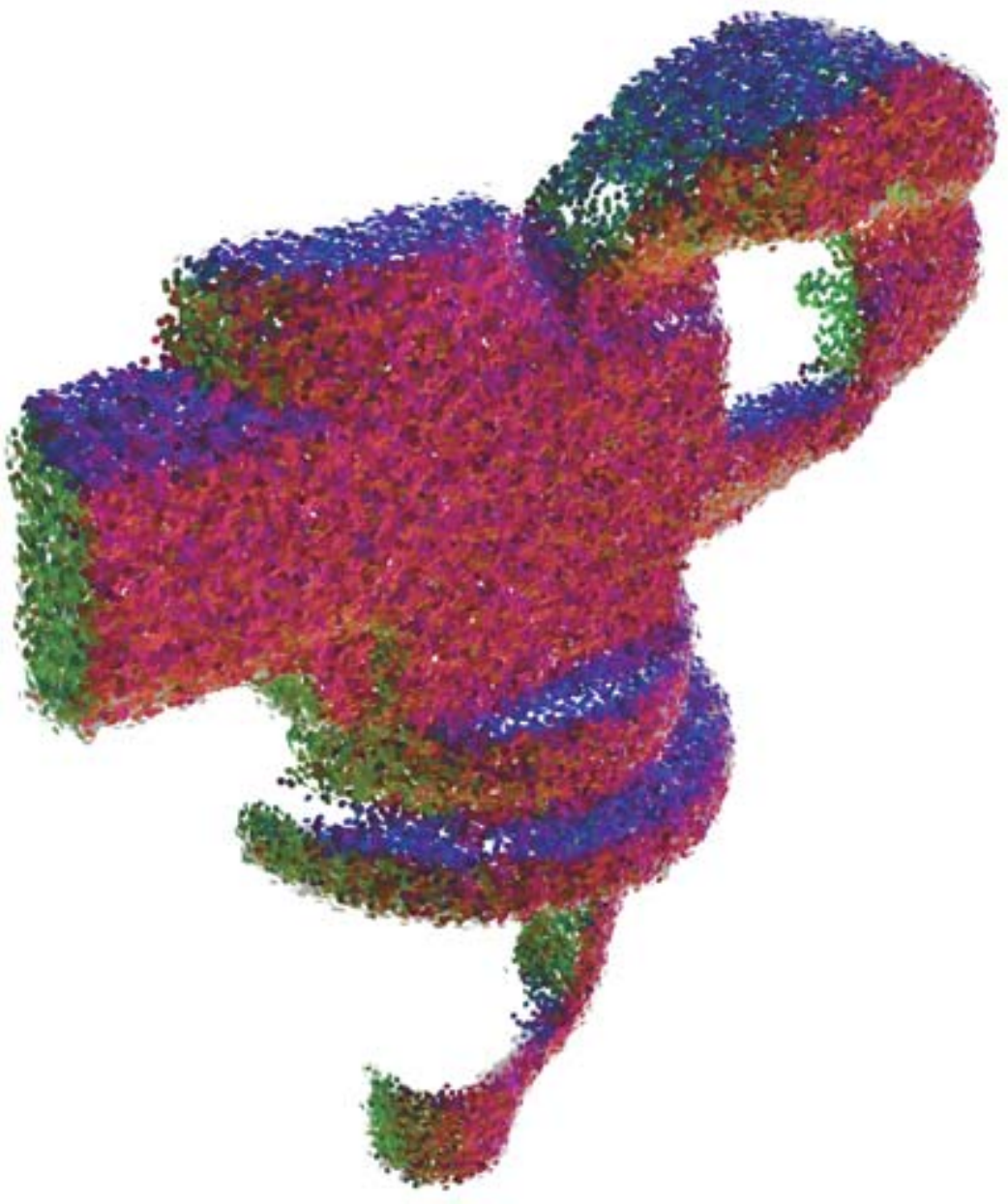}}
\end{minipage}
\begin{minipage}[b]{0.16\linewidth}
{\label{}\includegraphics[width=1\linewidth]{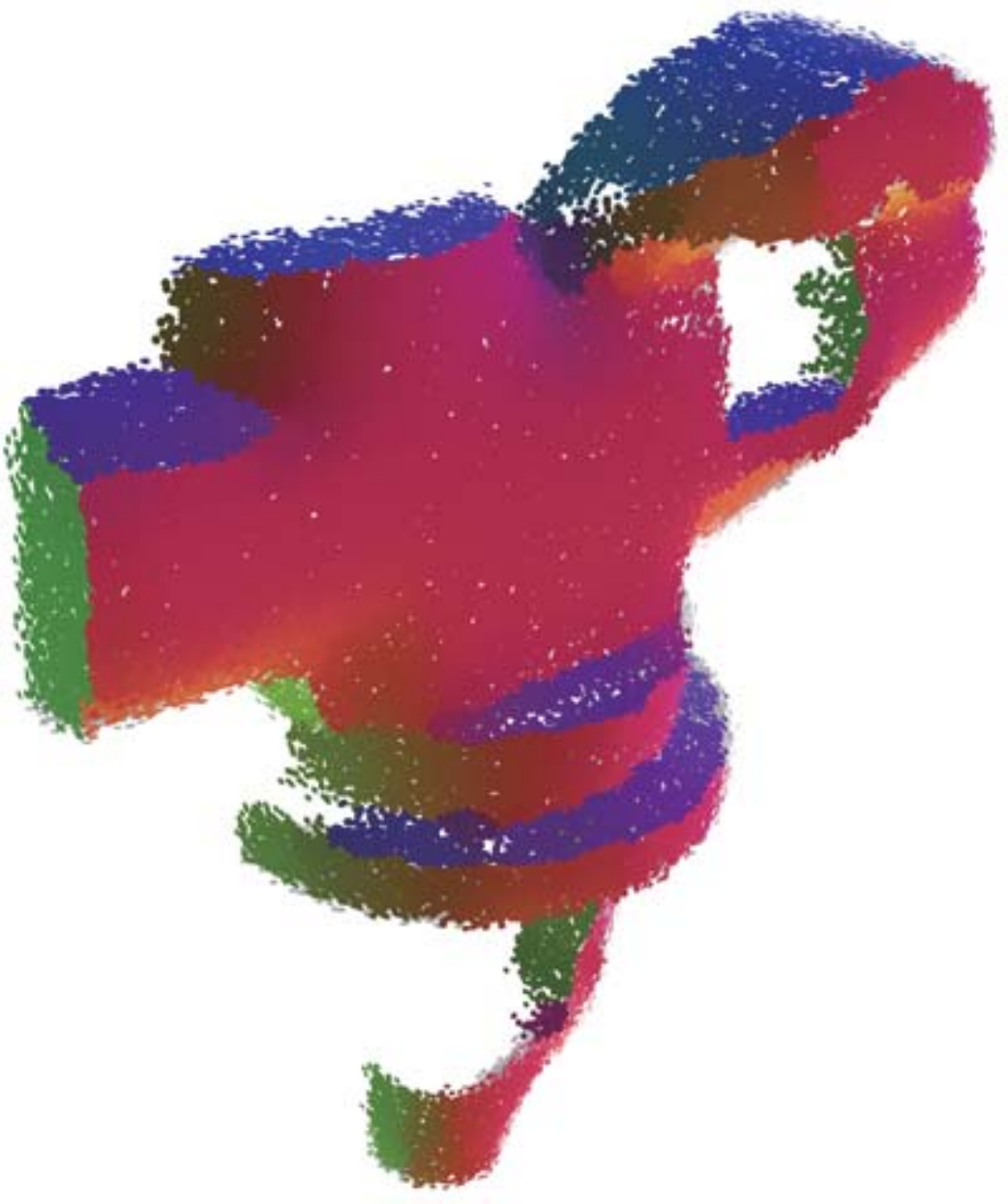}}
\end{minipage}	\\
\begin{minipage}[b]{0.16\linewidth}
{\label{}\includegraphics[width=1\linewidth]{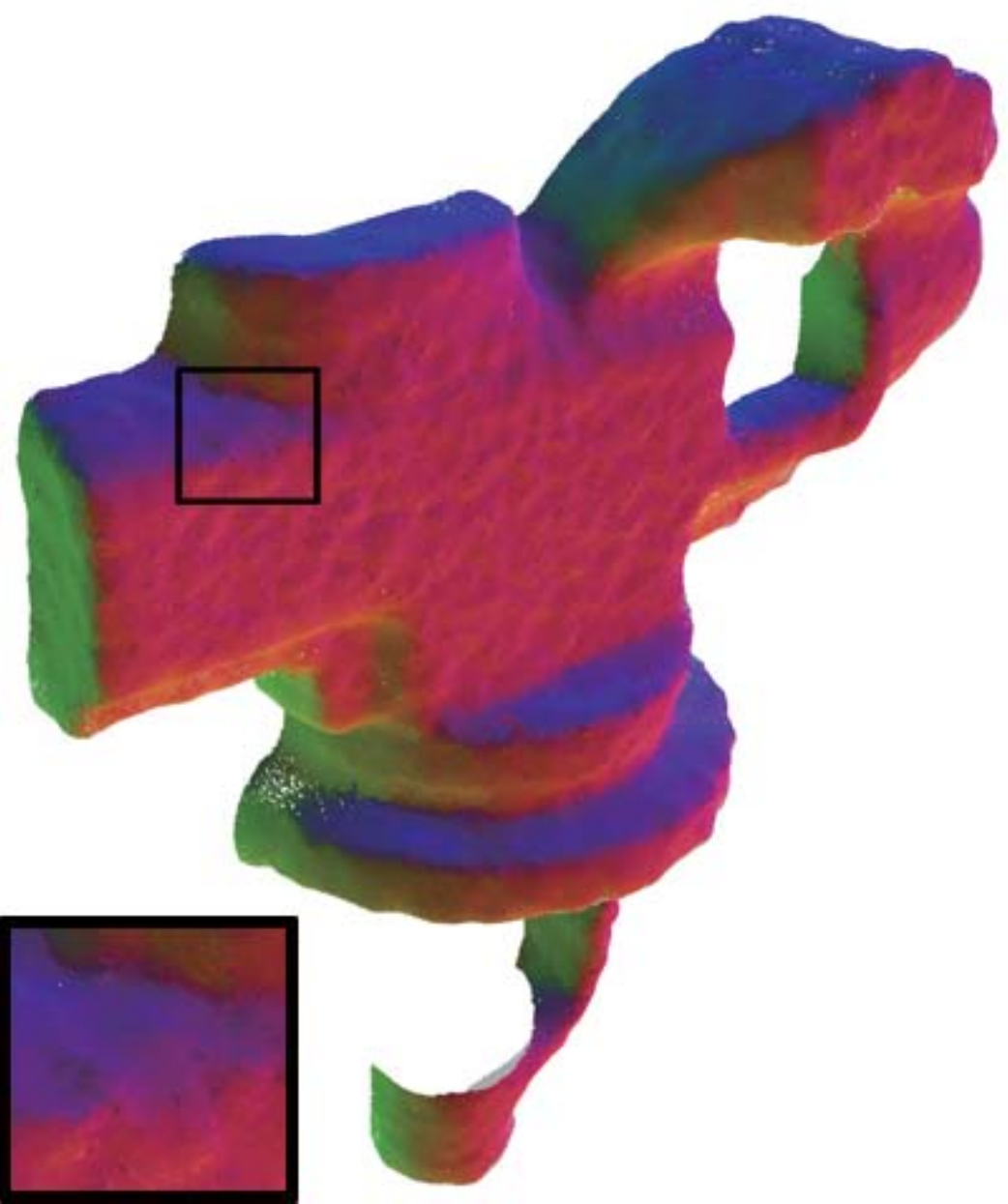}}
\end{minipage}
\begin{minipage}[b]{0.16\linewidth}
{\label{}\includegraphics[width=1\linewidth]{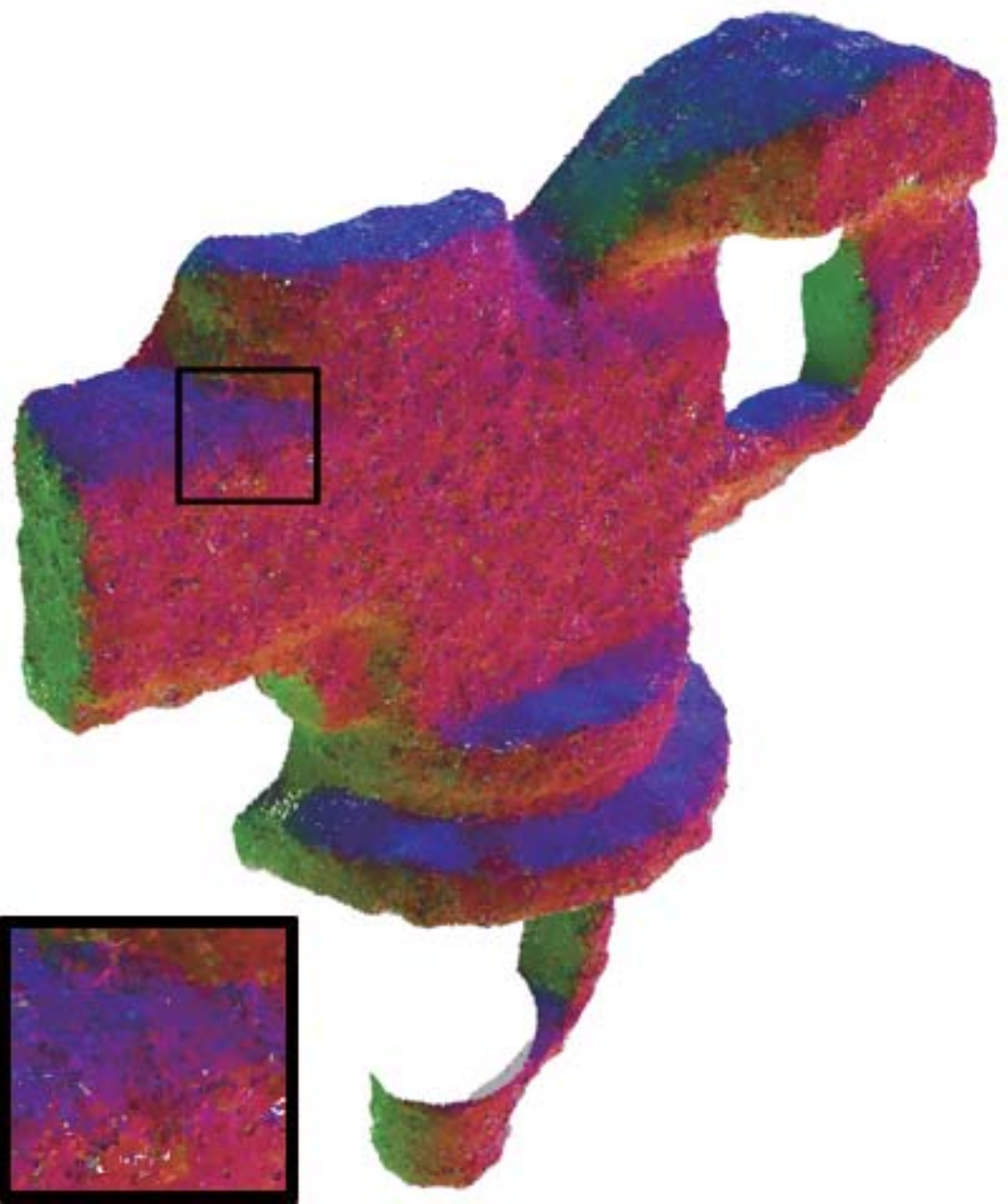}}
\end{minipage}
\begin{minipage}[b]{0.16\linewidth}
{\label{}\includegraphics[width=1\linewidth]{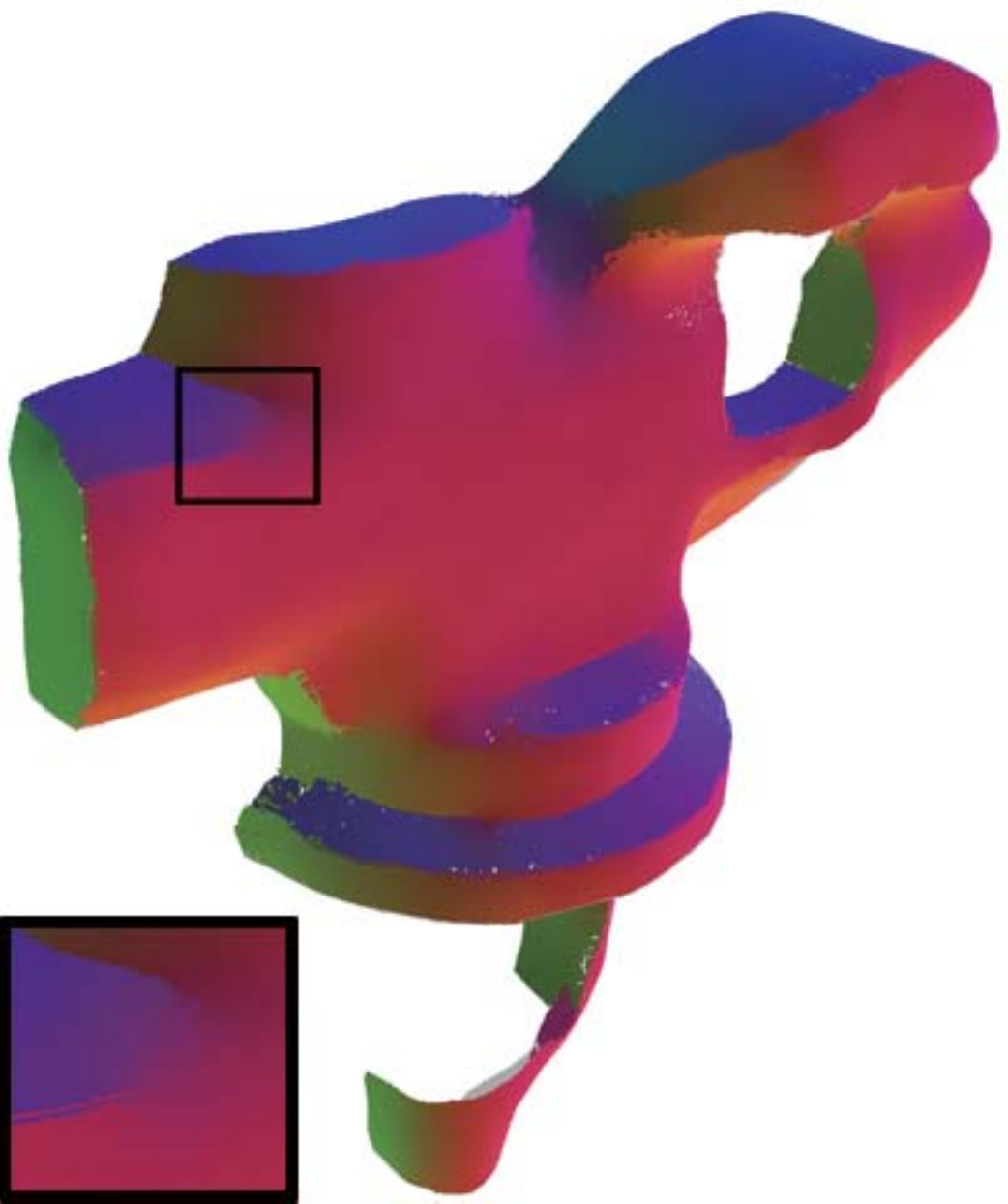}}
\end{minipage}
\begin{minipage}[b]{0.16\linewidth}
{\label{}\includegraphics[width=1\linewidth]{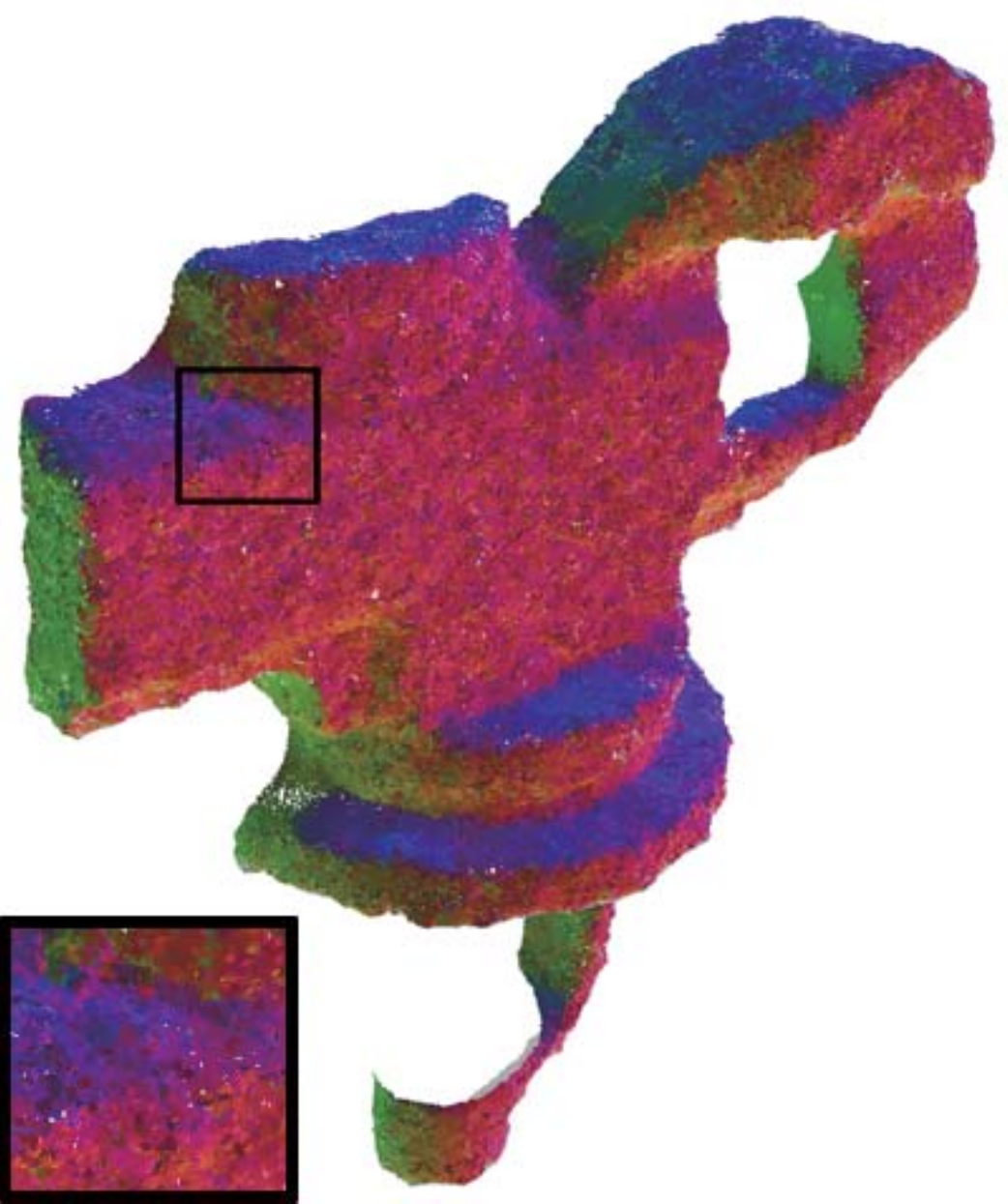}}
\end{minipage}
\begin{minipage}[b]{0.16\linewidth}
{\label{}\includegraphics[width=1\linewidth]{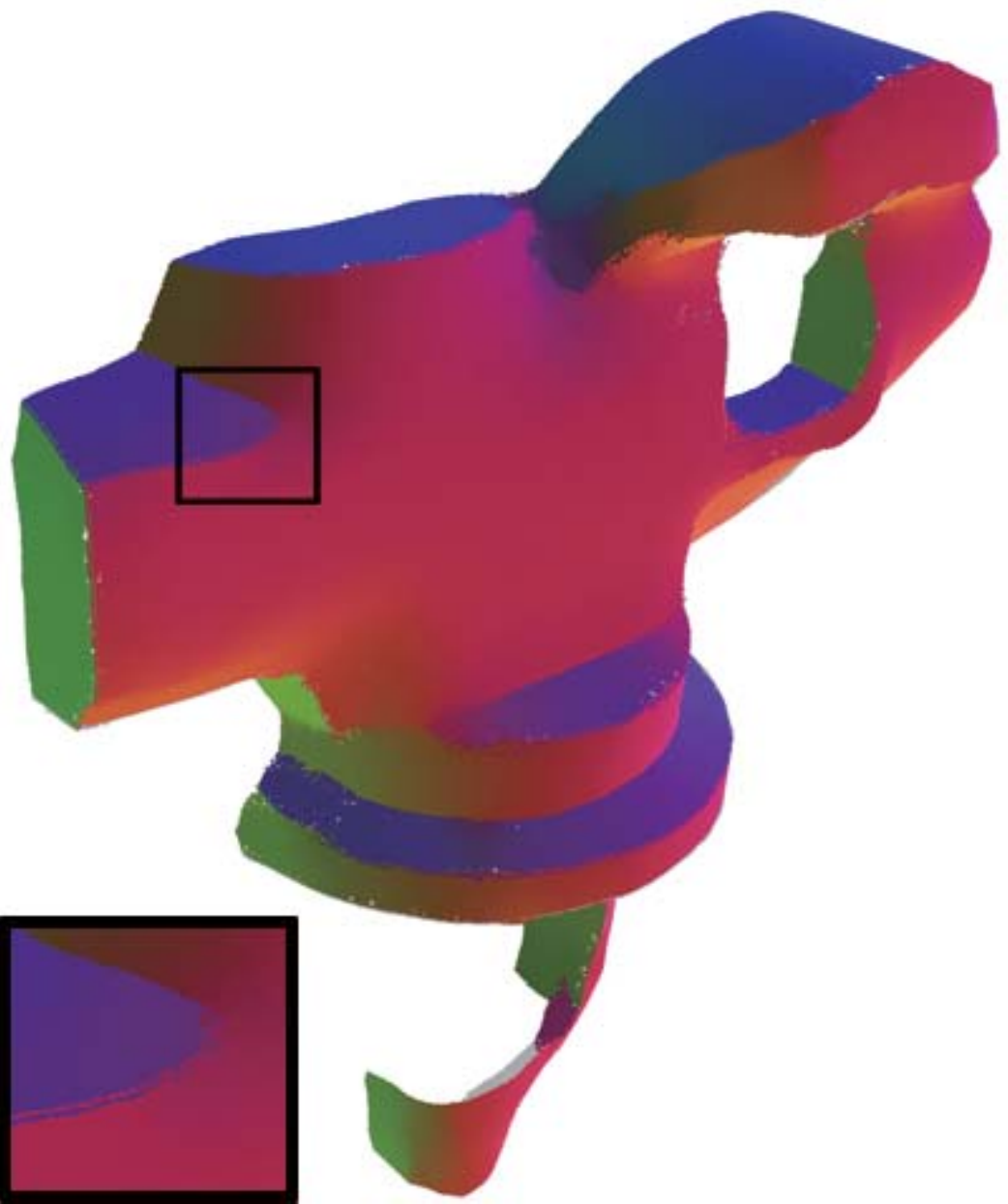}}
\end{minipage}	\\
\begin{minipage}[b]{0.16\linewidth}
\subfigure[\protect\cite{Hoppe1992}]{\label{}\includegraphics[width=1\linewidth]{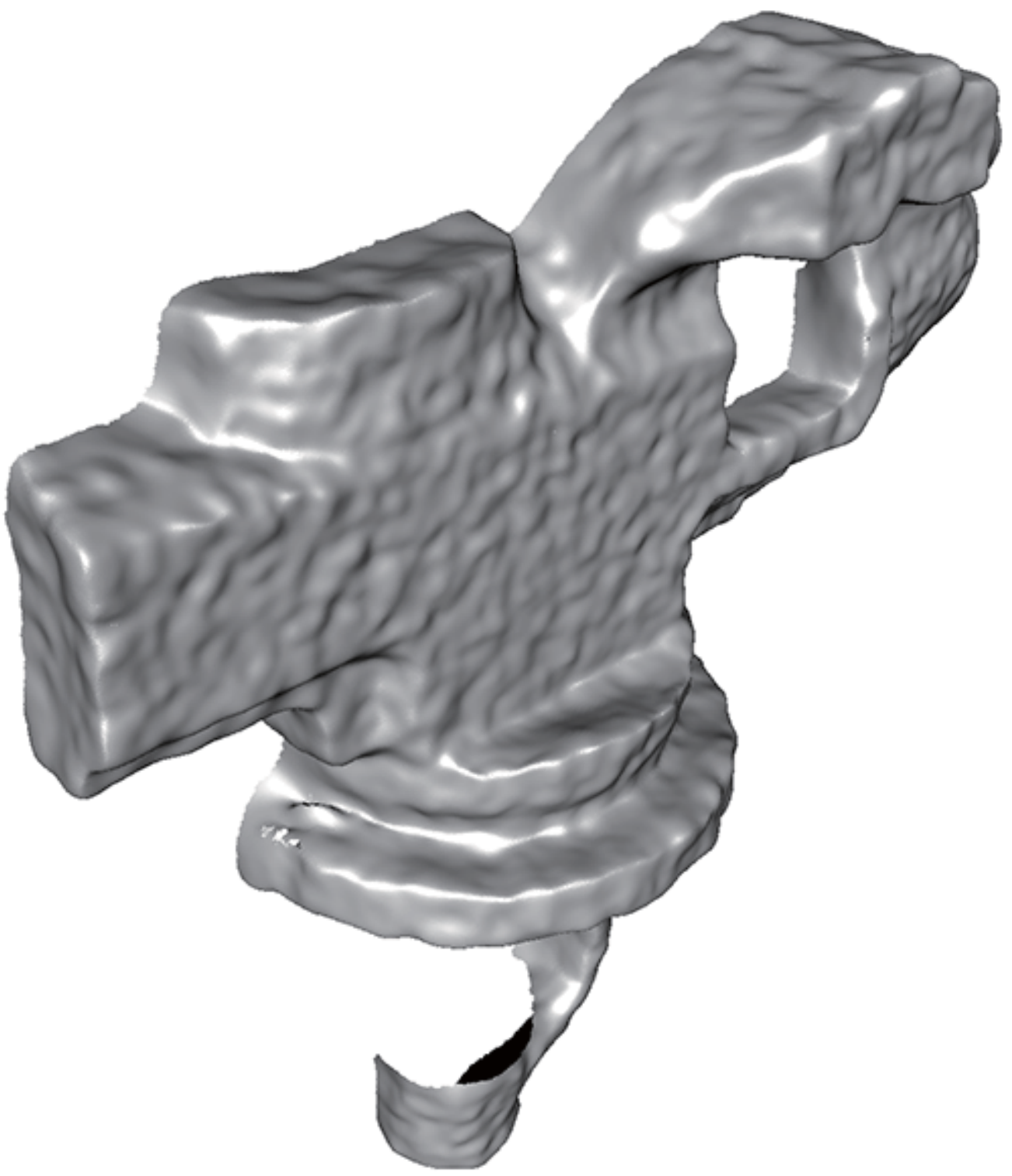}}
\end{minipage}
\begin{minipage}[b]{0.16\linewidth}
\subfigure[\protect\cite{Boulch2012}]{\label{}\includegraphics[width=1\linewidth]{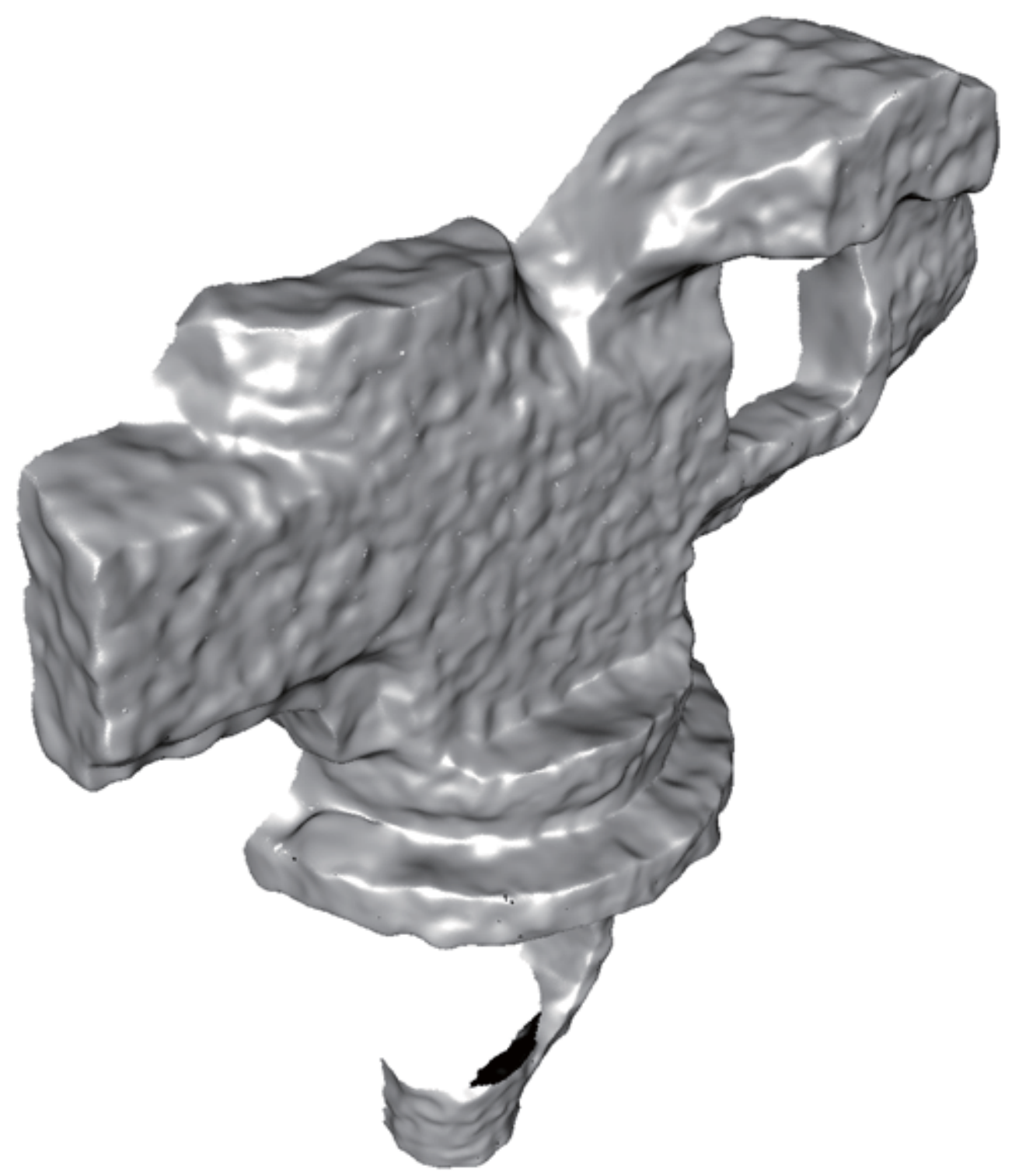}}
\end{minipage}
\begin{minipage}[b]{0.16\linewidth}
\subfigure[\protect\cite{Huang2013}]{\label{}\includegraphics[width=1\linewidth]{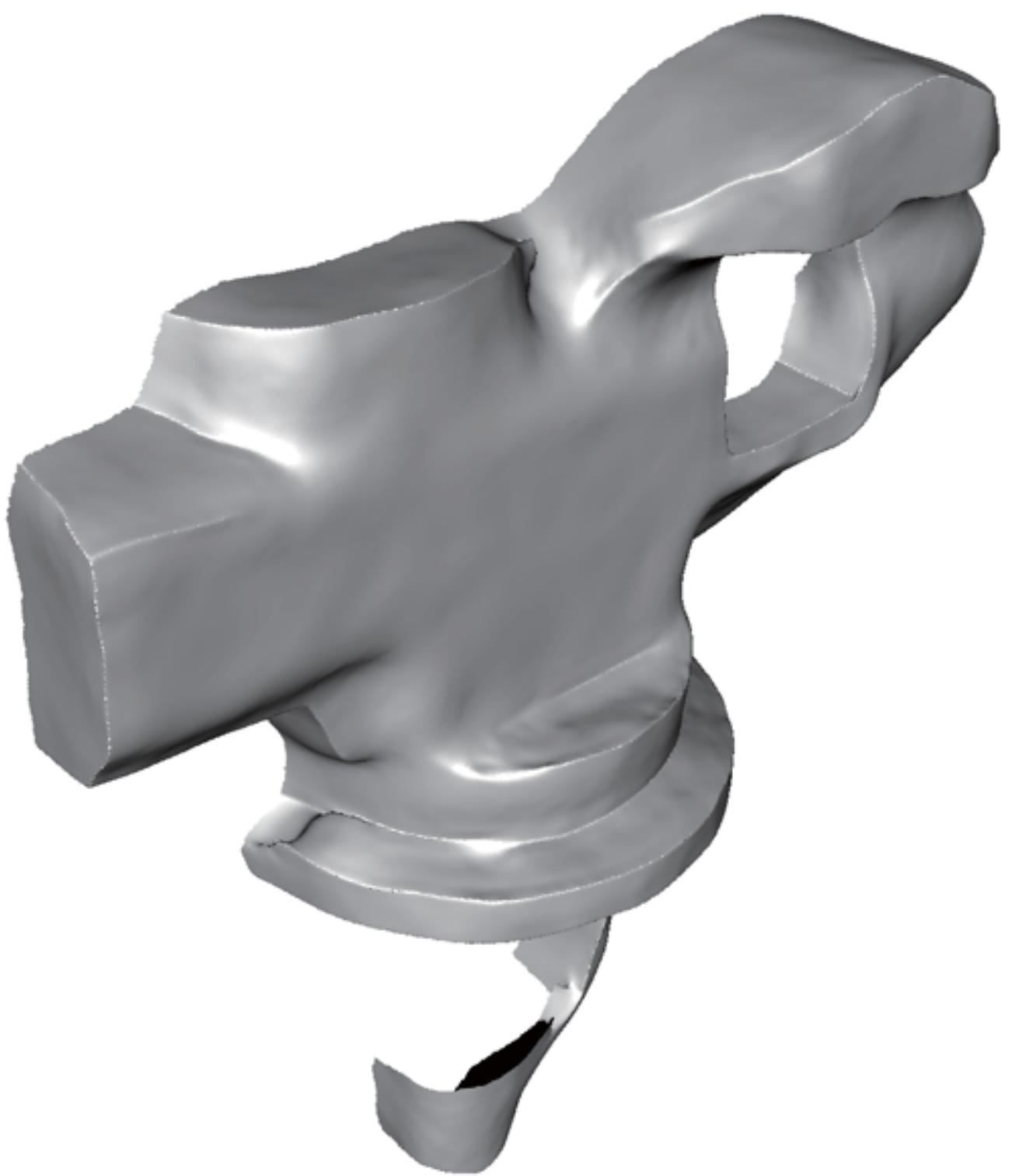}}
\end{minipage}
\begin{minipage}[b]{0.16\linewidth}
\subfigure[\protect\cite{Boulch2016}]{\label{}\includegraphics[width=1\linewidth]{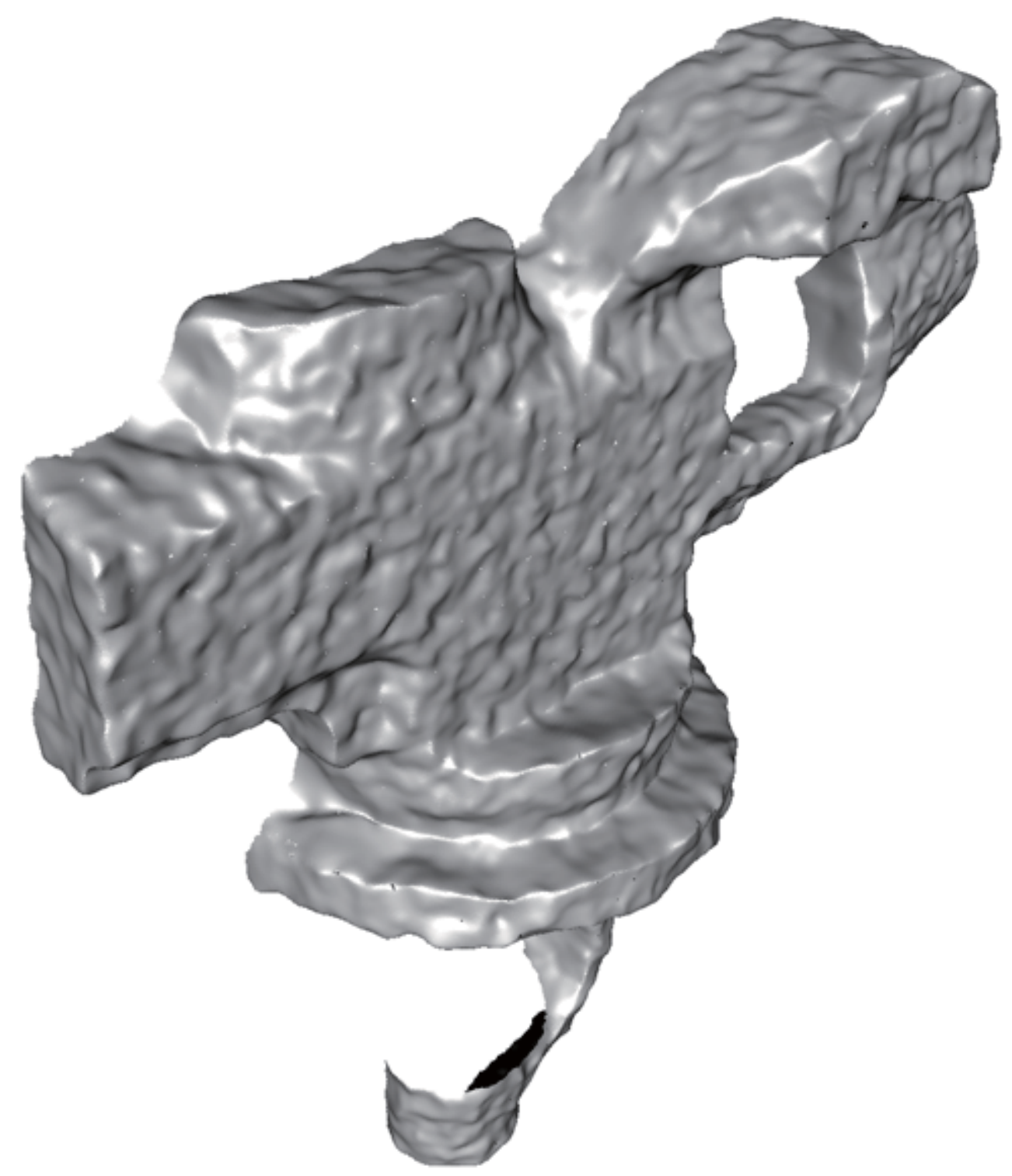}}
\end{minipage}
\begin{minipage}[b]{0.16\linewidth}
\subfigure[Ours]{\label{}\includegraphics[width=1\linewidth]{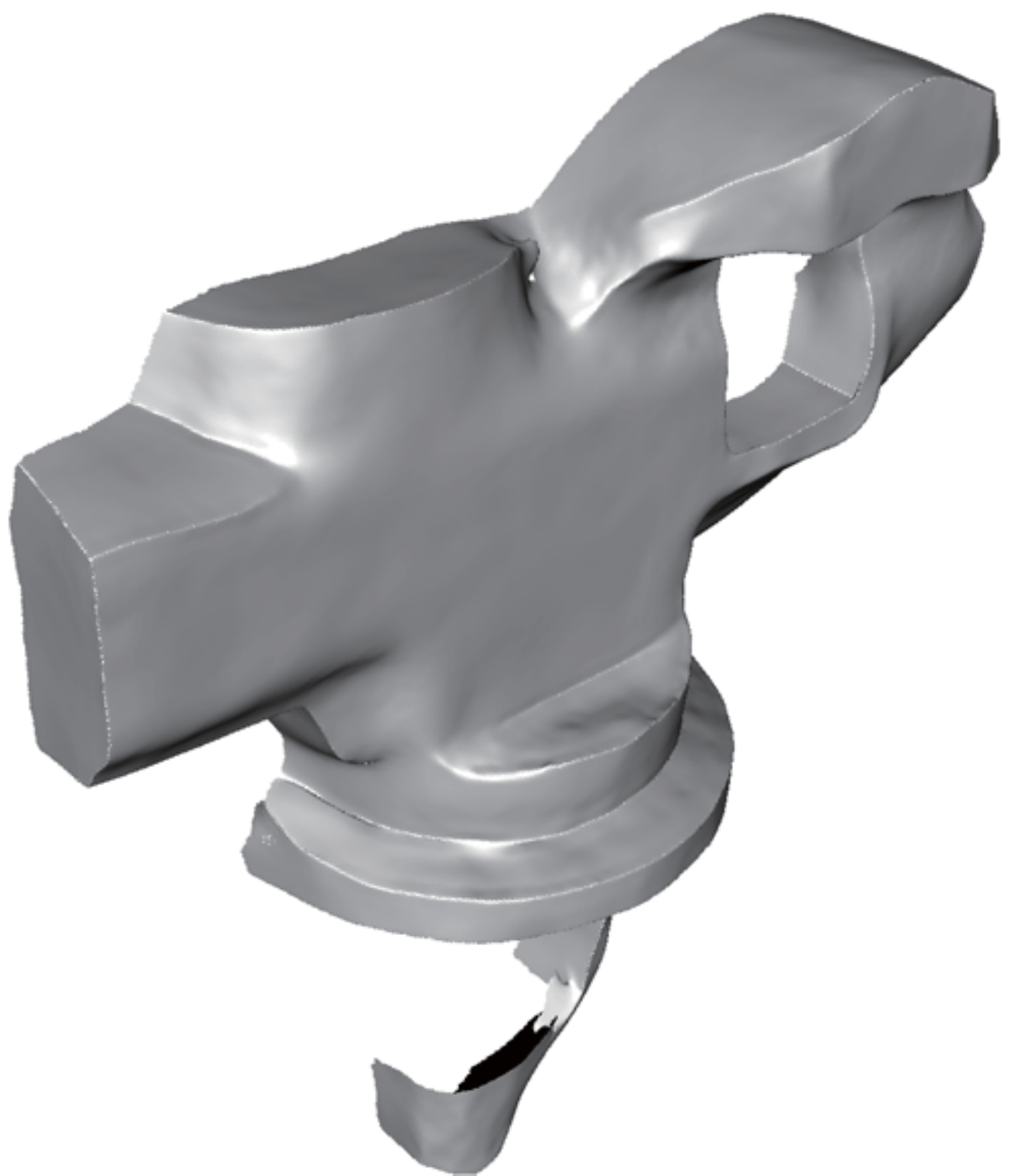}}
\end{minipage}
\caption{The first row: normal results of the scanned Iron point cloud. The second row: upsampling results of the filtered results by updating position with the normals in the first row. The third row: the corresponding surface reconstruction results.}
\label{fig:iron_point}
%\vspace{-0.65cm}
\end{figure*}

% parameter test: \theta_th, fixed and unfixed
\begin{figure}[thbp]
%\vspace{-0.0cm}
\centering
\begin{minipage}[b]{0.175\linewidth}
\subfigure[]{\label{}\includegraphics[width=1\linewidth]{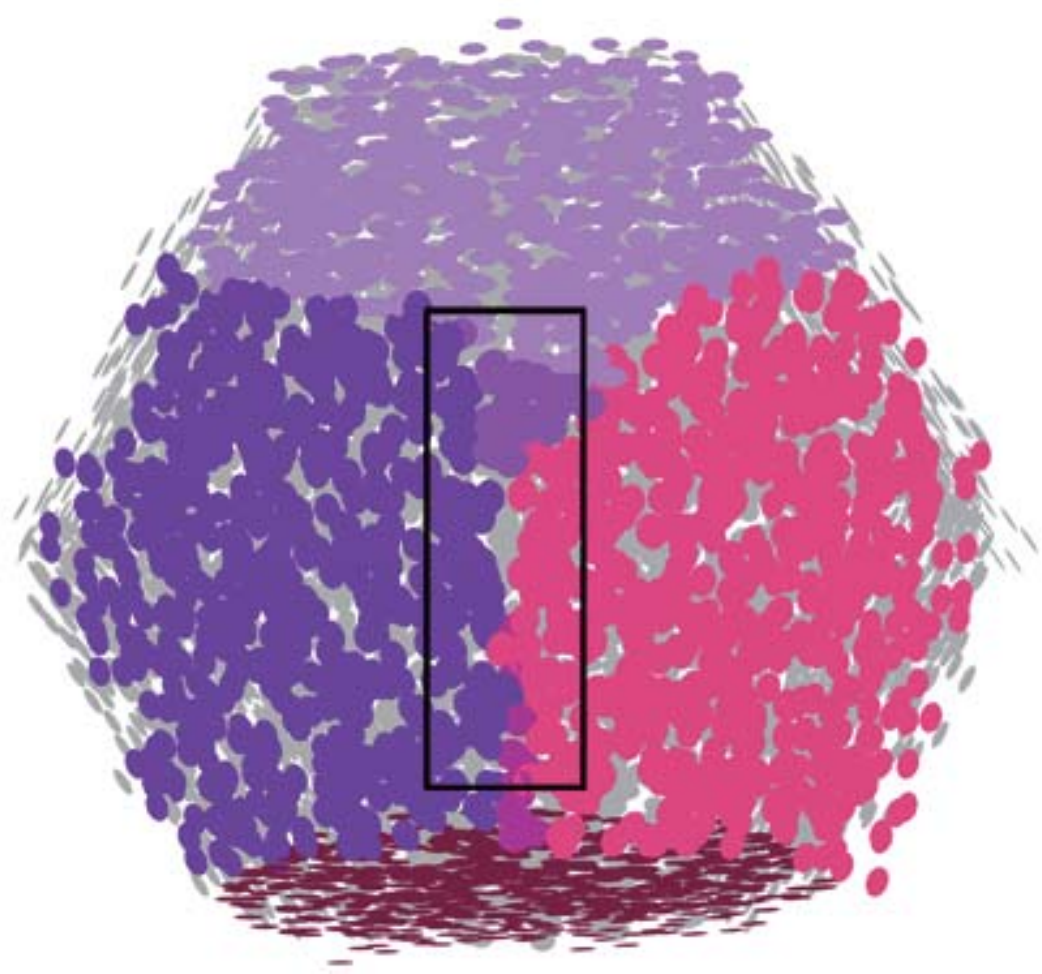}}
\end{minipage}
\begin{minipage}[b]{0.06\linewidth}
{\label{}\includegraphics[width=1\linewidth]{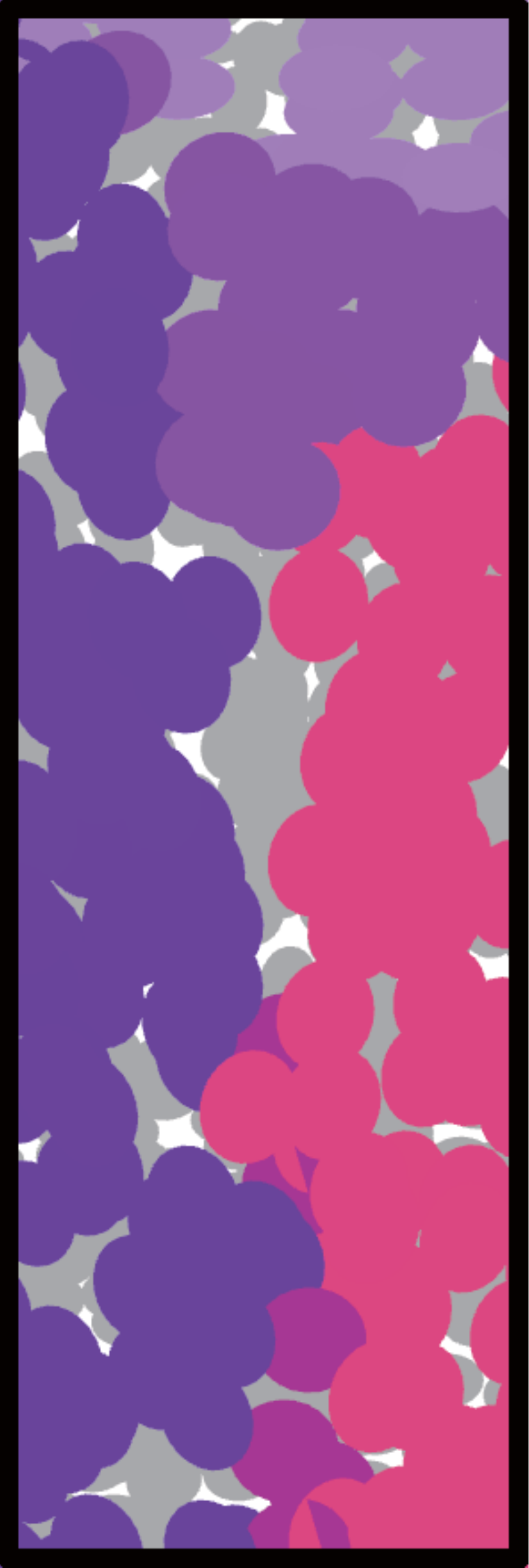}}
\end{minipage}
\begin{minipage}[b]{0.175\linewidth}
\subfigure[]{\label{}\includegraphics[width=1\linewidth]{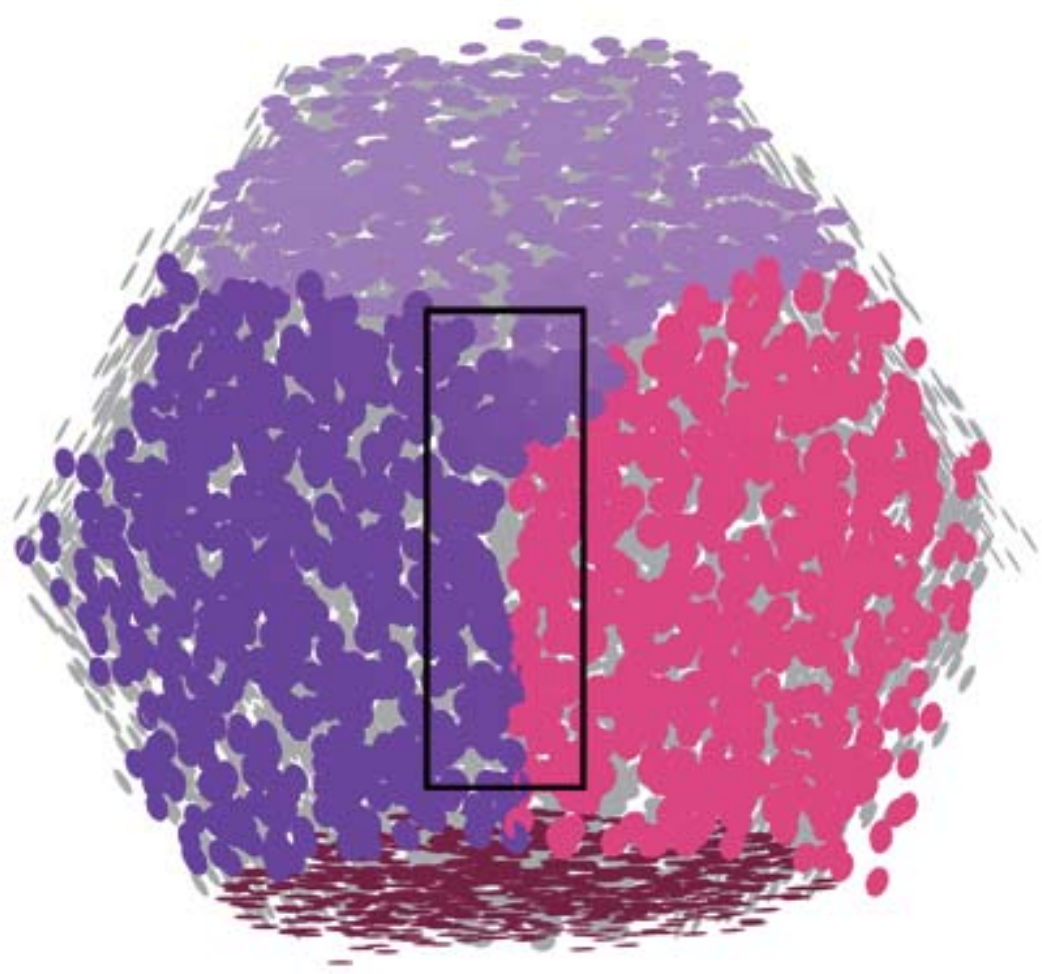}}
\end{minipage}
\begin{minipage}[b]{0.06\linewidth}
{\label{}\includegraphics[width=1\linewidth]{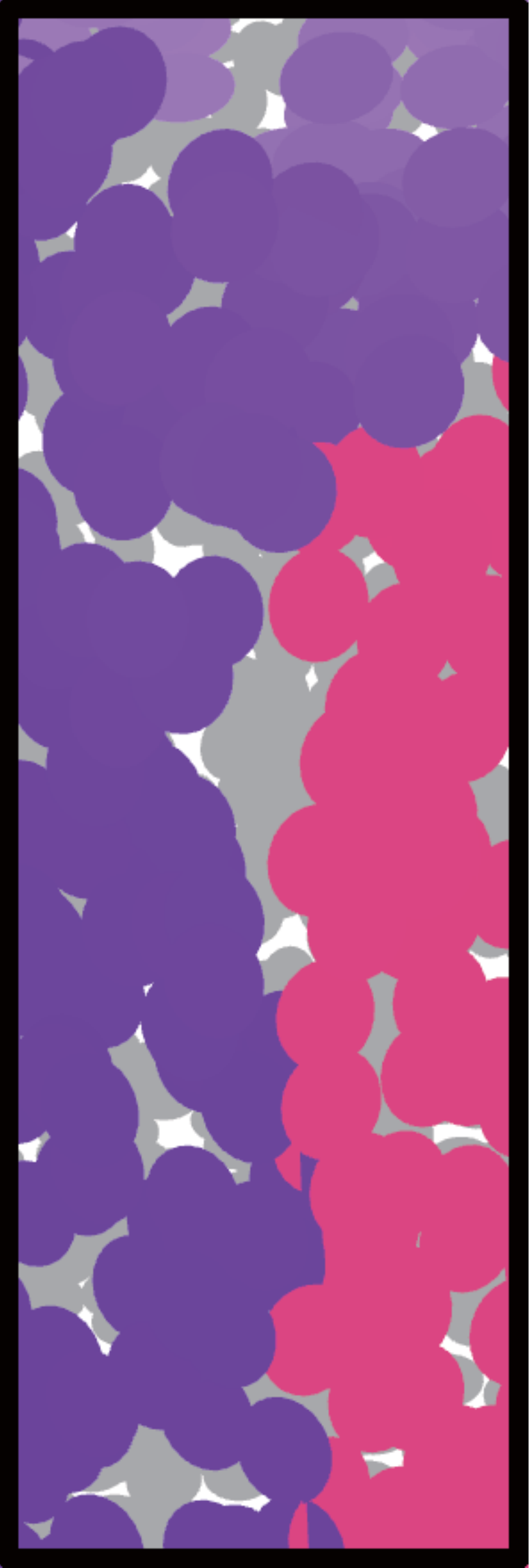}}
\end{minipage}
\begin{minipage}[b]{0.175\linewidth}
\subfigure[]{\label{}\includegraphics[width=1\linewidth]{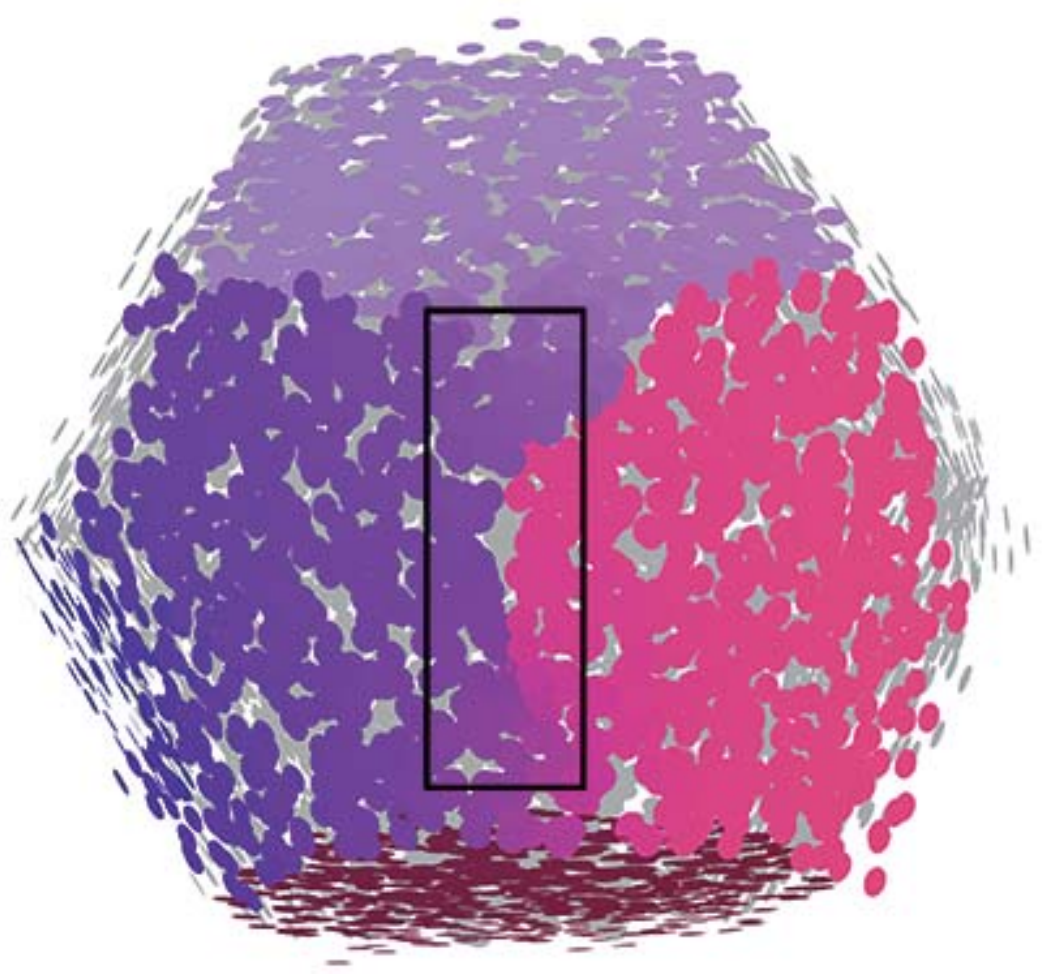}}
\end{minipage}
\begin{minipage}[b]{0.06\linewidth}
{\label{}\includegraphics[width=1\linewidth]{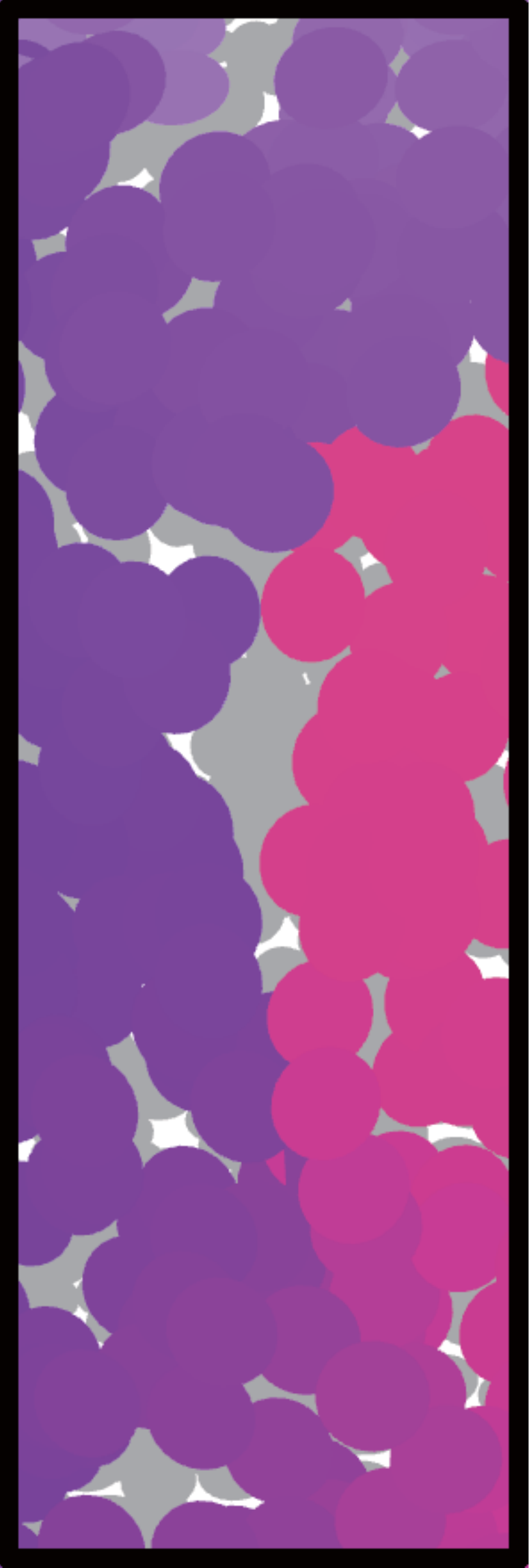}}
\end{minipage}
\begin{minipage}[b]{0.175\linewidth}
\subfigure[]{\label{}\includegraphics[width=1\linewidth]{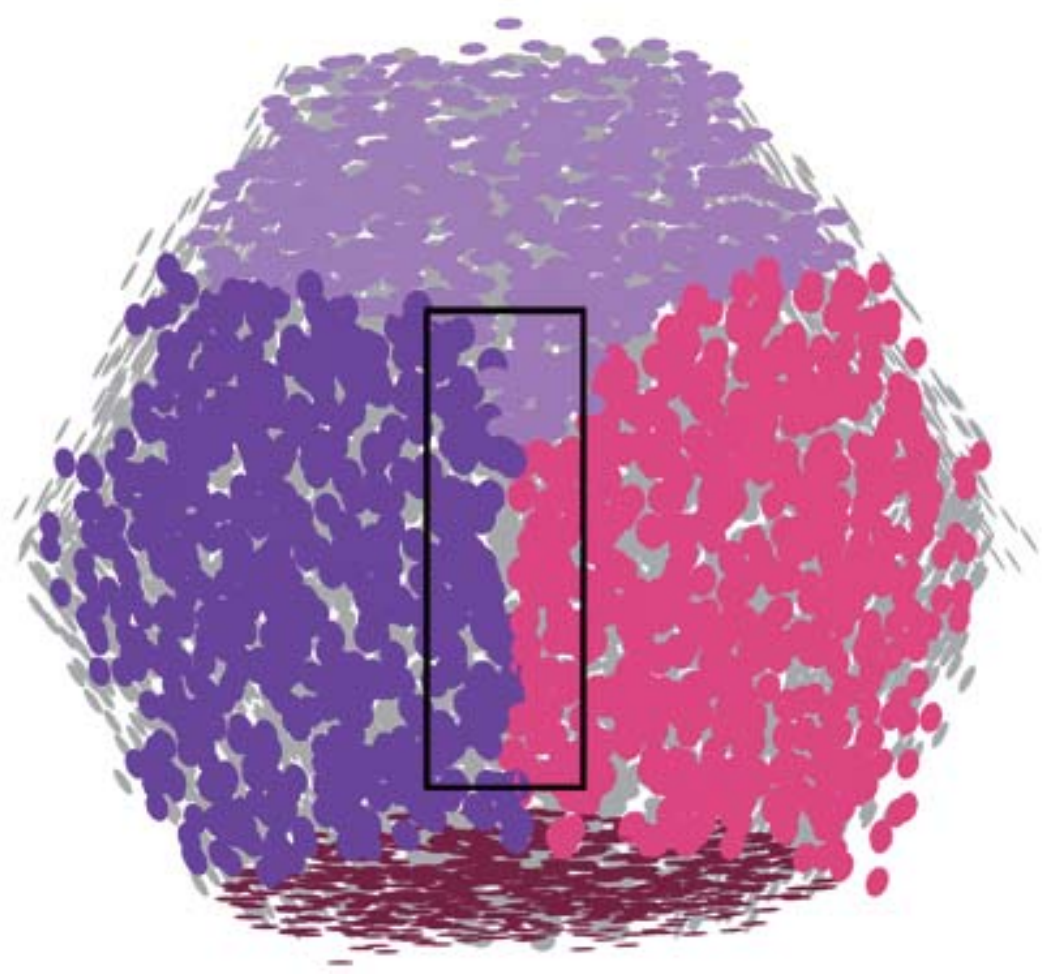}}
\end{minipage}
\begin{minipage}[b]{0.06\linewidth}
{\label{}\includegraphics[width=1\linewidth]{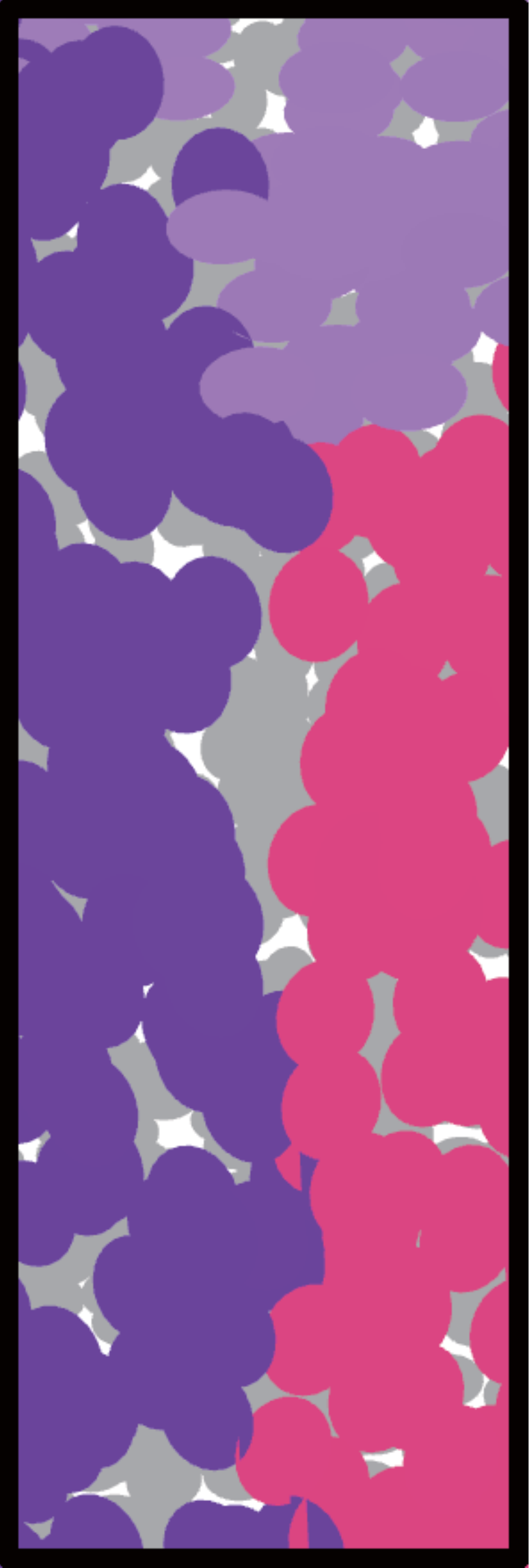}}
\end{minipage}
\caption{(a) and (b): two overly-sharpened results (more unique colors around the upper corner) by fixing $\theta_{th}$. (c) the smeared result (smoothly changed colors around the lower corner) by using a greater $\theta_{th}^{init}$. (d) The  result by using a smaller $\theta_{th}^{init}$. Zoom in to clearly observe the differences. }
\label{fig:dod_parametertest}
%\vspace{-0.65cm}
\end{figure}

%point cloud errors: synthetic noise
\begin{figure}[th]
%\vspace{-0.0cm}
\centering
\begin{minipage}[b]{0.47\linewidth}
\subfigure[Cube]{\label{}\includegraphics[width=1\linewidth]{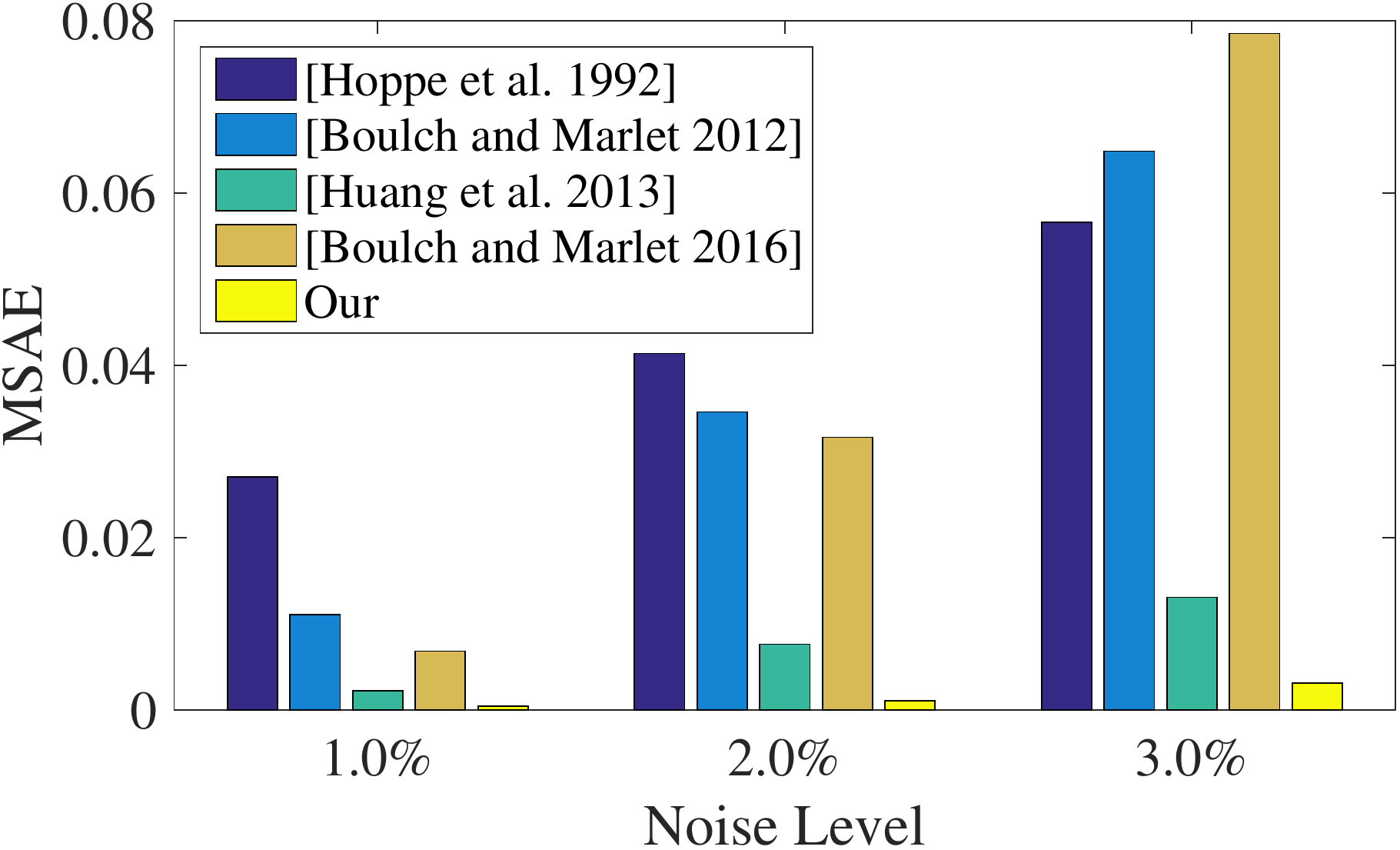}}
\end{minipage}
\begin{minipage}[b]{0.47\linewidth}
\subfigure[Dodecahedron]{\label{}\includegraphics[width=1\linewidth]{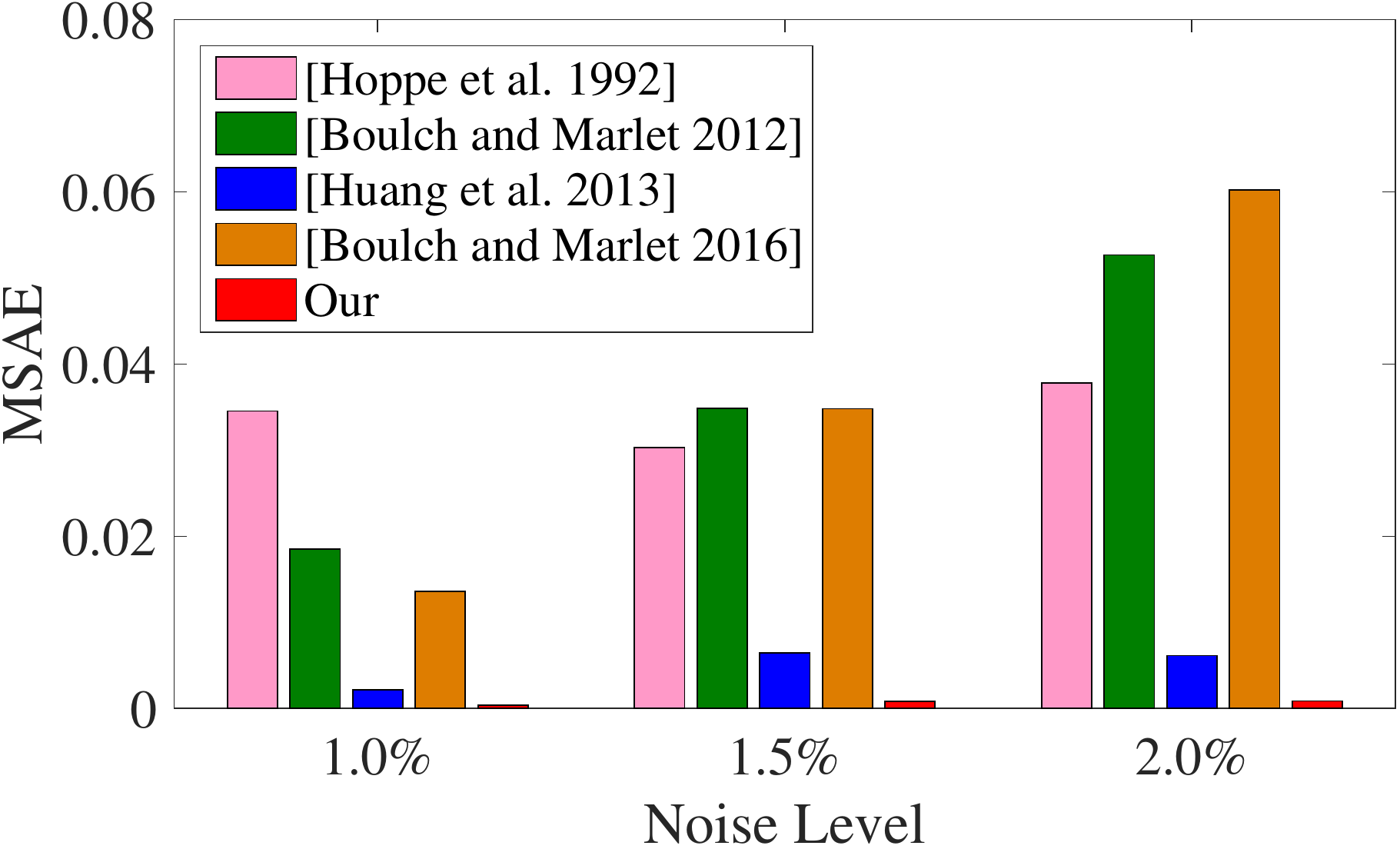}}
\end{minipage}
\caption{Normal errors (mean square angular error, in radians) of the Cube and Dodecahedron point sets corrupted with different levels of noise. }
\label{fig:syntheticnormalerror}
%\vspace{-0.65cm}
\end{figure}

%scanned: toy
\begin{figure*}[htbp]
%\vspace{-0.0cm}
\centering
\begin{minipage}[b]{0.16\linewidth}
{\label{}\includegraphics[width=1\linewidth]{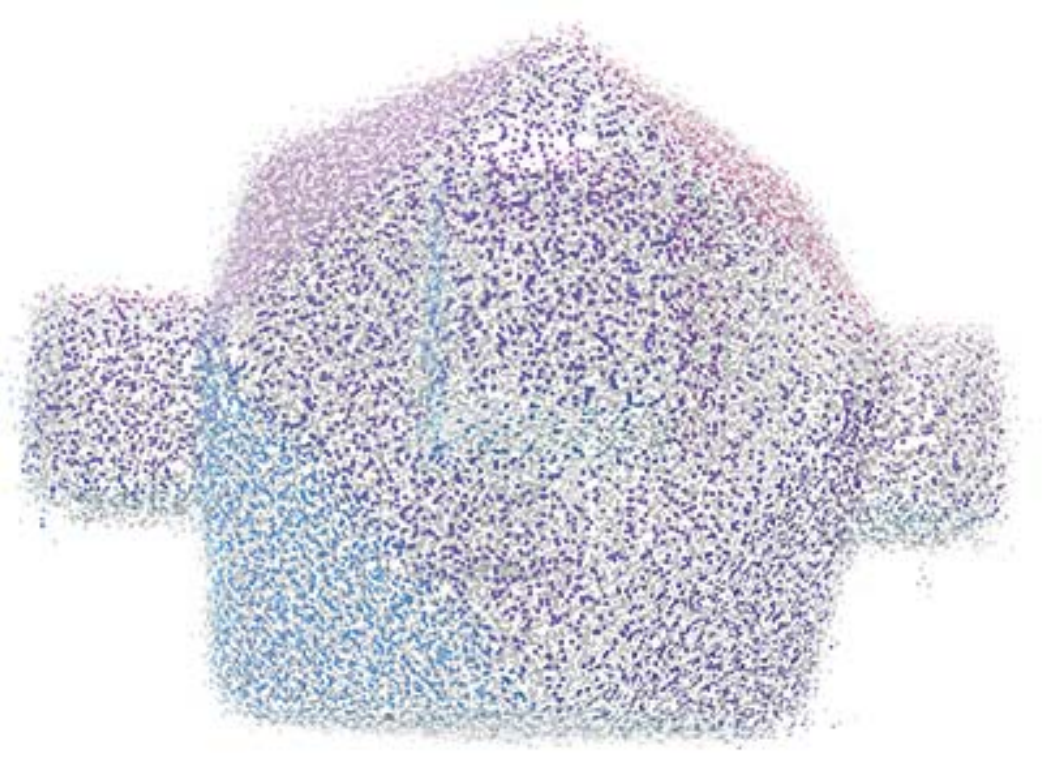}}
\end{minipage}
\begin{minipage}[b]{0.16\linewidth}
{\label{}\includegraphics[width=1\linewidth]{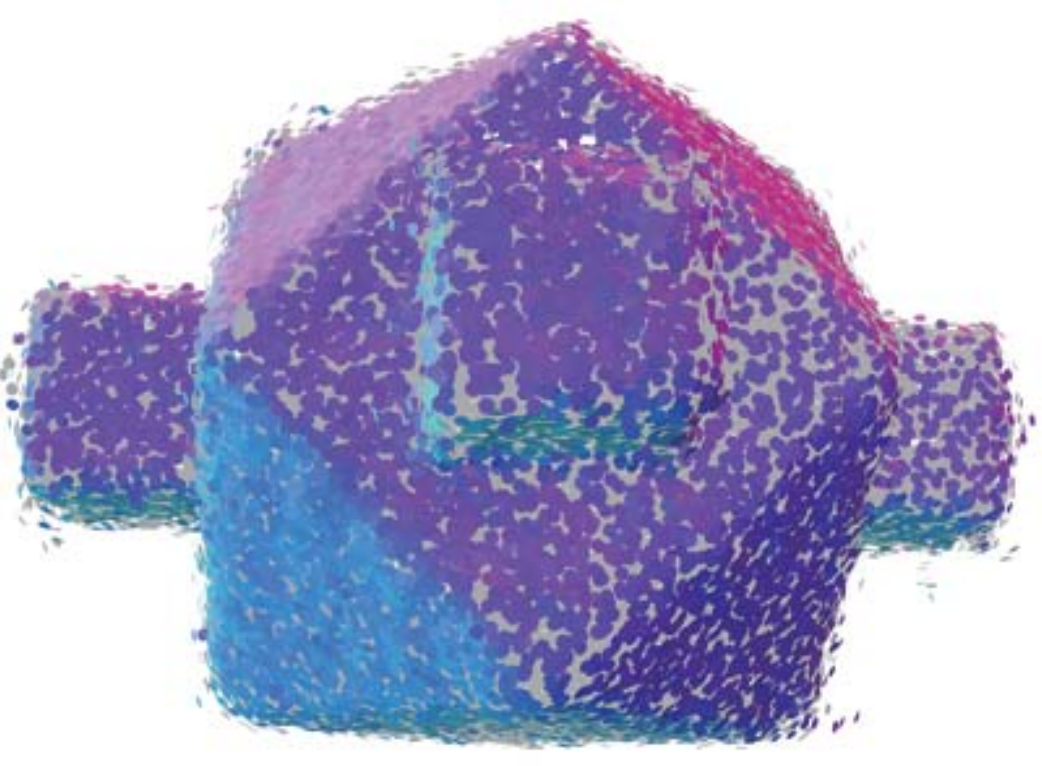}}
\end{minipage}
\begin{minipage}[b]{0.16\linewidth}
{\label{}\includegraphics[width=1\linewidth]{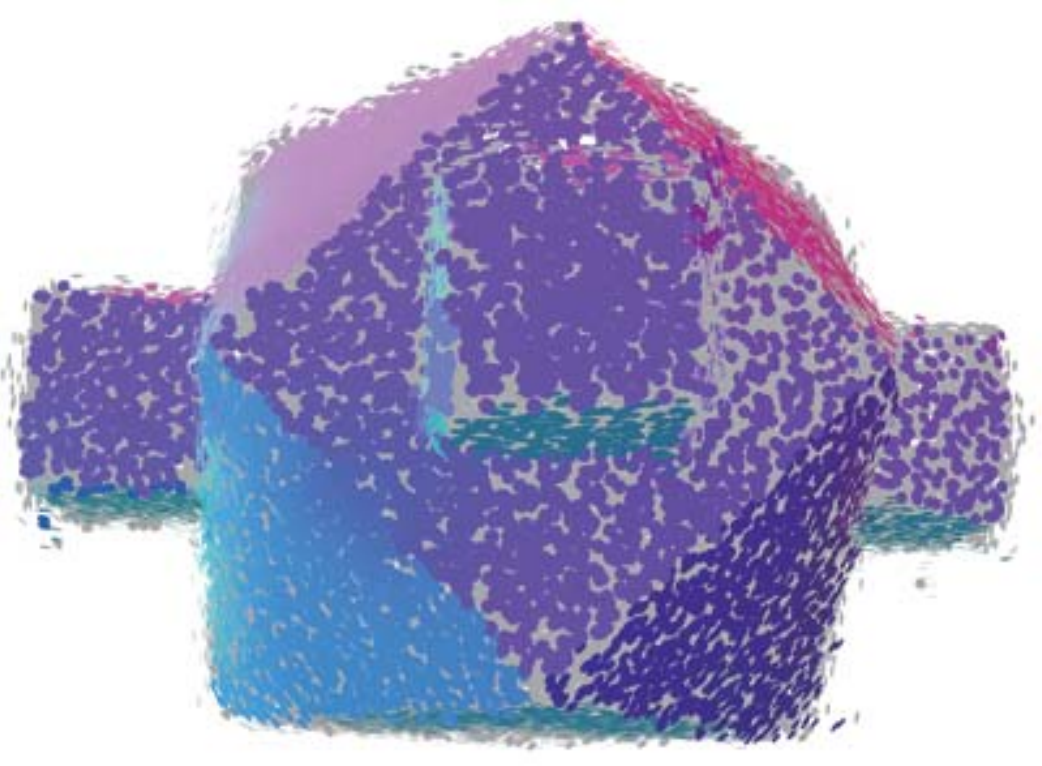}}
\end{minipage}
\begin{minipage}[b]{0.16\linewidth}
{\label{}\includegraphics[width=1\linewidth]{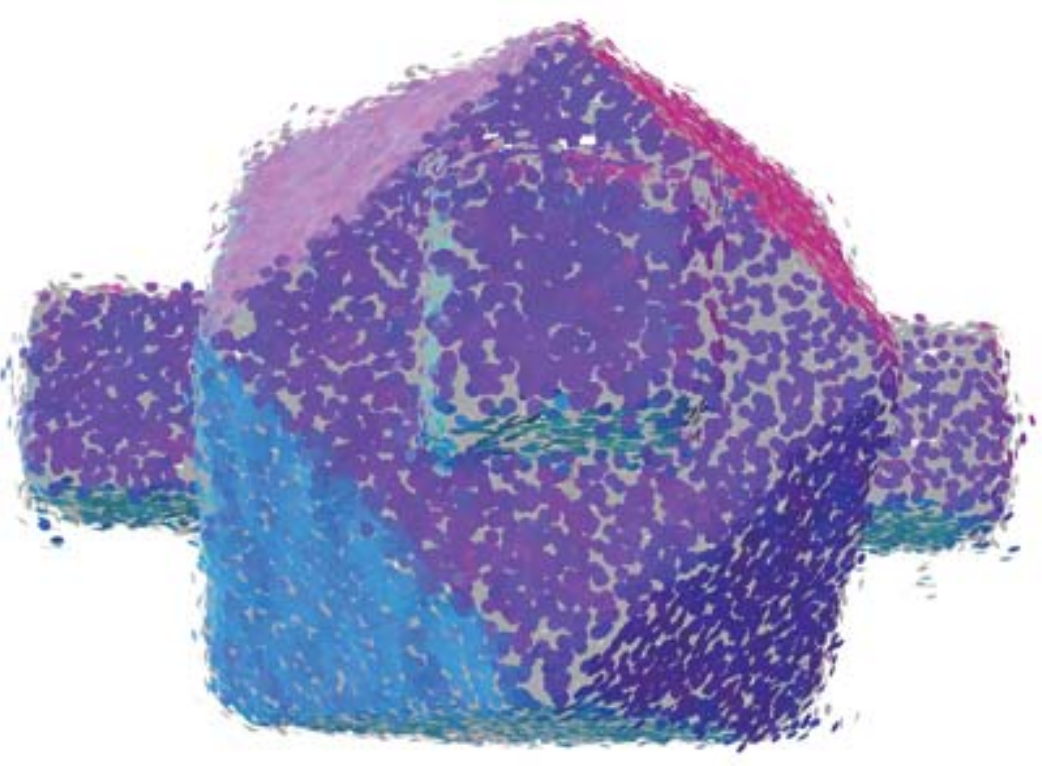}}
\end{minipage}
\begin{minipage}[b]{0.16\linewidth}
{\label{}\includegraphics[width=1\linewidth]{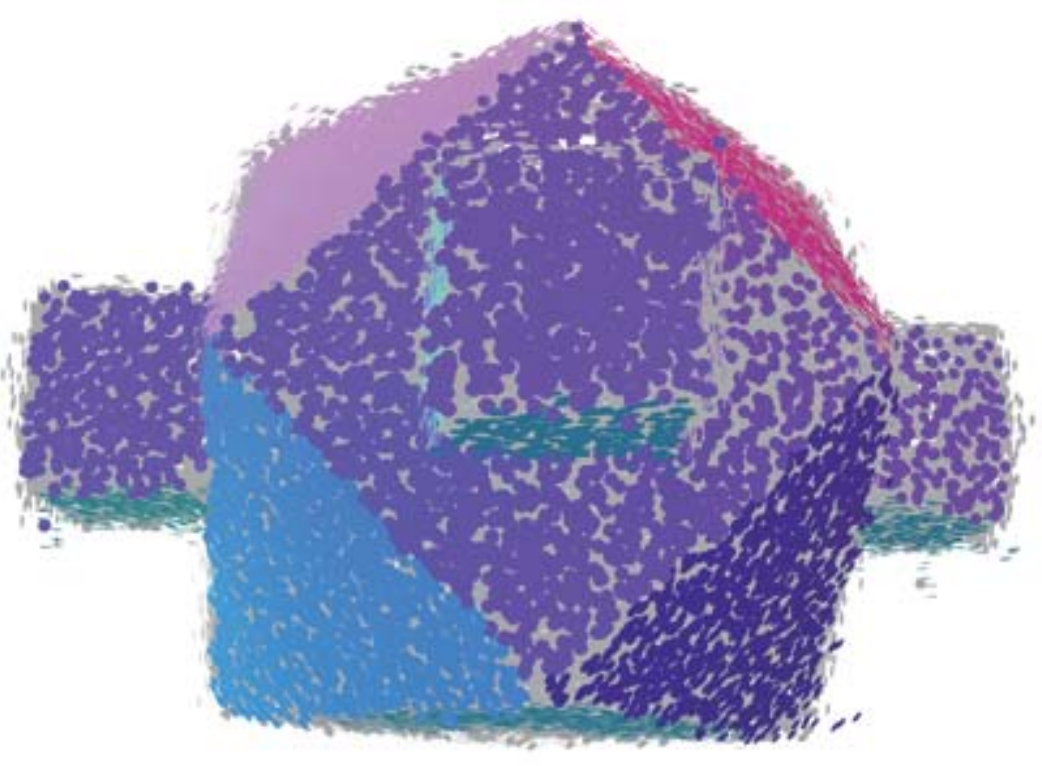}}
\end{minipage}	\\
\begin{minipage}[b]{0.16\linewidth}
{\label{}\includegraphics[width=1\linewidth]{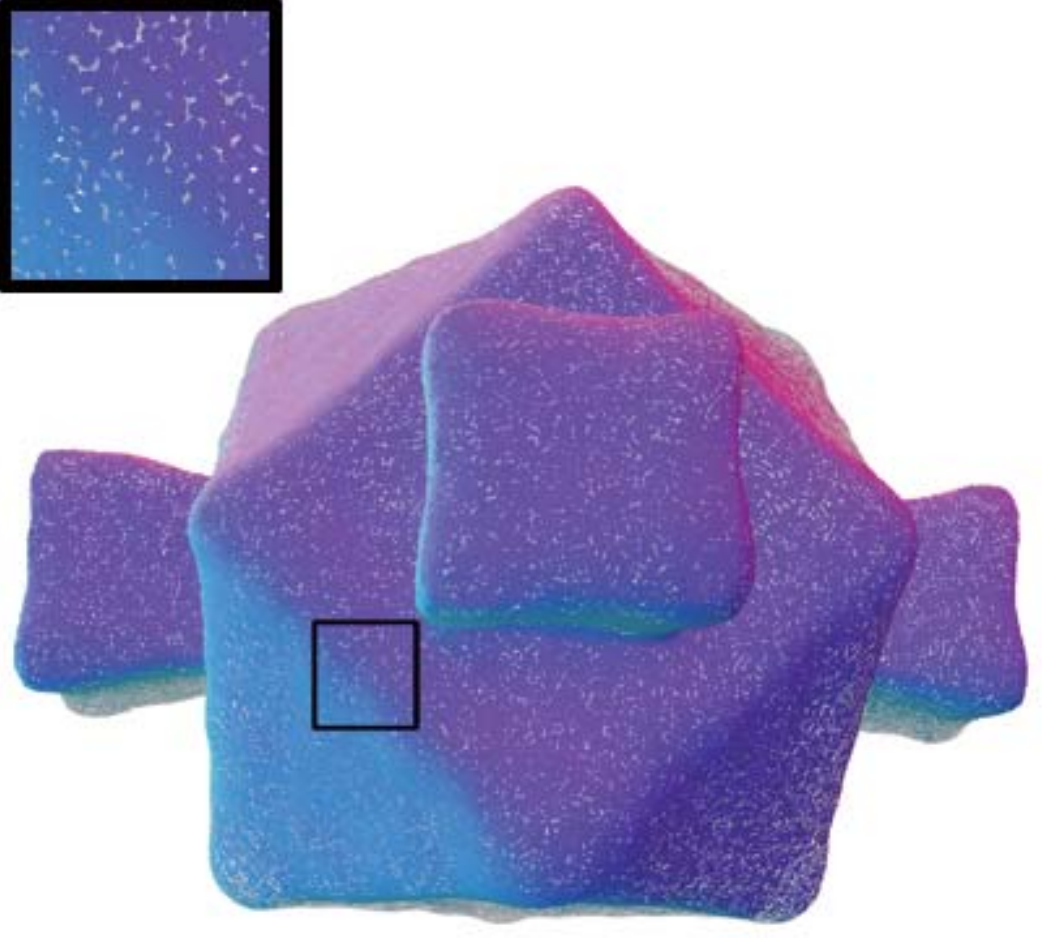}}
\end{minipage}
\begin{minipage}[b]{0.16\linewidth}
{\label{}\includegraphics[width=1\linewidth]{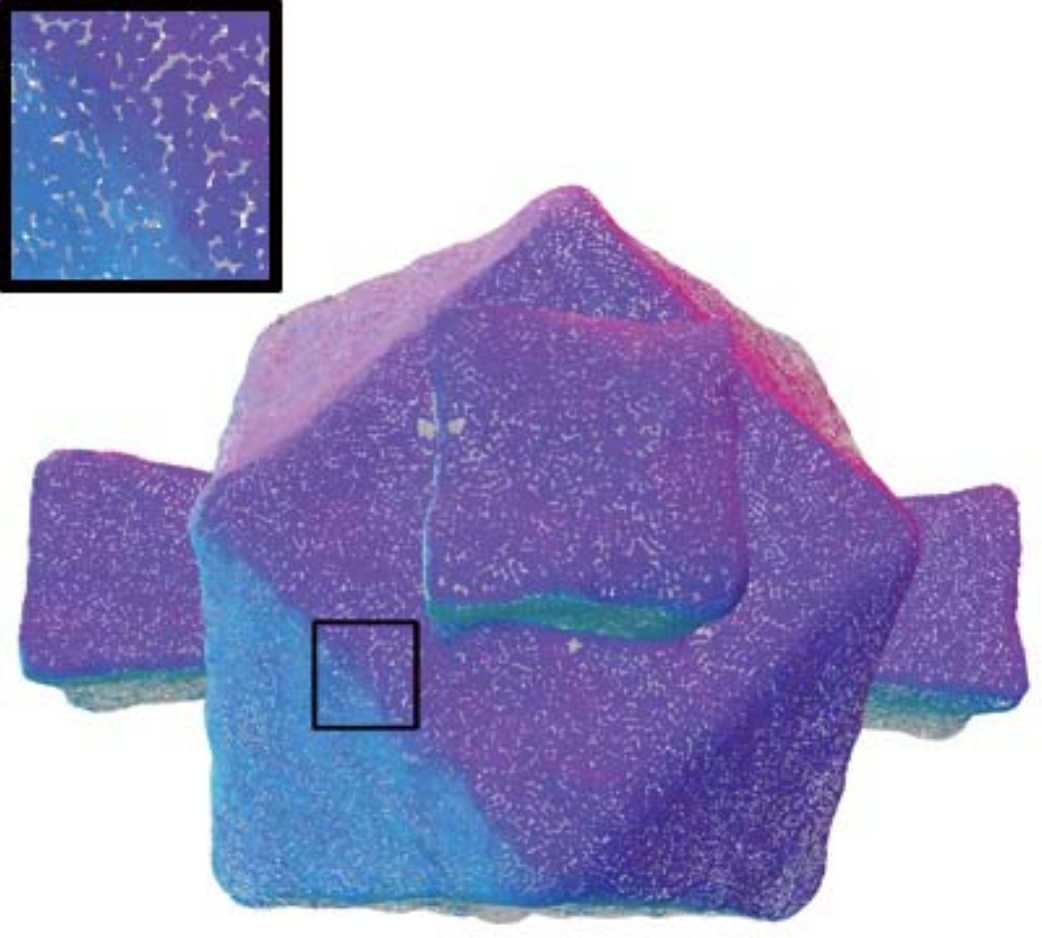}}
\end{minipage}
\begin{minipage}[b]{0.16\linewidth}
{\label{}\includegraphics[width=1\linewidth]{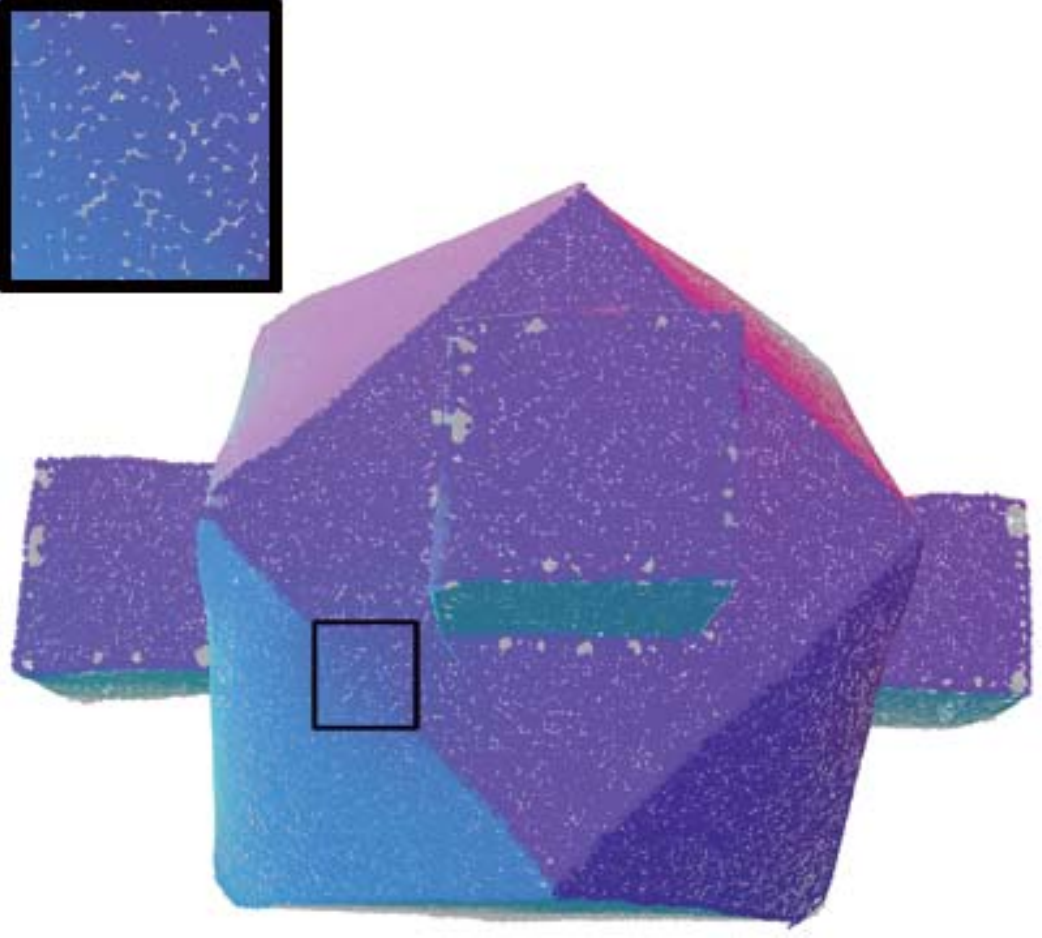}}
\end{minipage}
\begin{minipage}[b]{0.16\linewidth}
{\label{}\includegraphics[width=1\linewidth]{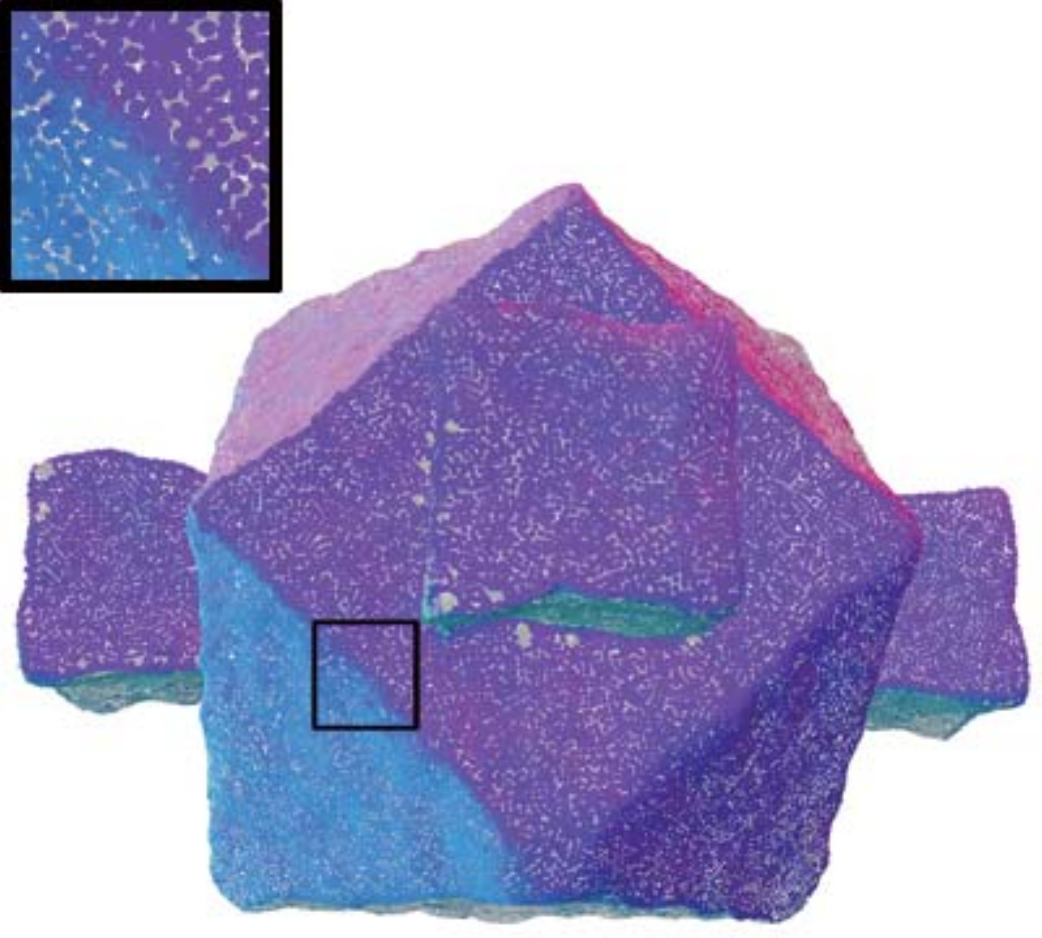}}
\end{minipage}
\begin{minipage}[b]{0.16\linewidth}
{\label{}\includegraphics[width=1\linewidth]{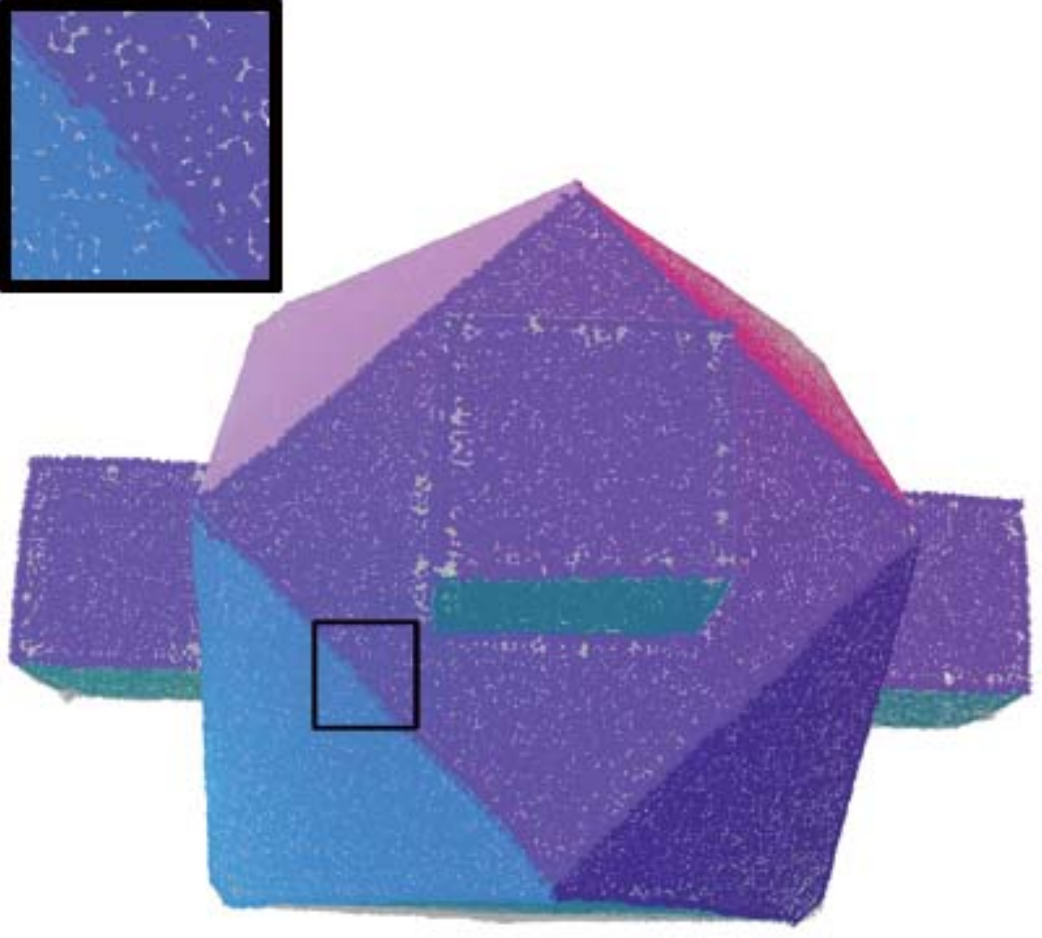}}
\end{minipage}	\\
\begin{minipage}[b]{0.16\linewidth}
\subfigure[\protect\cite{Hoppe1992}]{\label{}\includegraphics[width=1\linewidth]{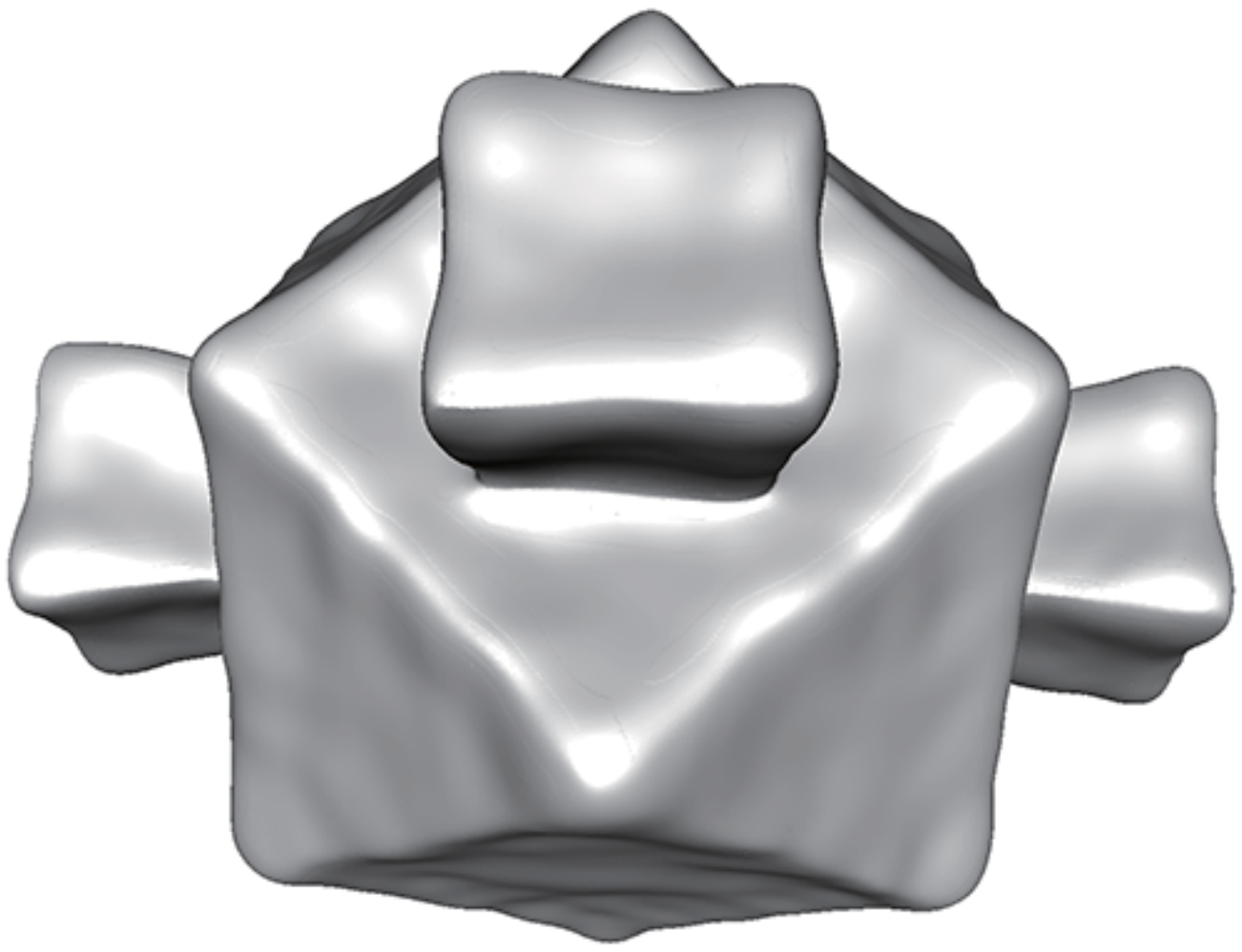}}
\end{minipage}
\begin{minipage}[b]{0.16\linewidth}
\subfigure[\protect\cite{Boulch2012}]{\label{}\includegraphics[width=1\linewidth]{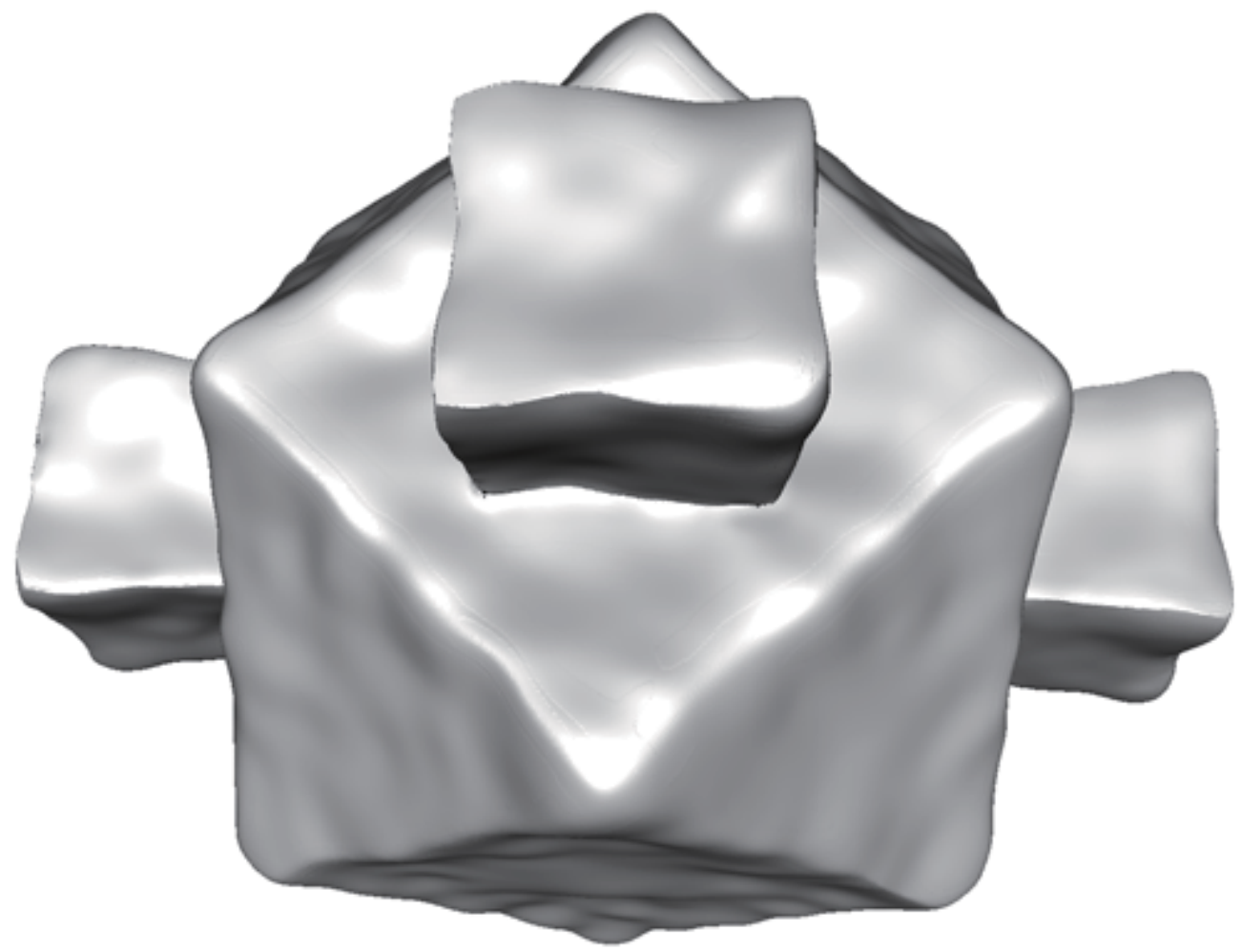}}
\end{minipage}
\begin{minipage}[b]{0.16\linewidth}
\subfigure[\protect\cite{Huang2013}]{\label{}\includegraphics[width=1\linewidth]{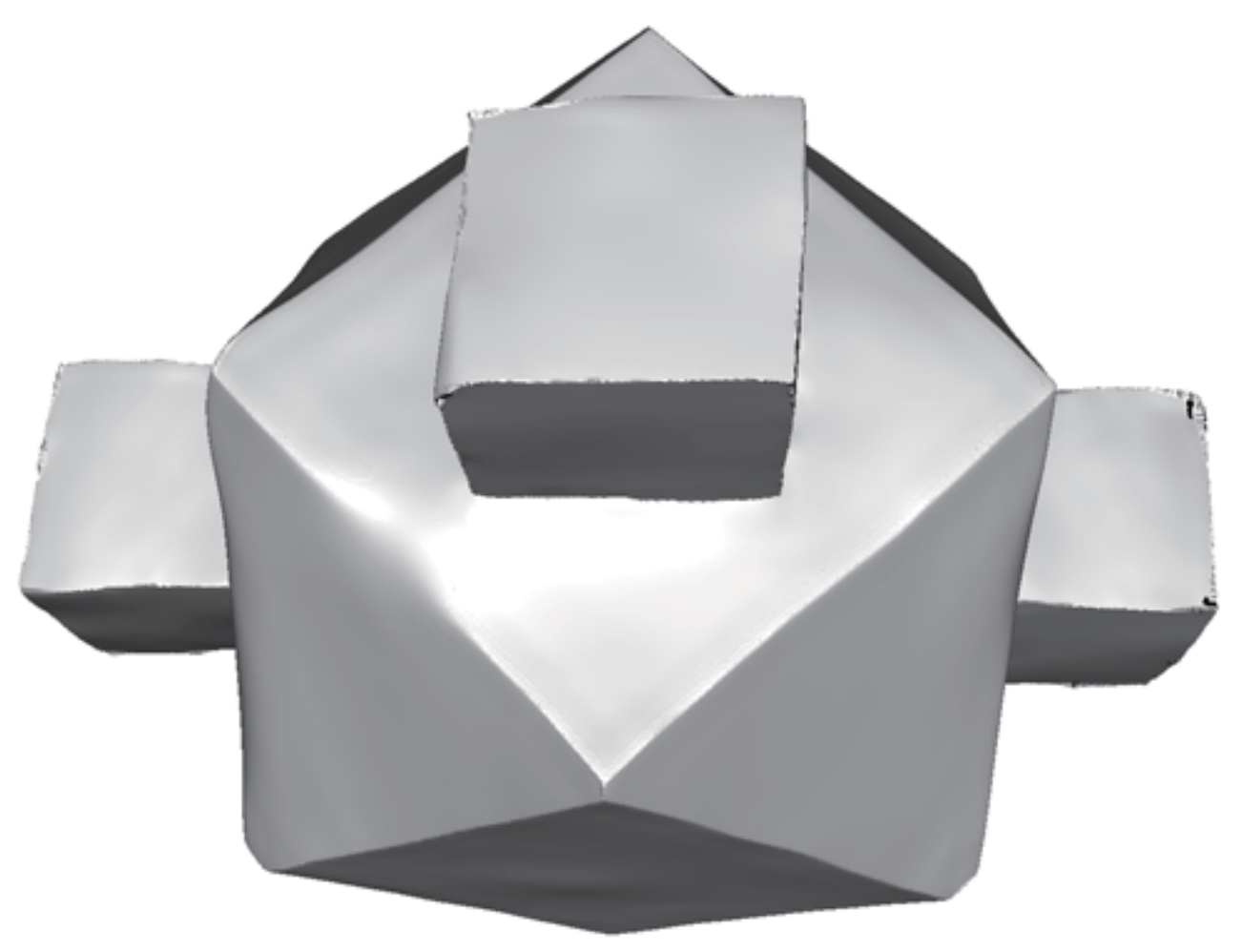}}
\end{minipage}
\begin{minipage}[b]{0.16\linewidth}
\subfigure[\protect\cite{Boulch2016}]{\label{}\includegraphics[width=1\linewidth]{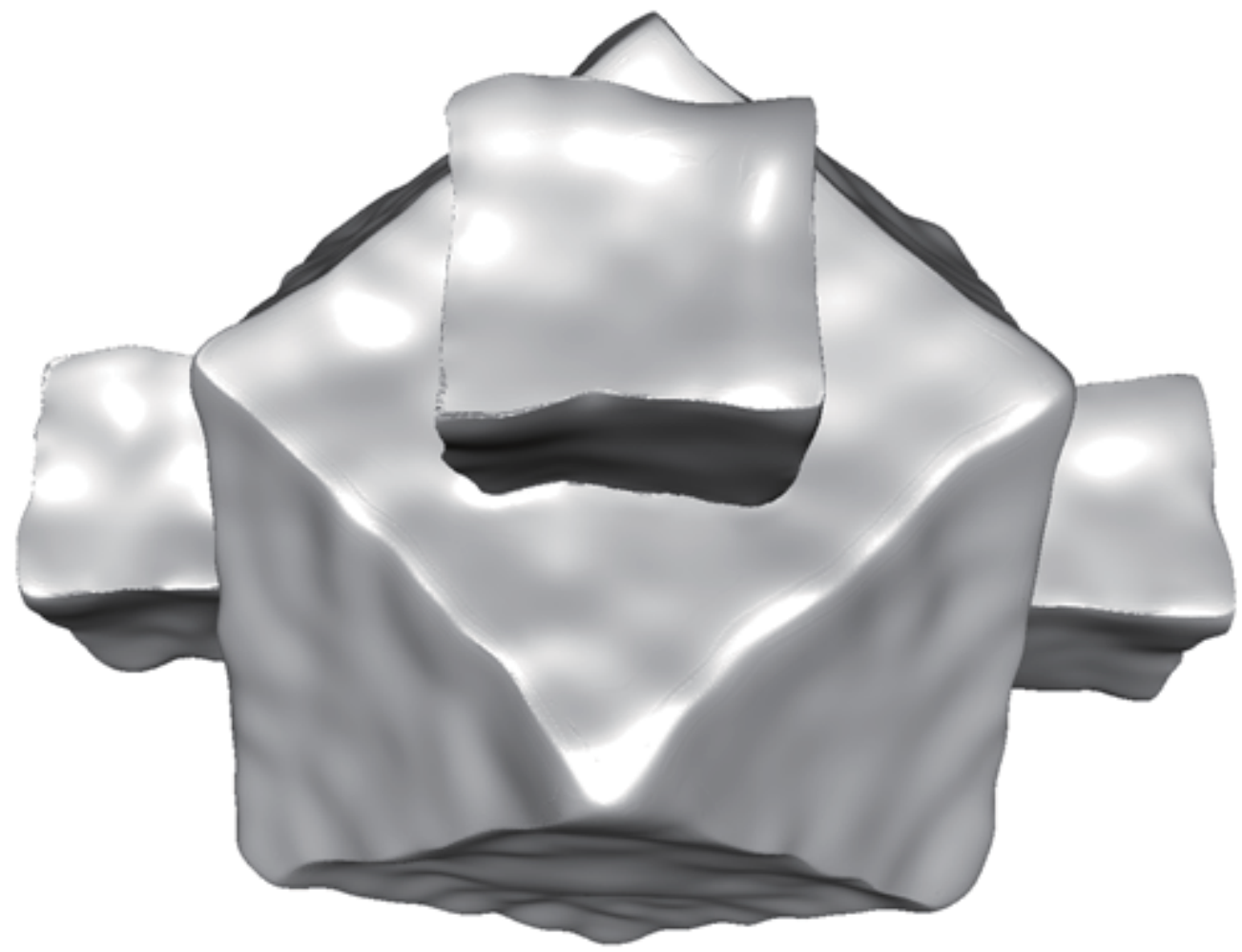}}
\end{minipage}
\begin{minipage}[b]{0.16\linewidth}
\subfigure[Ours]{\label{}\includegraphics[width=1\linewidth]{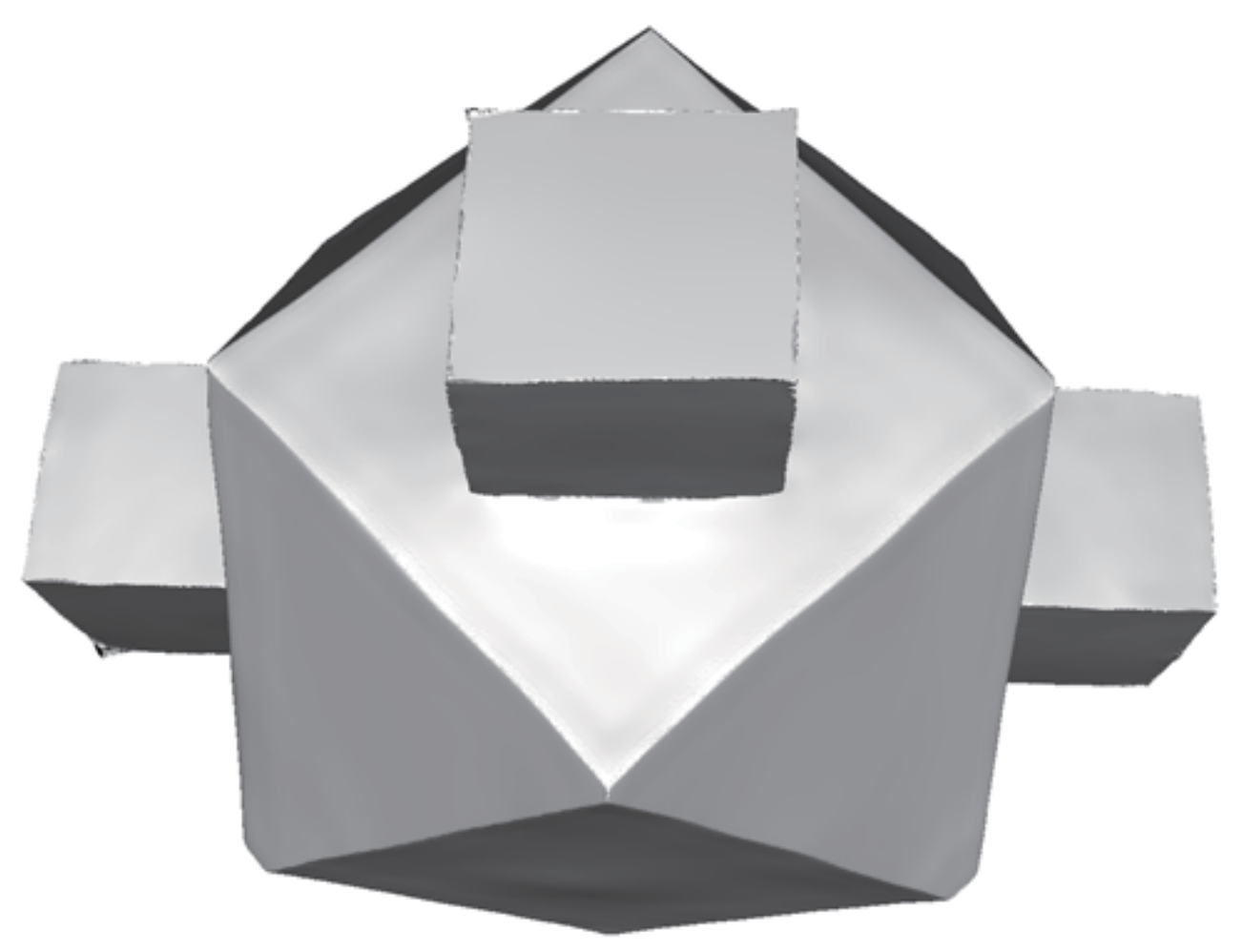}}
\end{minipage}
\caption{The first row: normal results of the scanned Toy point cloud. The second row: upsampling results of the filtered results by updating position with the normals in the first row. The third row: the corresponding surface reconstruction results.}
\label{fig:toy_point}
%\vspace{-0.65cm}
\end{figure*}

%bunnyhi
\begin{figure*}[htbp]
%\vspace{-0.0cm}
\centering
\begin{minipage}[b]{0.03\linewidth}
{\label{}\includegraphics[width=1\linewidth]{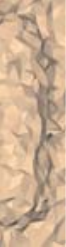}}
\end{minipage}
\begin{minipage}[b]{0.12\linewidth}
{\label{}\includegraphics[width=1\linewidth]{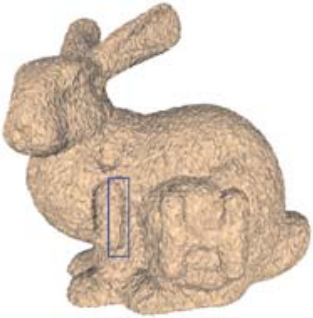}}
\end{minipage}
\begin{minipage}[b]{0.03\linewidth}
{\label{}\includegraphics[width=1\linewidth]{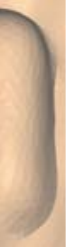}}
\end{minipage}
\begin{minipage}[b]{0.12\linewidth}
{\label{}\includegraphics[width=1\linewidth]{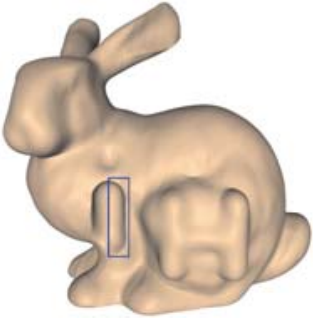}}
\end{minipage}
\begin{minipage}[b]{0.03\linewidth}
{\label{}\includegraphics[width=1\linewidth]{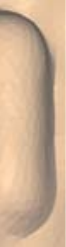}}
\end{minipage}
\begin{minipage}[b]{0.12\linewidth}
{\label{}\includegraphics[width=1\linewidth]{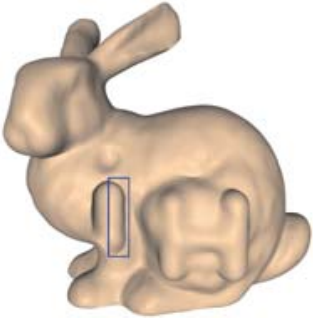}}
\end{minipage}
\begin{minipage}[b]{0.03\linewidth}
{\label{}\includegraphics[width=1\linewidth]{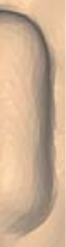}}
\end{minipage}
\begin{minipage}[b]{0.12\linewidth}
{\label{}\includegraphics[width=1\linewidth]{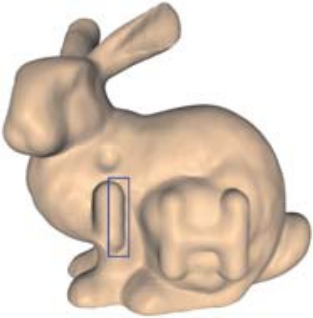}}
\end{minipage}
\begin{minipage}[b]{0.03\linewidth}
{\label{}\includegraphics[width=1\linewidth]{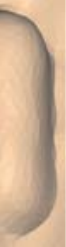}}
\end{minipage}
\begin{minipage}[b]{0.12\linewidth}
{\label{}\includegraphics[width=1\linewidth]{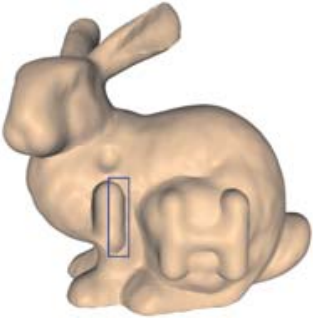}}
\end{minipage}
\begin{minipage}[b]{0.03\linewidth}
{\label{}\includegraphics[width=1\linewidth]{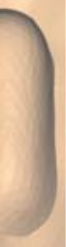}}
\end{minipage}
\begin{minipage}[b]{0.12\linewidth}
{\label{}\includegraphics[width=1\linewidth]{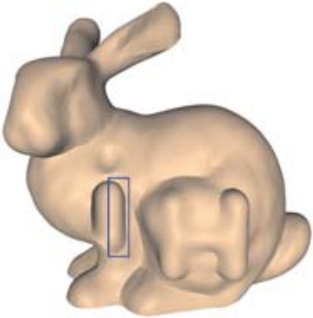}}
\end{minipage}	\\
\begin{minipage}[b]{0.115\linewidth}
{\label{}\includegraphics[width=1\linewidth]{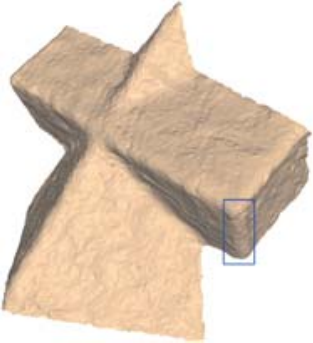}}
\end{minipage}
\begin{minipage}[b]{0.04\linewidth}
{\label{}\includegraphics[width=1\linewidth]{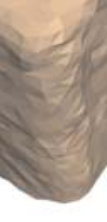}}
\end{minipage}
\begin{minipage}[b]{0.115\linewidth}
{\label{}\includegraphics[width=1\linewidth]{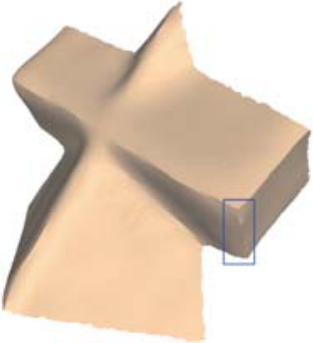}}
\end{minipage}
\begin{minipage}[b]{0.04\linewidth}
{\label{}\includegraphics[width=1\linewidth]{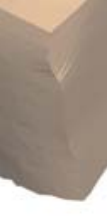}}
\end{minipage}
\begin{minipage}[b]{0.115\linewidth}
{\label{}\includegraphics[width=1\linewidth]{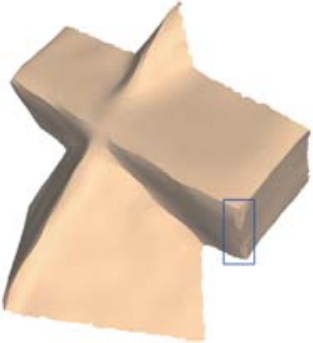}}
\end{minipage}
\begin{minipage}[b]{0.04\linewidth}
{\label{}\includegraphics[width=1\linewidth]{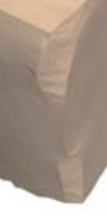}}
\end{minipage}
\begin{minipage}[b]{0.115\linewidth}
{\label{}\includegraphics[width=1\linewidth]{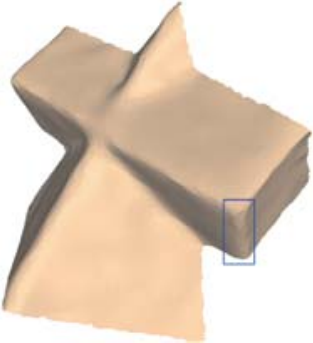}}
\end{minipage}
\begin{minipage}[b]{0.04\linewidth}
{\label{}\includegraphics[width=1\linewidth]{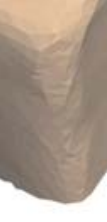}}
\end{minipage}
\begin{minipage}[b]{0.115\linewidth}
{\label{}\includegraphics[width=1\linewidth]{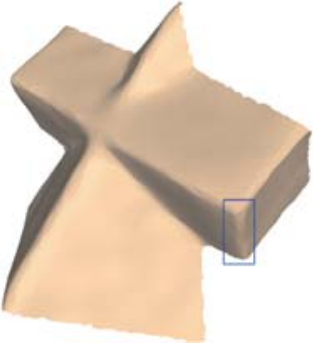}}
\end{minipage}
\begin{minipage}[b]{0.04\linewidth}
{\label{}\includegraphics[width=1\linewidth]{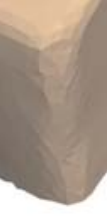}}
\end{minipage}
\begin{minipage}[b]{0.115\linewidth}
{\label{}\includegraphics[width=1\linewidth]{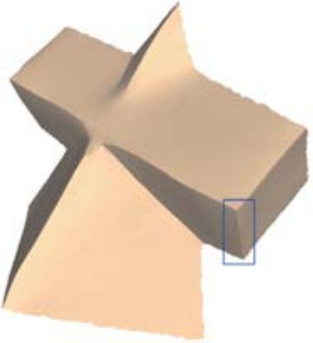}}
\end{minipage}
\begin{minipage}[b]{0.04\linewidth}
{\label{}\includegraphics[width=1\linewidth]{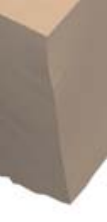}}
\end{minipage}	\\
\begin{minipage}[b]{0.16\linewidth}
{\label{}\includegraphics[width=1\linewidth]{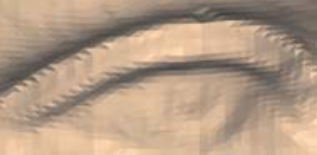}}
\end{minipage}
\begin{minipage}[b]{0.16\linewidth}
{\label{}\includegraphics[width=1\linewidth]{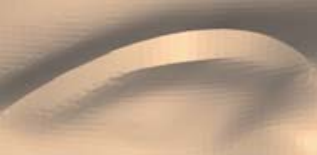}}
\end{minipage}
\begin{minipage}[b]{0.16\linewidth}
{\label{}\includegraphics[width=1\linewidth]{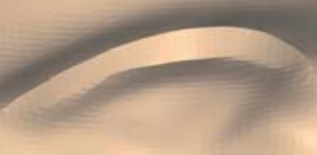}}
\end{minipage}
\begin{minipage}[b]{0.16\linewidth}
{\label{}\includegraphics[width=1\linewidth]{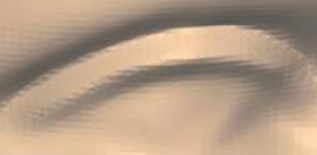}}
\end{minipage}
\begin{minipage}[b]{0.16\linewidth}
{\label{}\includegraphics[width=1\linewidth]{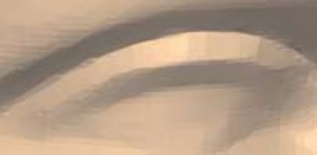}}
\end{minipage}
\begin{minipage}[b]{0.16\linewidth}
{\label{}\includegraphics[width=1\linewidth]{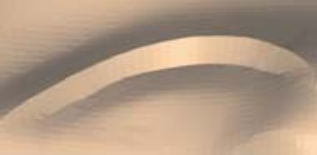}}
\end{minipage} \\
\begin{minipage}[b]{0.13\linewidth}
\subfigure[Noisy input]{\label{}\includegraphics[width=1\linewidth]{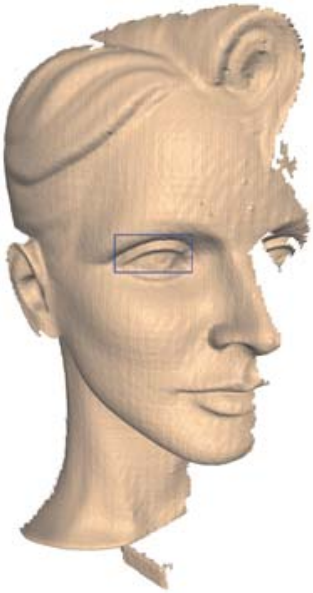}}
\end{minipage}\hfill
\begin{minipage}[b]{0.13\linewidth}
\subfigure[\protect\cite{Sun2007}]{\label{}\includegraphics[width=1\linewidth]{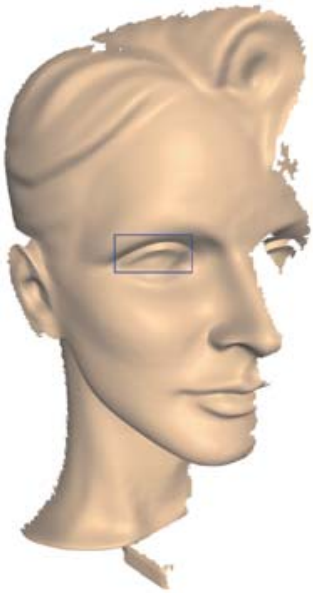}}
\end{minipage}\hfill
\begin{minipage}[b]{0.13\linewidth}
\subfigure[\protect\cite{Zheng2011} (local)]{\label{}\includegraphics[width=1\linewidth]{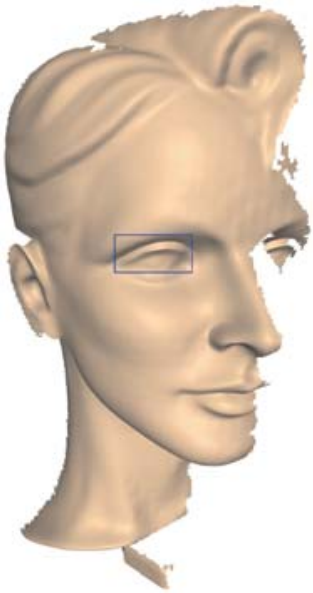}}
\end{minipage}\hfill
\begin{minipage}[b]{0.13\linewidth}
\subfigure[\protect\cite{Zheng2011} (global)]{\label{}\includegraphics[width=1\linewidth]{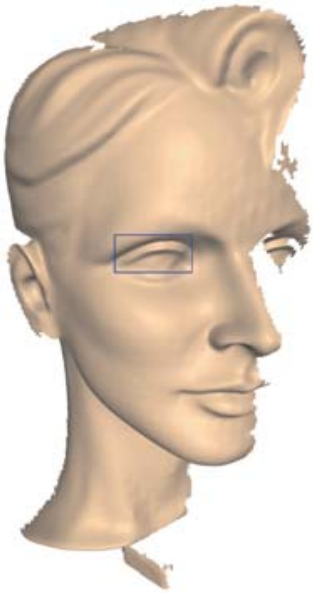}}
\end{minipage}\hfill
\begin{minipage}[b]{0.13\linewidth}
\subfigure[\protect\cite{Zhang2015}]{\label{}\includegraphics[width=1\linewidth]{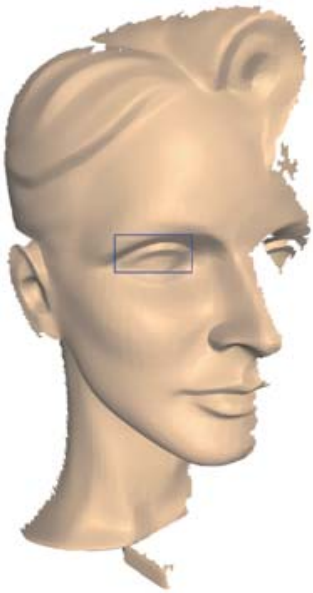}}
\end{minipage}\hfill
\begin{minipage}[b]{0.13\linewidth}
\subfigure[Ours]{\label{}\includegraphics[width=1\linewidth]{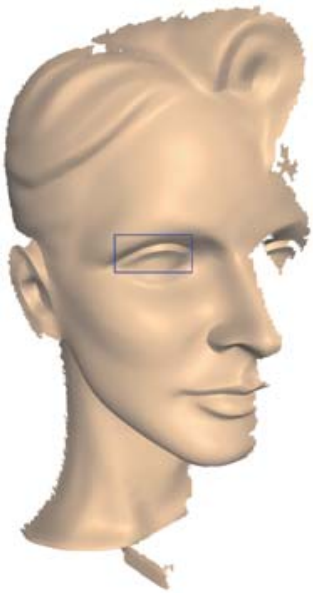}}
\end{minipage}
\caption{Denoised results of the Bunny (synthetic noise), the scanned Pyramid and  Wilhelm.}
\label{fig:meshdenoising}
%\vspace{-0.65cm}
\end{figure*}

\textbf{Convergence.} As with the reported results of previous techniques \cite{Zheng2011,Sun2007,Zhang2015}, we also cannot guarantee the convergence of our normal estimation method. Figure \ref{fig:normalnonconvergeandKNN}(a) shows a comparison between the bilateral filter \cite{Huang2013} and our method as the iteration count increases.  The figure indicates that our method is more accurate than the bilateral filter \cite{Huang2013}.

\section{Position Update}
\label{sec:positionupdate}
Besides normal estimation, we also present algorithms to update point or vertex positions to match the estimated normals, which is typically necessary before {applying other geometry processing algorithms.} 

\textbf{Vertex update for mesh models.} We use the algorithm \cite{Sun2007} to update vertices of mesh models, which minimizes the square of the dot product between the normal and the three edges of each face. 

\textbf{Point update for point clouds.} Compared to the vertex update for mesh models, updating point cloud positions is more difficult due to the absence of topological information. Furthermore, the local neighborhood information may vary during this position update. 
We propose a modification of the edge recovery algorithm in \cite{Sun2015} to update points in a feature-aware way and minimize
\begin{equation}\label{eq:pointminimization}
\begin{aligned}
\sum_i\sum_{j\in S_i} \normScalar{(\mathbf{p}_i-\mathbf{p}_j)\mathbf{n}_j^T}^2 + \normScalar{(\mathbf{p}_i-\mathbf{p}_j)\mathbf{n}_i^T}^2 .
\end{aligned}
\end{equation}

$\mathbf{p}_i$ and $\mathbf{p}_j$ are unknown, and $\mathbf{n}_i$ and $\mathbf{n}_j$ are computed by our normal estimation algorithm. Eq. \eqref{eq:pointminimization} encodes the sum of distances to the tangent planes defined by the neighboring points $\{\mathbf{p}_j\}$ and the corresponding normals $\{\mathbf{n}_j\}$, as well as the sum of distances to the tangent planes defined by $\{\mathbf{p}_i\}$ and $\{\mathbf{n}_i\}$. We use the gradient descent method to solve Eq. \eqref{eq:pointminimization}, by assuming the point $\mathbf{p}_i$ and its neighboring points $\{j\in S_i|\mathbf{p}_j\}$ in the previous iteration are known. Here we use ball neighbors instead of k nearest neighbors to ensure the convergence of our point update. Therefore, the new position of $\mathbf{p}_i$ can be computed by
\begin{equation}\label{eq:pointupdate}
\begin{aligned}
\mathbf{p}_i' = \mathbf{p}_i + \gamma_i \sum_{j\in S_i} (\mathbf{p}_j-\mathbf{p}_i) (\mathbf{n}_j^T\mathbf{n}_j+ \mathbf{n}_i^T\mathbf{n}_i),
\end{aligned}
\end{equation}
where $\mathbf{p}_i'$ is the new position. $\gamma_i$ is the step size, which is set to $\frac{1}{3|S_i|}$ to ensure the convergence (see Appendix).

\textbf{Remark 4.} While the neighboring information for point updating should be recomputed in each iteration, doing so can lead to artifacts.  As illustrated in Figure \ref{fig:dod_knnupdate}(a), our point update method enhances edges by automatically driving neighboring points to edges but also leads to obvious gaps near edges. As a result, the upsampling application could fail when the number of points is low (Figure \ref{fig:dod_knnupdate}(b)). To alleviate this issue, we simply keep the neighboring information unchanged in all iterations, which has the side-effect of reducing the computation in each iteration. Figure \ref{fig:dod_knnupdate} shows a comparison.  Though we cannot guarantee that our point update method preserves the volume of the shape, we found insignificant volume changes in our experiments. We show the \textit{proof of convergence} of our point update algorithm in the Appendix.

%% file: paper/results.tex
\section{Applications and Experimental Results}
\label{sec:results}
In this section, we demonstrate some geometry processing applications that benefit from our approach directly or indirectly including mesh denoising, point cloud filtering, point cloud upsampling, surface reconstruction, and geometric texture removal. Moreover, we also compared state of the art methods with our approach in each application. The source code (program) of the compared methods are available on the Internet or granted to us by the original authors. 

\textbf{Parameter setting.} The main parameters of our normal estimation method are the local neighborhood size $k_{local}$, the angle threshold $\theta_{th}$, the non-local searching range $k_{non}$, and the maximum iterations for normal estimation $n_{nor}$. For the position update procedure, our parameters are the local neighborhood size $k_{local}$ or 1-ring neighbors (mesh models) and the number of iterations for the position update $n_{pos}$. 

To more accurately find similar local isotropic structures, we set one initial value and one lower bound to $\theta_{th}$, namely $\theta_{th}^{init}$ and $\theta_{th}^{low}$. We reduce the start value $\theta_{th}^{init}$ towards $\theta_{th}^{low}$ at a rate of $1.1^n$ in the $n$-th iteration. We show the tests of our  parameters in Figure \ref{fig:normalnonconvergeandKNN} and \ref{fig:dod_parametertest}. In general, normal errors are decreased with the growing number of normal estimation iterations, but excessive iterations would cause greater normal errors (Figure \ref{fig:normalnonconvergeandKNN}(a)). The estimated normals of models with sharp features are more accurate with the increasing local neighborhood $k_{local}$ or non-local search range $k_{non}$ (Figure \ref{fig:normalnonconvergeandKNN}(b)). However, we need to compromise between accuracy and computation. Fixed $\theta_{th}$ are likely to inaccurately locate similar local isotropic structures and further generate erroneous normal estimations (Figure \ref{fig:dod_parametertest}(a-b)). Larger start values of $\theta_{th}^{init}$ smear geometric features (Figure \ref{fig:dod_parametertest}(c)). 

Based on our parameter tests and observations, for point clouds we empirically set: $k_{local}=60$, $k_{non}=150$, $\theta_{th}^{init}=30.0$, and $\theta_{th}^{low}=15.0$ for models with sharp features, but set $\theta_{th}^{init}=20.0$ and $\theta_{th}^{low}=8.0$ for models with low dihedral angle features.  
For mesh models, we replace the local neighborhood with the 2-ring of neighboring faces.  We use 4 to 10 iterations for normal estimation and 5 to 30 for the position update.

For fair comparisons, we set the same local neighborhood to other methods. The remaining parameters of other methods are tuned to achieve the best visual results. Regarding \cite{Boulch2012,Boulch2016}, the two methods have multiple solutions and the best results are selected for comparisons. For the position update, we set the same parameters for the compared normal estimation methods for each model.

%cube_upsample: error
\begin{figure}[htbp]
%\vspace{-0.0cm}
\centering
\begin{minipage}[b]{0.175\linewidth}
\subfigure[Hoppe et al. 1992]%\protect\cite{Hoppe1992}]
{\label{}\includegraphics[width=1\linewidth]{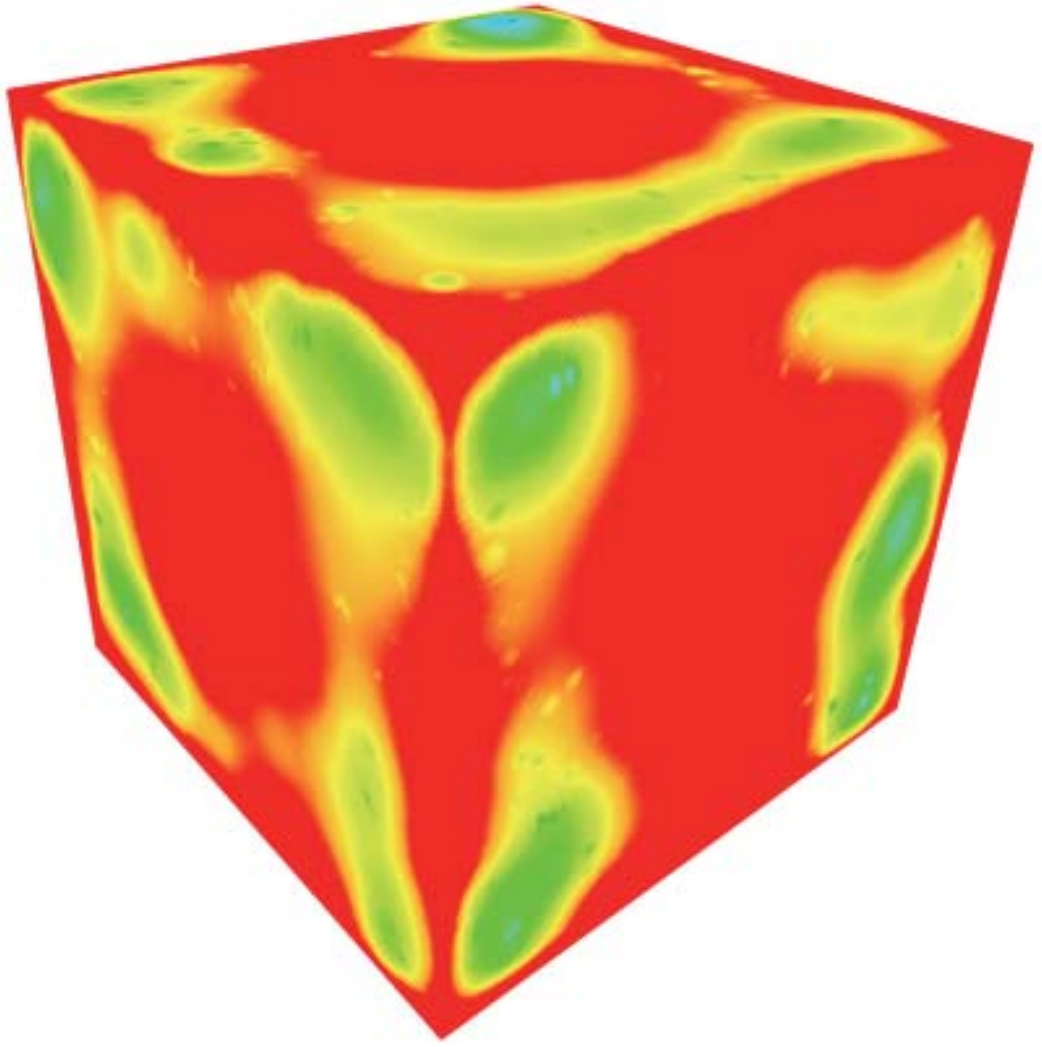}}
\end{minipage}
\begin{minipage}[b]{0.175\linewidth}
\subfigure[Boulch and Marlet 2012]%\protect\cite{Boulch2012}]
{\label{}\includegraphics[width=1\linewidth]{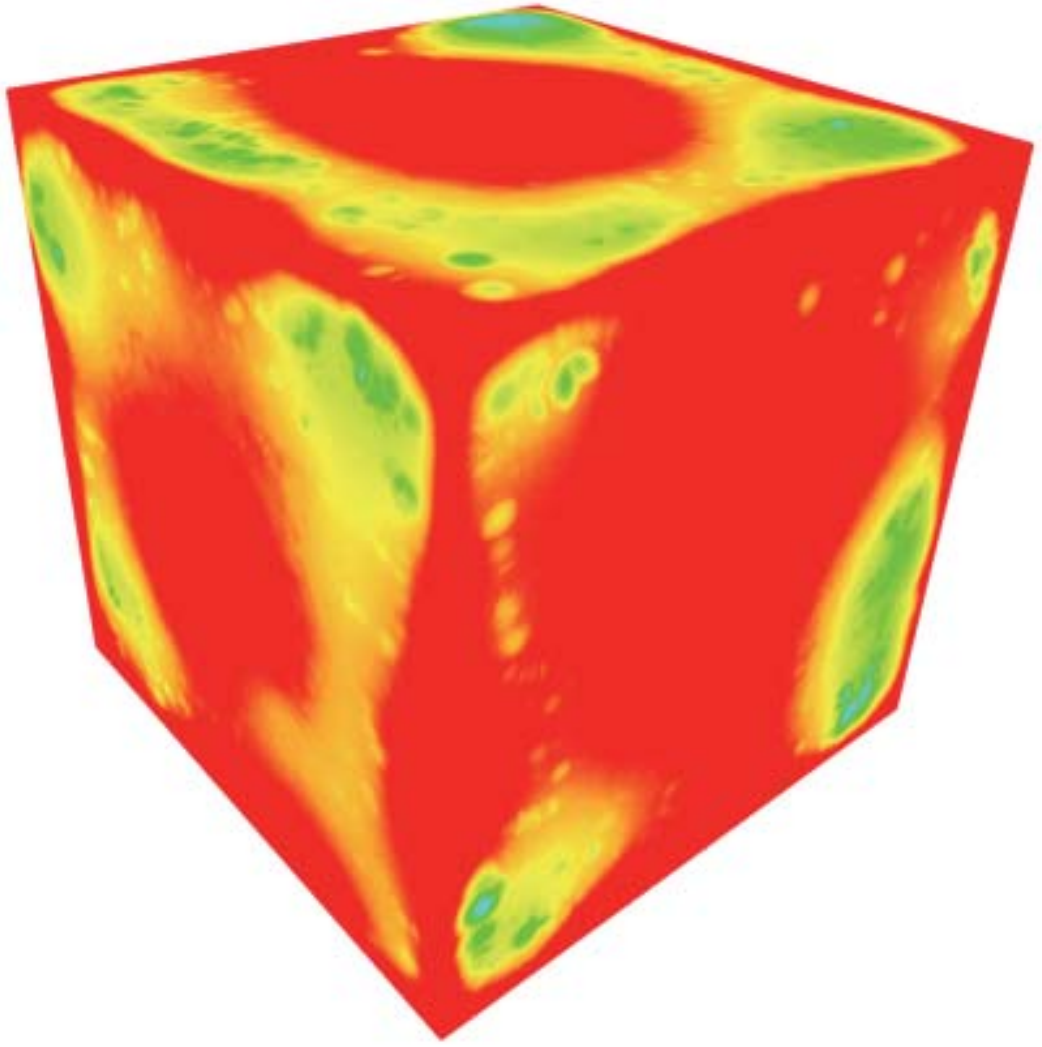}}
\end{minipage}
\begin{minipage}[b]{0.175\linewidth}
\subfigure[Huang et al. 2013]%\protect\cite{Huang2013}]
{\label{}\includegraphics[width=1\linewidth]{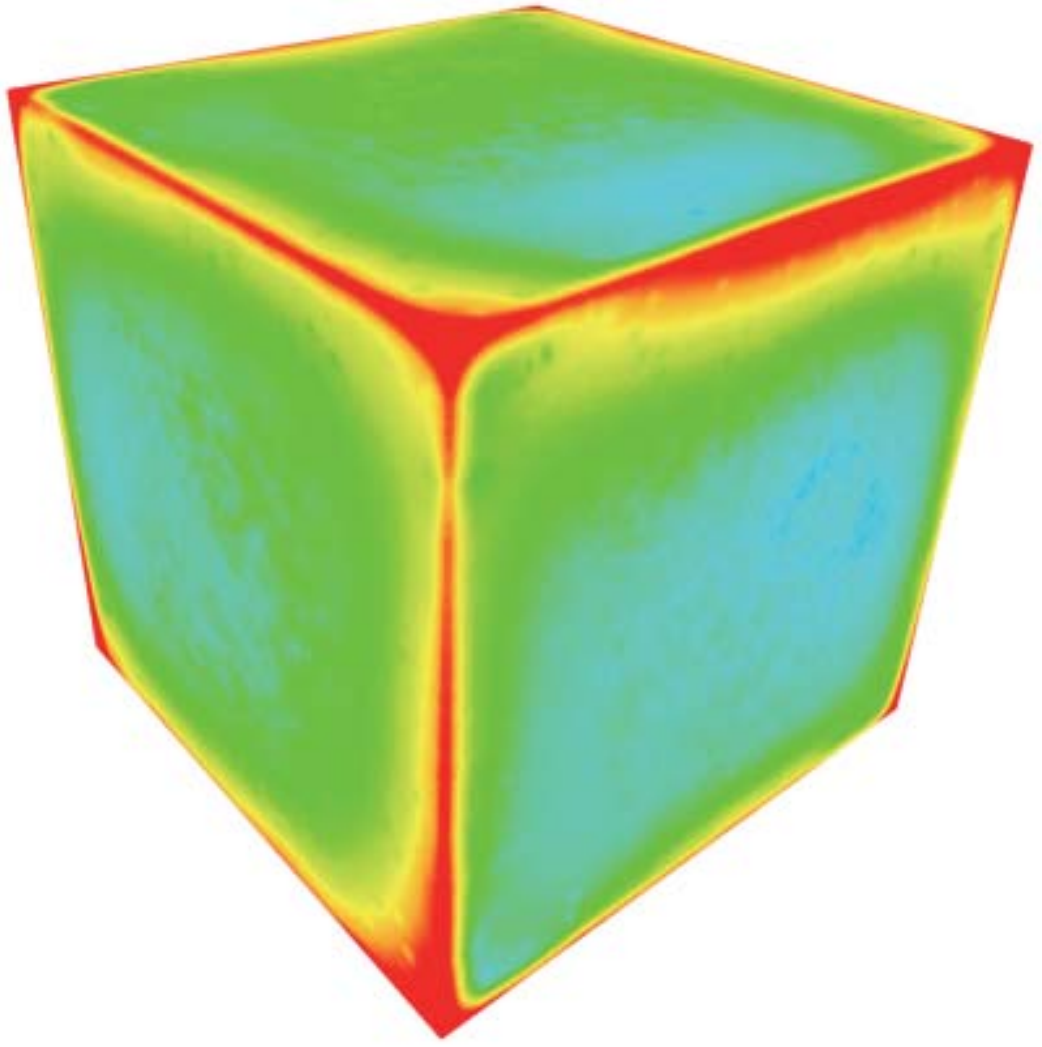}}
\end{minipage}
\begin{minipage}[b]{0.175\linewidth}
\subfigure[Boulch and Marlet 2016]%\protect\cite{Boulch2016}]
{\label{}\includegraphics[width=1\linewidth]{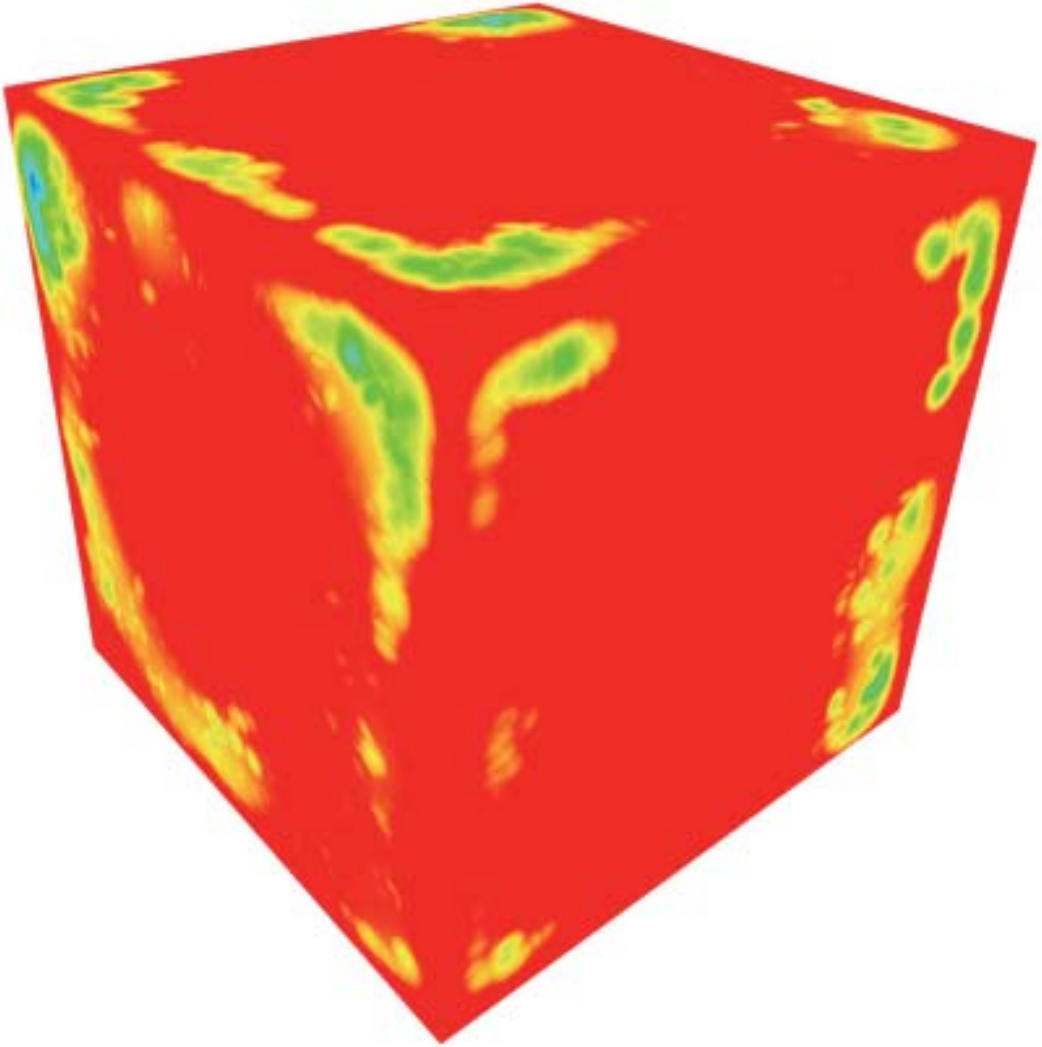}}
\end{minipage}
\begin{minipage}[b]{0.175\linewidth}
\subfigure[Ours]{\label{}\includegraphics[width=1\linewidth]{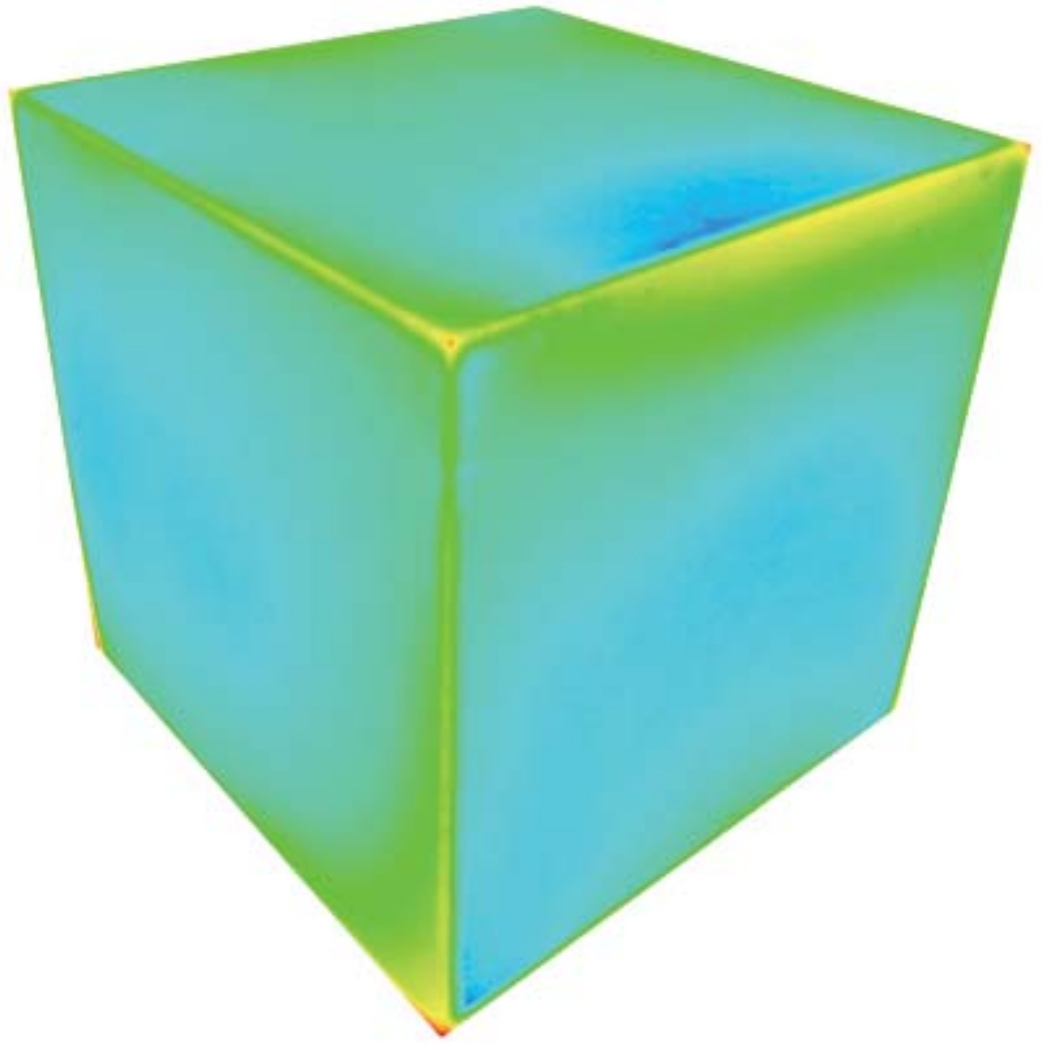}}
\end{minipage}
\begin{minipage}[b]{0.075\linewidth}
\subfigure[]{\label{}\includegraphics[width=1\linewidth]{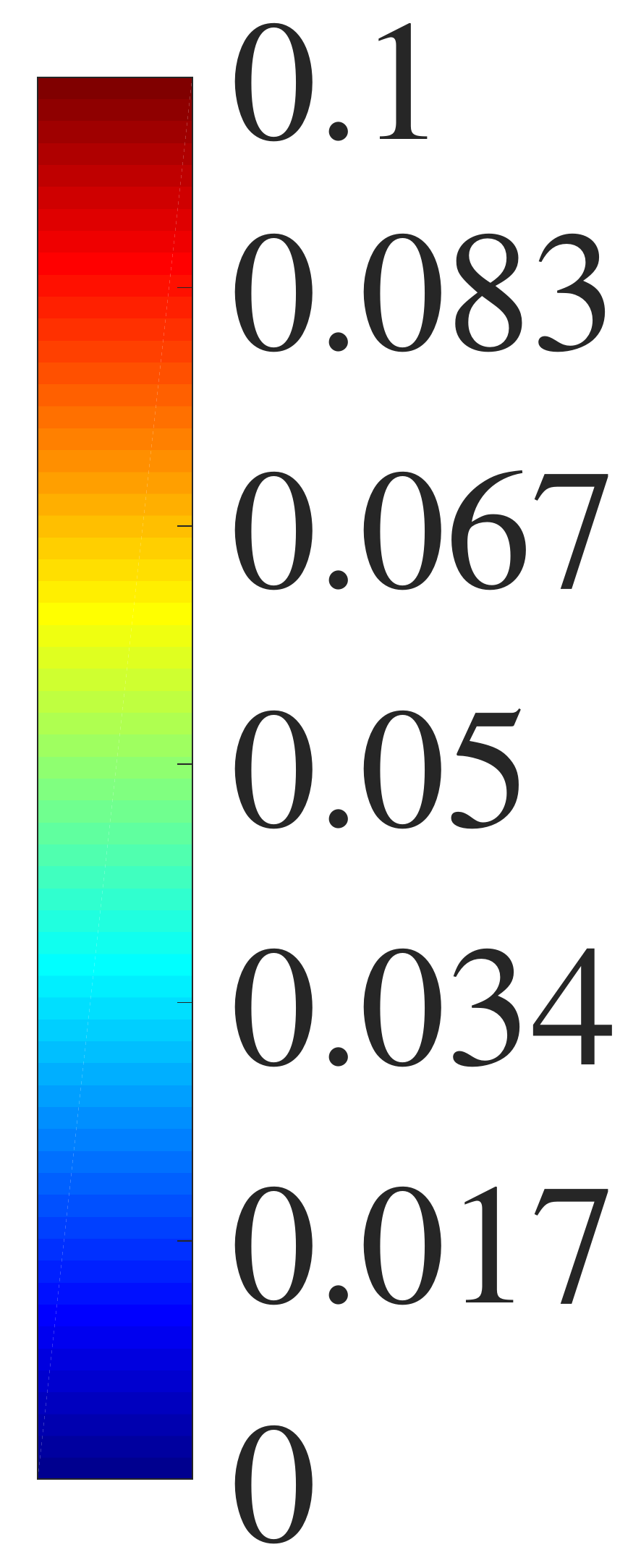}}
\end{minipage}
\caption{Position accuracies for Fig. \ref{fig:cube_point}. The root mean square errors are ($\times10^{-2}$): (a) $8.83$, (b) $9.05$, (c) $5.14$, (d) $9.64$, (e) $3.22$. The rmse of the corresponding surface reconstructions are ($\times 10^{-2}$): $7.73$, $6.45$, $3.28$, $7.71$ and $2.41$, respectively. (f) is the error bar for here %Fig. \ref{fig:girlposError} 
and \ref{fig:ironposError}. }
\label{fig:cubeposError}
%\vspace{-0.65cm}
\end{figure}

%iron_upsample: error
\begin{figure}[htbp]
%\vspace{-0.0cm}
\centering
\begin{minipage}[b]{0.19\linewidth}
\subfigure[Hoppe et al. 1992]{\label{}\includegraphics[width=1\linewidth]{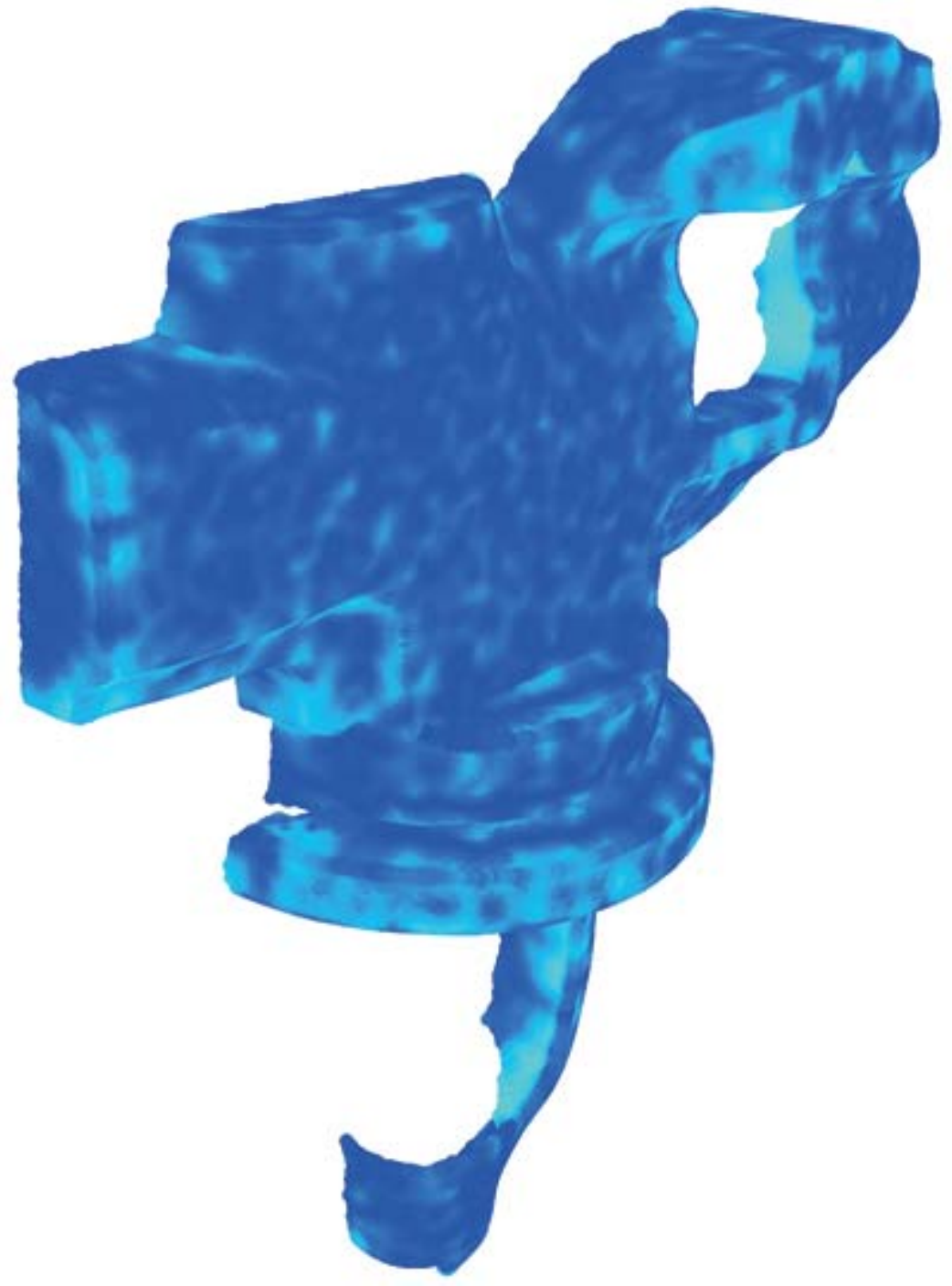}}
\end{minipage}
\begin{minipage}[b]{0.19\linewidth}
\subfigure[Boulch and Marlet 2012]{\label{}\includegraphics[width=1\linewidth]{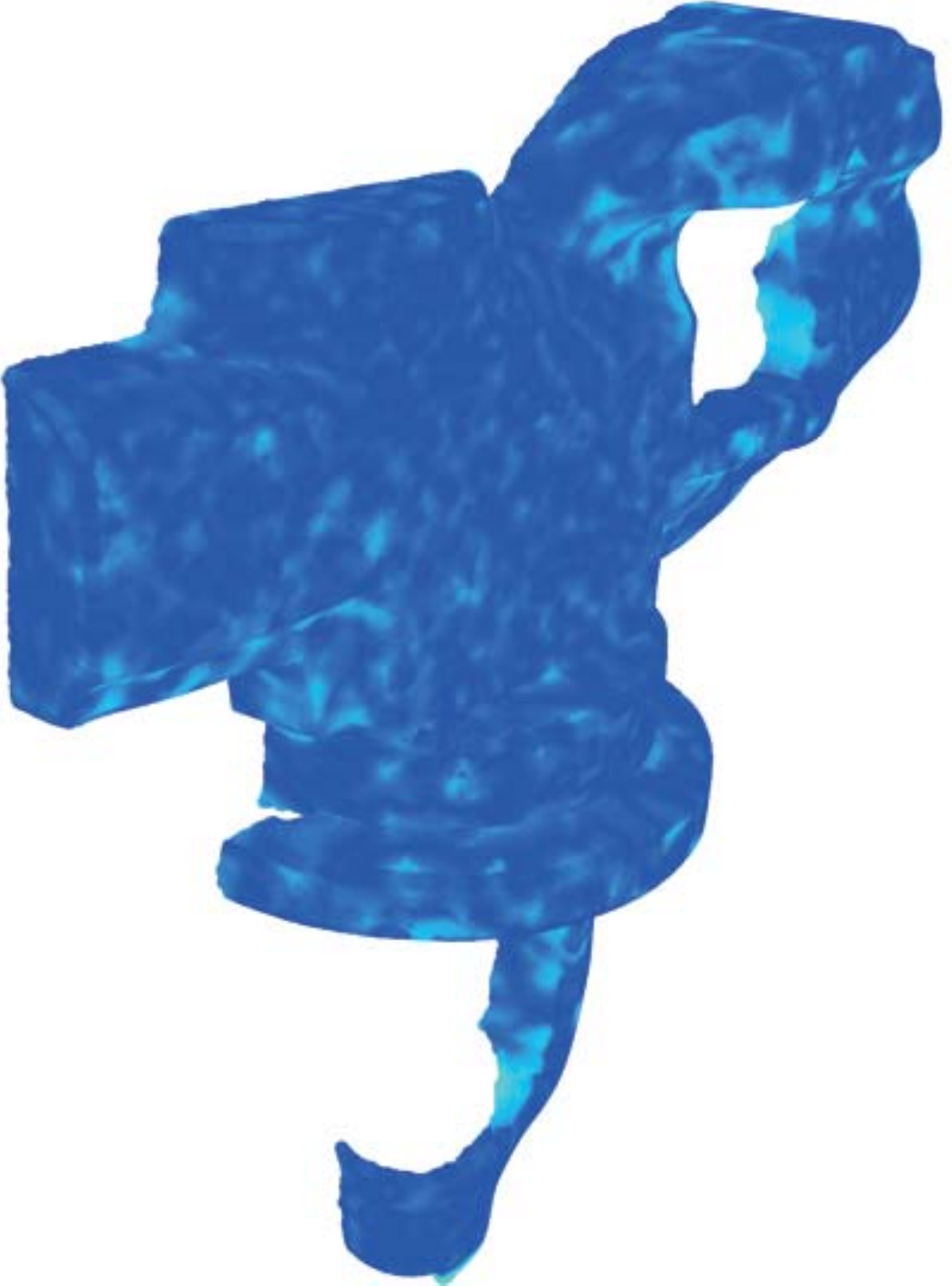}}
\end{minipage}
\begin{minipage}[b]{0.19\linewidth}
\subfigure[Huang et al. 2013]{\label{}\includegraphics[width=1\linewidth]{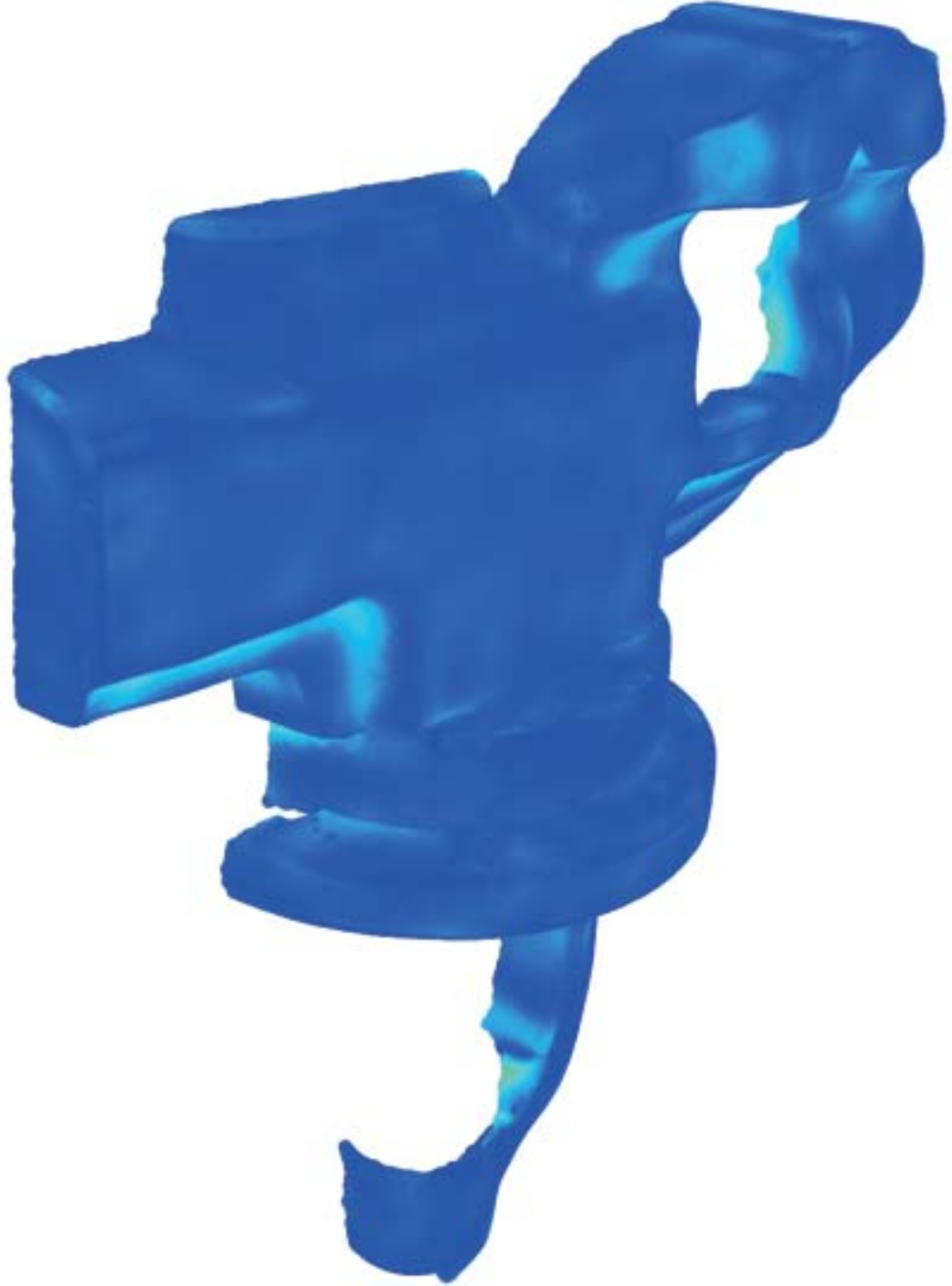}}
\end{minipage}
\begin{minipage}[b]{0.19\linewidth}
\subfigure[Boulch and Marlet 2016]{\label{}\includegraphics[width=1\linewidth]{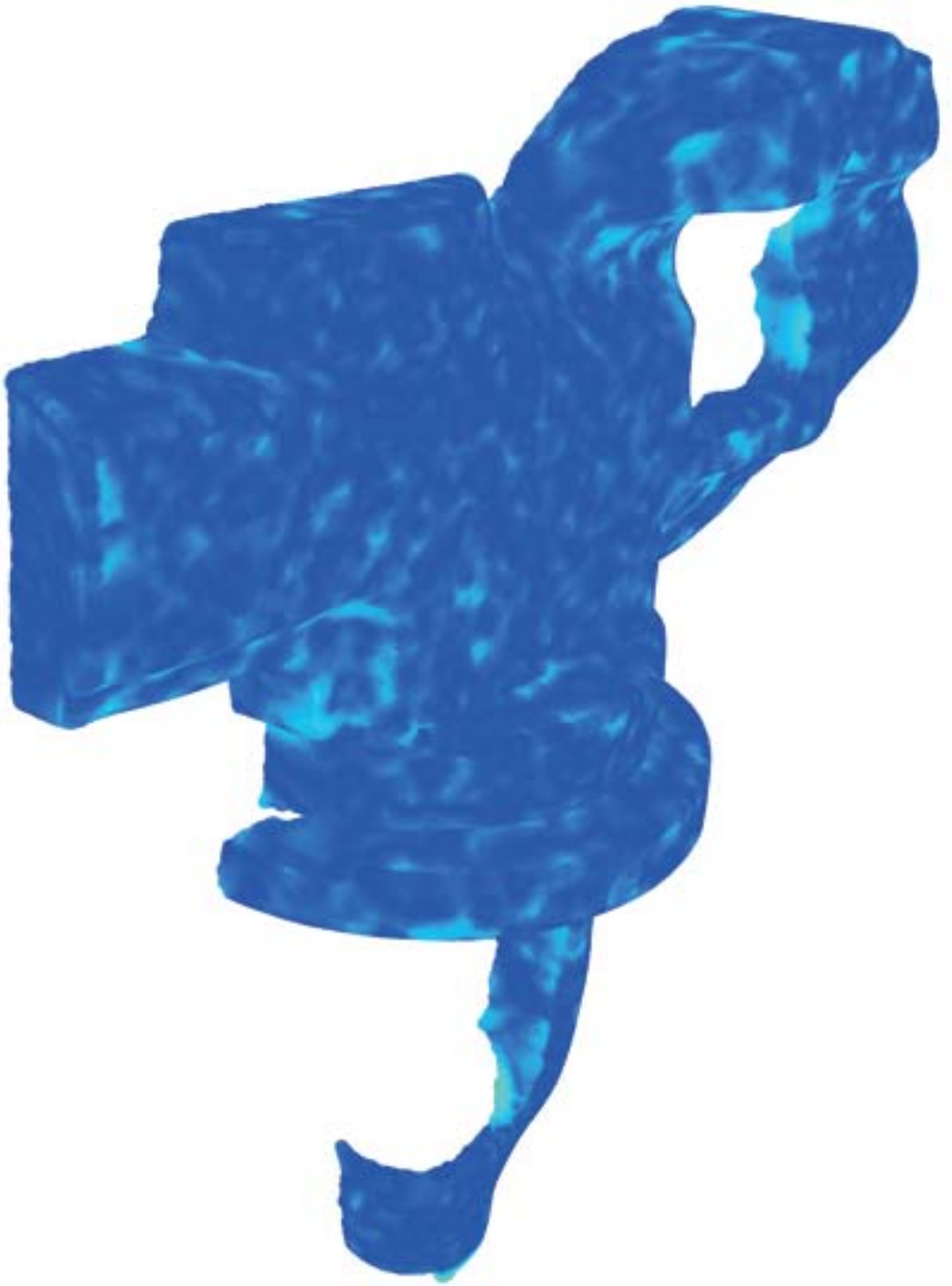}}
\end{minipage}
\begin{minipage}[b]{0.19\linewidth}
\subfigure[Ours]{\label{}\includegraphics[width=1\linewidth]{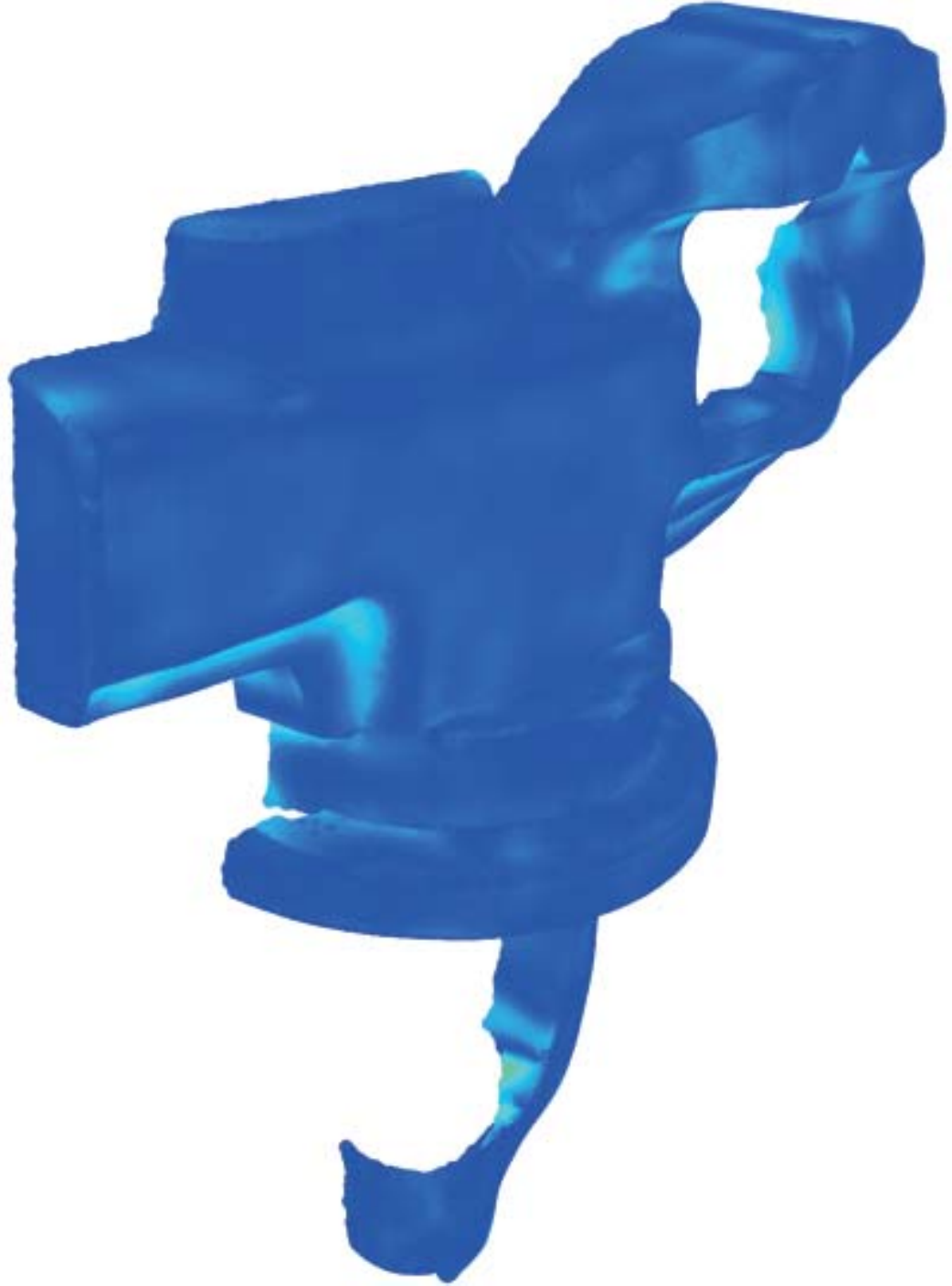}}
\end{minipage}
\caption{Position accuracies for Fig. \ref{fig:iron_point}. The root mean square errors are ($\times10^{-3}$): (a) $8.59$, (b) $6.84$, (c) $6.80$, (d) $6.82$, (e) $6.57$. The rmse of the corresponding surface reconstructions are ($\times 10^{-3}$): $8.60$, $6.75$, $6.68$, $6.74$ and $6.40$, respectively. }
\label{fig:ironposError}
%\vspace{-0.65cm}
\end{figure}

\textbf{Accuracy.} Since we used the pre-filter \cite{Lu2016} for meshes with large noise, 
there exist few flipped normals in the results so that different methods have limited difference in normal accuracy. However, the visual differences are easy to observe. Therefore, we compared the accuracies of normals and positions over point cloud shapes. Note that state of the art methods compute normals on edges differently: the normals on edges are either sideling (e.g., \cite{Hoppe1992,Boulch2016}) or perpendicular to one of the intersected tangent planes (e.g., \cite{Huang2013} and our method). The latter is more suitable for feature-aware position update. For fair comparisons, we have two ground truth models for each point cloud: the original ground truth for \cite{Hoppe1992,Boulch2016} and the other ground truth for \cite{Huang2013} and our method. The latter ground truth is generated by adapting normals on edges to be perpendicular to one of the intersected tangent planes. The ground truth model, which has the smaller mean square angular error (MSAE) \cite{Lu2016} among the two kinds of ground truth models, is selected as the ground truth for \cite{Boulch2012}. Figure \ref{fig:syntheticnormalerror} shows the normal errors of different levels of noise on the cube and dodecahedron models. We also compared our method with state of the art techniques in Table \ref{table:table1}. %In Table \ref{table:table1}, 
The ground truth for the \textit{Dod\_vir} model (Table~\ref{table:table1}) for \cite{Hoppe1992,Boulch2016} is achieved by averaging the neighboring face normals in the noise-free model. The other kind of ground truth for \cite{Huang2013} and our method is produced by further adapting normals on edges to one of the intersected tangent planes. We compute ground truth for Figure \ref{fig:iron_point} and \ref{fig:syntheticnormalerror} in a similar way.  
The normal error results demonstrate that our approach outperforms state of the art methods. We analyzed that this is probably due to more exhibited useful information of non-local similar structures than local techniques. 

\begin{table}[thbp]\tablefont
    \centering
    \caption{Normal errors (mean square angular error, in radians) of two scanned models. Dod\_vir is a virtual scan of a noise-free model rather than Figure \ref{fig:dod_point} corrupted with synthetic noise. }\label{table:table1} 
    \begin{tabular}{l c c c c c c}
    \toprule
    Methods & \tabincell{l}{
    Hoppe et\\al.1992\\
    } & \tabincell{l}{
    Boulch and\\Marlet 2012
    } & \tabincell{l}{
    Huang et \\al.2015
    } & \tabincell{l}{Boulch and\\Marlet 2016} & \tabincell{l}{
    Ours} \\ 
    \midrule
    Dod\_vir & 0.0150 & 0.0465 & 0.0054 & 0.0553 & \textbf{0.0023}
    \\
    Fig. \protect\ref{fig:iron_point} & 0.0118 & 0.1274 & 0.0060 & 0.1208 & \textbf{0.0036}
    \\
    \bottomrule
    \end{tabular}
\end{table}

In addition, we compared the position errors of different techniques, see Figure \ref{fig:cubeposError} %\ref{fig:girlposError}, 
and \ref{fig:ironposError}. The position error is measured using the average distance between points of the ground truth and their closest points of the reconstructed point set \cite{Lu2017tvcg}. For visualization purpose, we rendered the colors of position errors on the upsampling results. The root mean square error (RMSE) of both the upsampling and reconstruction results show that our approach is more accurate than state of the art methods.

% compare with clop and wlop
\begin{figure}[htbp]
%\vspace{-0.0cm}
\centering
\begin{minipage}[b]{0.24\linewidth}
\subfigure[Huang et al. 2009]%\protect\cite{Huang2009}]
{\label{}\includegraphics[width=1\linewidth]{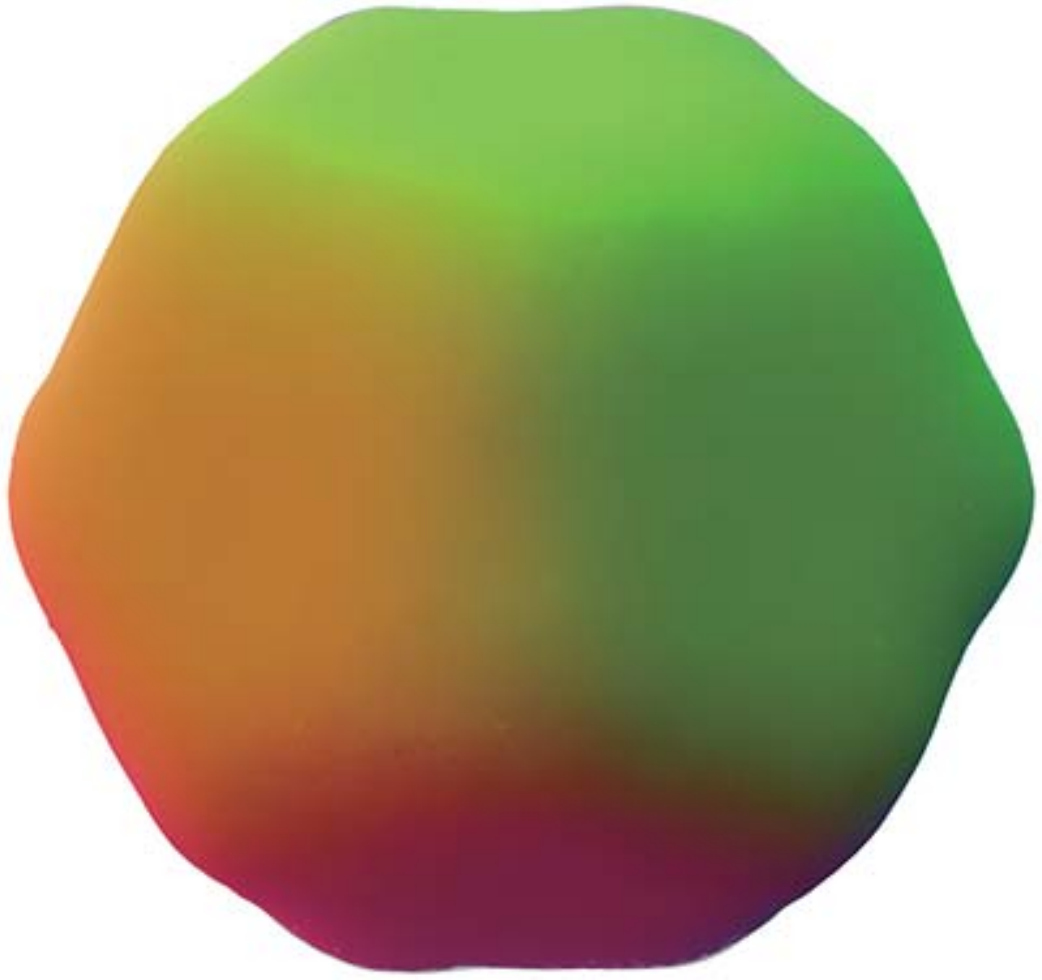}}
\end{minipage}
\begin{minipage}[b]{0.23\linewidth}
\subfigure[RIMLS over (a)]{\label{}\includegraphics[width=1\linewidth]{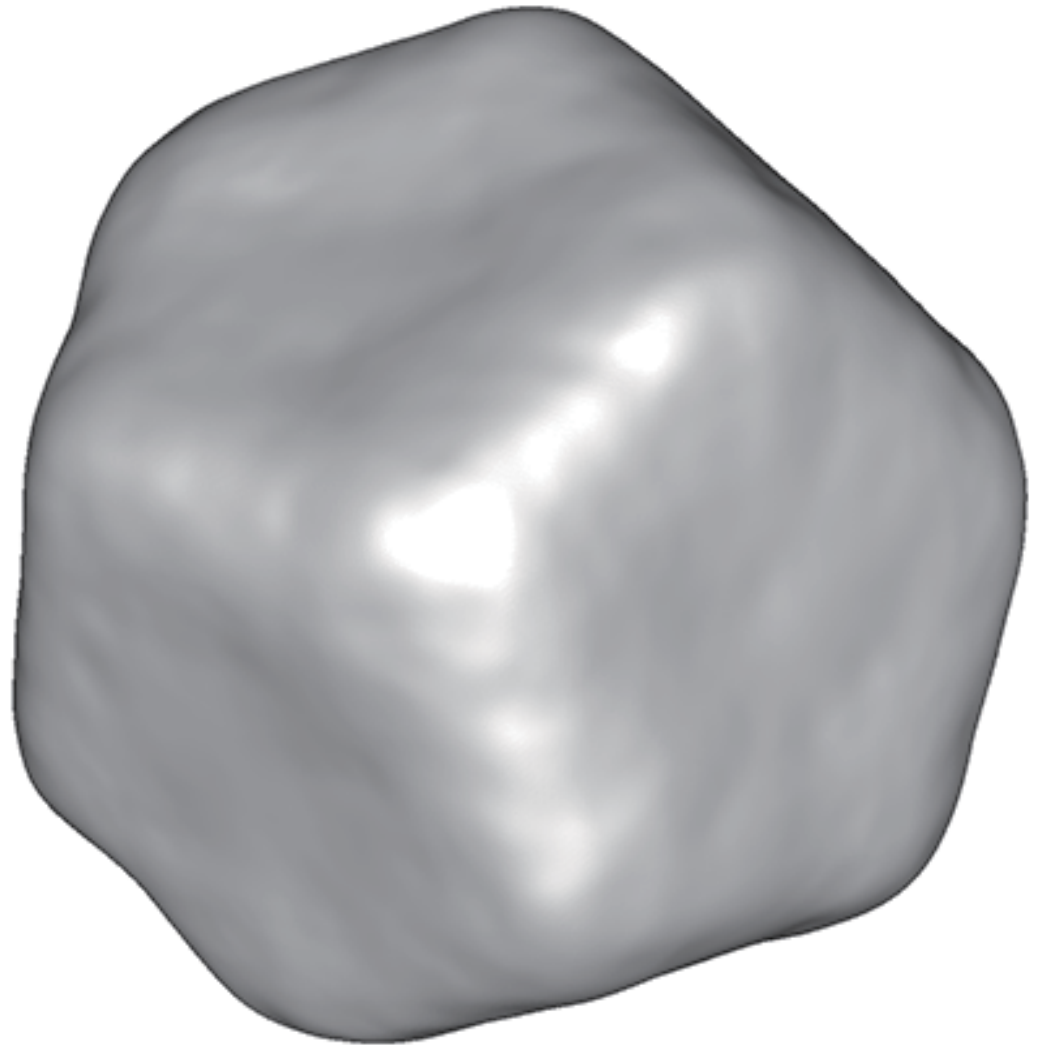}}
\end{minipage}
\begin{minipage}[b]{0.24\linewidth}
\subfigure[Preiner et al. 2014]%\protect\cite{Preiner2014}]
{\label{}\includegraphics[width=1\linewidth]{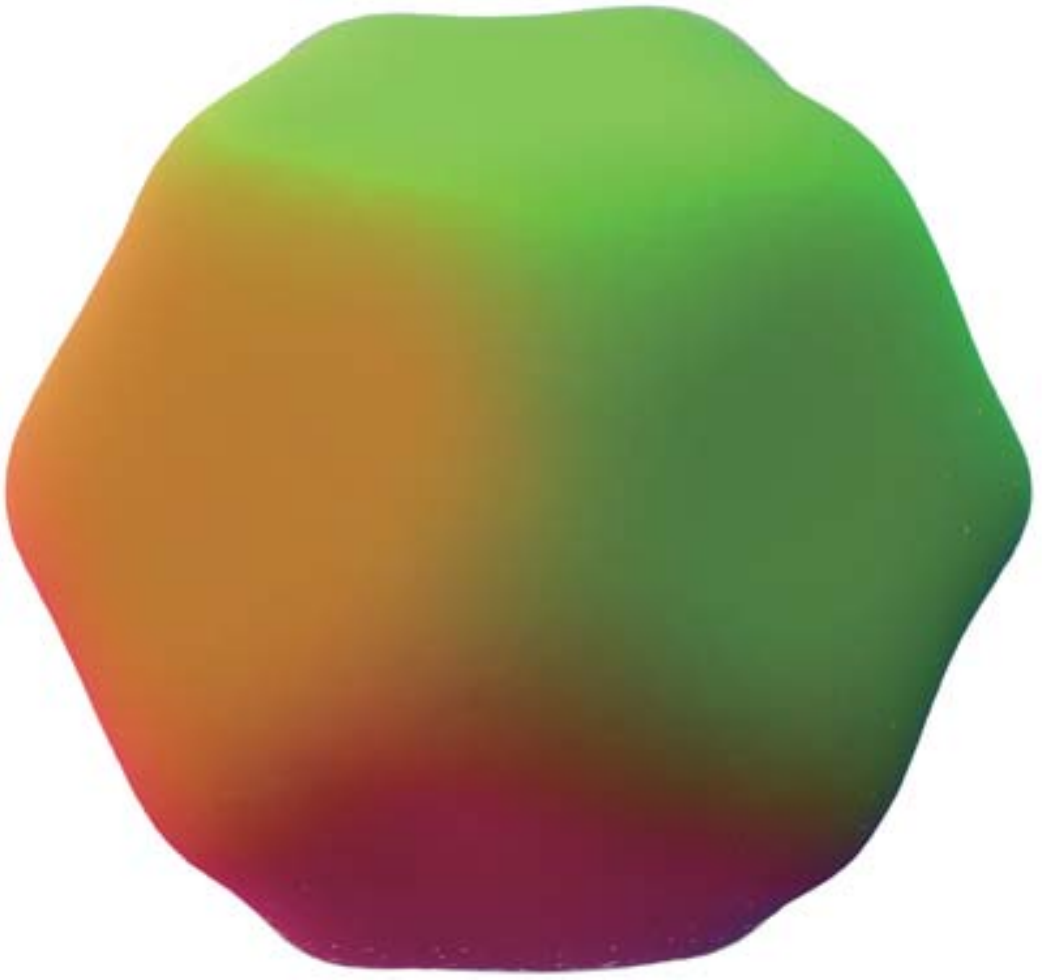}}
\end{minipage}
\begin{minipage}[b]{0.23\linewidth}
\subfigure[RIMLS over (c)]{\label{}\includegraphics[width=1\linewidth]{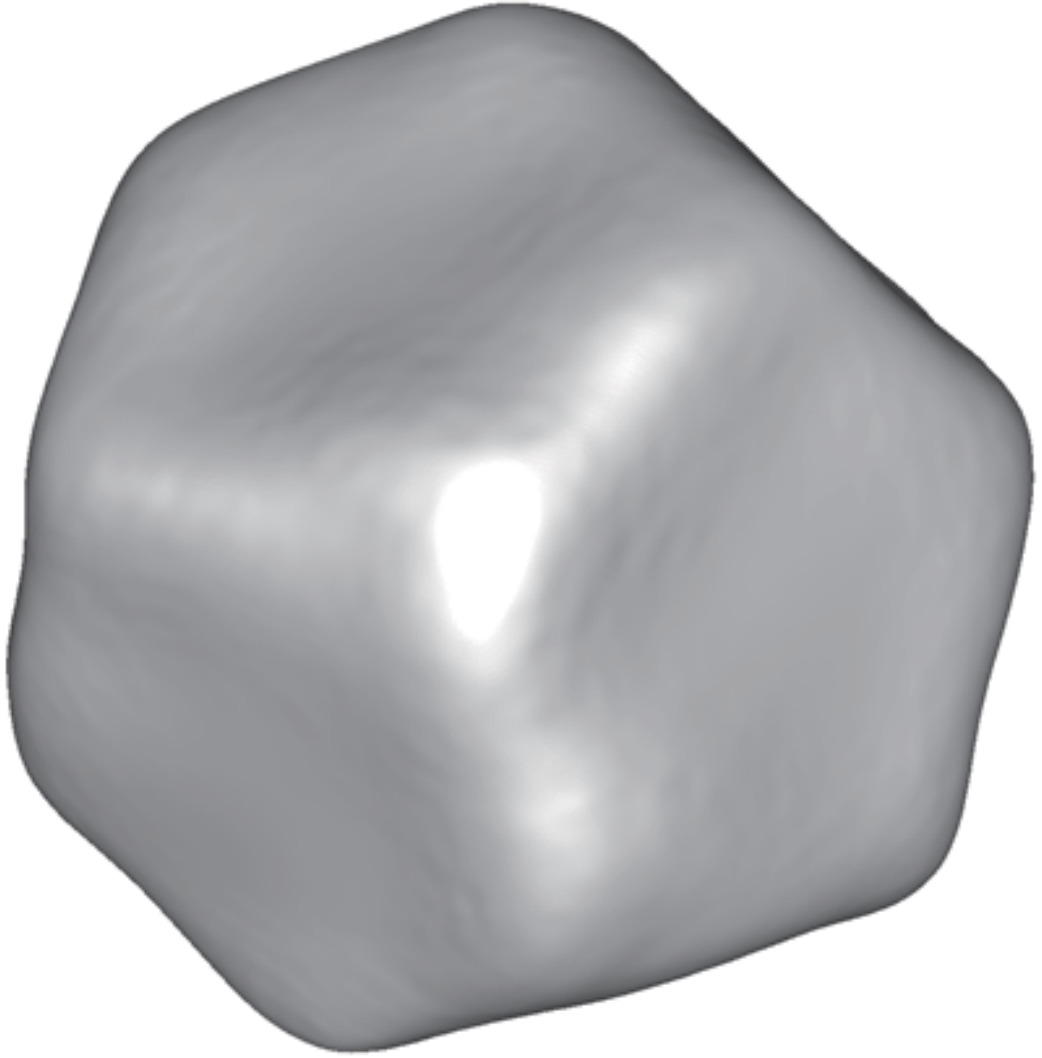}}
\end{minipage}
\caption{Upsampling and reconstruction results over \protect\cite{Huang2009,Preiner2014}. The input is the same as Figure \ref{fig:dod_point}. }
\label{fig:wlopclop_dod}
%\vspace{-0.2cm}
\end{figure}

%point clouds with complicated features
\begin{figure}[htbp]
%\vspace{-0.0cm}
\centering
\begin{minipage}[b]{0.26\linewidth}
\subfigure[Huang et al. 2013]
{\label{}\includegraphics[width=1\linewidth]{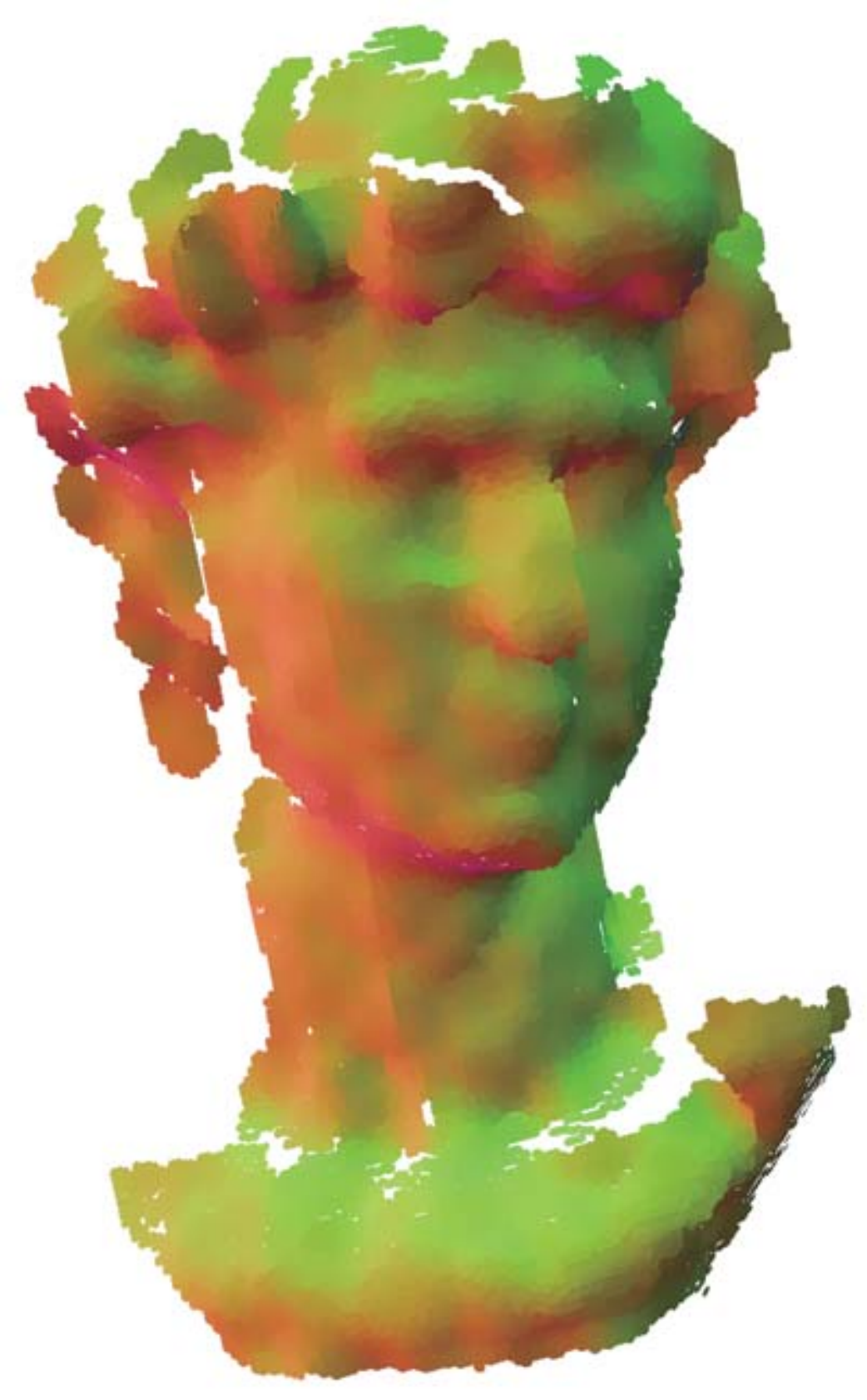}}
\end{minipage}
\begin{minipage}[b]{0.26\linewidth}
\subfigure[Ours]{\label{}\includegraphics[width=1\linewidth]{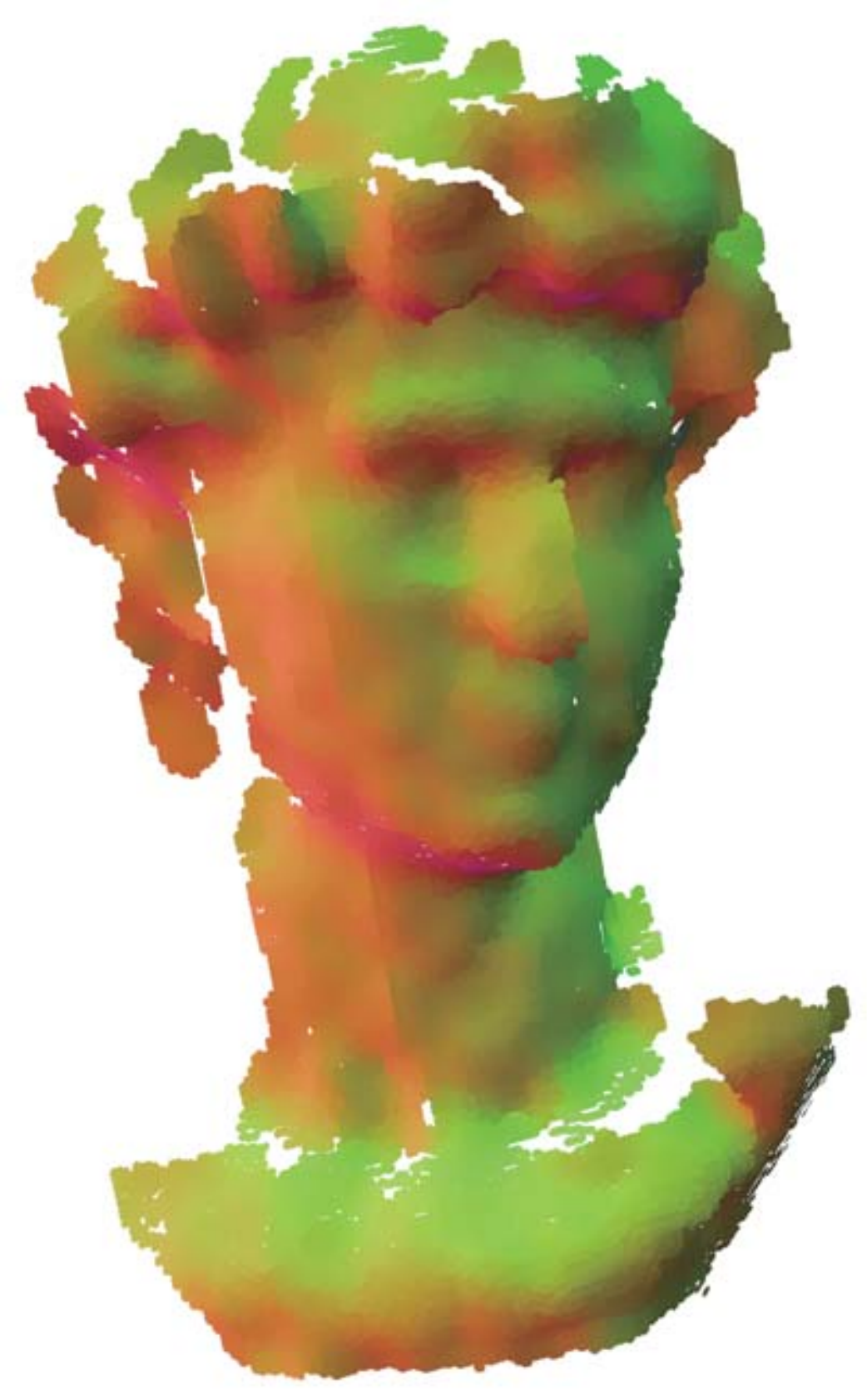}}
\end{minipage}
\begin{minipage}[b]{0.185\linewidth}
\subfigure[Huang et al. 2013]
{\label{}\includegraphics[width=1\linewidth]{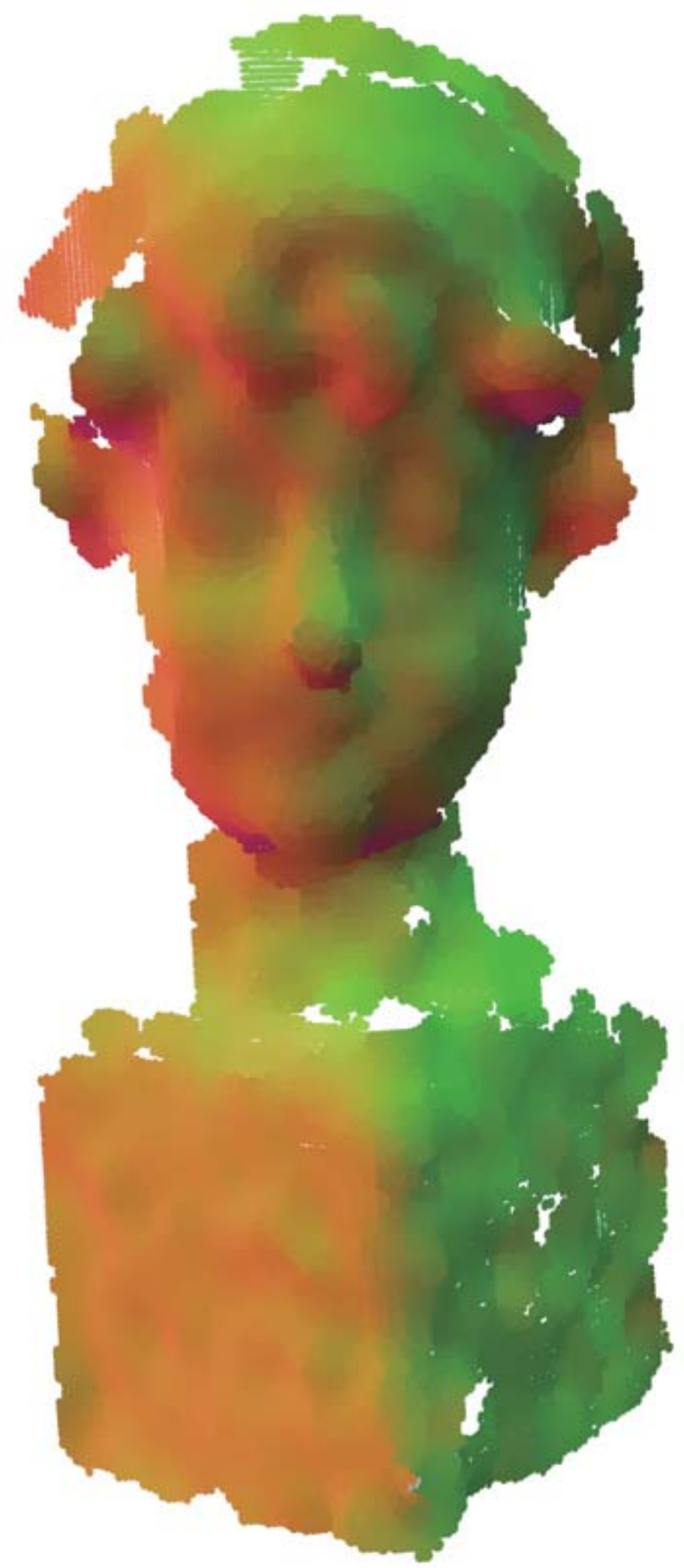}}
\end{minipage}
\begin{minipage}[b]{0.185\linewidth}
\subfigure[Ours]{\label{}\includegraphics[width=1\linewidth]{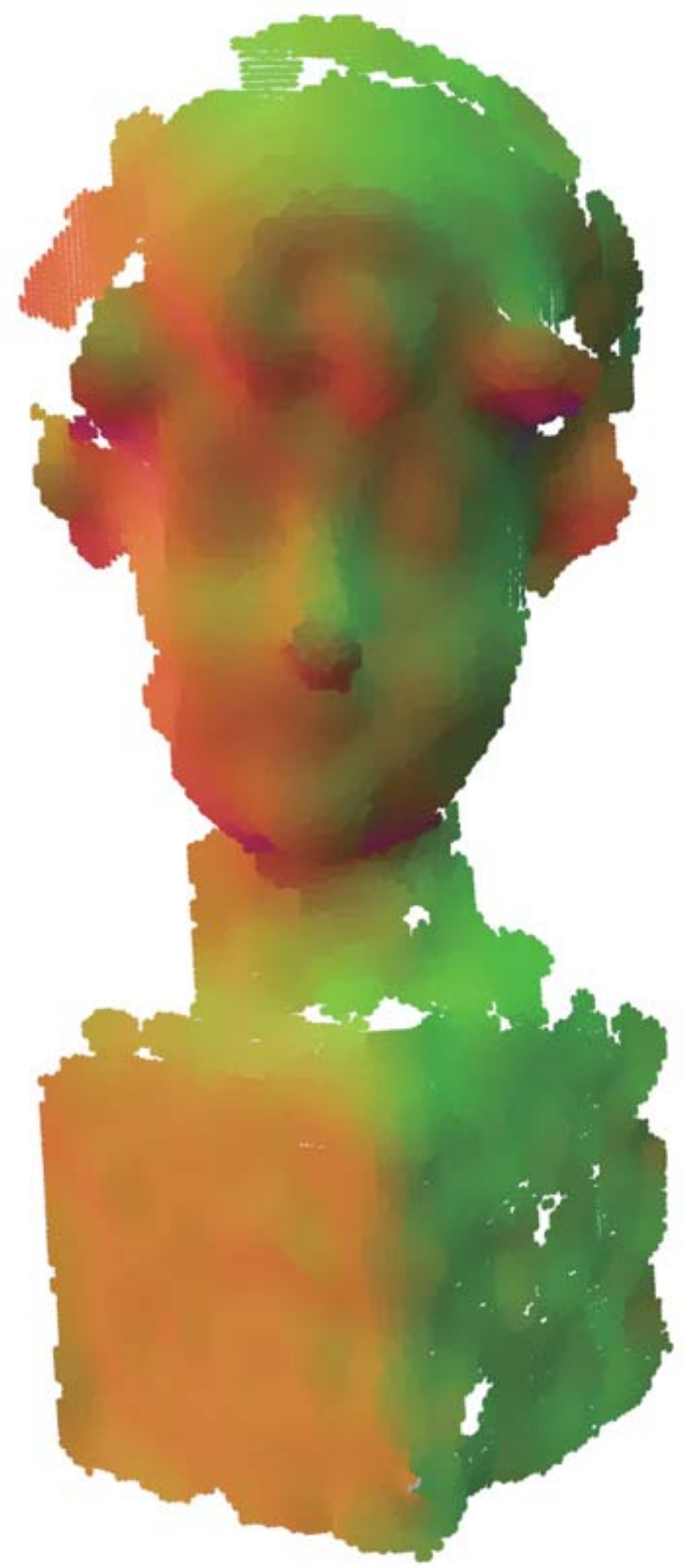}}
\end{minipage} \\
\begin{minipage}[b]{0.32\linewidth}
\subfigure[Huang et al. 2013]
{\label{}\includegraphics[width=1\linewidth]{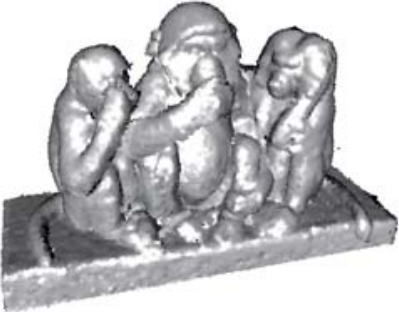}}
\end{minipage}
\begin{minipage}[b]{0.32\linewidth}
\subfigure[Ours]{\label{}\includegraphics[width=1\linewidth]{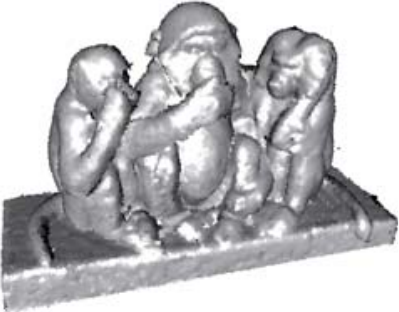}}
\end{minipage}
\begin{minipage}[b]{0.32\linewidth}
\subfigure[Ours]
{\label{}\includegraphics[width=1\linewidth]{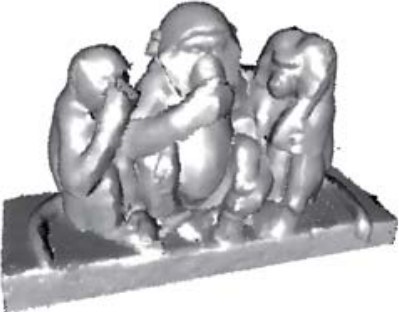}}
\end{minipage}
\caption{Normal estimation results on David (a,b), a female statue (c,d) and monkeys (e,f,g). The mean square angular errors of (a-g) are respectively ($\times 10^{-2}$): $10.684$, $10.636$, $9.534$, $9.423$, $5.004$, $4.853$ and $4.893$. (b,d,f) used smaller $k_{non}$ and $k_{local}$, and (g) used the default $k_{non}$ and $k_{local}$.  }
\label{fig:complexfeatures}
%\vspace{-0.2cm}
\end{figure}

\subsection{Point Cloud Filtering}
\label{sec:pointcloudfiltering}
We compare our normal estimation method with several state of the art normal estimation techniques. We then perform the same number of iterations of our position update algorithm with the estimated normals of all methods. 

Figure \ref{fig:cube_point} and \ref{fig:dod_point} show two point cloud models corrupted with heavy, synthetic noise. The results demonstrate that our method performs better than state of the art approaches in terms of feature preservation and non-feature smoothness. Figure \ref{fig:car_point}, \ref{fig:house_point}, 
\ref{fig:iron_point}, and \ref{fig:toy_point} show the methods applied to a variety of real scanned point cloud models. Our approach outperforms other methods in terms of the quality of the estimated normals. We demonstrate our technique on point clouds with more complicated features. Figure \ref{fig:complexfeatures} shows that our method produces slightly lower normal errors than \cite{Huang2013}.  Figure \ref{fig:complexfeatures} (f) and (g) show our method with different parameters, which leads to a less/more sharpened version of the input.    
We also show some results using \cite{Huang2009,Preiner2014}, which do not preserve sharp features (Figure \ref{fig:wlopclop_dod}).

\subsection{Point Cloud Upsampling}
\label{sec:pointcloudupsample}
As described in Section \ref{sec:pointcloudfiltering}, the point cloud filtering also consists of a two-step procedure: normal estimation and point update. However, unlike mesh shapes, point cloud models often need to be resampled to enhance point density after filtering operations have been applied. 

We apply the edge-aware point set resampling technique \cite{Huang2013} to all the results after point cloud filtering and contrast the different upsampling results. For fair comparisons, we upsample the filtered point clouds of each model to reach a similar number of points. Figure \ref{fig:cube_point} to \ref{fig:toy_point} display various upsampling results on state of the art normal estimation methods and different point cloud models. The figures show that the upsampling results on our filtered point clouds are substantially better than those filtered by other methods in preserving geometric features.  
Bilateral normal smoothing \cite{Huang2013} usually produces good results, but this method sometimes blur edges with low dihedral angles.

\subsection{Surface Reconstruction}
\label{sec:surfacereconstruct}
One common application for point cloud models is to reconstruct surfaces from the upsampled point clouds in Section \ref{sec:pointcloudupsample} before use in other applications.  Here, we select the edge-aware surface reconstruction technique--RIMLS \cite{Oztireli2009}. For fair comparisons, we use the same parameters for all the upsampled point clouds of each model.

Figure \ref{fig:cube_point} to \ref{fig:toy_point} show a variety of surface reconstruction results on different point cloud models. The comparison results demonstrate that the RIMLS technique over our method produces the best surface reconstruction results, in terms of edge preservation.

\subsection{Mesh Denoising}
\label{sec:meshdenoisingapp}
Many state of the art mesh denoising methods involve a two-step procedure which first estimates normals and then updates vertex positions. We selected several of these methods \cite{Sun2007,Zheng2011,Zhang2015} for comparisons in Figure \ref{fig:meshdenoising}.  Note that \cite{Zheng2011} provides both a local and global solution, and we provide comparisons for both.

When the noise level is high, many of these methods produce flipped face normals. For the Bunny model (Figure \ref{fig:meshdenoising}), which involves frequent flipped triangles, we utilize the technique in \cite{Lu2016} to estimate a starting mesh from the original noisy mesh input for all methods.
The comparison results show that our method outperforms state of the art mesh denoising methods in terms of feature preservation.

\subsection{Geometric Texture Removal}
\label{sec:textureremoval}
\begin{figure*}[htbp]
%\vspace{-0.0cm}
\centering
\begin{minipage}[b]{0.15\linewidth}
{\label{}\includegraphics[width=1\linewidth]{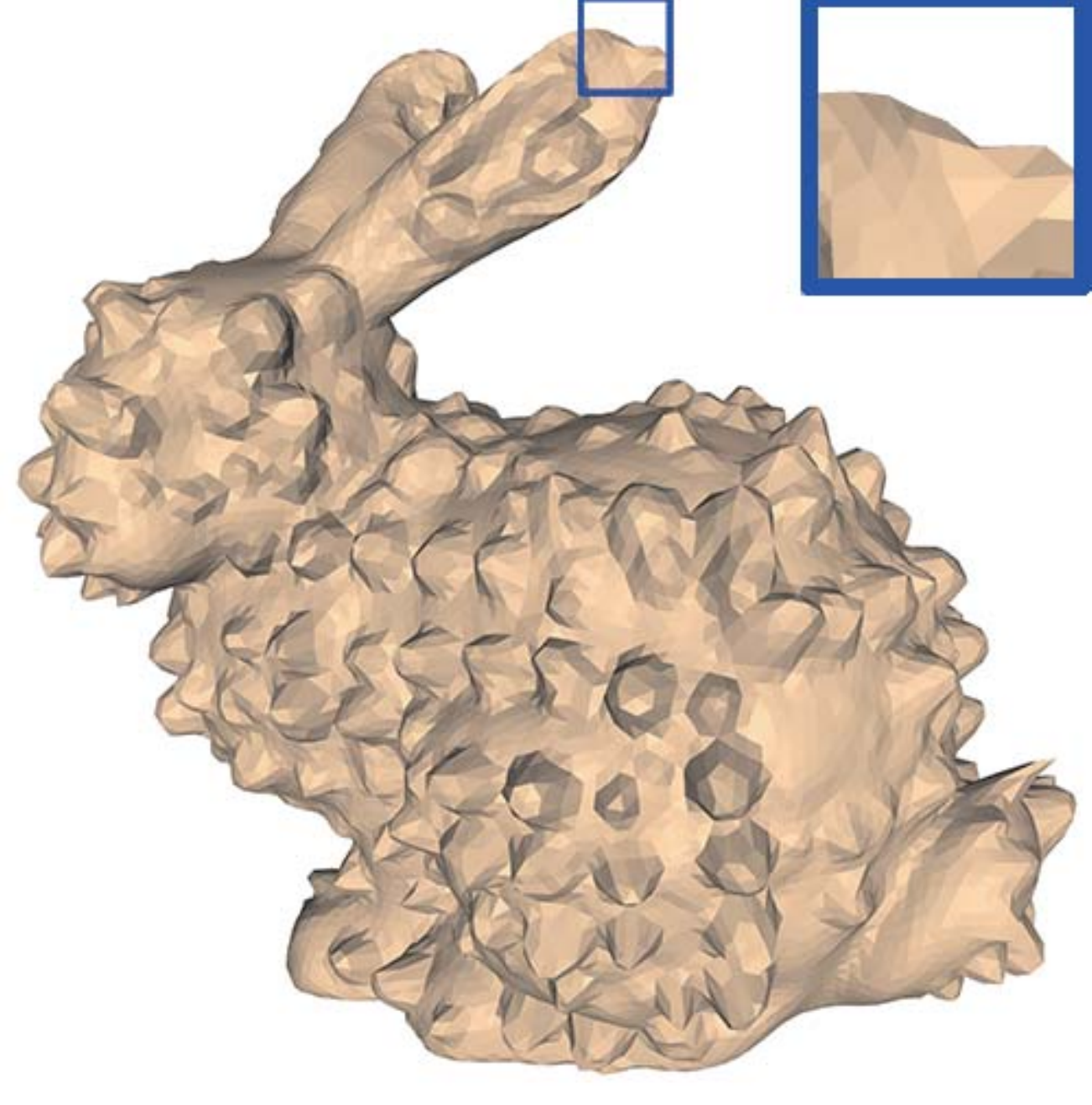}}
\end{minipage}
\begin{minipage}[b]{0.15\linewidth}
{\label{}\includegraphics[width=1\linewidth]{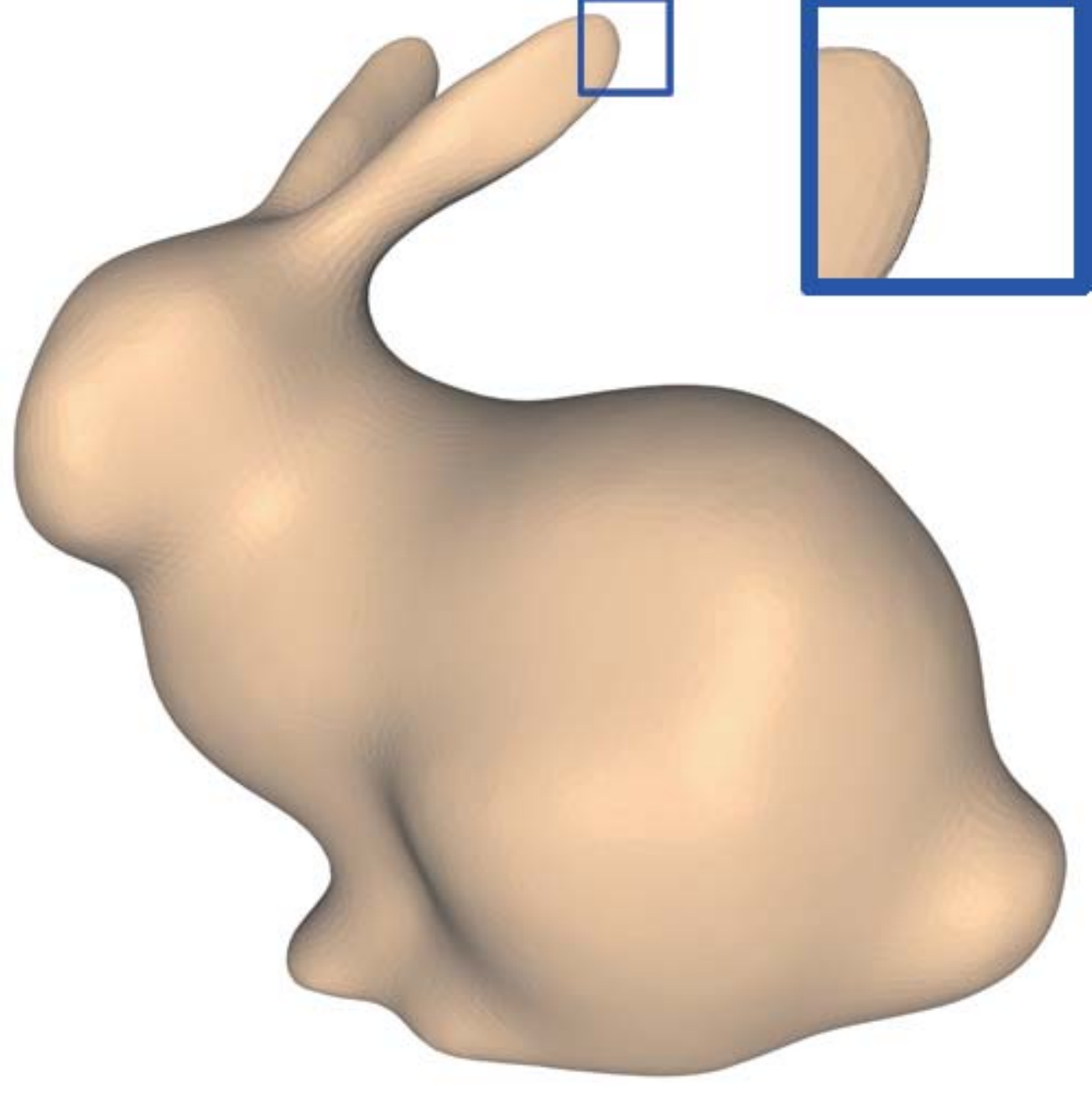}}
\end{minipage}
\begin{minipage}[b]{0.15\linewidth}
{\label{}\includegraphics[width=1\linewidth]{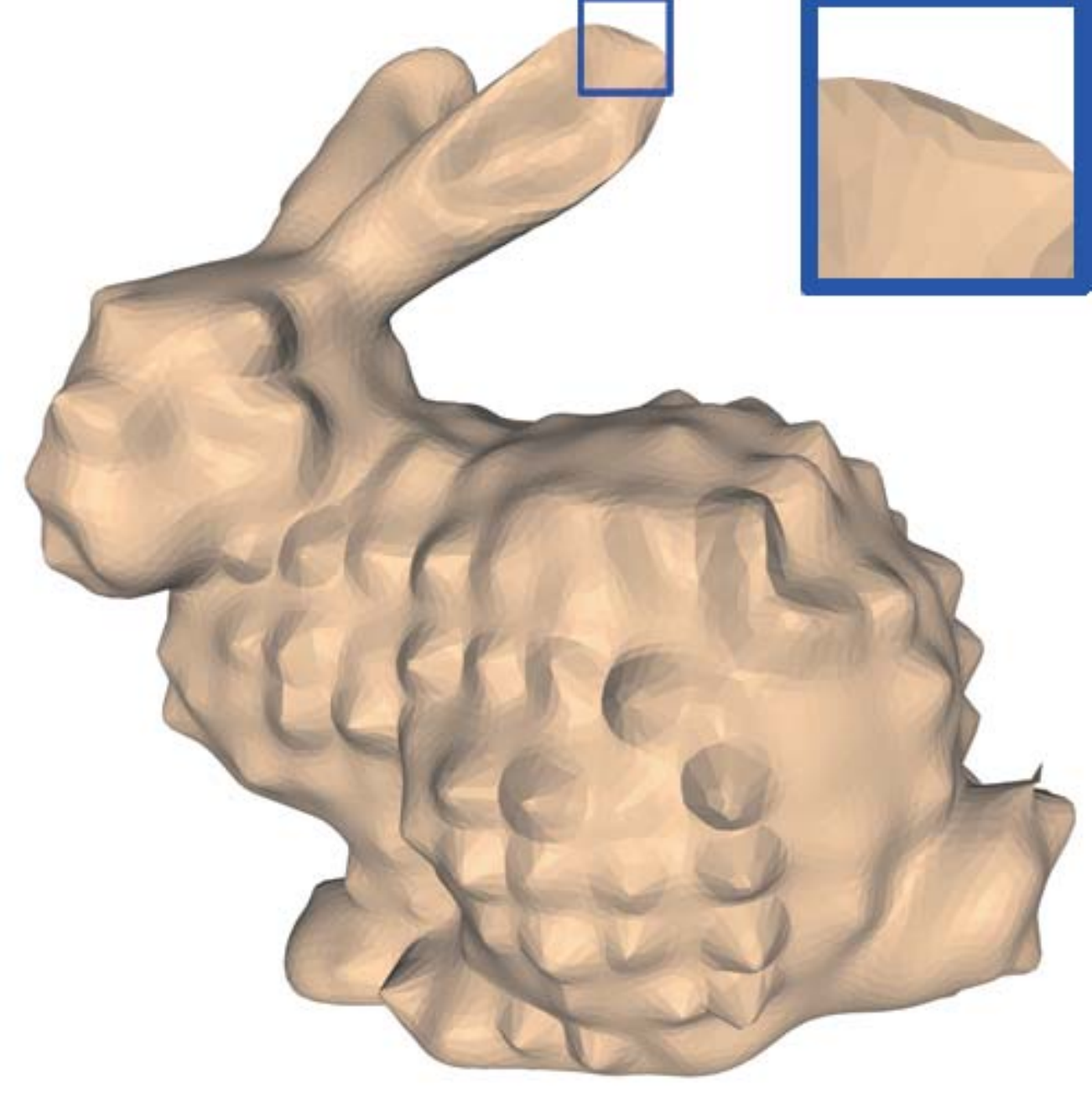}}
\end{minipage}
\begin{minipage}[b]{0.15\linewidth}
{\label{}\includegraphics[width=1\linewidth]{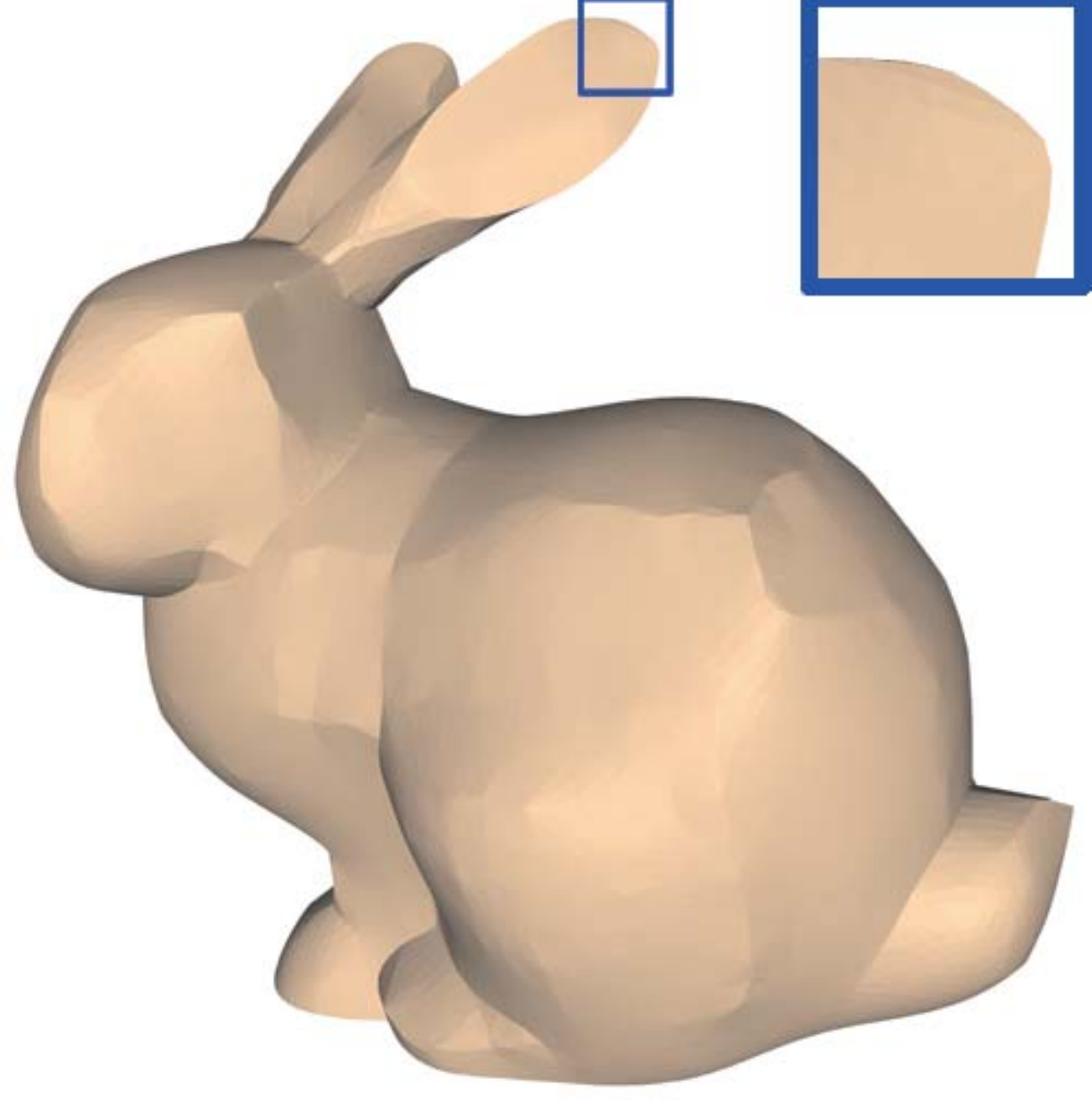}}
\end{minipage}
\begin{minipage}[b]{0.15\linewidth}
{\label{}\includegraphics[width=1\linewidth]{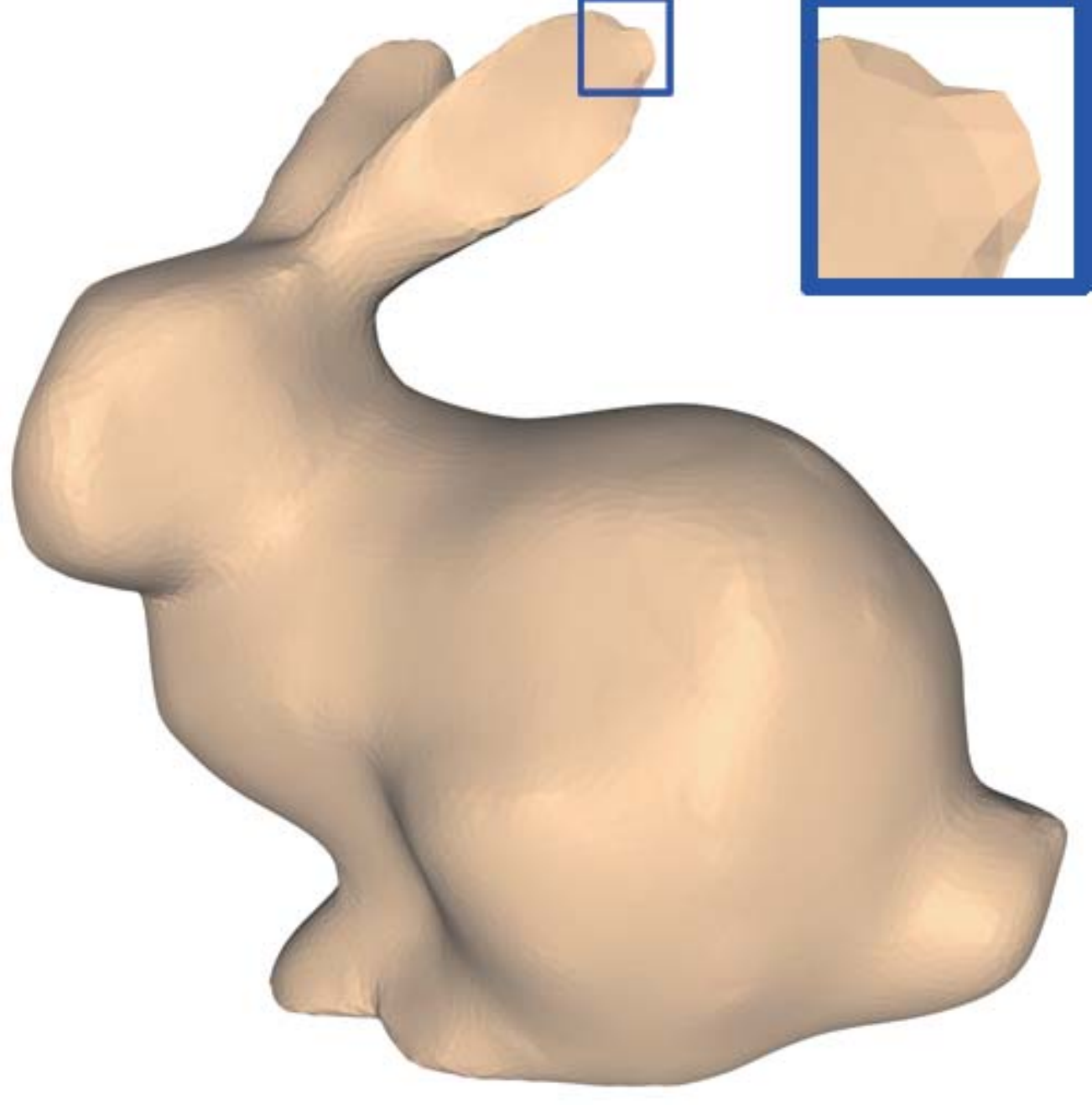}}
\end{minipage}
\begin{minipage}[b]{0.15\linewidth}
{\label{}\includegraphics[width=1\linewidth]{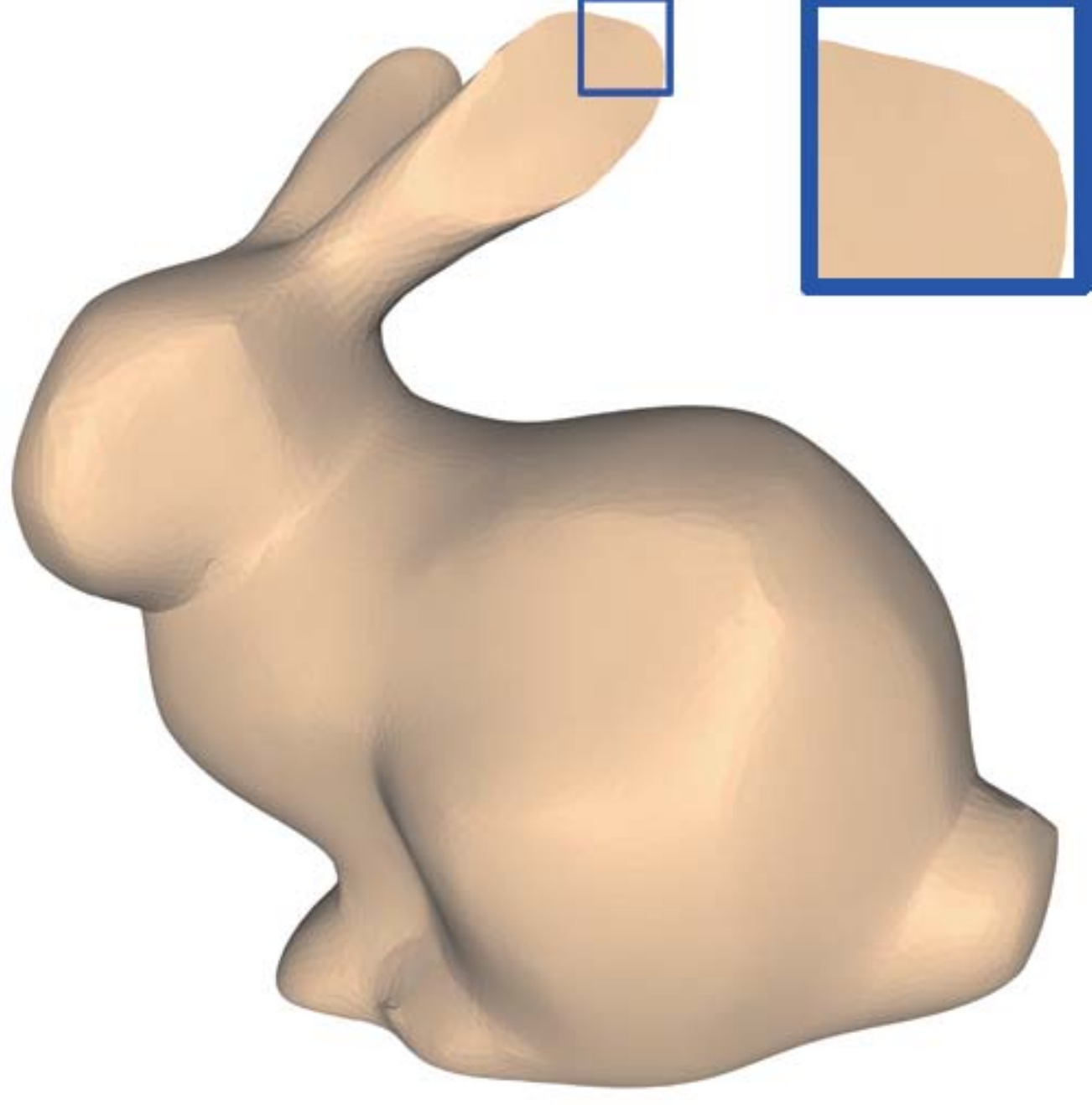}}
\end{minipage}	\\
\begin{minipage}[b]{0.15\linewidth}
\subfigure[Input]{\label{}\includegraphics[width=1\linewidth]{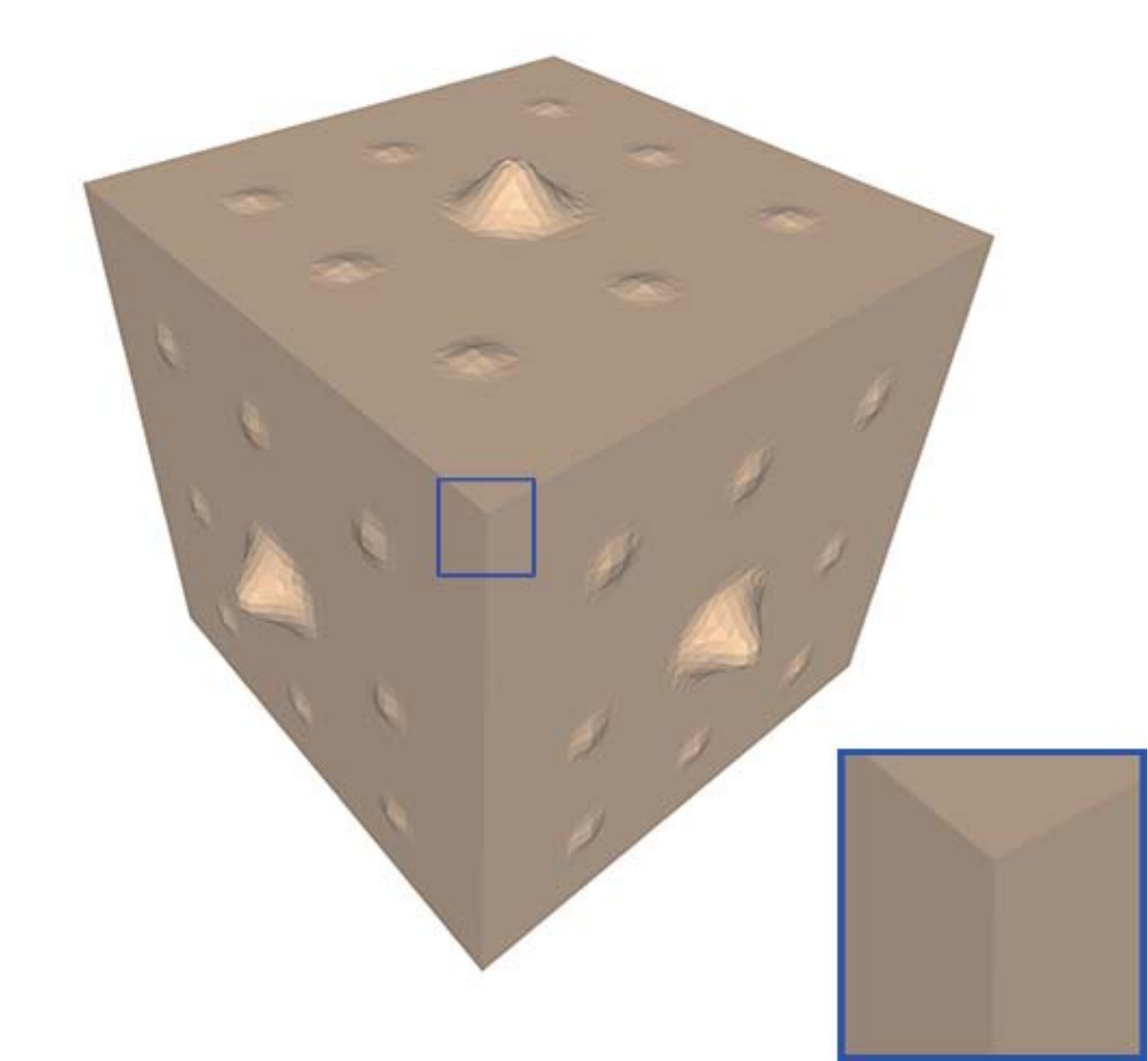}}
\end{minipage}
\begin{minipage}[b]{0.15\linewidth}
\subfigure[Laplacian]{\label{}\includegraphics[width=1\linewidth]{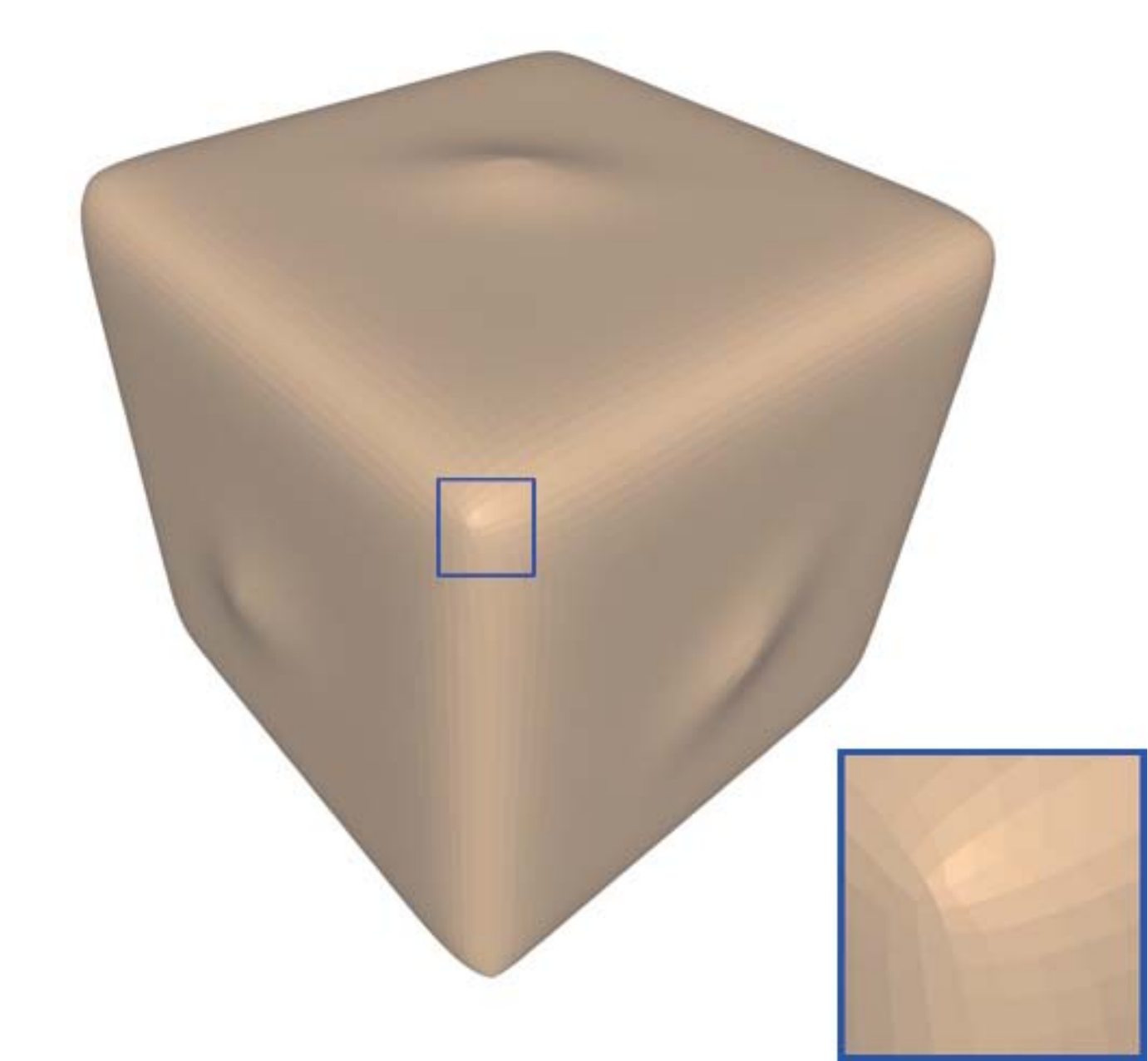}}
\end{minipage}
\begin{minipage}[b]{0.15\linewidth}
\subfigure[\protect\cite{Zheng2011} (local)]{\label{}\includegraphics[width=1\linewidth]{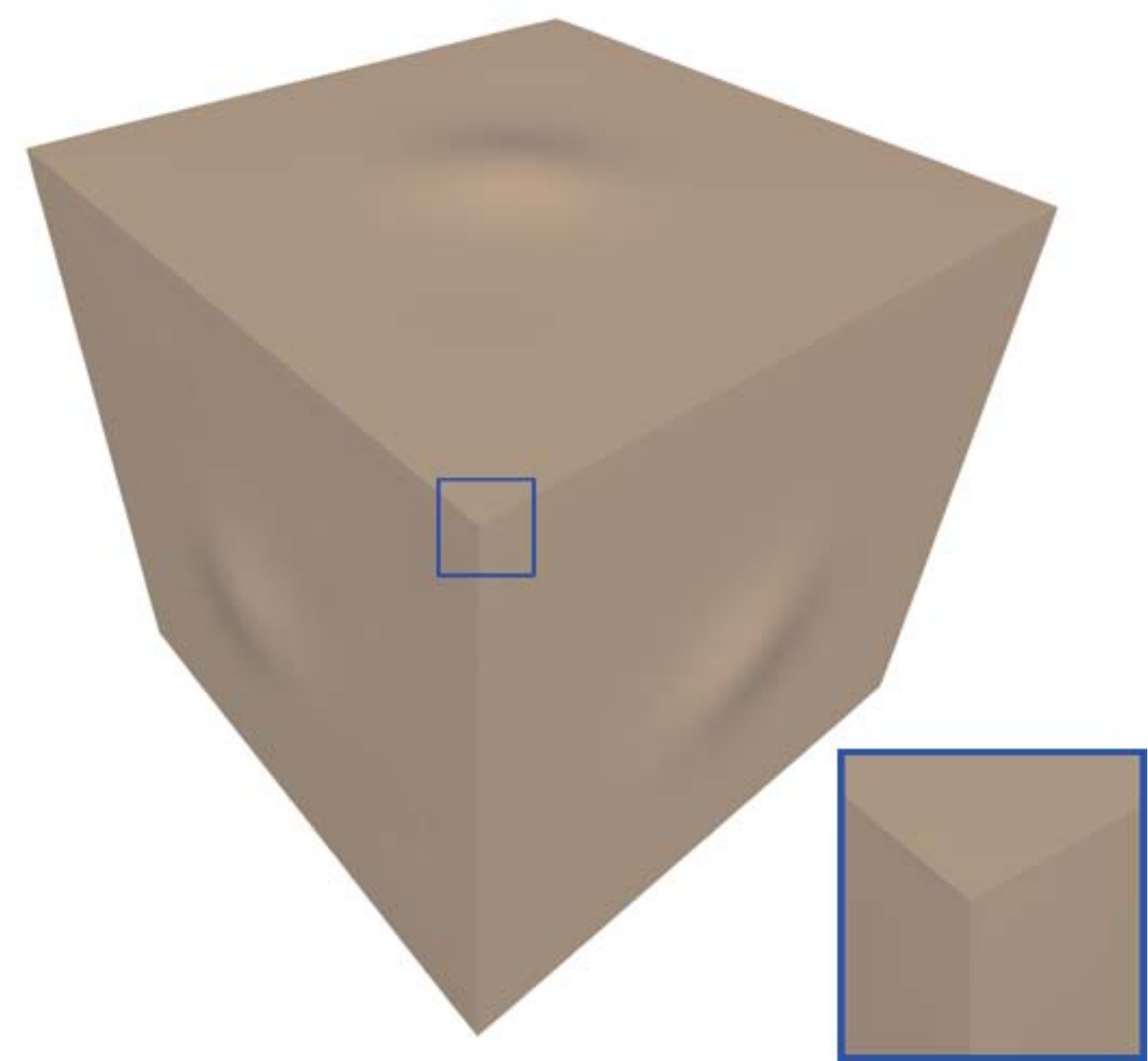}}
\end{minipage}
\begin{minipage}[b]{0.15\linewidth}
\subfigure[\protect\cite{He2013}]{\label{}\includegraphics[width=1\linewidth]{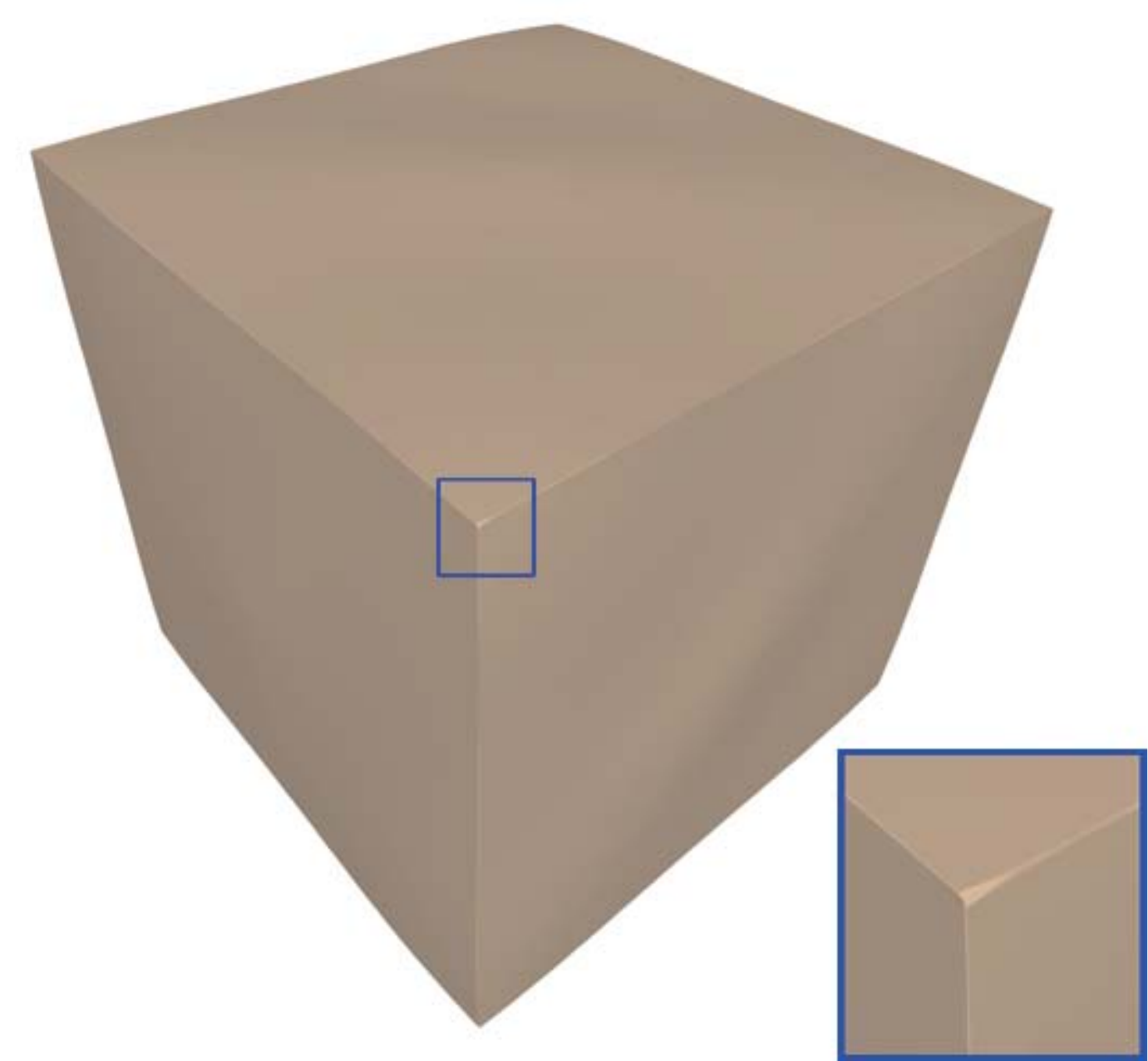}}
\end{minipage}
\begin{minipage}[b]{0.15\linewidth}
\subfigure[\protect\cite{Wang2015}]{\label{}\includegraphics[width=1\linewidth]{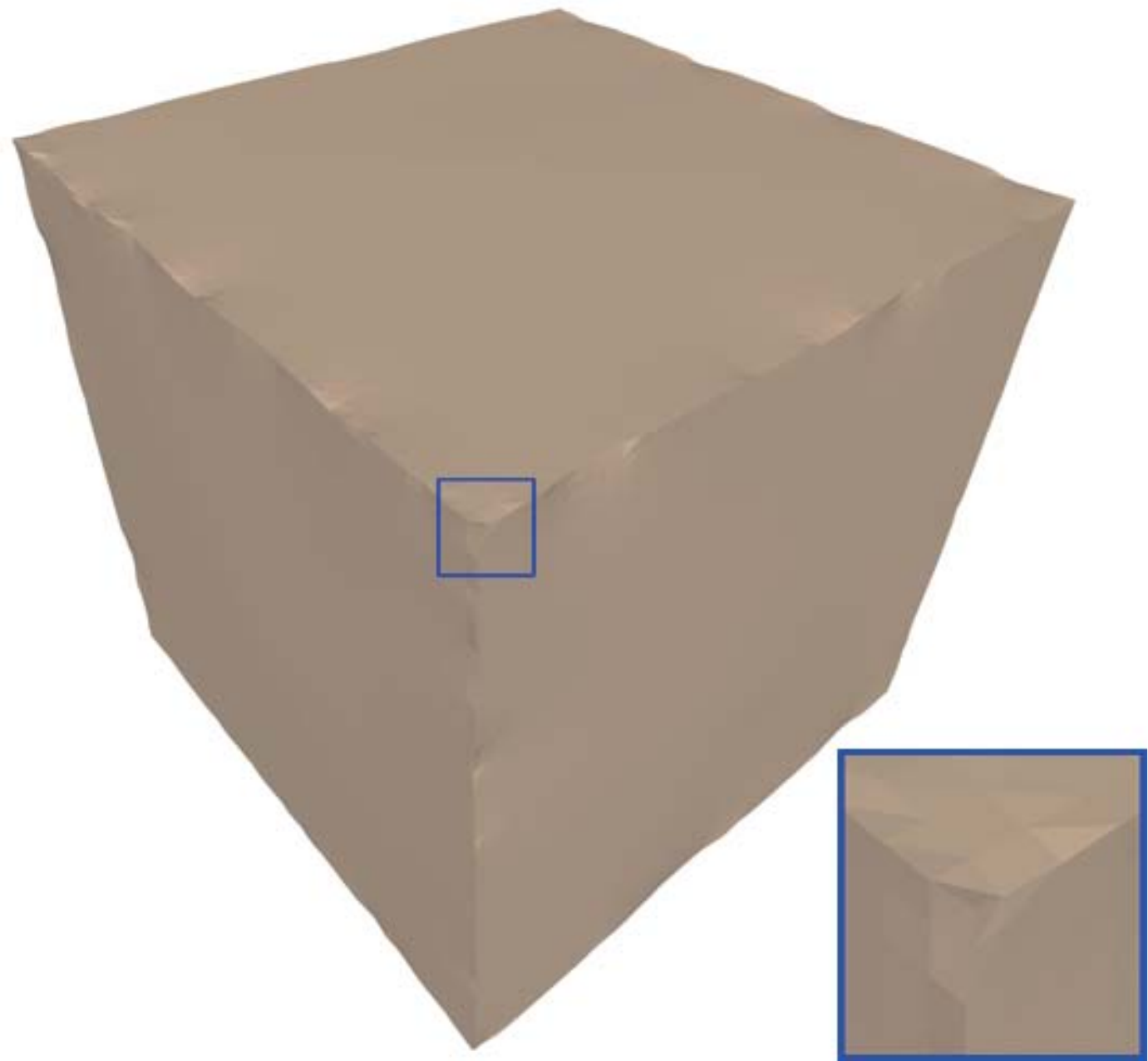}}
\end{minipage}
\begin{minipage}[b]{0.15\linewidth}
\subfigure[Ours]{\label{}\includegraphics[width=1\linewidth]{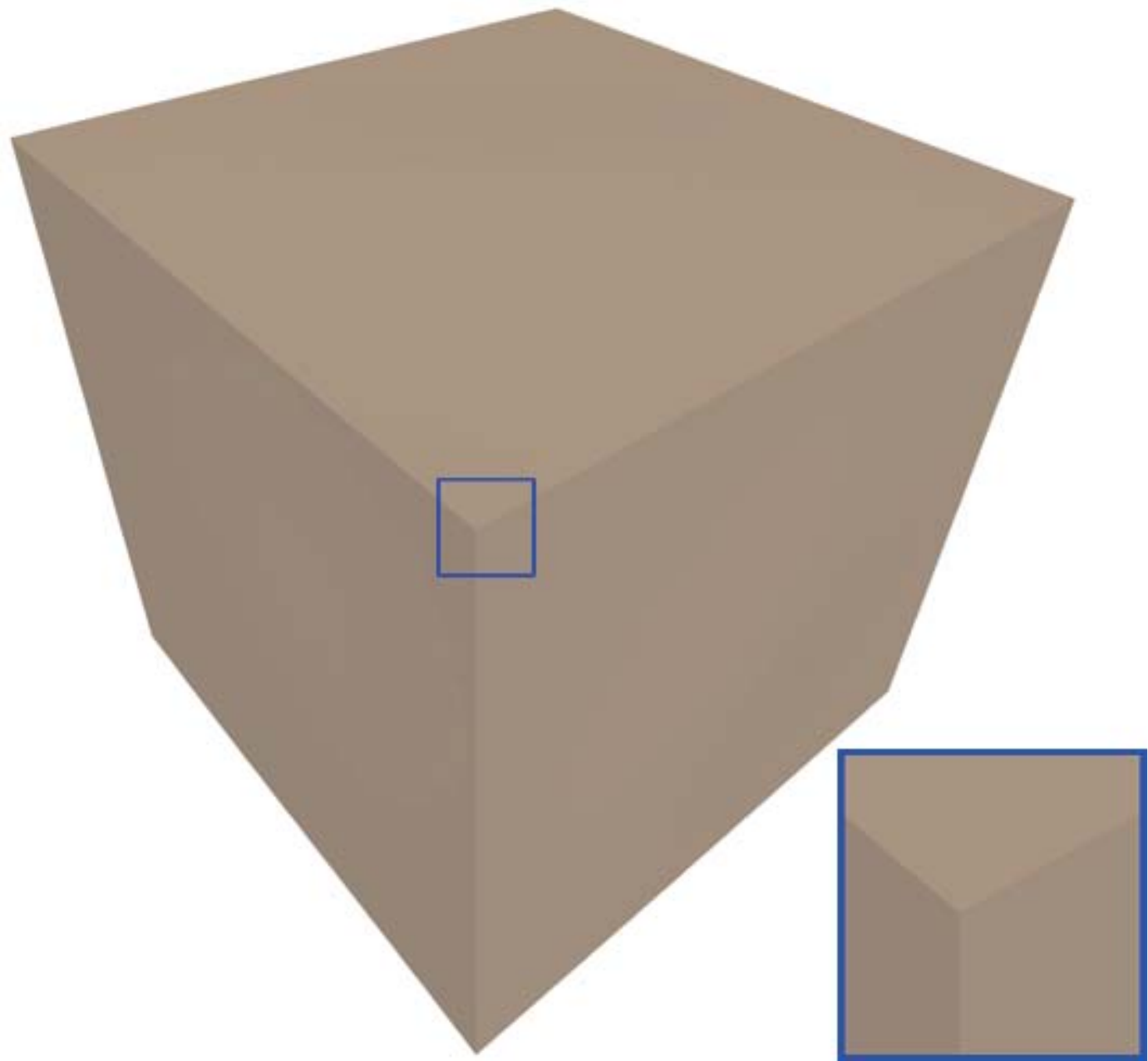}}
\end{minipage}
\caption{Geometric texture removal results of the Bunny and Cube. Please refer to the zoomed rectangular windows. }
\label{fig:meshtexture}
%\vspace{-0.65cm}
\end{figure*}

\begin{figure}[htbp]
%\vspace{-0.0cm}
\centering
\begin{minipage}[b]{0.24\linewidth}
\subfigure[Input]{\label{}\includegraphics[width=1\linewidth]{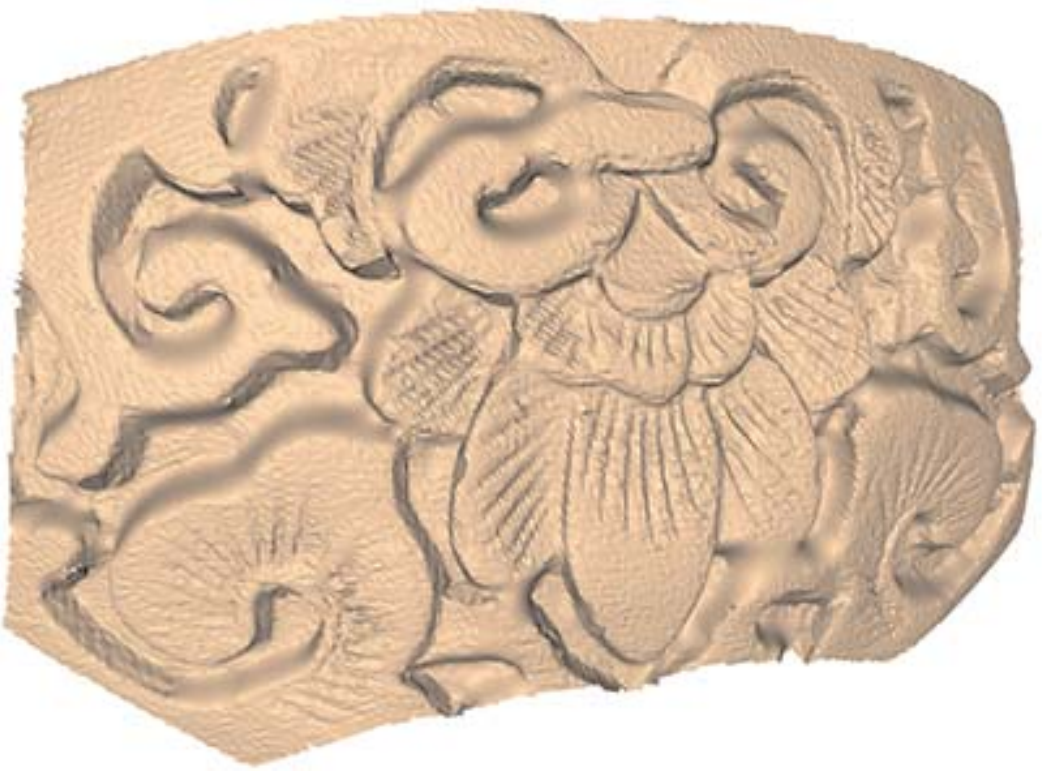}}
\end{minipage}
\begin{minipage}[b]{0.24\linewidth}
\subfigure[Small texture removal]{\label{}\includegraphics[width=1\linewidth]{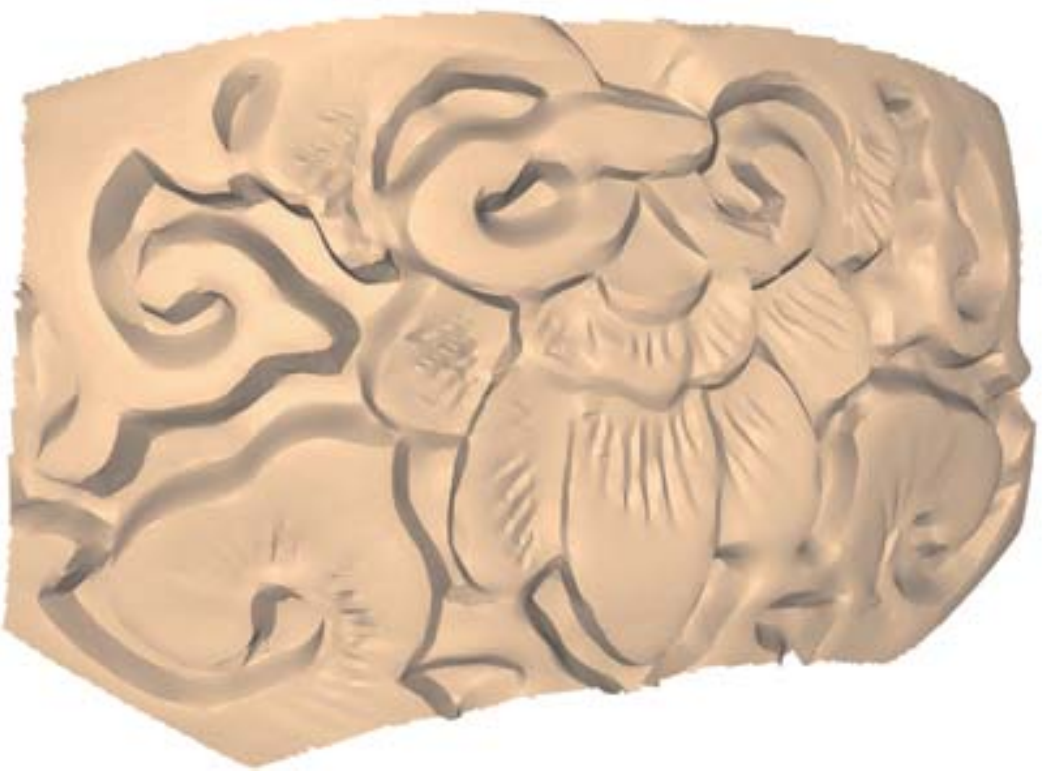}}
\end{minipage}
\begin{minipage}[b]{0.24\linewidth}
\subfigure[Medium texture removal]{\label{}\includegraphics[width=1\linewidth]{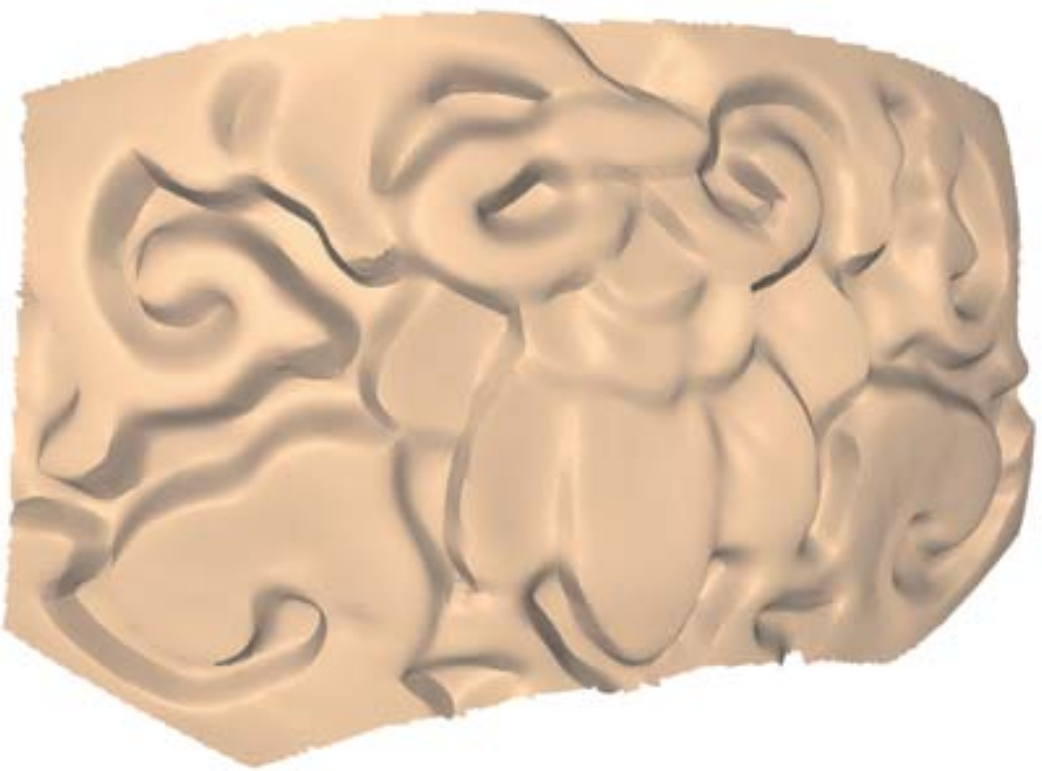}}
\end{minipage}
\begin{minipage}[b]{0.24\linewidth}
\subfigure[Large texture removal]{\label{}\includegraphics[width=1\linewidth]{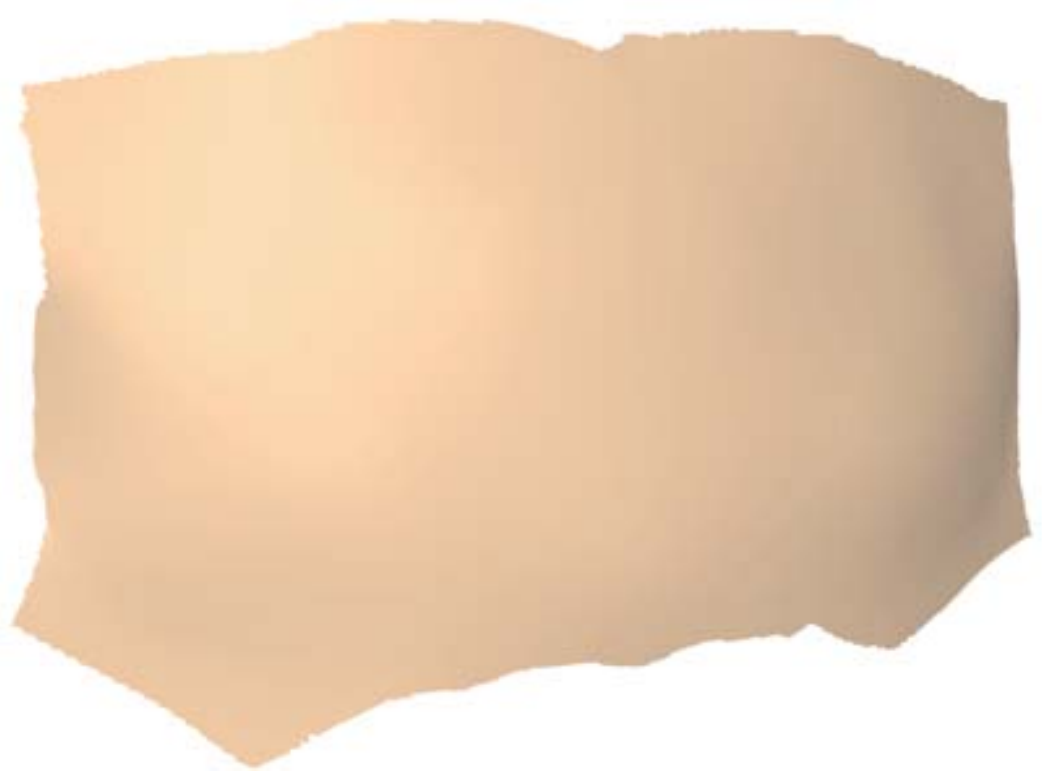}}
\end{minipage}
\caption{Different scales of geometric texture removal results of the Circular model. }
\label{fig:circularfeature}
%\vspace{-0.65cm}
\end{figure}

\begin{figure}[htbp]
%\vspace{-0.0cm}
\centering
\begin{minipage}[b]{0.26\linewidth}
\subfigure[Input]{\label{}\includegraphics[width=1\linewidth]{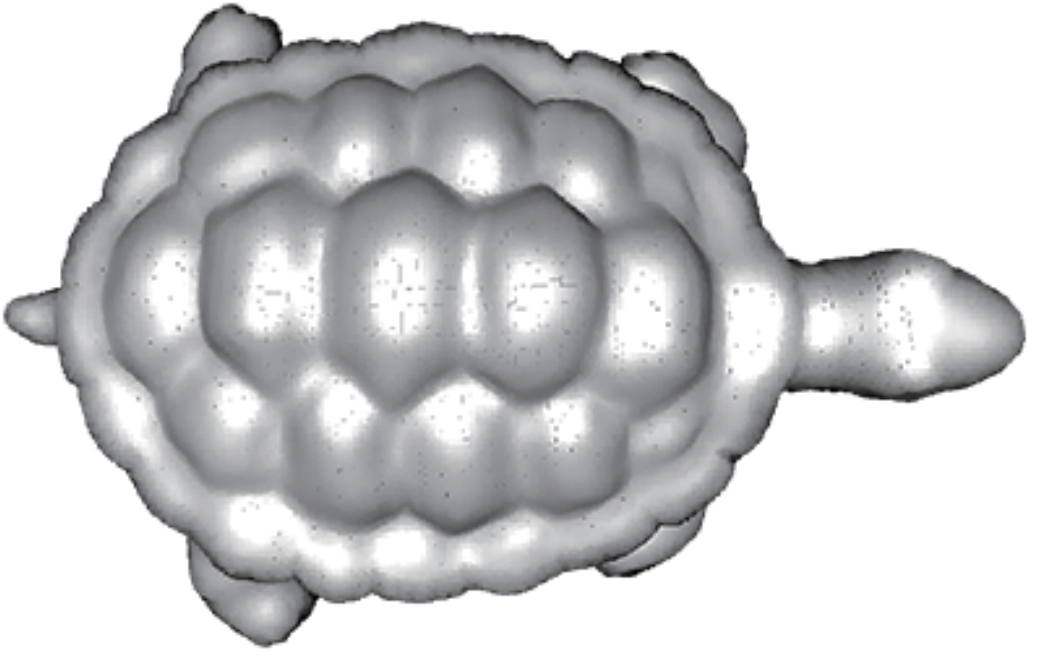}}
\end{minipage}
\begin{minipage}[b]{0.26\linewidth}
\subfigure[Ours]{\label{}\includegraphics[width=1\linewidth]{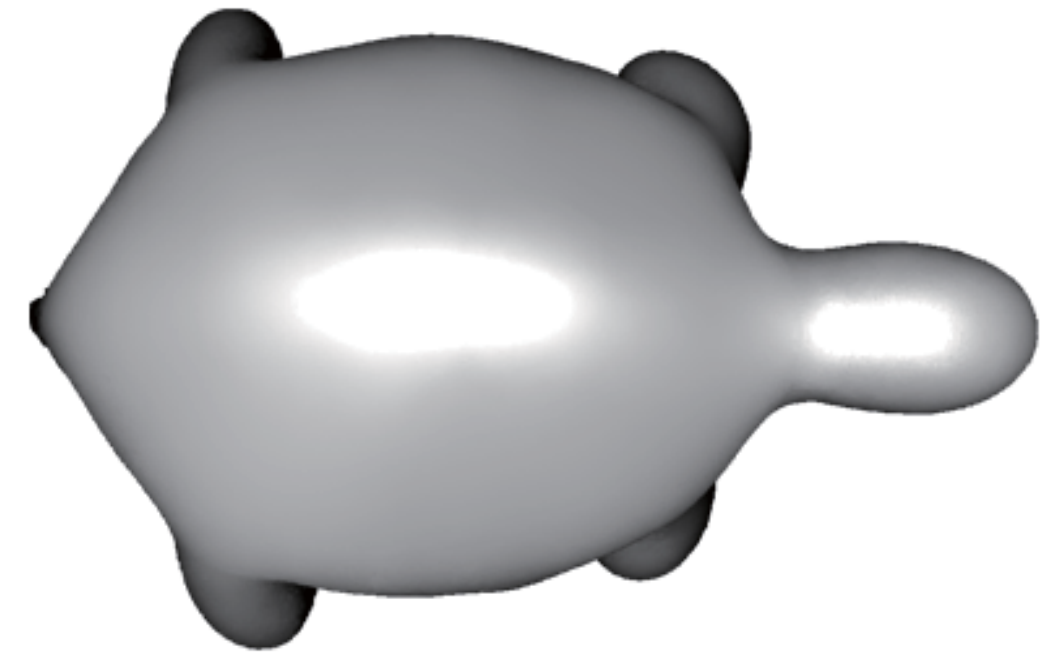}}
\end{minipage}
\begin{minipage}[b]{0.21\linewidth}
\subfigure[Input]{\label{}\includegraphics[width=1\linewidth]{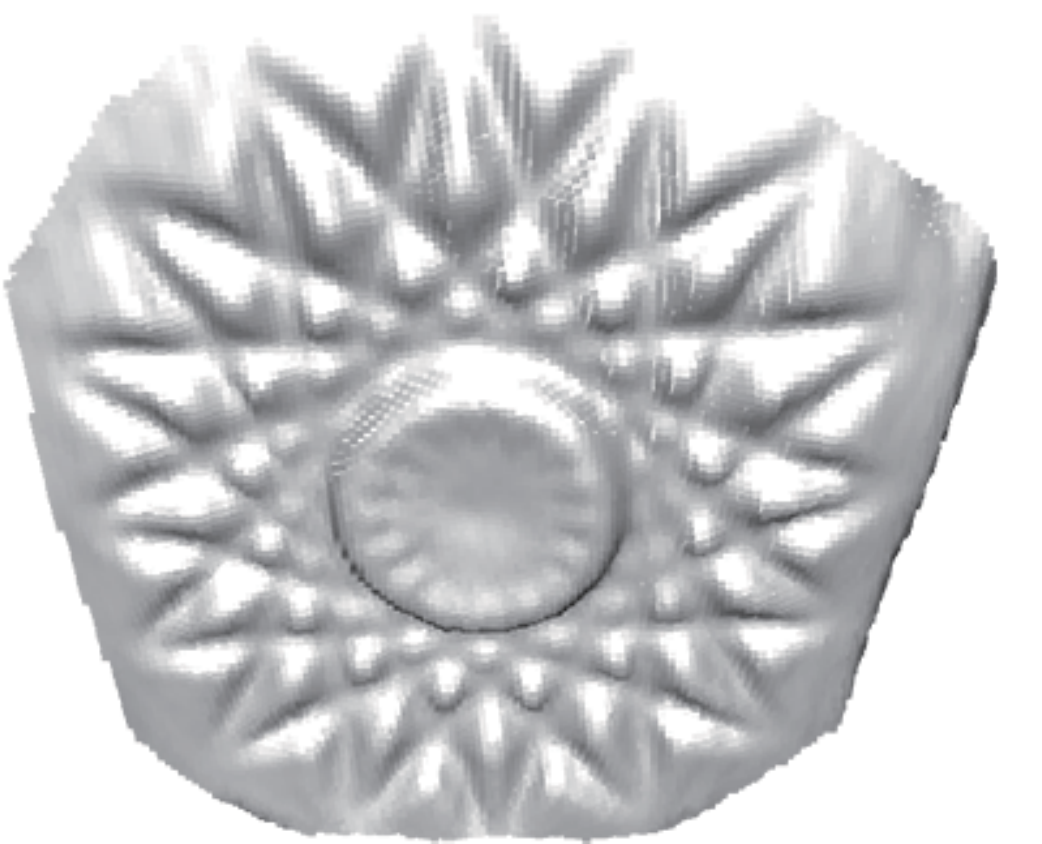}}
\end{minipage}
\begin{minipage}[b]{0.21\linewidth}
\subfigure[Ours]{\label{}\includegraphics[width=1\linewidth]{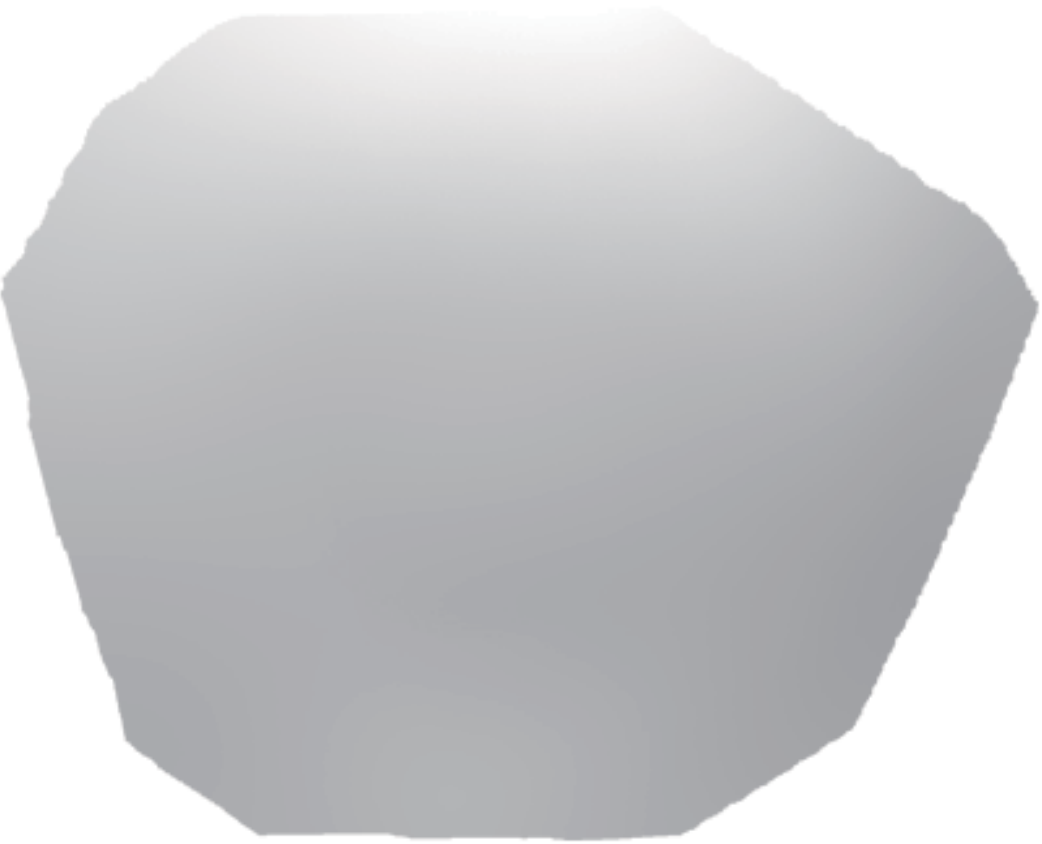}}
\end{minipage}
\caption{Geometric texture removal results of the Turtle point cloud and embossed point cloud. We render point set surfaces of each point cloud for visualization. }
\label{fig:pointcloudfeature}
%\vspace{-0.65cm}
\end{figure}

We also successfully applied our method to geometric de-texturing where we remove high frequency features of different scales. Figure \ref{fig:meshtexture} shows comparisons of different methods that demonstrate that our method outperforms other approaches. Note that \cite{Wang2015} is specifically designed for geometric texture removal. However, that method cannot preserve sharp edges well. Figure \ref{fig:circularfeature} shows the results of removing different scales of geometric features on a mesh. We produced Figure \ref{fig:circularfeature} (d) by applying the pre-filtering technique \cite{Lu2016} in advance, since the vertex update algorithm \cite{Zheng2011} could generate frequent flipped triangles when dealing with such large and steep geometric features. As an alternative, our normal estimation method can be combined with the vertex update in \cite{Wang2015} to handle such challenging mesh models. Figure \ref{fig:pointcloudfeature} shows the geometric texture removal on two different point clouds, which are particularly challenging due to a lack of topology.

%bunny feature removal

%circular mesh self-test

%point cloud feature removal

%% file: paper/discussion.tex
%timing
\begin{table}[thbp]\eqfont
    \centering
    \caption{Timing statistics for different normal estimation techniques over point clouds.}\label{table:table2}
    \begin{tabular}{l c c c c c c}
    \toprule
    Methods & \tabincell{l}{
    Hoppe et\\al.1992\\
    } & \tabincell{l}{
    Boulch and\\Marlet 2012
    } & \tabincell{l}{
    Huang et\\al.2015
    } & \tabincell{l}{Boulch and\\Marlet 2016} & \tabincell{l}{
    Ours\\SVD} & \tabincell{l}{
    Ours\\RSVD} \\ 
    \midrule
    \tabincell{l}{Fig. \protect\ref{fig:cube_point}\\\#6146} & 0.57 & 141.5 & 0.48 & 8 & 95.6 & 65.1
    \\
    \tabincell{l}{Fig. \protect\ref{fig:iron_point}\\\#100k} & 18.7 & 2204 & 17.2 & 115 & 3147 & 2458
    \\
    \tabincell{l}{Fig. \protect\ref{fig:house_point}\\\#127k} & 10.8 & 3769 & 12.5 & 141 & 3874 & 2856
    \\
    \bottomrule
    \end{tabular}
\end{table}

\subsection{Timings}
\label{sec:discussion}
Table \ref{table:table2} summarizes the timings of different normal estimation methods on several point clouds.  While our method produces high quality output, the algorithm takes a long time to run due to the svd operation for each normal estimation. Therefore, our method is more suitable for offline geometry processing. However, it is possible to accelerate our method using specific svd decomposition algorithms, such as the randomized svd (RSVD) decomposition algorithm \cite{Halko2011} as shown in Table \ref{table:table2}. In addition, many parts of the algorithm could benefit from parallelization.

%% file: paper/conclusion.tex
\section{Conclusion}
\label{sec:conclusion}
In this paper, we have presented an approach consisting of two steps: normal estimation and position update. Our method can handle both mesh shapes and point cloud models. We also show various geometry processing applications that benefit from our approach directly or indirectly. The extensive experiments demonstrate that our method performs substantially better than state of the art techniques, in terms of both visual quality and accuracy. 

While our method works well, speed is an issue if online processing speeds are required.  In addition, though we mitigate issues associated with the point distribution in the position update procedure, the point distribution could still be improved. One way to do so is to re-distribute points after our position update through a ``repulsion force'' from each point to its neighbors. We could potentially accomplish this effect by adding this repulsion force directly to Eq. \eqref{eq:pointminimization}.

%% file: paper/appendix.tex
\section*{Appendix}
\label{sec:appendix}
In this section, we show how to prove the convergence of the point update algorithm. The point update is convergent in the sense that the total energy $E = \sum_i\sum_{j\in S_i} \normScalar{(\mathbf{p}_i-\mathbf{p}_j)\mathbf{n}_j^T}_2^2 + \normScalar{(\mathbf{p}_i-\mathbf{p}_j)\mathbf{n}_i^T}_2^2$ decreases in each iteration. With the assumption of ball neighbors search, we can obtain:
\begin{equation}\label{eq:E}
\begin{aligned}
E = \mathbf{P}\mathbf{Q}\mathbf{P}^T,
\end{aligned}
\end{equation}
where $\mathbf{P}$ is a $1\times3|i|$ vector concatenated by all point positions and $|i|$ is the number of points. $\mathbf{Q}$ is a $3|i|\times 3|i|$ matrix which consists of $|i|\times|i|$ blocks ($3\times3$). We use $\mathbf{Q}_{i,i}$ and $\mathbf{Q}_{i,j}$ to respectively denote the diagonal and other blocks: $\mathbf{Q}_{i,i}=2\sum_{j}(\mathbf{n}_i^T\mathbf{n}_i+\mathbf{n}_j^T\mathbf{n}_j)$ and $\mathbf{Q}_{i,j} = \mathbf{Q}_{j,i} = -2(\mathbf{n}_i^T\mathbf{n}_i+\mathbf{n}_j^T\mathbf{n}_j)$. 

From Eq. \eqref{eq:pointupdate}, we can achieve the new positions $\mathbf{P}' = \mathbf{P}(\mathbf{I}-\mathbf{O}\mathbf{G})$. $\mathbf{I}$ is the identity matrix, $\mathbf{O}=\frac{1}{2}\mathbf{Q}$ and $\mathbf{G}$ is a diagonal block matrix with diagonal elements $g_{3i,3i}= g_{3i+1,3i+1}=g_{3i+2,3i+2}= \gamma_i$. Here, the lowercase indicates elements while uppercase denotes block matrices. Based on $\mathbf{P}'$, we have $E' = \mathbf{P}'\mathbf{Q}\mathbf{P}'^T = \mathbf{P}(\mathbf{Q}-\mathbf{QGO} - \mathbf{OGQ} + \mathbf{OGQGO}) \mathbf{P}^T$. Thus, we obtain
\begin{equation}\label{eq:differenceE}
\begin{aligned}
E - E' &= \mathbf{P}\mathbf{Q}\mathbf{P}^T - \mathbf{P}(\mathbf{Q}-\mathbf{QGO} - \mathbf{OGQ} + \mathbf{OGQGO}) \mathbf{P}^T \\
&= 2\mathbf{P} \mathbf{OG}(2\mathbf{G}^{-1}-\mathbf{O})\mathbf{GO} \mathbf{P}^T
\end{aligned}
\end{equation}

To demonstrate the convergence of the point update, $E-E'$ should be non-negative. $\mathbf{O}$ and $\mathbf{G}$ are both symmetric positive semidefinite matrices, and we should prove $2\mathbf{G}^{-1}-\mathbf{O}$ is a symmetric positive semidefinite matrix.

Obviously, $2\mathbf{G}^{-1}-\mathbf{O}$ is a symmetric matrix. $2\mathbf{G}^{-1}-\mathbf{O}$ is a positive semidefinite matrix in the sense that its eigenvalues are non-negative. We denote $\lambda$ as one of its eigenvalues, and $\mathbf{X}$ the corresponding eigenvector. Without loss of generality, we assume $|x_l|$ (i.e., $|x_l| \geq |x_k|$) is the greatest in $\mathbf{X}$. $\lambda$ can be computed by
\begin{equation}\label{eq:matrixprove}
\begin{aligned}
\lambda &= \frac{\sum_k (2g'_{l,k} - o_{l,k})x_k} {x_l} \\
&= \sum_k 2g'_{l,k}\frac{x_k}{x_l} - \sum_k o_{l,k}\frac{x_k}{x_l} \\
&= 2g'_{l,l} - \sum_k o_{l,k}\frac{x_k}{x_l} \\
\end{aligned}
\end{equation}
where $\mathbf{G}^{-1}$ is the inverse matrix of $\mathbf{G}$, with elements $g'_{l,l} = \frac{1}{\gamma_l}$ and $g'_{l,k\neq l}=0$. We can easily demonstrate that the sum of the absolute values of each row in $\mathbf{n}_i^T\mathbf{n}_i$ or $\mathbf{n}_j^T\mathbf{n}_j$ is equal or less than $\frac{1+\sqrt{3}}{2}$. Since $g'_{l,l} = \frac{1}{\gamma_l} = 3|S_l|$ and $\sum_k o_{l,k}\frac{x_k}{x_l} \leq \sum_k |o_{l,k}|\frac{|x_k|}{|x_l|} \leq \sum_k |o_{l,k}| \leq 2(1+\sqrt{3})|S_l|$, we can achieve $\lambda \geq 2g'_{l,l} - 2(1+\sqrt{3})|S_l| \geq 0$. Consequently, $2\mathbf{G}^{-1}-\mathbf{O}$ is a symmetric positive semidefinite matrix and $E-E'\geq 0$.

%% file: sample-journal.bbl

\begin{thebibliography}{51}

%%% the Web address.

\ifx \showCODEN    \undefined \def \showCODEN     #1{\unskip}     \fi
\ifx \showDOI      \undefined \def \showDOI       #1{#1}\fi
\ifx \showISBNx    \undefined \def \showISBNx     #1{\unskip}     \fi
\ifx \showISBNxiii \undefined \def \showISBNxiii  #1{\unskip}     \fi
\ifx \showISSN     \undefined \def \showISSN      #1{\unskip}     \fi
\ifx \showLCCN     \undefined \def \showLCCN      #1{\unskip}     \fi
\ifx \shownote     \undefined \def \shownote      #1{#1}          \fi
\ifx \showarticletitle \undefined \def \showarticletitle #1{#1}   \fi
\ifx \showURL      \undefined \def \showURL       {\relax}        \fi
% invisible to TeX
\providecommand\bibfield[2]{#2}
\providecommand\bibinfo[2]{#2}
\providecommand\natexlab[1]{#1}
\providecommand\showeprint[2][]{arXiv:#2}

\bibitem[\protect\citeauthoryear{Alexa, Behr, Cohen-Or, Fleishman, Levin, and
  Silva}{Alexa et~al\mbox{.}}{2001}]%
        {Alexa2001}
\bibfield{author}{\bibinfo{person}{Marc Alexa}, \bibinfo{person}{Johannes
  Behr}, \bibinfo{person}{Daniel Cohen-Or}, \bibinfo{person}{Shachar
  Fleishman}, \bibinfo{person}{David Levin}, {and} \bibinfo{person}{Claudio~T.
  Silva}.} \bibinfo{year}{2001}\natexlab{}.
\newblock \showarticletitle{Point Set Surfaces}. In \bibinfo{booktitle}{{\em
  Proceedings of the Conference on Visualization '01}} {\em
  (\bibinfo{series}{VIS '01})}. \bibinfo{publisher}{IEEE Computer Society},
  \bibinfo{address}{Washington, DC, USA}, \bibinfo{pages}{21--28}.
\newblock
\showISBNx{0-7803-7200-X}
\showURL{%
\url{http://dl.acm.org/citation.cfm?id=601671.601673}}


\bibitem[\protect\citeauthoryear{Alliez, Cohen-Steiner, Tong, and
  Desbrun}{Alliez et~al\mbox{.}}{2007}]%
        {Alliez2007}
\bibfield{author}{\bibinfo{person}{P. Alliez}, \bibinfo{person}{D.
  Cohen-Steiner}, \bibinfo{person}{Y. Tong}, {and} \bibinfo{person}{M.
  Desbrun}.} \bibinfo{year}{2007}\natexlab{}.
\newblock \showarticletitle{Voronoi-based Variational Reconstruction of
  Unoriented Point Sets}. In \bibinfo{booktitle}{{\em Proceedings of the Fifth
  Eurographics Symposium on Geometry Processing}} {\em (\bibinfo{series}{SGP
  '07})}. \bibinfo{publisher}{Eurographics Association},
  \bibinfo{address}{Aire-la-Ville, Switzerland, Switzerland},
  \bibinfo{pages}{39--48}.
\newblock
\showISBNx{978-3-905673-46-3}
\showURL{%
\url{http://dl.acm.org/citation.cfm?id=1281991.1281997}}


\bibitem[\protect\citeauthoryear{Avron, Sharf, Greif, and Cohen-Or}{Avron
  et~al\mbox{.}}{2010}]%
        {Avron2010}
\bibfield{author}{\bibinfo{person}{Haim Avron}, \bibinfo{person}{Andrei Sharf},
  \bibinfo{person}{Chen Greif}, {and} \bibinfo{person}{Daniel Cohen-Or}.}
  \bibinfo{year}{2010}\natexlab{}.
\newblock \showarticletitle{L1-Sparse Reconstruction of Sharp Point Set
  Surfaces}.
\newblock \bibinfo{journal}{{\em ACM Trans. Graph.\/}} \bibinfo{volume}{29},
  \bibinfo{number}{5}, Article \bibinfo{articleno}{135} (\bibinfo{date}{Nov.}
  \bibinfo{year}{2010}), \bibinfo{numpages}{12}~pages.
\newblock
\showISSN{0730-0301}
\showDOI{%
\url{https://doi.org/10.1145/1857907.1857911}}


\bibitem[\protect\citeauthoryear{Boulch and Marlet}{Boulch and Marlet}{2012}]%
        {Boulch2012}
\bibfield{author}{\bibinfo{person}{Alexandre Boulch} {and}
  \bibinfo{person}{Renaud Marlet}.} \bibinfo{year}{2012}\natexlab{}.
\newblock \showarticletitle{Fast and Robust Normal Estimation for Point Clouds
  with Sharp Features}.
\newblock \bibinfo{journal}{{\em Comput. Graph. Forum\/}} \bibinfo{volume}{31},
  \bibinfo{number}{5} (\bibinfo{date}{Aug.} \bibinfo{year}{2012}),
  \bibinfo{pages}{1765--1774}.
\newblock
\showISSN{0167-7055}
\showDOI{%
\url{https://doi.org/10.1111/j.1467-8659.2012.03181.x}}


\bibitem[\protect\citeauthoryear{Boulch and Marlet}{Boulch and Marlet}{2016}]%
        {Boulch2016}
\bibfield{author}{\bibinfo{person}{Alexandre Boulch} {and}
  \bibinfo{person}{Renaud Marlet}.} \bibinfo{year}{2016}\natexlab{}.
\newblock \showarticletitle{Deep Learning for Robust Normal Estimation in
  Unstructured Point Clouds}.
\newblock \bibinfo{journal}{{\em Computer Graphics Forum\/}}
  \bibinfo{volume}{35}, \bibinfo{number}{5} (\bibinfo{year}{2016}),
  \bibinfo{pages}{281--290}.
\newblock
\showISSN{1467-8659}
\showDOI{%
\url{https://doi.org/10.1111/cgf.12983}}


\bibitem[\protect\citeauthoryear{Cai, Cand\`{e}s, and Shen}{Cai
  et~al\mbox{.}}{2010}]%
        {Cai2010}
\bibfield{author}{\bibinfo{person}{Jian-Feng Cai}, \bibinfo{person}{Emmanuel~J.
  Cand\`{e}s}, {and} \bibinfo{person}{Zuowei Shen}.}
  \bibinfo{year}{2010}\natexlab{}.
\newblock \showarticletitle{A Singular Value Thresholding Algorithm for Matrix
  Completion}.
\newblock \bibinfo{journal}{{\em SIAM J. on Optimization\/}}
  \bibinfo{volume}{20}, \bibinfo{number}{4} (\bibinfo{date}{March}
  \bibinfo{year}{2010}), \bibinfo{pages}{1956--1982}.
\newblock
\showISSN{1052-6234}
\showDOI{%
\url{https://doi.org/10.1137/080738970}}


\bibitem[\protect\citeauthoryear{Cand{\`e}s and Recht}{Cand{\`e}s and
  Recht}{2009}]%
        {Candes2009}
\bibfield{author}{\bibinfo{person}{Emmanuel~J. Cand{\`e}s} {and}
  \bibinfo{person}{Benjamin Recht}.} \bibinfo{year}{2009}\natexlab{}.
\newblock \showarticletitle{Exact Matrix Completion via Convex Optimization}.
\newblock \bibinfo{journal}{{\em Foundations of Computational Mathematics\/}}
  \bibinfo{volume}{9}, \bibinfo{number}{6} (\bibinfo{date}{03 Apr}
  \bibinfo{year}{2009}), \bibinfo{pages}{717}.
\newblock
\showISSN{1615-3383}
\showDOI{%
\url{https://doi.org/10.1007/s10208-009-9045-5}}


\bibitem[\protect\citeauthoryear{Dey and Goswami}{Dey and Goswami}{2004}]%
        {Dey2004}
\bibfield{author}{\bibinfo{person}{Tamal~K. Dey} {and} \bibinfo{person}{Samrat
  Goswami}.} \bibinfo{year}{2004}\natexlab{}.
\newblock \showarticletitle{Provable Surface Reconstruction from Noisy
  Samples}. In \bibinfo{booktitle}{{\em Proceedings of the Twentieth Annual
  Symposium on Computational Geometry}} {\em (\bibinfo{series}{SCG '04})}.
  \bibinfo{publisher}{ACM}, \bibinfo{address}{New York, NY, USA},
  \bibinfo{pages}{330--339}.
\newblock
\showISBNx{1-58113-885-7}
\showDOI{%
\url{https://doi.org/10.1145/997817.997867}}


\bibitem[\protect\citeauthoryear{Digne}{Digne}{2012}]%
        {Digne2012}
\bibfield{author}{\bibinfo{person}{J. Digne}.} \bibinfo{year}{2012}\natexlab{}.
\newblock \showarticletitle{Similarity based filtering of point clouds}. In
  \bibinfo{booktitle}{{\em 2012 IEEE Computer Society Conference on Computer
  Vision and Pattern Recognition Workshops}}. \bibinfo{pages}{73--79}.
\newblock
\showISSN{2160-7508}
\showDOI{%
\url{https://doi.org/10.1109/CVPRW.2012.6238917}}


\bibitem[\protect\citeauthoryear{Digne, Valette, and Chaine}{Digne
  et~al\mbox{.}}{2018}]%
        {Digne2018}
\bibfield{author}{\bibinfo{person}{J. Digne}, \bibinfo{person}{S. Valette},
  {and} \bibinfo{person}{R. Chaine}.} \bibinfo{year}{2018}\natexlab{}.
\newblock \showarticletitle{Sparse Geometric Representation Through Local Shape
  Probing}.
\newblock \bibinfo{journal}{{\em IEEE Transactions on Visualization and
  Computer Graphics\/}} \bibinfo{volume}{24}, \bibinfo{number}{7}
  (\bibinfo{date}{July} \bibinfo{year}{2018}), \bibinfo{pages}{2238--2250}.
\newblock
\showISSN{1077-2626}
\showDOI{%
\url{https://doi.org/10.1109/TVCG.2017.2719024}}


\bibitem[\protect\citeauthoryear{Fleishman, Drori, and Cohen-Or}{Fleishman
  et~al\mbox{.}}{2003}]%
        {Fleishman2003}
\bibfield{author}{\bibinfo{person}{Shachar Fleishman}, \bibinfo{person}{Iddo
  Drori}, {and} \bibinfo{person}{Daniel Cohen-Or}.}
  \bibinfo{year}{2003}\natexlab{}.
\newblock \showarticletitle{Bilateral Mesh Denoising}.
\newblock \bibinfo{journal}{{\em ACM Trans. Graph.\/}} \bibinfo{volume}{22},
  \bibinfo{number}{3} (\bibinfo{date}{July} \bibinfo{year}{2003}),
  \bibinfo{pages}{950--953}.
\newblock
\showISSN{0730-0301}
\showDOI{%
\url{https://doi.org/10.1145/882262.882368}}


\bibitem[\protect\citeauthoryear{Gu, Xie, Meng, Zuo, Feng, and Zhang}{Gu
  et~al\mbox{.}}{2017}]%
        {Gu2017}
\bibfield{author}{\bibinfo{person}{Shuhang Gu}, \bibinfo{person}{Qi Xie},
  \bibinfo{person}{Deyu Meng}, \bibinfo{person}{Wangmeng Zuo},
  \bibinfo{person}{Xiangchu Feng}, {and} \bibinfo{person}{Lei Zhang}.}
  \bibinfo{year}{2017}\natexlab{}.
\newblock \showarticletitle{Weighted Nuclear Norm Minimization and Its
  Applications to Low Level Vision}.
\newblock \bibinfo{journal}{{\em International Journal of Computer Vision\/}}
  \bibinfo{volume}{121}, \bibinfo{number}{2} (\bibinfo{date}{01 Jan}
  \bibinfo{year}{2017}), \bibinfo{pages}{183--208}.
\newblock
\showISSN{1573-1405}
\showDOI{%
\url{https://doi.org/10.1007/s11263-016-0930-5}}


\bibitem[\protect\citeauthoryear{Gu, Zhang, Zuo, and Feng}{Gu
  et~al\mbox{.}}{2014}]%
        {Gu2014}
\bibfield{author}{\bibinfo{person}{Shuhang Gu}, \bibinfo{person}{Lei Zhang},
  \bibinfo{person}{Wangmeng Zuo}, {and} \bibinfo{person}{Xiangchu Feng}.}
  \bibinfo{year}{2014}\natexlab{}.
\newblock \showarticletitle{Weighted Nuclear Norm Minimization with Application
  to Image Denoising}. In \bibinfo{booktitle}{{\em Proceedings of the 2014 IEEE
  Conference on Computer Vision and Pattern Recognition}} {\em
  (\bibinfo{series}{CVPR '14})}. \bibinfo{publisher}{IEEE Computer Society},
  \bibinfo{address}{Washington, DC, USA}, \bibinfo{pages}{2862--2869}.
\newblock
\showISBNx{978-1-4799-5118-5}
\showDOI{%
\url{https://doi.org/10.1109/CVPR.2014.366}}


\bibitem[\protect\citeauthoryear{Halko, Martinsson, and Tropp}{Halko
  et~al\mbox{.}}{2011}]%
        {Halko2011}
\bibfield{author}{\bibinfo{person}{N. Halko}, \bibinfo{person}{P.~G.
  Martinsson}, {and} \bibinfo{person}{J.~A. Tropp}.}
  \bibinfo{year}{2011}\natexlab{}.
\newblock \showarticletitle{Finding Structure with Randomness: Probabilistic
  Algorithms for Constructing Approximate Matrix Decompositions}.
\newblock \bibinfo{journal}{{\it SIAM Rev.}} \bibinfo{volume}{53},
  \bibinfo{number}{2} (\bibinfo{year}{2011}), \bibinfo{pages}{217--288}.
\newblock
\showDOI{%
\url{https://doi.org/10.1137/090771806}}
\showeprint{https://doi.org/10.1137/090771806}


\bibitem[\protect\citeauthoryear{He and Schaefer}{He and Schaefer}{2013}]%
        {He2013}
\bibfield{author}{\bibinfo{person}{Lei He} {and} \bibinfo{person}{Scott
  Schaefer}.} \bibinfo{year}{2013}\natexlab{}.
\newblock \showarticletitle{Mesh Denoising via L0 Minimization}.
\newblock \bibinfo{journal}{{\em ACM Trans. Graph.\/}} \bibinfo{volume}{32},
  \bibinfo{number}{4}, Article \bibinfo{articleno}{64} (\bibinfo{date}{July}
  \bibinfo{year}{2013}), \bibinfo{numpages}{8}~pages.
\newblock
\showISSN{0730-0301}
\showDOI{%
\url{https://doi.org/10.1145/2461912.2461965}}


\bibitem[\protect\citeauthoryear{Hoppe, DeRose, Duchamp, McDonald, and
  Stuetzle}{Hoppe et~al\mbox{.}}{1992}]%
        {Hoppe1992}
\bibfield{author}{\bibinfo{person}{Hugues Hoppe}, \bibinfo{person}{Tony
  DeRose}, \bibinfo{person}{Tom Duchamp}, \bibinfo{person}{John McDonald},
  {and} \bibinfo{person}{Werner Stuetzle}.} \bibinfo{year}{1992}\natexlab{}.
\newblock \showarticletitle{Surface Reconstruction from Unorganized Points}.
\newblock \bibinfo{journal}{{\em SIGGRAPH Comput. Graph.\/}}
  \bibinfo{volume}{26}, \bibinfo{number}{2} (\bibinfo{date}{July}
  \bibinfo{year}{1992}), \bibinfo{pages}{71--78}.
\newblock
\showISSN{0097-8930}
\showDOI{%
\url{https://doi.org/10.1145/142920.134011}}


\bibitem[\protect\citeauthoryear{Huang, Li, Zhang, Ascher, and Cohen-Or}{Huang
  et~al\mbox{.}}{2009}]%
        {Huang2009}
\bibfield{author}{\bibinfo{person}{Hui Huang}, \bibinfo{person}{Dan Li},
  \bibinfo{person}{Hao Zhang}, \bibinfo{person}{Uri Ascher}, {and}
  \bibinfo{person}{Daniel Cohen-Or}.} \bibinfo{year}{2009}\natexlab{}.
\newblock \showarticletitle{Consolidation of Unorganized Point Clouds for
  Surface Reconstruction}.
\newblock \bibinfo{journal}{{\em ACM Trans. Graph.\/}} \bibinfo{volume}{28},
  \bibinfo{number}{5}, Article \bibinfo{articleno}{176} (\bibinfo{date}{Dec.}
  \bibinfo{year}{2009}), \bibinfo{numpages}{7}~pages.
\newblock
\showISSN{0730-0301}
\showDOI{%
\url{https://doi.org/10.1145/1618452.1618522}}


\bibitem[\protect\citeauthoryear{Huang, Wu, Gong, Cohen-Or, Ascher, and
  Zhang}{Huang et~al\mbox{.}}{2013}]%
        {Huang2013}
\bibfield{author}{\bibinfo{person}{Hui Huang}, \bibinfo{person}{Shihao Wu},
  \bibinfo{person}{Minglun Gong}, \bibinfo{person}{Daniel Cohen-Or},
  \bibinfo{person}{Uri Ascher}, {and} \bibinfo{person}{Hao~(Richard) Zhang}.}
  \bibinfo{year}{2013}\natexlab{}.
\newblock \showarticletitle{Edge-aware Point Set Resampling}.
\newblock \bibinfo{journal}{{\em ACM Trans. Graph.\/}} \bibinfo{volume}{32},
  \bibinfo{number}{1}, Article \bibinfo{articleno}{9} (\bibinfo{date}{Feb.}
  \bibinfo{year}{2013}), \bibinfo{numpages}{12}~pages.
\newblock
\showISSN{0730-0301}
\showDOI{%
\url{https://doi.org/10.1145/2421636.2421645}}


\bibitem[\protect\citeauthoryear{Jones, Durand, and Desbrun}{Jones
  et~al\mbox{.}}{2003}]%
        {Jones2003}
\bibfield{author}{\bibinfo{person}{Thouis~R. Jones}, \bibinfo{person}{Fr{\'e}do
  Durand}, {and} \bibinfo{person}{Mathieu Desbrun}.}
  \bibinfo{year}{2003}\natexlab{}.
\newblock \showarticletitle{Non-iterative, Feature-preserving Mesh Smoothing}.
\newblock \bibinfo{journal}{{\em ACM Trans. Graph.\/}} \bibinfo{volume}{22},
  \bibinfo{number}{3} (\bibinfo{date}{July} \bibinfo{year}{2003}),
  \bibinfo{pages}{943--949}.
\newblock
\showISSN{0730-0301}
\showDOI{%
\url{https://doi.org/10.1145/882262.882367}}


\bibitem[\protect\citeauthoryear{Lange and Polthier}{Lange and
  Polthier}{2005}]%
        {Lange2005}
\bibfield{author}{\bibinfo{person}{Carsten Lange} {and} \bibinfo{person}{Konrad
  Polthier}.} \bibinfo{year}{2005}\natexlab{}.
\newblock \showarticletitle{Anisotropic smoothing of point sets}.
\newblock \bibinfo{journal}{{\em Computer Aided Geometric Design\/}}
  \bibinfo{volume}{22}, \bibinfo{number}{7} (\bibinfo{year}{2005}),
  \bibinfo{pages}{680 -- 692}.
\newblock
\showISSN{0167-8396}
\showDOI{%
\url{https://doi.org/10.1016/j.cagd.2005.06.010}}
\newblock
\shownote{Geometric Modelling and Differential Geometry.}


\bibitem[\protect\citeauthoryear{Lee and Wang}{Lee and Wang}{2005}]%
        {Lee2005}
\bibfield{author}{\bibinfo{person}{Kai-Wah Lee} {and} \bibinfo{person}{Wen-Ping
  Wang}.} \bibinfo{year}{2005}\natexlab{}.
\newblock \showarticletitle{Feature-preserving mesh denoising via bilateral
  normal filtering}. In \bibinfo{booktitle}{{\em Proc. of Int'l Conf. on
  Computer Aided Design and Computer Graphics 2005}}.
\newblock


\bibitem[\protect\citeauthoryear{Li, Schnabel, Klein, Cheng, Dang, and Jin}{Li
  et~al\mbox{.}}{2010}]%
        {Li2010}
\bibfield{author}{\bibinfo{person}{Bao Li}, \bibinfo{person}{Ruwen Schnabel},
  \bibinfo{person}{Reinhard Klein}, \bibinfo{person}{Zhiquan Cheng},
  \bibinfo{person}{Gang Dang}, {and} \bibinfo{person}{Shiyao Jin}.}
  \bibinfo{year}{2010}\natexlab{}.
\newblock \showarticletitle{Robust normal estimation for point clouds with
  sharp features}.
\newblock \bibinfo{journal}{{\em Computers \& Graphics\/}}
  \bibinfo{volume}{34}, \bibinfo{number}{2} (\bibinfo{year}{2010}),
  \bibinfo{pages}{94 -- 106}.
\newblock
\showISSN{0097-8493}
\showDOI{%
\url{https://doi.org/10.1016/j.cag.2010.01.004}}


\bibitem[\protect\citeauthoryear{Liu, Lin, and Yu}{Liu et~al\mbox{.}}{2010}]%
        {Liu2010}
\bibfield{author}{\bibinfo{person}{Guangcan Liu}, \bibinfo{person}{Zhouchen
  Lin}, {and} \bibinfo{person}{Yong Yu}.} \bibinfo{year}{2010}\natexlab{}.
\newblock \showarticletitle{Robust Subspace Segmentation by Low-rank
  Representation}. In \bibinfo{booktitle}{{\em Proceedings of the 27th
  International Conference on International Conference on Machine Learning}}
  {\em (\bibinfo{series}{ICML'10})}. \bibinfo{publisher}{Omnipress},
  \bibinfo{address}{USA}, \bibinfo{pages}{663--670}.
\newblock
\showISBNx{978-1-60558-907-7}
\showURL{%
\url{http://dl.acm.org/citation.cfm?id=3104322.3104407}}


\bibitem[\protect\citeauthoryear{Liu, Zhang, Cao, Li, and Liu}{Liu
  et~al\mbox{.}}{2015}]%
        {Liu2015}
\bibfield{author}{\bibinfo{person}{Xiuping Liu}, \bibinfo{person}{Jie Zhang},
  \bibinfo{person}{Junjie Cao}, \bibinfo{person}{Bo Li}, {and}
  \bibinfo{person}{Ligang Liu}.} \bibinfo{year}{2015}\natexlab{}.
\newblock \showarticletitle{Quality point cloud normal estimation by guided
  least squares representation}.
\newblock \bibinfo{journal}{{\em Computers \& Graphics\/}}
  \bibinfo{volume}{51}, \bibinfo{number}{Supplement C} (\bibinfo{year}{2015}),
  \bibinfo{pages}{106 -- 116}.
\newblock
\showISSN{0097-8493}
\showDOI{%
\url{https://doi.org/10.1016/j.cag.2015.05.024}}
\newblock
\shownote{International Conference Shape Modeling International.}


\bibitem[\protect\citeauthoryear{Lu, Chen, and Schaefer}{Lu
  et~al\mbox{.}}{2017a}]%
        {Lu2017}
\bibfield{author}{\bibinfo{person}{Xuequan Lu}, \bibinfo{person}{Wenzhi Chen},
  {and} \bibinfo{person}{Scott Schaefer}.} \bibinfo{year}{2017}\natexlab{a}.
\newblock \showarticletitle{Robust mesh denoising via vertex pre-filtering and
  L1-median normal filtering}.
\newblock \bibinfo{journal}{{\em Computer Aided Geometric Design\/}}
  \bibinfo{volume}{54}, \bibinfo{number}{Supplement C} (\bibinfo{year}{2017}),
  \bibinfo{pages}{49 -- 60}.
\newblock
\showISSN{0167-8396}
\showDOI{%
\url{https://doi.org/10.1016/j.cagd.2017.02.011}}


\bibitem[\protect\citeauthoryear{Lu, Deng, and Chen}{Lu et~al\mbox{.}}{2016}]%
        {Lu2016}
\bibfield{author}{\bibinfo{person}{Xuequan Lu}, \bibinfo{person}{Zhigang Deng},
  {and} \bibinfo{person}{Wenzhi Chen}.} \bibinfo{year}{2016}\natexlab{}.
\newblock \showarticletitle{A Robust Scheme for Feature-Preserving Mesh
  Denoising}.
\newblock \bibinfo{journal}{{\em {IEEE} Trans. Vis. Comput. Graph.\/}}
  \bibinfo{volume}{22}, \bibinfo{number}{3} (\bibinfo{year}{2016}),
  \bibinfo{pages}{1181--1194}.
\newblock


\bibitem[\protect\citeauthoryear{Lu, Wu, Chen, Yeung, Chen, and Zwicker}{Lu
  et~al\mbox{.}}{2017b}]%
        {Lu2017tvcg}
\bibfield{author}{\bibinfo{person}{X. Lu}, \bibinfo{person}{S. Wu},
  \bibinfo{person}{H. Chen}, \bibinfo{person}{S.~K. Yeung}, \bibinfo{person}{W.
  Chen}, {and} \bibinfo{person}{M. Zwicker}.} \bibinfo{year}{2017}\natexlab{b}.
\newblock \showarticletitle{GPF: GMM-inspired Feature-preserving Point Set
  Filtering}.
\newblock \bibinfo{journal}{{\em IEEE Transactions on Visualization and
  Computer Graphics\/}} \bibinfo{volume}{PP}, \bibinfo{number}{99}
  (\bibinfo{year}{2017}), \bibinfo{pages}{1--1}.
\newblock
\showISSN{1077-2626}
\showDOI{%
\url{https://doi.org/10.1109/TVCG.2017.2725948}}


\bibitem[\protect\citeauthoryear{Mitra and Nguyen}{Mitra and Nguyen}{2003}]%
        {Mitra2003}
\bibfield{author}{\bibinfo{person}{Niloy~J. Mitra} {and} \bibinfo{person}{An
  Nguyen}.} \bibinfo{year}{2003}\natexlab{}.
\newblock \showarticletitle{Estimating Surface Normals in Noisy Point Cloud
  Data}. In \bibinfo{booktitle}{{\em Proceedings of the Nineteenth Annual
  Symposium on Computational Geometry}} {\em (\bibinfo{series}{SCG '03})}.
  \bibinfo{publisher}{ACM}, \bibinfo{address}{New York, NY, USA},
  \bibinfo{pages}{322--328}.
\newblock
\showISBNx{1-58113-663-3}
\showDOI{%
\url{https://doi.org/10.1145/777792.777840}}


\bibitem[\protect\citeauthoryear{\"{O}ztireli, Guennebaud, and
  Gross}{\"{O}ztireli et~al\mbox{.}}{2009}]%
        {Oztireli2009}
\bibfield{author}{\bibinfo{person}{A.~C. \"{O}ztireli}, \bibinfo{person}{G.
  Guennebaud}, {and} \bibinfo{person}{M. Gross}.}
  \bibinfo{year}{2009}\natexlab{}.
\newblock \showarticletitle{Feature Preserving Point Set Surfaces based on
  Non-Linear Kernel Regression}.
\newblock \bibinfo{journal}{{\em Computer Graphics Forum\/}}
  \bibinfo{volume}{28}, \bibinfo{number}{2} (\bibinfo{year}{2009}),
  \bibinfo{pages}{493--501}.
\newblock
\showISSN{1467-8659}
\showDOI{%
\url{https://doi.org/10.1111/j.1467-8659.2009.01388.x}}


\bibitem[\protect\citeauthoryear{Pauly, Gross, and Kobbelt}{Pauly
  et~al\mbox{.}}{2002}]%
        {Pauly2002}
\bibfield{author}{\bibinfo{person}{Mark Pauly}, \bibinfo{person}{Markus Gross},
  {and} \bibinfo{person}{Leif~P. Kobbelt}.} \bibinfo{year}{2002}\natexlab{}.
\newblock \showarticletitle{Efficient Simplification of Point-sampled
  Surfaces}. In \bibinfo{booktitle}{{\em Proceedings of the Conference on
  Visualization '02}} {\em (\bibinfo{series}{VIS '02})}.
  \bibinfo{publisher}{IEEE Computer Society}, \bibinfo{address}{Washington, DC,
  USA}, \bibinfo{pages}{163--170}.
\newblock
\showISBNx{0-7803-7498-3}
\showURL{%
\url{http://dl.acm.org/citation.cfm?id=602099.602123}}


\bibitem[\protect\citeauthoryear{Preiner, Mattausch, Arikan, Pajarola, and
  Wimmer}{Preiner et~al\mbox{.}}{2014}]%
        {Preiner2014}
\bibfield{author}{\bibinfo{person}{Reinhold Preiner}, \bibinfo{person}{Oliver
  Mattausch}, \bibinfo{person}{Murat Arikan}, \bibinfo{person}{Renato
  Pajarola}, {and} \bibinfo{person}{Michael Wimmer}.}
  \bibinfo{year}{2014}\natexlab{}.
\newblock \showarticletitle{Continuous Projection for Fast L1 Reconstruction}.
\newblock \bibinfo{journal}{{\em ACM Trans. Graph.\/}} \bibinfo{volume}{33},
  \bibinfo{number}{4}, Article \bibinfo{articleno}{47} (\bibinfo{date}{July}
  \bibinfo{year}{2014}), \bibinfo{numpages}{13}~pages.
\newblock
\showISSN{0730-0301}
\showDOI{%
\url{https://doi.org/10.1145/2601097.2601172}}


\bibitem[\protect\citeauthoryear{Shen and Barner}{Shen and Barner}{2004}]%
        {Shen2004}
\bibfield{author}{\bibinfo{person}{Yuzhong Shen} {and} \bibinfo{person}{K.E.
  Barner}.} \bibinfo{year}{2004}\natexlab{}.
\newblock \showarticletitle{Fuzzy vector median-based surface smoothing}.
\newblock \bibinfo{journal}{{\em IEEE Transactions on Visualization and
  Computer Graphics\/}} \bibinfo{volume}{10}, \bibinfo{number}{3}
  (\bibinfo{date}{May} \bibinfo{year}{2004}), \bibinfo{pages}{252--265}.
\newblock


\bibitem[\protect\citeauthoryear{Solomon, Crane, Butscher, and Wojtan}{Solomon
  et~al\mbox{.}}{2014}]%
        {Solomon2014}
\bibfield{author}{\bibinfo{person}{Justin Solomon}, \bibinfo{person}{Keenan
  Crane}, \bibinfo{person}{Adrian Butscher}, {and} \bibinfo{person}{Chris
  Wojtan}.} \bibinfo{year}{2014}\natexlab{}.
\newblock \showarticletitle{A General Framework for Bilateral and Mean Shift
  Filtering}.
\newblock \bibinfo{journal}{{\em CoRR\/}}  \bibinfo{volume}{abs/1405.4734}
  (\bibinfo{year}{2014}).
\newblock
\showeprint[arxiv]{1405.4734}
\showURL{%
\url{http://arxiv.org/abs/1405.4734}}


\bibitem[\protect\citeauthoryear{Sun, Rosin, Martin, and Langbein}{Sun
  et~al\mbox{.}}{2007}]%
        {Sun2007}
\bibfield{author}{\bibinfo{person}{Xianfang Sun}, \bibinfo{person}{P.L. Rosin},
  \bibinfo{person}{R.R. Martin}, {and} \bibinfo{person}{F.C. Langbein}.}
  \bibinfo{year}{2007}\natexlab{}.
\newblock \showarticletitle{Fast and Effective Feature-Preserving Mesh
  Denoising}.
\newblock \bibinfo{journal}{{\em IEEE Transactions on Visualization and
  Computer Graphics\/}} \bibinfo{volume}{13}, \bibinfo{number}{5}
  (\bibinfo{date}{Sept} \bibinfo{year}{2007}), \bibinfo{pages}{925--938}.
\newblock


\bibitem[\protect\citeauthoryear{Sun, Rosin, Martin, and Langbein}{Sun
  et~al\mbox{.}}{2008}]%
        {Sun2008}
\bibfield{author}{\bibinfo{person}{Xianfang Sun}, \bibinfo{person}{Paul~L.
  Rosin}, \bibinfo{person}{Ralph~R. Martin}, {and} \bibinfo{person}{Frank~C.
  Langbein}.} \bibinfo{year}{2008}\natexlab{}.
\newblock \showarticletitle{Random walks for feature-preserving mesh
  denoising}.
\newblock \bibinfo{journal}{{\em Computer Aided Geometric Design\/}}
  \bibinfo{volume}{25}, \bibinfo{number}{7} (\bibinfo{year}{2008}),
  \bibinfo{pages}{437 -- 456}.
\newblock
\showISSN{0167-8396}
\showDOI{%
\url{https://doi.org/10.1016/j.cagd.2007.12.008}}
\newblock
\shownote{Solid and Physical Modeling Selected papers from the Solid and
  Physical Modeling and Applications Symposium 2007 (SPM 2007) Solid and
  Physical Modeling and Applications Symposium 2007.}


\bibitem[\protect\citeauthoryear{Sun, Schaefer, and Wang}{Sun
  et~al\mbox{.}}{2015}]%
        {Sun2015}
\bibfield{author}{\bibinfo{person}{Yujing Sun}, \bibinfo{person}{Scott
  Schaefer}, {and} \bibinfo{person}{Wenping Wang}.}
  \bibinfo{year}{2015}\natexlab{}.
\newblock \showarticletitle{Denoising point sets via $L_0$ minimization}.
\newblock \bibinfo{journal}{{\em Computer Aided Geometric Design\/}}
  \bibinfo{volume}{35-36} (\bibinfo{year}{2015}), \bibinfo{pages}{2 -- 15}.
\newblock
\showISSN{0167-8396}
\showDOI{%
\url{https://doi.org/10.1016/j.cagd.2015.03.011}}
\newblock
\shownote{Geometric Modeling and Processing 2015.}


\bibitem[\protect\citeauthoryear{Wang}{Wang}{2006}]%
        {Wang2006}
\bibfield{author}{\bibinfo{person}{Charlie C~L Wang}.}
  \bibinfo{year}{2006}\natexlab{}.
\newblock \showarticletitle{Bilateral recovering of sharp edges on
  feature-insensitive sampled meshes}.
\newblock \bibinfo{journal}{{\em Visualization and Computer Graphics, IEEE
  Transactions on\/}} \bibinfo{volume}{12}, \bibinfo{number}{4}
  (\bibinfo{date}{July} \bibinfo{year}{2006}), \bibinfo{pages}{629--639}.
\newblock
\showISSN{1077-2626}
\showDOI{%
\url{https://doi.org/10.1109/TVCG.2006.60}}


\bibitem[\protect\citeauthoryear{Wang, Fu, Liu, Tong, Liu, and Guo}{Wang
  et~al\mbox{.}}{2015}]%
        {Wang2015}
\bibfield{author}{\bibinfo{person}{Peng-Shuai Wang}, \bibinfo{person}{Xiao-Ming
  Fu}, \bibinfo{person}{Yang Liu}, \bibinfo{person}{Xin Tong},
  \bibinfo{person}{Shi-Lin Liu}, {and} \bibinfo{person}{Baining Guo}.}
  \bibinfo{year}{2015}\natexlab{}.
\newblock \showarticletitle{Rolling Guidance Normal Filter for Geometric
  Processing}.
\newblock \bibinfo{journal}{{\em ACM Trans. Graph.\/}} \bibinfo{volume}{34},
  \bibinfo{number}{6}, Article \bibinfo{articleno}{173} (\bibinfo{date}{Oct.}
  \bibinfo{year}{2015}), \bibinfo{numpages}{9}~pages.
\newblock
\showISSN{0730-0301}
\showDOI{%
\url{https://doi.org/10.1145/2816795.2818068}}


\bibitem[\protect\citeauthoryear{Wang, Liu, and Tong}{Wang
  et~al\mbox{.}}{2016}]%
        {Wang2016}
\bibfield{author}{\bibinfo{person}{Peng-Shuai Wang}, \bibinfo{person}{Yang
  Liu}, {and} \bibinfo{person}{Xin Tong}.} \bibinfo{year}{2016}\natexlab{}.
\newblock \showarticletitle{Mesh Denoising via Cascaded Normal Regression}.
\newblock \bibinfo{journal}{{\em ACM Trans. Graph.\/}} \bibinfo{volume}{35},
  \bibinfo{number}{6}, Article \bibinfo{articleno}{232} (\bibinfo{date}{Nov.}
  \bibinfo{year}{2016}), \bibinfo{numpages}{12}~pages.
\newblock
\showISSN{0730-0301}
\showDOI{%
\url{https://doi.org/10.1145/2980179.2980232}}


\bibitem[\protect\citeauthoryear{Wright, Ganesh, Rao, Peng, and Ma}{Wright
  et~al\mbox{.}}{2009}]%
        {Wright2009}
\bibfield{author}{\bibinfo{person}{John Wright}, \bibinfo{person}{Arvind
  Ganesh}, \bibinfo{person}{Shankar Rao}, \bibinfo{person}{Yigang Peng}, {and}
  \bibinfo{person}{Yi Ma}.} \bibinfo{year}{2009}\natexlab{}.
\newblock \showarticletitle{Robust Principal Component Analysis: Exact Recovery
  of Corrupted Low-Rank Matrices via Convex Optimization}.
\newblock In \bibinfo{booktitle}{{\em Advances in Neural Information Processing
  Systems 22}}, \bibfield{editor}{\bibinfo{person}{Y.~Bengio},
  \bibinfo{person}{D.~Schuurmans}, \bibinfo{person}{J.~D. Lafferty},
  \bibinfo{person}{C.~K.~I. Williams}, {and} \bibinfo{person}{A.~Culotta}}
  (Eds.). \bibinfo{publisher}{Curran Associates, Inc.},
  \bibinfo{pages}{2080--2088}.
\newblock


\bibitem[\protect\citeauthoryear{Wu, Yeung, Jia, Tang, and Medioni}{Wu
  et~al\mbox{.}}{2012}]%
        {Wu2012}
\bibfield{author}{\bibinfo{person}{T.~P. Wu}, \bibinfo{person}{S.~K. Yeung},
  \bibinfo{person}{J. Jia}, \bibinfo{person}{C.~K. Tang}, {and}
  \bibinfo{person}{G. Medioni}.} \bibinfo{year}{2012}\natexlab{}.
\newblock \showarticletitle{A Closed-Form Solution to Tensor Voting: Theory and
  Applications}.
\newblock \bibinfo{journal}{{\em IEEE Transactions on Pattern Analysis and
  Machine Intelligence\/}} \bibinfo{volume}{34}, \bibinfo{number}{8}
  (\bibinfo{date}{Aug} \bibinfo{year}{2012}), \bibinfo{pages}{1482--1495}.
\newblock
\showISSN{0162-8828}
\showDOI{%
\url{https://doi.org/10.1109/TPAMI.2011.250}}


\bibitem[\protect\citeauthoryear{Yadav, Reitebuch, and Polthier}{Yadav
  et~al\mbox{.}}{2017}]%
        {Yadav2017}
\bibfield{author}{\bibinfo{person}{S.~K. Yadav}, \bibinfo{person}{U.
  Reitebuch}, {and} \bibinfo{person}{K. Polthier}.}
  \bibinfo{year}{2017}\natexlab{}.
\newblock \showarticletitle{Mesh Denoising based on Normal Voting Tensor and
  Binary Optimization}.
\newblock \bibinfo{journal}{{\em IEEE Transactions on Visualization and
  Computer Graphics\/}} \bibinfo{volume}{PP}, \bibinfo{number}{99}
  (\bibinfo{year}{2017}), \bibinfo{pages}{1--1}.
\newblock
\showISSN{1077-2626}
\showDOI{%
\url{https://doi.org/10.1109/TVCG.2017.2740384}}


\bibitem[\protect\citeauthoryear{Yagou, Ohtake, and Belyaev}{Yagou
  et~al\mbox{.}}{2002}]%
        {Yagou2002}
\bibfield{author}{\bibinfo{person}{H. Yagou}, \bibinfo{person}{Y. Ohtake},
  {and} \bibinfo{person}{A. Belyaev}.} \bibinfo{year}{2002}\natexlab{}.
\newblock \showarticletitle{Mesh smoothing via mean and median filtering
  applied to face normals}. In \bibinfo{booktitle}{{\em Geometric Modeling and
  Processing, 2002. Proceedings}}. \bibinfo{pages}{124--131}.
\newblock
\showDOI{%
\url{https://doi.org/10.1109/GMAP.2002.1027503}}


\bibitem[\protect\citeauthoryear{{Y}agou, Ohtake, and Belyaev}{{Y}agou
  et~al\mbox{.}}{2003}]%
        {Yagou2003}
\bibfield{author}{\bibinfo{person}{H. {Y}agou}, \bibinfo{person}{Y. Ohtake},
  {and} \bibinfo{person}{A.G. Belyaev}.} \bibinfo{year}{2003}\natexlab{}.
\newblock \showarticletitle{Mesh denoising via iterative alpha-trimming and
  nonlinear diffusion of normals with automatic thresholding}. In
  \bibinfo{booktitle}{{\em Computer Graphics International, 2003.
  Proceedings}}. \bibinfo{pages}{28--33}.
\newblock
\showISSN{1530-1052}
\showDOI{%
\url{https://doi.org/10.1109/CGI.2003.1214444}}


\bibitem[\protect\citeauthoryear{Zhang, Wu, Zhang, and Deng}{Zhang
  et~al\mbox{.}}{2015b}]%
        {Zhang2015}
\bibfield{author}{\bibinfo{person}{Huayan Zhang}, \bibinfo{person}{Chunlin Wu},
  \bibinfo{person}{Juyong Zhang}, {and} \bibinfo{person}{Jiansong Deng}.}
  \bibinfo{year}{2015}\natexlab{b}.
\newblock \showarticletitle{Variational Mesh Denoising Using Total Variation
  and Piecewise Constant Function Space}.
\newblock \bibinfo{journal}{{\em Visualization and Computer Graphics, IEEE
  Transactions on\/}} \bibinfo{volume}{21}, \bibinfo{number}{7}
  (\bibinfo{date}{July} \bibinfo{year}{2015}), \bibinfo{pages}{873--886}.
\newblock
\showISSN{1077-2626}
\showDOI{%
\url{https://doi.org/10.1109/TVCG.2015.2398432}}


\bibitem[\protect\citeauthoryear{Zhang, Cao, Liu, He, Li, and Liu}{Zhang
  et~al\mbox{.}}{2018}]%
        {Zhang2018}
\bibfield{author}{\bibinfo{person}{J. Zhang}, \bibinfo{person}{J. Cao},
  \bibinfo{person}{X. Liu}, \bibinfo{person}{C. He}, \bibinfo{person}{B. Li},
  {and} \bibinfo{person}{L. Liu}.} \bibinfo{year}{2018}\natexlab{}.
\newblock \showarticletitle{Multi-Normal Estimation via Pair Consistency
  Voting}.
\newblock \bibinfo{journal}{{\em IEEE Transactions on Visualization and
  Computer Graphics\/}} (\bibinfo{year}{2018}), \bibinfo{pages}{1--1}.
\newblock
\showISSN{1077-2626}
\showDOI{%
\url{https://doi.org/10.1109/TVCG.2018.2827998}}


\bibitem[\protect\citeauthoryear{Zhang, Cao, Liu, Wang, Liu, and Shi}{Zhang
  et~al\mbox{.}}{2013}]%
        {Zhang2013}
\bibfield{author}{\bibinfo{person}{Jie Zhang}, \bibinfo{person}{Junjie Cao},
  \bibinfo{person}{Xiuping Liu}, \bibinfo{person}{Jun Wang},
  \bibinfo{person}{Jian Liu}, {and} \bibinfo{person}{Xiquan Shi}.}
  \bibinfo{year}{2013}\natexlab{}.
\newblock \showarticletitle{Point cloud normal estimation via low-rank subspace
  clustering}.
\newblock \bibinfo{journal}{{\em Computers \& Graphics\/}}
  \bibinfo{volume}{37}, \bibinfo{number}{6} (\bibinfo{year}{2013}),
  \bibinfo{pages}{697 -- 706}.
\newblock
\showISSN{0097-8493}
\showDOI{%
\url{https://doi.org/10.1016/j.cag.2013.05.008}}
\newblock
\shownote{Shape Modeling International (SMI) Conference 2013.}


\bibitem[\protect\citeauthoryear{Zhang, Deng, Zhang, Bouaziz, and Liu}{Zhang
  et~al\mbox{.}}{2015a}]%
        {Zhang2015cgf}
\bibfield{author}{\bibinfo{person}{Wangyu Zhang}, \bibinfo{person}{Bailin
  Deng}, \bibinfo{person}{Juyong Zhang}, \bibinfo{person}{Sofien Bouaziz},
  {and} \bibinfo{person}{Ligang Liu}.} \bibinfo{year}{2015}\natexlab{a}.
\newblock \showarticletitle{Guided Mesh Normal Filtering}.
\newblock \bibinfo{journal}{{\em Comput. Graph. Forum\/}} \bibinfo{volume}{34},
  \bibinfo{number}{7} (\bibinfo{date}{Oct.} \bibinfo{year}{2015}),
  \bibinfo{pages}{23--34}.
\newblock
\showISSN{0167-7055}
\showDOI{%
\url{https://doi.org/10.1111/cgf.12742}}


\bibitem[\protect\citeauthoryear{Zhang, Ganesh, Liang, and Ma}{Zhang
  et~al\mbox{.}}{2012}]%
        {Zhang2012}
\bibfield{author}{\bibinfo{person}{Zhengdong Zhang}, \bibinfo{person}{Arvind
  Ganesh}, \bibinfo{person}{Xiao Liang}, {and} \bibinfo{person}{Yi Ma}.}
  \bibinfo{year}{2012}\natexlab{}.
\newblock \showarticletitle{TILT: Transform Invariant Low-Rank Textures}.
\newblock \bibinfo{journal}{{\em International Journal of Computer Vision\/}}
  \bibinfo{volume}{99}, \bibinfo{number}{1} (\bibinfo{date}{01 Aug}
  \bibinfo{year}{2012}), \bibinfo{pages}{1--24}.
\newblock
\showISSN{1573-1405}
\showDOI{%
\url{https://doi.org/10.1007/s11263-012-0515-x}}


\bibitem[\protect\citeauthoryear{Zheng, Sharf, Wan, Li, Mitra, Cohen-Or, and
  Chen}{Zheng et~al\mbox{.}}{2010}]%
        {Zheng2010}
\bibfield{author}{\bibinfo{person}{Qian Zheng}, \bibinfo{person}{Andrei Sharf},
  \bibinfo{person}{Guowei Wan}, \bibinfo{person}{Yangyan Li},
  \bibinfo{person}{Niloy~J. Mitra}, \bibinfo{person}{Daniel Cohen-Or}, {and}
  \bibinfo{person}{Baoquan Chen}.} \bibinfo{year}{2010}\natexlab{}.
\newblock \showarticletitle{Non-local Scan Consolidation for 3D Urban Scenes}.
\newblock \bibinfo{journal}{{\em ACM Trans. Graph.\/}} \bibinfo{volume}{29},
  \bibinfo{number}{4}, Article \bibinfo{articleno}{94} (\bibinfo{date}{July}
  \bibinfo{year}{2010}), \bibinfo{numpages}{9}~pages.
\newblock
\showISSN{0730-0301}
\showDOI{%
\url{https://doi.org/10.1145/1778765.1778831}}


\bibitem[\protect\citeauthoryear{Zheng, Fu, Au, and Tai}{Zheng
  et~al\mbox{.}}{2011}]%
        {Zheng2011}
\bibfield{author}{\bibinfo{person}{Youyi Zheng}, \bibinfo{person}{Hongbo Fu},
  \bibinfo{person}{O.K.-C. Au}, {and} \bibinfo{person}{Chiew-Lan Tai}.}
  \bibinfo{year}{2011}\natexlab{}.
\newblock \showarticletitle{Bilateral Normal Filtering for Mesh Denoising}.
\newblock \bibinfo{journal}{{\em IEEE Transactions on Visualization and
  Computer Graphics\/}} \bibinfo{volume}{17}, \bibinfo{number}{10}
  (\bibinfo{date}{Oct} \bibinfo{year}{2011}), \bibinfo{pages}{1521--1530}.
\newblock
\showISSN{1077-2626}
\showDOI{%
\url{https://doi.org/10.1109/TVCG.2010.264}}


\end{thebibliography}
